\documentclass[pdftex,twocolumn,epjc3]{svjour3}          

\RequirePackage[T1]{fontenc}

\smartqed  

\RequirePackage{graphicx}
\RequirePackage{mathptmx}      
\RequirePackage{flushend}
\RequirePackage[numbers,sort&compress]{natbib}
\RequirePackage[colorlinks,citecolor=blue,urlcolor=blue,linkcolor=blue]{hyperref}
\usepackage{bookmark} 
\usepackage{multirow}
\usepackage{amssymb}
\usepackage[numbers,sort&compress]{natbib}
\usepackage{amsfonts,amsmath,amssymb}
\usepackage{mathrsfs}		
\usepackage{subfigure}
\usepackage{ifthen}
\usepackage{float}
\usepackage{etoolbox}
\usepackage{lipsum}

\usepackage{amsmath}
\numberwithin{equation}{section}
\usepackage[titletoc]{appendix}

\renewcommand{\bf}{\normalfont \bfseries}
\renewcommand{\tt}{\normalfont \ttfamily}

\newcommand{\Jbar}{\bar{J}}

\newcommand{\bra}[1]{\langle #1 |}
\newcommand{\ket}[1]{| #1 \rangle}

\newcommand{\aav}[1]{\langle #1 \rangle}

\newcommand{\etapi}{{\eta \to 3 \pi}}
\newcommand{\chpt}{$\chi$PT}

\newcommand{\lib}{{\cal L}_\text{QCD}^{\Delta \hspace{-0.2mm}m}}

\renewcommand{\Im}{\text{Im}\,}
\renewcommand{\Re}{\text{Re}\,}
\newcommand{\disc}{\text{disc}\,}

\newcommand{\mpi}{M_\pi}

\newcommand{\meta}{M_\eta}

\newcommand{\DMKQCD}{(M_{K^0}^2-M_{K^+}^2)\rule[-0.4em]{0em}{0em}_{\mathrm{QCD}}}

\newcommand{\GeV}{\text{GeV}}
\newcommand{\MeV}{\text{MeV}}

\newcommand{\comment}[1]{}

\newcommand{\eolp}{\,.}
\newcommand{\eolc}{\,,}

\newcommand{\entryTextPos}{}
\newcounter{entrynum}

\newcommand{\entryFont}{\small}
\newcommand{\entry}[4]{	\addtocounter{entrynum}{-1}
								\ifthenelse{\equal{#2}{\empty}}{
									\psline{|-|}(#3,\theentrynum)(#4,\theentrynum)
								}{
									\pscircle*(#2,\theentrynum){2.5pt}
									\ifthenelse{\equal{#3}{\empty}}{}{
										\SpecialCoor
										\psline{|-|}(!#2 #3 sub \theentrynum)(!#2 #4 add \theentrynum)
										\NormalCoor
									}
								}
								\rput[l](\entryTextPos,\theentrynum){\entryFont #1}
							}

\newlength{\mylw}
\newlength{\mydotlw}
\newlength{\myslw}
\newlength{\mysdotlw}
\def\query#1{}%

\newcommand{\lsim}{\,\raisebox{-0.3em}{$\stackrel{\raisebox{-0.1em}{$<$}}{\sim}$
}\,}

\newcommand{\al}{&}
\newcommand{\DeltaG}{\Delta}
\newcommand{\Deltapipi}{\delta}
\newcommand{\Mtilde}{M\rule{0em}{1em}^{\hspace{-0.8em}\sim}\hspace{0.15em} }

\journalname{Eur. Phys. J. C}

\begin{document}
\sloppy 
\title{Dispersive analysis of $\mathbf\eta \rightarrow 3 \pi$}

\author{Gilberto Colangelo\thanksref{e1,addr1} \and Stefan Lanz\thanksref{e2,addr1} \and Heinrich Leutwyler\thanksref{e3,addr1} 
        \and
        Emilie Passemar\thanksref{e4,addr2,addr3} 
}

\thankstext{e1}{gilberto@itp.unibe.ch}
\thankstext{e2}{lanz.stefan@gmx.ch}
\thankstext{e3}{leutwyler@itp.unibe.ch}
\thankstext{e4}{epassema@indiana.edu}

\institute{Albert Einstein Center for Fundamental Physics, Institute for Theoretical Physics, University of Bern, Sidlerstrasse 5, 3012 Bern, Switzerland\label{addr1}
          \and
          Department of Physics, Indiana University, Bloomington, IN 47405, USA\newline
Center for Exploration of Energy and Matter, Indiana University, Bloomington, IN 47408, USA\label{addr2}
          \and
          Theory Center, Thomas Jefferson National Accelerator Facility, Newport News, VA 23606, USA\label{addr3}
}

\date{Received: date / Accepted: date}

\maketitle

\begin{abstract}

The  dispersive analysis of the decay $\eta\to3\pi$ is reviewed and thoroughly updated with 
the aim of determining the quark mass ratio ~$Q^2=(m_s^2-m_{ud}^2)/(m_d^2-m_u^2)$.
With the number of subtractions we are using, the effects generated by the
final state interaction are dominated by low energy $\pi\pi$ scattering. Since the
corresponding phase shifts are now accurately known, causality and unitarity 
determine the decay amplitude within small uncertainties -- except for the
values of the subtraction constants. Our determination of 
these constants relies on the Dalitz plot distribution 
of the charged channel, which  is now measured with good accuracy. The theoretical 
constraints that follow from the fact that the particles involved 
in the transition represent Nambu-Goldstone bosons of a hidden approximate 
symmetry play an equally important role. The ensuing predictions for the Dalitz plot
distribution of the neutral channel and for the branching ratio $\Gamma_{\eta\to3\pi^0}/
\Gamma_{\eta\to\pi^+\pi^-\pi^0}$ are in very good agreement with experiment.
Relying on a known low-energy theorem that relates the meson masses to
the masses of the three lightest quarks, our analysis leads to $Q=22.1(7)$,
where the error covers all of the uncertainties encountered in the
course of the calculation: experimental uncertainties in decay rates and Dalitz  
plot distributions, noise in the input used for the phase shifts, as well as theoretical
uncertainties in the constraints imposed by chiral symmetry and in the evaluation of isospin breaking effects.
Our result indicates that the current algebra formulae for the meson masses only receive small 
corrections from higher orders of the chiral expansion, but not all of the recent lattice results are consistent with this conclusion. 
\end{abstract}

\setcounter{tocdepth}{2}
\tableofcontents

\section{Introduction}
Our world is almost isospin symmetric: The up and the down quarks can be
freely interchanged (or replaced by any linear combination of them) inside
hadrons almost without any observable consequence. Of course the charge of
the two quarks is different, so that after an isospin transformation the
charge of the hadronic state might change, but since the electromagnetic
interactions are much weaker than the strong ones, we can classify this as
a small effect. Besides the charge, the only difference between the two
quarks is their mass. In relative terms their mass difference is large, but
very small when compared to the mass of a typical hadron: If we interchange
the up and down quarks inside a hadron, the mass of the latter barely
changes. Observables which are sensitive to isospin violations are
therefore particularly interesting, as they offer us rare insights into the
sector of the Standard Model Lagrangian which breaks the isospin symmetry.
One of them is the decay of the $\eta$-meson into three pions. This decay
would be forbidden by isospin symmetry and moreover it is mainly due to
purely strong isospin violations~\cite{Sutherland1966,Bell+1968}: Among the
already rare observables sensitive to isospin breaking, this is even more
special as it allows to clearly separate the two sources, which are
otherwise mostly present at a similar level. To a good approximation the
decay rate is proportional to the square of the up and down mass
difference. If one were able to accurately calculate the proportionality
factor -- the modulus squared of the transition amplitude between the $\eta$
and a three-pion state mediated by the third component of the scalar
isovector quark bilinear -- a measurement of the decay rate would provide a
determination of this quark mass difference. This approach has been adopted
before, but both, recent improved measurements of the differential decay
rates as well as progress on the theory side call for an updated and
improved analysis.  This is the aim of the present paper, where we give a 
detailed account of the work reported in Ref.~\cite{Colangelo:2016jmc}. 

The calculation of hadronic matrix elements is not an easy task, especially
if the aim is high precision. Several methods are available and can be
applied with varying degree of success, depending on the circumstances: They
range from lattice QCD to Chiral Perturbation Theory (\chpt), to dispersive
approaches. Decays into three particles are not accessible to
lattice calculations yet,\footnote{The formalism for carrying out such
  calculations on the lattice is being developed, however,
  see~\cite{Polejaeva:2012ut,Briceno:2012rv,
Hansen:2014eka,Hansen:2015zga,Hammer:2017uqm,Hammer:2017kms,Mai:2017bge}.}
but both the effective field theory approach and
dispersion relations can be and have been used to analyze these processes.
As it turns out, the main difficulty concerns the evaluation of rescattering
effects among the pions in the final state.
In particular, the lowest resonance occurring in QCD, the $f_0(500)$, strongly amplifies the final 
state interaction in the $S$-wave with $I=0$. For this reason, the first few terms of the
chiral pertubation series do not provide a good description of the momentum 
dependence of the amplitude, even if the one-loop representation \cite{Gasser+1985a} is
extended to two loops \cite{Bijnens+2007}. We will discuss the limitations of the effective
theory in the present case in Sec.~\ref{sec:Anatomy}. 
Dispersion relations, on the other hand, are perfectly suited to evaluate rescattering effects to 
all orders \cite{Anisovich1995,Kambor+1996,Anisovich+1996}. They express the amplitude in terms
of a few subtraction constants, which play a role analogous to the low-energy constants (LEC) of \chpt. 
Those relevant for the momentum dependence of the amplitude can be determined very well 
on the basis of the experimental information on the Dalitz plot distribution. Theory is needed 
only for the analogs of those LECs that describe the dependence on the quark masses.

In the literature there are already a few papers which follow essentially the same approach, but there are
several compelling reasons for redoing this analysis:  
\begin{enumerate}
\item 
  Until recently, the dispersive analyses relied on a rather crude input for
  the $\pi \pi$ phase shifts, which is the essential ingredient in
  the dispersive calculation. Today a much more accurate representation for
  this amplitude is available~\cite{Colangelo2001,Kaminski:2006qe}.
\item
  Improved calculations of the electromagnetic effects in this decay are
  available~\cite{Ditsche+2009} and it is impossible to use these in
  combination with old dispersive calculations.
\item 
  There have been recent, more accurate experimental measurements of the Dalitz
  plot in the charged channel~\cite{Ambrosino:2008ht,Adlarson:2014aks,Ablikim:2015cmz,KLOE:2016qvh},
  which challenge the theory to correctly describe this momentum
  dependence.
  \item
  The experimental information concerning the momentum dependence in the 
  neutral channel also improved very significantly \cite{Unverzagt+2009,Prakhov+2009,Prakhov+2018,
  Ambrosino+2010}, but represents a theoretical puzzle, because Chiral
  Perturbation Theory does not predict the slope correctly, in fact, not
  even the sign.
\end{enumerate}
In the following we take up this challenge and apply and combine all
theoretical improvements listed above to come up with a representation for
the $\eta \to 3 \pi$ amplitude which can be used to describe the data. The
most challenging aspects concern:
\begin{itemize}
\item[{\em i)}] obtaining numerical solutions of the
integral equations which follow from the dispersion relations;
\item[{\em ii)}] the dispersion relations are analyzed in the isospin limit -- isospin breaking effects
must be accounted for;
\item[{\em iii)}] formulate and impose the constraints that follow from the fact that the
particles involved in this decay are Nambu-Goldstone bosons of a hidden approximate symmetry. 
\end{itemize}
As we will show, we have been able to successfully address all these
challenges and have set up a framework which allows us to describe the data
well with values of the subtraction constants -- the input
parameters in the dispersion relations -- which agree well with the
prediction of \chpt. A proper treatment of isospin breaking corrections is
essential, at the current level of precision, to simultaneously describe
experimental data in both the charged and the neutral channel of the
decay. 

The plan of the paper is as follows.  
We set up our dispersive framework in Sec.~\ref{sec:Theoretical framework}
and review \chpt~calculations and predictions on this process in
Sec.~\ref{sec:Chiral perturbation theory}. Our dispersive analysis is
performed in the isospin limit -- the approach used to account for
isospin breaking effects is discussed in Sec.~\ref{sec:Isospin breaking corrections}.
In Sec.~\ref{sec:Dalitz plot}, we describe our fits to the KLOE
measurements of the Dalitz plot for $\eta \to \pi^+\pi^- \pi^0$ and discuss
the importance of the theoretical constraints in this context. 
The results of the dispersive  analysis are compared with the \chpt~two-loop 
representation of the decay amplitude in Sec.~\ref{sec:Anatomy}, whereas, 
in Sec.~\ref{sec:etato3pi0}, we analyze the consequences for the decay
$\eta\to3\pi^0$. In Sec.~\ref{sec:Fits to MAMI}, the results are compared with 
the recent update of the MAMI data on this decay \cite{Prakhov+2018}. 
Sec.~\ref{sec:MKQ} discusses our determination of the kaon mass
difference in QCD and of the quark mass ratios $Q$ and $m_u/m_d$. 
Finally, in Sec.~\ref{sec:Comparison}, we compare our analysis with
related work.  Our conclusions in
Sec.~\ref{sec:Conclusions} are followed by a number of appendices containing
details of our calculation.

\section{Theoretical framework}\label{sec:Theoretical framework}

\subsection{Isospin}\label{sec:Isospin}
The transition $\etapi$ proceeds exclusively through isospin breaking operators since three pions
cannot be in a state where isospin and angular momentum vanish at the same time. Indeed, the
three-pion isoscalar state
has odd (and therefore non-zero) angular momentum according to Bose
statistics. In the Standard Model, isospin breaking contributions can arise
either from the electromagnetic or the strong interaction. However,
according to a theorem by Sutherland~\cite{Bell+1968,Sutherland1966}, the
electromagnetic (e.m.) contribution  to the decay $\etapi$ vanishes at leading order
of the chiral perturbation series: The transition is  mainly due to the fact that
QCD does not conserve isospin. The isospin breaking part of the QCD Lagrangian, 
\begin{equation}
	\lib  =-\mbox{$\frac{1}{2}$}(m_u - m_d)\, ( \bar{u} u - \bar{d}d )
	 \eolc
	\label{eq:LIB}
\end{equation}
carries $I=1$ and can indeed generate transitions
between the $\eta$ and three-pion states with $I=1$. Up to contributions from the e.m.~interaction
and higher orders in $m_u-m_d$, the transition 
amplitude is given by the matrix element of the perturbation $\lib$ between the unperturbed, 
stable initial and final states,\footnote{The relative phase of the amplitudes for the charged and neutral channels depends on the convention used to specify the phase of the one-particle states. We are working with $|\pi^\pm\rangle =(|\pi^1\rangle \pm i\hspace{0.05em} |\pi^2\rangle) /\sqrt{2}$, $|\pi^0\rangle=|\pi^3\rangle$.}
\begin{equation}\label{eq:Ac}A_c(s,t,u)=
\langle \pi^+ \pi^-  \pi^0 \,
\mathrm{out}|\lib|\eta \,\mathrm{in}\rangle\,.\end{equation}
The Mandelstam variables stand for  
\begin{eqnarray}
	s &= (p_{\pi^+} + p_{\pi^-})^2 = (p_\eta - p_{\pi^0})^2\eolc\nonumber\\
	t &= (p_{\pi^-} + p_{\pi^0})^2 = (p_\eta - p_{\pi^+})^2\eolc\\
	u &= (p_{\pi^+} + p_{\pi^0})^2 = (p_\eta - p_{\pi^-})^2 \eolp\nonumber
	\label{eq:eta3piMandelstam}
\end{eqnarray}
The quantity $A_c(s,t,u)$ is  dimensionless like the amplitude 
of $\pi\pi$ scattering and is proportional to the quark mass difference $m_d-m_u$. As pointed 
out in~\cite{Gasser+1985a}, it is convenient to (i) decompose the amplitude into a momentum-independent 
term $N$ that breaks isospin symmetry times a remainder $M_c(s,t,u)$ that is isospin-invariant
and (ii) define $N$ in terms of the kaon mass difference in QCD and the pion decay constant $F_\pi$:
\begin{equation}\label{eq:normalization}A_c(s,t,u) =-N M_c(s,t,u)\,,\quad
N\equiv \frac{\hat{M}_{K^0}^2-\hat{M}_{K^+}^2}{3\sqrt{3}\,F_\pi^2}\,.\end{equation}
We follow the notation used by FLAG: $\hat{M}_{K^0}$ and $\hat{M}_{K^+}$ stand for the masses of the kaons in QCD \cite{Aoki:2016frl}. The amplitude $M_c(s,t,u)$ concerns the isospin limit of QCD, where the charged and neutral pions and kaons carry the common mass $M_\pi$ and $M_K$, respectively.  The normalization~\eqref{eq:normalization} implies that, in current algebra approximation~\cite{Cronin1967, Osborn+1970}, the amplitude $M_c$ exclusively involves the meson masses:\footnote{The mass of the $\eta$ is protected from isospin breaking: The e.m.~self-energy vanishes at leading order of the chiral expansion and the expansion of $M_\eta$ in powers of the difference $m_d-m_u$ only starts at $O(m_d-m_u)^2$. The difference between the physical mass of the $\eta$ and its value in the isospin limit is beyond the accuracy of our calculation.} $M_c(s,t,u)=(3s-4M_\pi^2)/(M_\eta^2-M_\pi^2)$. 

In this notation, the rate of the decay $\eta\to\pi^+\pi^-\pi^0$ is given by 
\begin{eqnarray}
\Gamma_{\eta\to\pi^+\pi^-\pi^0}&=&\frac{(2\pi)^4N^2}{2M_\eta}\hspace{-0.3em}\int\hspace{-0.3em} d\mu(p_{\pi^+})d\mu(p_{\pi^-})d\mu(p_{\pi^0})  \\
&& \times \delta^4(p_\eta-p_{\pi^+}-p_{\pi^-}-p_{\pi^0})|M_c(s,t,u)|^2, \nonumber
\end{eqnarray}
with $d\mu(p)=d^3p/(2p^0)/(2\pi)^3.$ Since only two of the Mandelstam variables are independent, the rate can be expressed as an integral over two of these: 
 \begin{equation}\label{eq:Gammac}\Gamma_{\eta\to\pi^+\pi^-\pi^0}=\frac{N^2J_c}{256\pi^3 M_\eta^3}\;,\quad J_c\equiv \int\hspace{-0.3em} 
ds \,dt\, |M_c(s,t,u)|^2\,.\end{equation} 

In the entire first part of the present paper, we will limit ourselves to an analysis of the transition amplitude $M_c(s,t,u)$
in the isospin limit. The neglected contributions of order $e^2$ and $(m_u-m_d)^2$ do not respect isospin symmetry 
and are referred to as {\em isospin breaking corrections}. We will analyze
these in detail in Sec.~\ref{sec:Isospin breaking corrections}.

Charge conjugation symmetry requires the amplitude to be invariant under the exchange of the two
charged pions,
\begin{equation}
  M_c(s,t,u) = M_c(s,u,t) \eolc
\end{equation}
and isospin symmetry implies that the amplitude for the transition  $\eta\to\pi^i\pi^j\pi^k$ is determined by the one relevant for the charged decay mode:
\begin{eqnarray}
M^{ijk}(s,t,u)&=& M_c(s,t,u)\, \delta^{ij} \delta^{k3} + M_c(t,u,s)\, \delta^{ik} \delta^{j3} \nonumber\\
			&& + M_c(u,s,t)\,\delta^{jk}  \delta^{i3} \eolp\end{eqnarray}
%
In particular, the transition amplitude for the decay $\eta\to 3\pi^0$, which we denote by $M_n(s,t,u)$, is represented as:  
\begin{equation}\label{eq:Mn(s,t,u)}M_n(s,t,u)=
 M_c(s,t,u)+M_c(t,u,s)+ M_c(u,s,t)   \eolp \end{equation}
The formula explicitly shows that the amplitude for the neutral mode is symmetric in all three Mandelstam variables. 

Note that the indistinguishability of the pions generated in the decay \mbox{$\eta\to 3\pi^0$} implies that the corresponding Mandelstam variables are not unique. While an event occurring in the decay $\eta\to \pi^+\pi^-\pi^0$ corresponds to a unique set of values for $s,t,u$, the six different permutations of $s,t,u$ belonging to a configuration of three neutral pions correspond to six different points in the physical region, but describe the same event. If the phase space integral is extended over the entire physical region, the result must be divided by six:
\begin{equation}\label{eq:Gamman}
\Gamma_{\eta\to 3\pi^0}=\frac{N^2J_n}{256\pi^3 M_\eta^3}\;,\quad J_n\equiv\frac{1}{6}\int\hspace{-0.4em} 
ds \,dt\, |M_n(s,t,u)|^2\,.\end{equation} 
%

\subsection{Branch cuts, discontinuities}\label{sec:Branch cuts}
The consequences  of causality and unitarity for transitions with three particles in the final state were investigated long ago~\cite{Gribov:1958ez,Khuri+1960,Anisovich+1966,Neveu+1970,Roiesnel+1981} and many papers concerning the decays $K\to3\pi$ and $\eta\to 3\pi$ have appeared since then. In particular, as shown in~\cite{Anisovich1995,Anisovich:2013gha,Anisovich+1996,Kambor+1996}, 
the final state interaction can reliably be accounted for
with dispersion relations. Since the publication of these papers, the $\pi\pi$ phase shifts
have been determined to remarkable 
precision~\cite{Colangelo2001,Descotes-Genon+2002,Kaminski:2006qe} and the
quality of the experimental information about these decays is now also much
better. Moreover, the nonrelativistic effective field theory has been set up for these transitions. The application of this method to $K\to 3\pi$ turned out to be very successful~\cite{Colangelo+2006a,Bissegger:2007yq,Bissegger:2008ff,Gasser:2011ju}.  These developments have triggered renewed interest in theoretical studies of $\eta\to3\pi$~\cite{Gullstrom:2008sy,Schneider+2011,Kampf+2011,Lanz:2013ku,Guo:2014vya,
Guo:2015zqa,Guo:2016wsi,Albaladejo+2015,Albaladejo+2017,Kolesar:2016iyz,Kolesar:2016jwe,Kolesar:2017xrl,Kolesar:2017jdx}.

We briefly summarize the main properties of the transition amplitude at low energies. On account of causality, the function $M_c(s,t,u)$ is analytic in the Mandelstam variables $s,t,u$. At low energies, the final state interaction among the pions generates the most important singularities. The branch cut due to the interaction between $\pi^+$ and $\pi^-$ starts at $s=4M_\pi^2$ (`$s$-channel'), while the cuts associated with the interactions in the $t$- and $u$-channels stem from the pairs $\pi^+\pi^0$ and $\pi^-\pi^0$ and start at $t=4M_\pi^2$ and $u=4M_\pi^2$, respectively. The strength of these singularities can be characterized with the discontinuity across the cut, that is with the difference between the values of the amplitude at the upper and lower rim of the cuts. The discontinuity across the branch cut in the $s$-channel, for instance, is defined by
\begin{equation} \mathrm{disc}_{s}\,M_c(s,t,u)=\frac{1}{2i}\{M_c(s+i\epsilon,t,u)-M_c(s-i \epsilon,t,u)\}\,.\end{equation}
Since the angular momentum barrier strongly suppresses the discontinuities due to the D- and higher partial waves, the low-energy structure is dominated by those from the S- and P-waves.  This also manifests itself in \chpt: Discontinuities due to partial waves with $\ell\geq 2$ start showing up only at $O(p^8)$ of the chiral expansion. 

The discontinuity generated by the S-wave with isospin $I=0$ only shows up in the $s$-channel, with a term that does not depend on the scattering angle, i.e.~exclusively involves the variable $s$. We denote the discontinuity due to this partial wave by $\disc M_0(s)$:
\begin{equation}\label{eq:discS0}\mathrm{disc}_{\mbox{\scriptsize S}_0}\,M_c(s,t,u)=\disc M_0(s)\end{equation}

In the $t$-channel, the interaction in the S-wave with $I = 2$ generates a discontinuity that only depends on $t$:  $\disc M_2(t)$. Since the transition amplitude is symmetric with respect to the exchange of $t$ and $u$, the corresponding discontinuity in the $u$-channel is determined by the same function: $\disc M_2(u)$. The interaction in the exotic wave also manifests itself in the $s$-channel, with a discontinuity proportional to $\disc M_2(s)$. The proportionality factor must be such that the projection onto the isoscalar S-wave vanishes. This projection is given by the sum over $i=j$ of the matrix element  $\bra{\pi^i \pi^j \pi^k\mathrm{out}} \bar{q}\lambda^3 q \ket{\eta}$, i.e.~by $3f(s,t,u)+f(t,u,s)+f(u,s,t)$. With $f(s,t,u)\propto \disc M_2(t)+\disc M_2(u)+\lambda \,\disc M_2(s)$, this reduces to $(3\lambda+2)\, \disc M_2(s) +\cdots\,$, where the ellipsis stands for terms that only depend on $t$ or $u$. Hence $\lambda=-\frac{2}{3}$, so that: 
\begin{equation}\label{eq:discS2}\mathrm{disc}_{\mbox{\scriptsize S}_2}M_c(s,t,u)=\disc M_2(t)+\disc M_2(u)-\mbox{$\frac{2}{3}$}\disc M_2(s)\,.\end{equation}

Since the P-wave carries $I=1$, it cannot show up in the $s$-channel, but generates a $t$-channel contribution of the form $f(t)\cos \theta_t$, where $\theta_t$ is the scattering angle. Expressed in terms of the Mandelstam variables, $\cos \theta_t$ is proportional to $s-u$. Together with the analogous term in the $u$-channel the P-wave discontinuity thus takes the form
\begin{equation}\mathrm{disc}_{\mbox{\scriptsize P}}\,M_c(s,t,u)=  (s-u)\,\disc M_1(t)+(s-t)\,\disc M_1(u) \,.\end{equation}

This shows that the suppression of the higher partial waves simplifies the analytic structure of the transition amplitude considerably: Retaining only the discontinuities due to the leading partial waves with isospin $I=0,1,2$, those of the full amplitude can be decomposed into three functions of a single variable:
\begin{eqnarray}
\label{eq:RT1}
\disc M_c(s, t, u) \al=\al\disc M_0(s) \\
\al\al + (s-u)\, \disc M_1(t) + (s-t) \, \disc M_1(u) \nonumber\\
\al\al +\; \disc M_2(t) + \disc M_2(u) - \mbox{$\frac{2}{3}$} \disc M_2(s) \; \; . \nonumber
\end{eqnarray}
The functions $\disc M_0(x), \disc M_1(x)$ and $\disc M_2(x)$ describe the discontinuities in the lowest partial waves with $I=0,1$ and 2, respectively.  
\subsection{Dispersion relations, subtractions} \label{sec:Dispersion relations}
We denote the contribution to the transition amplitude generated by the discontinuity from the leading partial wave with isospin $I$ by $M_I(s)$ and refer to the functions $M_0(s)$, $M_1(s)$, $M_2(s)$ as the isospin components of the amplitude. These functions only have a right hand cut for $4M_\pi^2<s<\infty$ and, as suggested by the notation, the discontinuity of $M_I(s)$ across this cut is given by $\disc M_I(s)$. Accordingly, $M_I(s)$ obeys a dispersion relation of the form
\begin{equation}\label{eq:MI}M_I(s)=P_I(s)+\frac{s^{\hspace{0.05em}n_I}}{\pi}\int_{4M_\pi^2}^\infty \frac{ds'}{s'^{\hspace{0.1em}n_I}}\frac{\disc M_I(s')}{(s'-s-i\epsilon)}\,,\quad I=0,1,2\,,\end{equation}
where we have allowed for subtractions, collecting the subtraction constants in the polynomial $P_I(s)$. The representation illustrates the fact that analytic functions are fully determined by their singularities. In the present context, not only those occurring at finite values of the Mandelstam variables, but also those at infinity matter. Although we are not interested in the asymptotic behaviour of the amplitude as such, it provides a convenient handle on the subtractions: The singularities unambiguously determine the amplitude provided the asymptotic behaviour is known. 

The Mandelstam variables are not independent, but obey the constraint
$s+t+u=M_\eta^2+3M_\pi^2$. We use the two independent variables $s$ and
$\tau \equiv t-u$ (the constraint then fixes all three variables in terms
of these two). The condition that the amplitude $M_c(s,t,u)$ does not grow
more rapidly than with the square of $\lambda$ if $s$ and $\tau$ grow in
proportion to $\lambda$ turns out to lead to a suitable framework that
allows sufficiently many subtractions, so that the poorly known high
energy behaviour of the amplitude and inelastic contributions do not
play a significant role. The general polynomial that is even in $\tau$ and
obeys this asymptotic condition is of the form $p_0+p_1 s+ p_2 s^2 + p_3
\tau^2$ and it is easy to see that a polynomial of this form can be
absorbed in the functions $M_0(s),M_1(s),M_2(s)$. Hence, if the discontinuities 
are of the form~\eqref{eq:RT1}, then the asymptotic condition ensures that 
the amplitude itself can be decomposed into three
functions of a single variable,
\begin{eqnarray}
\label{eq:RT}
M_c(s, t, u) \al=\al M_0(s) + (s-u) M_1(t) + (s-t) M_1(u) \nonumber\\
&&+ M_2(t) + M_2(u) - \mbox{$\frac{2}{3}$} M_2(s) \; \; .
\end{eqnarray}
Inserting this in~\eqref{eq:Mn(s,t,u)}, the analogous decomposition of the neutral transition takes the remarkably simple form:
\begin{equation}\label{eq:RTn}M_n(s,t,u)= M_n(s)+M_n(t)+M_n(u) \;.\end{equation}
In the approximation we are using, only the combination 
\begin{equation}\label{eq:Mn}M_n(s)\equiv M_0(s)+\mbox{$\frac{4}{3}$}M_2(s)\end{equation}
 of the S-waves is relevant for the neutral decay mode -- the P-wave drops out altogether.

We expect that, in the physical region of the decay, the representations~\eqref{eq:RT}, \eqref{eq:RTn} constitute an excellent approximation to the isospin limit of the transition amplitudes. In \chpt, the approximation holds up to and including next-to-next-to-leading order (NNLO) -- in that framework, the decomposition~\eqref{eq:RT} is referred to as the `reconstruction theorem'~\cite{Stern+1993}.

\subsection{Polynomial ambiguities}\label{sec:Polynomial}

There is a problem of technical nature with the approximation~\eqref{eq:RT}: The decomposition is unique only modulo polynomials. Indeed, one readily checks that the functions
\begin{eqnarray}\label{eq:gauge a} \Mtilde_1(s)\al=\al M_1(s)+3\,a\, s^2
+b\, s+c\hspace{5em}\\
  \Mtilde_2(s)\al=\al M_2(s)+a\,
s^3-9\,a\, s_0s^2-b\, s^2+d\,s+e\,,\nonumber\end{eqnarray} 
with $s_0=\frac{1}{3}M_\eta^2+M_\pi^2$, yield the same
total amplitude as $M_1(s)$, $M_2(s)$, except for a
contribution which is independent of $t,u$ and may thus be absorbed in
$M_0(s)$,
\begin{eqnarray}\label{eq:gauge b}
\Mtilde_0(s)\al=\al M_0(s)-\mbox{$\frac{4}{3}$}\,a\, s^3
+12\,a\, s_0s^2-54\,a\, s_0^2(s-s_0) \nonumber\\ 
 &\!\!&\!\!+\mbox{$\frac{4}{3}$}\, b\, s^2-9\,b\, s_0(s-s_0)-3\,c\,
(s-s_0) +\mbox{$\frac{5}{3}$}\,d\, s \nonumber\\
 &\!\!&\!\!-3\,d\, s_0
-\mbox{$\frac{4}{3}$},\end{eqnarray} Conversely, the two sets
$\Mtilde_0(s)$, $\Mtilde_1(s)$, $\Mtilde_2(s)$ and
$M_0(s)$, $M_1(s)$, $M_2(s)$ give rise to the same sum only if they are related in this manner.
To verify this statement, eliminate $s$ in favour of the two independent variables $t,u$ and
consider the derivative
$(\partial_t-\partial_u)\partial_t^2\partial_u$ of the function $M_c(s,t,u)$. The 
operation eliminates all of the isospin components except for $M_1$ -- the result is proportional to 
the third derivative, $M_1^{'''}(t)$. Accordingly, for the two decompositions to have the same sum,
the third derivative of $\Mtilde_1(s)-M_1(s)$ must vanish. Hence this difference is
a second order polynomial -- the first line of Eq.~\eqref{eq:gauge a} is verified. Once the polynomial ambiguity in $M_1$ is
determined, those in $M_0$ and $M_2$ readily follow.

This demonstrates that the decomposition~\eqref{eq:RT} is unique up to a five-parameter family of polynomials. The transformations specified in~\eqref{eq:gauge a}, \eqref{eq:gauge b} form a Lie group, which we denote by $G_5$. Under this group, the isospin components $M_0(s)$, $M_1(s)$ and $M_2(s)$ transform in a non-trivial manner, but their sum, $M_c(s,t,u)$ is invariant. 

The above calculation also shows that the component $M_1(t)$ cannot grow more rapidly than with the square of $t$: Otherwise, the function $M_1^{'''}(t)$ would not tend to zero when $t$ is sent to infinity, as required by the asymptotic condition.  We exploit the freedom inherent in the polynomial ambiguities as follows. First, we choose the parameter $a$ in~\eqref{eq:gauge a}, \eqref{eq:gauge b} such that the term in $M_1(t)$ which asymptotically grows with $t^2$ is cancelled, such that $M_1(t)\propto t$. For large values of $t$, the derivative $(\partial_t-\partial_u)\partial_t^2  M_c(s,t,u)$ is then dominated by the contribution from $M_2(t)$, which is proportional to $M_2^{'''}(t)$. The asymptotic condition on $M_c(s,t,u)$ thus implies that $M_2^{'''}(t)$ must tend to zero when $t\to\infty$, so that $M_2(t)$ grows at most quadratically. The leading term can again be removed: With a suitable choice of the parameter $b$, we arrive at a decomposition for which both $M_1$ and $M_2$ at most grow linearly. The ambiguities in the decomposition then reduce to a three-parameter family of polynomials, labeled with $c, d, e$. We fix $c$ with the condition $M_1(0)=0$ and, finally, choose $d,e$ such that $M_2(0)=M_2'(0)=0$. This shows that the decomposition can be made unique by 
imposing the five constraints
%
\begin{eqnarray}\label{eq:M asymptotics}
&& M_1(0)=0\;,\quad M_1(s)\propto s\;, \nonumber\\
&& M_2(0)=0\;,\quad M_2'(0)=0\;,\quad M_2(s)\propto s\;.\end{eqnarray}
With this choice, the asymptotic condition is obeyed by the individual isospin components, not only by their sum. In particular, $M_0(s)$ then grows at most quadratically: $M_0(s)\propto s^2$.
\subsection{Elastic unitarity}\label{sec:Elastic unitarity}

The occurrence of $\pi\pi$ branch cuts is a consequence of unitarity, but an amplitude of the simple form~\eqref{eq:RT} can obey the unitarity condition only approximately. The relevant approximation is referred to as elastic unitarity. For $\pi\pi$ scattering, the Roy equations~\cite{Roy1971} provide a rigorous framework, within which the singularities due to the final state interaction in the S- and P-waves can be sorted out explicitly. For the decay of an $\eta$ or a kaon into three pions, however, the constraints imposed by elastic unitarity are more subtle. For a detailed discussion, we refer to the literature quoted above. In the following, we rely on the framework developed in~\cite{Anisovich+1966,Anisovich1995,Anisovich+1996}, where the final state interaction effects are analyzed by means of analytic continuation in $M_\eta$. The net result of that analysis is the following expression for the leading discontinuities:
\begin{equation}
\disc M_I(s)  = \theta(s-4M_\pi^2)\left\{ M_I(s) + \hat{M}_I(s) \right\} \sin \delta_I(s) e^{-i \delta_I(s)}\,,
  \label{eq:discMI}
\end{equation}
with $I = 0,1,2$.
The first term in the curly bracket stems from collisions in the $s$-channel, the second accounts for those in the $t$- and $u$-channels and $\delta_0(s), \delta_1(s),\delta_2(s)$ denote the phase shifts  of the leading partial waves of $\pi\pi$ scattering with isospin $I=0,1,2$, respectively (in the standard notation, the phase shifts are denoted by $\delta_I^\ell(s)$, where $I$ and $\ell$ indicate the isospin and angular momentum quantum numbers of the partial wave, respectively; as only the lowest value of $\ell$ is relevant in our approximation, we drop the upper index).  
The contributions from the $t$- and $u$-channels are given by averages over the functions $M_0(s)$, $M_1(s)$, $M_2(s)$:
\begin{eqnarray}\label{eq:mhat}
  \hat M_0 \al= \al\mbox{$\frac{2}{3}$} \aav{M_0} + 2 (s-s_0) \aav{M_1} +\mbox{$ \frac{2}{3}$} \kappa \aav{z M_1}  + \mbox{$\frac{20}{9}$} \aav{M_2} \; ,\nonumber \\
  \hat M_1 \al=\al \kappa^{-1} \left\{ 3 \aav{z M_0} + \mbox{$\frac{9}{2} $}(s-s_0) \aav{z M_1} - 5 \aav{z M_2} + \mbox{$\frac{3}{2}$} \kappa \aav{z^2 M_1} \right\}, \nonumber\\
  \hat M_2 \al=\al \aav{M_0} - \mbox{$\frac{3}{2}$} (s-s_0) \aav{M_1} - \mbox{$\frac{1}{2}$} \kappa \aav{z M_1} + \mbox{$\frac{1}{3}$} \aav{M_2} \; ,
\end{eqnarray}
with $\hat M_0=\hat M_0(s)$, $\aav{M_0}=\aav{M_0}(s)$, etc. The quantities $s_0$ and $\kappa=\kappa(s)$ stand for
\begin{eqnarray}\label{eq:kappa}
&& s_0=\mbox{$\frac{1}{3}$}M_\eta^2+M_\pi^2\;, \\
&& \kappa(s)= \sqrt{1 - 4 \mpi^2/s} \sqrt{(\meta-\mpi)^2-s} \sqrt{(\meta+\mpi)^2-s} \nonumber
\end{eqnarray}
and the averages are defined by
\begin{equation}
\aav{z^n  M_I}(s) = \mbox{$\frac{1}{2}$} \int_{-1}^1\!\! dz \, z^n M_I( \mbox{$\frac{3}{2}$} s_0 -\mbox{$\frac{1}{2}$} s +\mbox{$\frac{1}{2}$} z \hspace{0.05em}\kappa(s)) \; ,
  \label{eq:angularAverage}
\end{equation}
with $I=0,1,2$ and $n=0,1,\ldots$
The complications occurring with elastic unitarity in the decay into three pions concern the specification of these averages. They arise because the $\eta$ is an unstable particle. 

We  use the standard method proposed in the pioneering papers on the subject and define the angular averages by means of analytic continuation in the square of the mass of the $\eta$. Reserving the symbol $M_\eta$ for the physical value of the mass, we denote the corresponding complex variable by $M$. Starting with a real value of $M^2$ below  $9M_\pi^2$, where the $\eta$ is stable, the physical mass is approached with $M^2=M_\eta^2+i\delta$, where $\delta$ is positive and tends to zero.  For $\Re M^2<9M_\pi^2$, the integral over $z$ in~\eqref{eq:angularAverage} runs over values that are in the analyticity domain of the integrand, so that the integral is meaningful as it stands. Since the integrand is an analytic function of $z$, the path of integration can be deformed without changing the value of the integral, as long as the path stays within the domain of analyticity. Indeed, if $\Re M^2$ is increased above $9M_\pi^2$, such a deformation is necessary to avoid the singularities of the integrand. The matter is discussed in some detail in~Appendix~\ref{sec:Angular averages}.

Gasser and Rusetsky \cite{Gasser+2018} very recently found a more efficient method for the solution of the integral equations. Their approach relies on a formulation of these equations for complex values of the Mandelstam variables and avoids the numerical problems altogether, which are encountered in the method we are using to evaluate the angular averages and are described in 
Appendix~\ref{sec:Angular averages}. They kindly made their numerical results for the fundamental solutions available to us prior to publication -- see the ancillary files in \cite{Gasser+2018}. In the vicinity of the critical points, their solutions are significantly more accurate than those obtained with our numerical procedure, while away from these points, their results offer a very welcome check. The numerical results given in the present paper are based on their fundamental solutions -- some of our numerical results differ from those quoted in the letter version  \cite{Colangelo:2016jmc}, but in all cases, the difference amounts to a small fraction of the quoted error.

Analytic continuation in the mass of the $\eta$ fully specifies the elastic unitarity approximation used in the present work. As mentioned in Sec.~\ref{sec:Branch cuts}, the approximation~\eqref{eq:RT}, which represents the amplitude in terms of three functions of a single variable, is valid in \chpt, up to and including NNLO. This statement holds within the effective theory based on SU(3)$\times$SU(3), i.e.~includes loops involving kaons or $\eta$-mesons. Our treatment of elastic unitarity, however, only accounts for the discontinuities generated by elastic collisions among the pions and does not include intermediate states containing heavy members of the Nambu-Goldstone octet.

Albaladejo and Moussallam~\cite{Albaladejo+2015,Albaladejo+2017} have set up a dispersive framework for the analysis of the decay $\eta\to 3\pi$ which extends elastic unitarity to the quasi-elastic collisions among the members of the pseudoscalar octet. We compare our approach with theirs in Sec.~\ref{sec:Dispersion theory}.
In the range of energies of interest to us and in view of the fact that we use dispersion relations with many subtractions, the polynomial approximation for the contributions from the heavy intermediate states is perfectly adequate. What is important, however, is that the singularities generated by the final state interaction among the pions are properly accounted for and we have checked that this is the case: The elastic unitarity approximation specified above does account for the pionic singularities contained in the chiral representation of the transition amplitude, up to and including two loops.  

\subsection{Phase shifts}\label{sec:Phase shifts}

The Roy equations~\cite{Roy1971} very strongly constrain the behaviour of the $\pi\pi$ scattering amplitude at low energies.  In particular, these equations fully determine the amplitude in terms of its imaginary part, up to the two S-wave scattering lengths, which enter as subtraction constants. Together with the predictions for the scattering lengths obtained on the basis of \chpt, this framework offers a remarkably precise representation for the scattering amplitude at low energies~\cite{Ananthanarayan+2001,Colangelo2001}. In the meantime, the experimental work on kaon decays
\cite{Pislak:2001bf,Batley:2000zz,Batley:2010zza,Batley:2012rf} and pionic or kaonic atoms~\cite{Adeva:2011tc, Adeva:2014xtx} has tested the predictions for the scattering lengths to high accuracy and the dispersive analysis is also confirmed within errors~\cite{Kaminski:2006qe,Pelaez:2015qba}.  
\begin{figure}[t]\centering
\includegraphics[width=8cm]{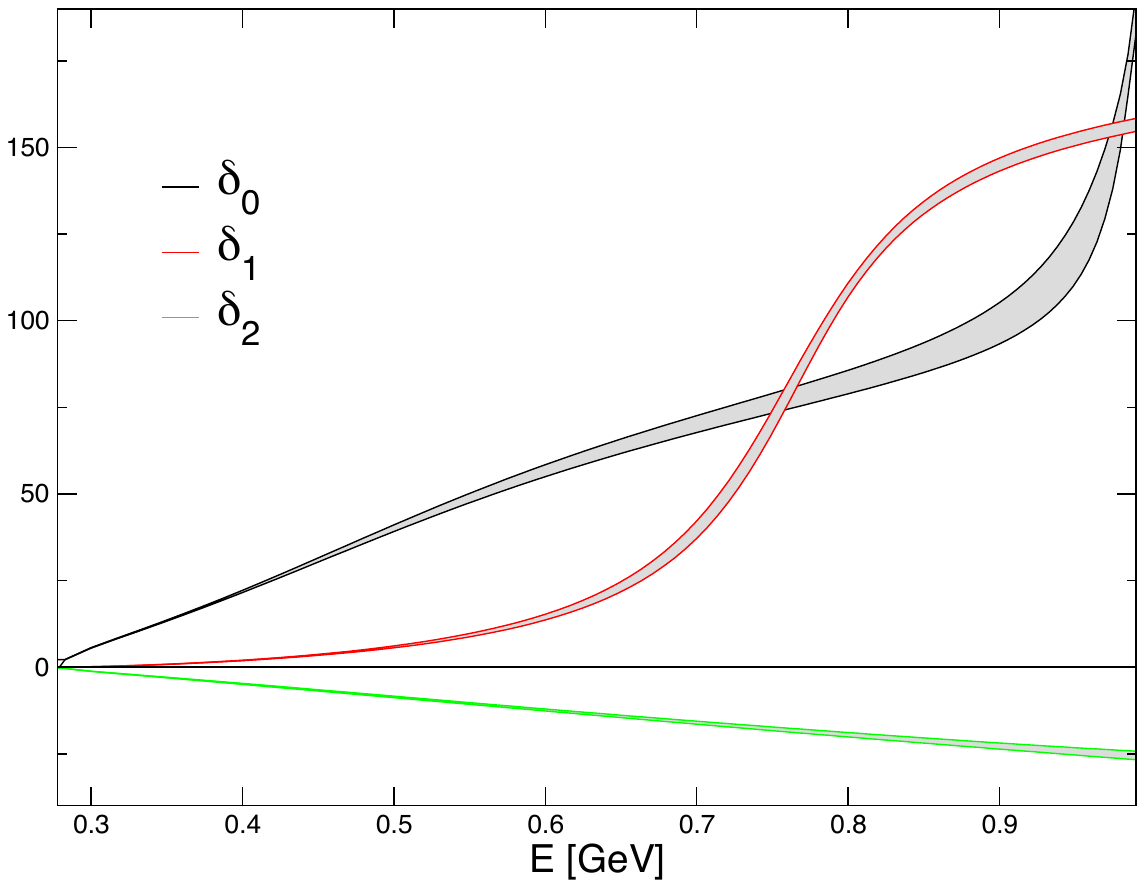}
\caption{Phase shifts of the leading $\pi\pi$ partial waves.\label{fig:phaseshifts}}
\end{figure}

We use the representations for the three phase shifts $\delta_0(s)$, $\delta_1(s)$, $\delta_2(s)$ given in~\cite{Colangelo2001}. In that analysis, the values of the phase shifts at $\sqrt{s_1} = 0.8\,\GeV$ are used to control the uncertainties in the low-energy region. We vary these in the range 
\begin{eqnarray}
\label{eq:deltanum}
&& \delta_0(s_1)=82.3^\circ(3.4^\circ)\,,\quad\delta_1(s_1)=108.9^\circ(2.0^\circ)\,, \nonumber \\
&& \delta_2(s_1)=-19.5^\circ(0.6^\circ)\;.
\end{eqnarray}
Fig.~\ref{fig:phaseshifts} shows the energy dependence below $K\bar{K}$-threshold.
Above that energy, dispersion theory does not impose strong constraints on the behaviour of the phase shifts, but since we are using dispersion relations with many subtractions, the uncertainties in the input used there do not play a significant role. For definiteness, we use a parametrization where, above 1.7 GeV, $\delta_0(s)$ and $\delta_1(s)$ are set equal to $180^\circ$, while the exotic phase $\delta_2(s)$ is set equal to zero. By far the most important contribution stems from $\delta_0(s)$. In order to test the sensitivity to the behaviour of this phase shift  in the region between $K\bar{K}$-threshold  and 1.7 GeV, we generously varied the parametrization used in that region, but found that this barely affects any of the results (see the detailed discussion of our numerical results in Appendix~\ref{sec:Sensitivity to phase shifts}). 
 
\subsection{Integral equations}

For our method it is crucial that the dispersion relations
used uniquely determine the amplitude in terms of the subtraction constants. With the form~\eqref{eq:MI} of these relations,
that is not the case, however. There, the subtraction constants are collected in the polynomials $P_I(s)$. The problem
is that the homogeneous equations obtained if these polynomials are set equal to zero admit non-trivial solutions.

In its simplest form, the problem shows up if the
contributions to the discontinuities from
the crossed channels are dropped. The elastic unitarity relation~\eqref{eq:discMI} then reduces
to three independent constraints of the form
$\disc M_I(s)=\sin\delta_I(s)\, e^{-i\delta_I(s)}
M_I(s)$, or, equivalently, $M_I(s+i\epsilon)
=e^{2i\delta_I(s)}$ \mbox{$M_I(s-i\epsilon)$}. This condition
is well-known from the dispersive analysis of form factors and
can be
solved explicitly: The Omn\`{e}s function \cite{Omnes:1958hv}, defined by 
\begin{equation}\label{eq:Omega} \Omega_I(s)=\exp\left\{\frac{s}{\pi}\int_{4M_\pi^2}^\infty
\!\frac{ds'}{s'}\frac{\delta_I(s')}{(s'-s-i\epsilon)}\right\}\;,\end{equation}
obeys
$\Omega_I(s+i\epsilon)=
e^{2i\delta_I(s)}\,\Omega_I(s-i\epsilon)$, so that the ratio
$m_I(s)=M_I(s)/\Omega_I(s)$ is continuous
across the cut. Since $\Omega_I(s)$ does not have any zeros,
$m_I(s)$ is an entire function. With the asymptotic behaviour
of the phase shifts specified in the preceding section, $\Omega_0(s),\Omega_1(s)$
tend to zero in inverse proportion to $s$, while $\Omega_2(s)$ approaches
a constant:
\begin{equation}\label{eq:Omega asymptotics}\Omega_0(s)\propto\frac{1}{s}\;,\quad\Omega_1(s)\propto\frac{1}{s}\;,\quad\Omega_2(s)\propto \mathrm{constant}\;.\end{equation}
As shown in Sec.~\ref{sec:Polynomial}, the asymptotic condition we are imposing ensures that the functions
$M_I(s)$ do not grow faster than a power of $s$. Hence this also holds for the functions
$m_I(s)$. Being entire,
$m_0(s),m_1(s)$ and $m_2(s)$ thus represent polynomials:
The general solution of the simplified unitarity conditions
is of the form $M_I(s)=m_I(s)\,\Omega_I(s)$, where
$m_I(s)$ is a polynomial. 

Bookkeeping then shows, however, that the dispersion relation~\eqref{eq:MI} cannot determine the solution uniquely: The asymptotic behaviour $M_0(s)\propto s^2$ allows a cubic polynomial for $m_0(s)$, but only a quadratic one for $P_0(s)$. Hence the general solution involves four free parameters while the dispersion relation only contains three subtraction constants. Evidently, the phenomenon occurs because the Omn\`{e}s factor $\Omega_0(s)$ tends to zero if $s$ becomes large. This is the case also for $\Omega_1(s)$, while the solution of the dispersion relation for $M_2(s)$ is determined uniquely by the subtraction constants.  

The problem also occurs if the functions $\hat M_I(s)$ are retained.
The preceding discussion points the way towards
a solution of the problem: It suffices to
replace the dispersion relation for $M_I(s)$ with the one for the ratio
\mbox{$m_I(s)\equiv M_I(s)/\Omega_I(s)$}. The corresponding discontinuity
is given by 
\begin{eqnarray} 
&&m_I(s+i\epsilon)-m_I(s-i\epsilon) = \nonumber\\
&&\{
M_I(s+i\epsilon)e^{\!-i\delta_I(s)}\!-\!M_I(s-i\epsilon) e^{i\delta_I(s)}\}/
|\Omega_I(s)|\;.\end{eqnarray}
With the relation $M_I(s-i\epsilon)=M_I(s+i\epsilon)-2 i \,\disc M_I(s)$
and the expression~\eqref{eq:discMI} for the discontinuity, this becomes
\begin{equation} m_I(s+i\epsilon)-m_I(s-i\epsilon)  =2 i\,\frac{\sin\delta_I(s) \hat M_I(s)}{|\Omega_I(s)|}\;.\end{equation}
Since the functions $M_I(s)$ and $\Omega_I(s)$ only have a right hand cut and $\Omega_I(s)$ does not have a zero, the dispersion relations can be rewritten in the form
\begin{equation}\label{eq:DROmnes}
 M_I(s)=\Omega_I(s)\left\{\tilde{P}_I(s)+\frac{s^{\hspace{0.05em}n_I}}{\pi}
 \!\int_{4M_\pi^2}^\infty
\frac{ds'}{s^{\prime\,n_I}}\,\frac{\sin\delta_I(s')\,\hat M_I(s')}
{|\Omega_I(s')|\,(s'-s-i\epsilon)}\right\}\;.\end{equation}
In the simplified situation considered above, these equations
indeed unambiguously fix the solution in terms of the
polynomials $\tilde{P}_I(s)$. Our numerical results indicate that the same
is true also for the full set of coupled integral equations, but we do
not have an analytic proof of this statement.
\begin{figure*}[thb]
\begin{center}
\includegraphics[width=7.8cm]{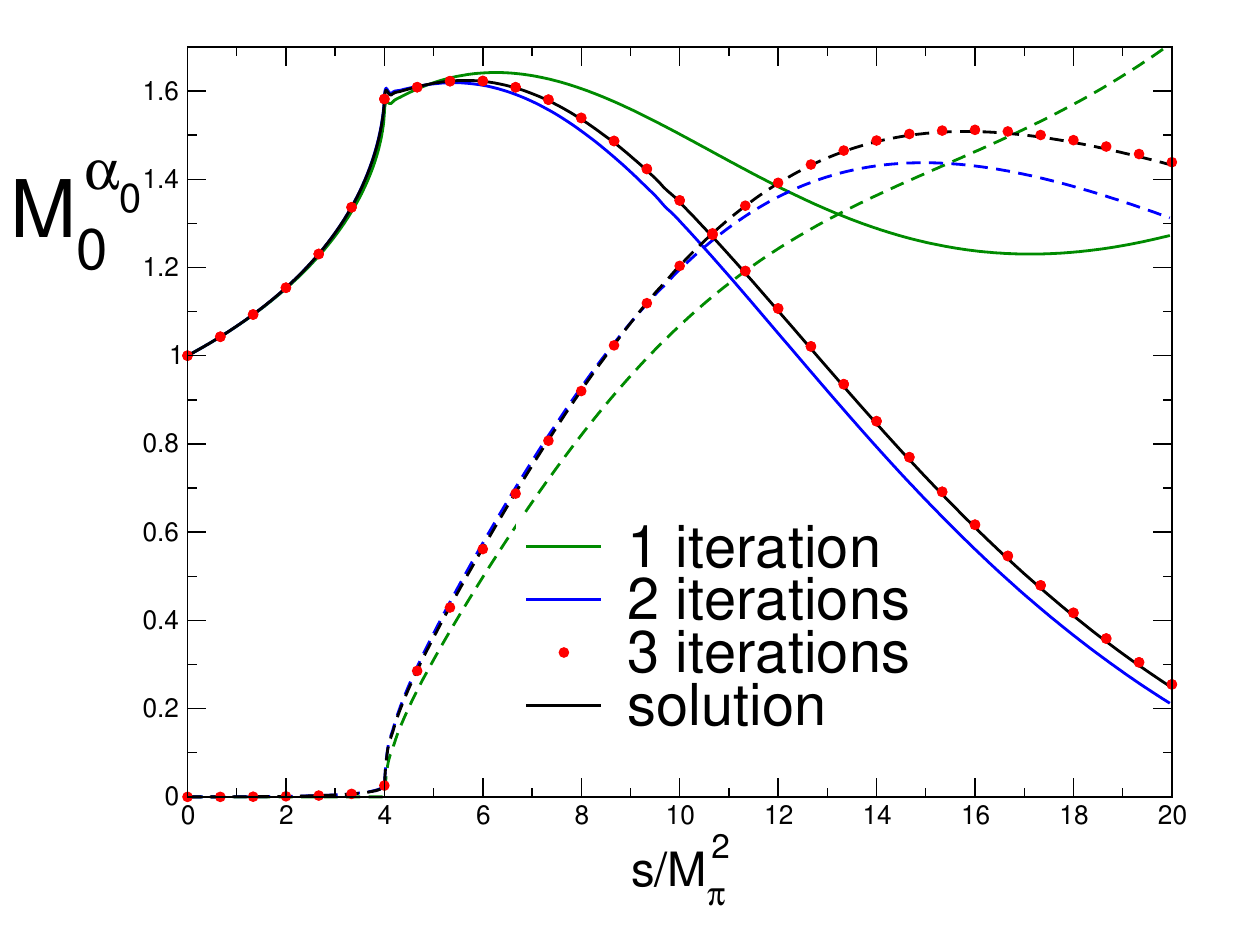}\includegraphics[width=7.8cm]{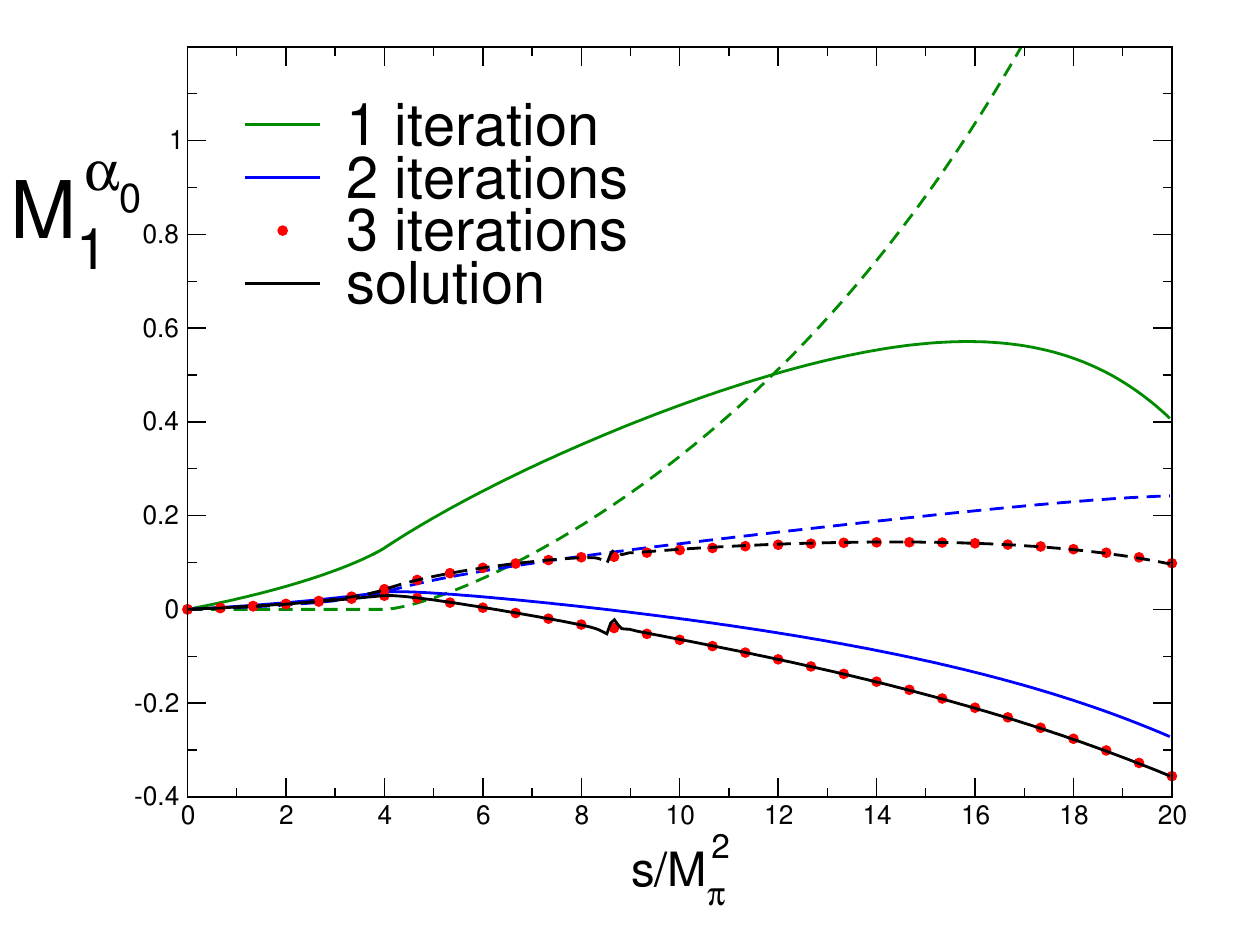}\\
\includegraphics[width=7.8cm]{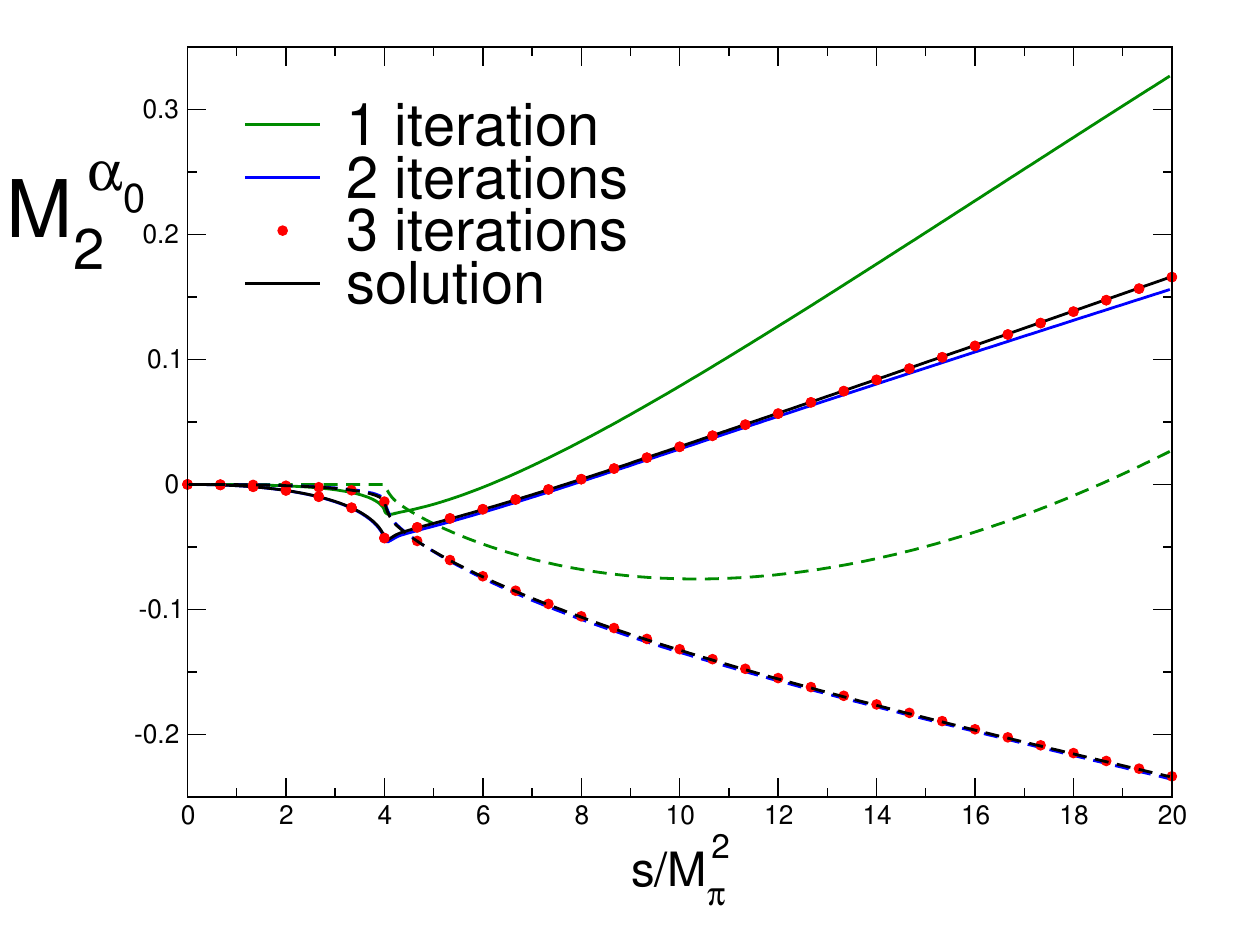}\includegraphics[width=7.8cm]{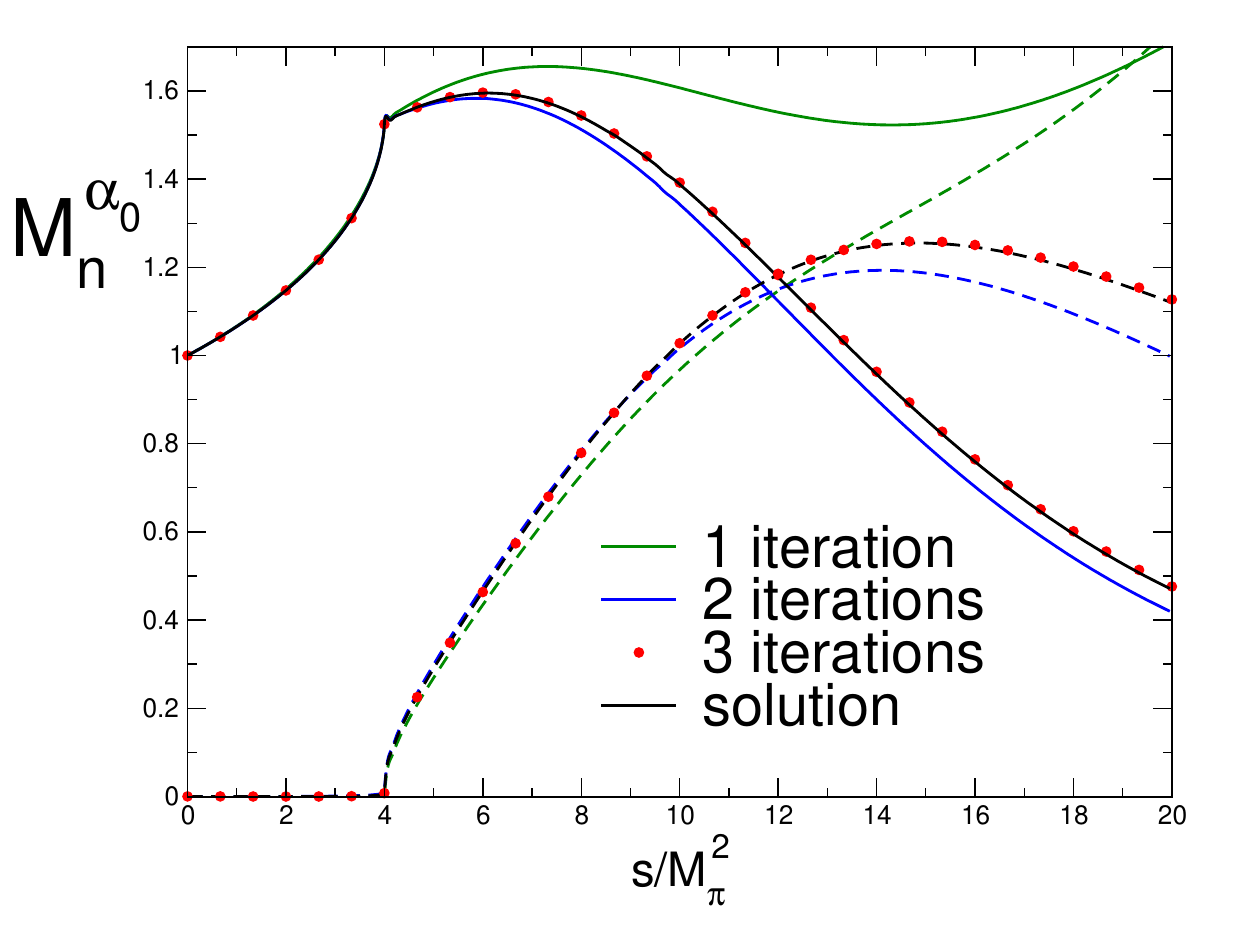}
\caption{Isospin components and neutral channel amplitude of the fundamental solution belonging to $\alpha_0$ (real and imaginary parts are shown as full and dashed lines, respectively). The plot illustrates the convergence of the iterative procedure. The result of the third iteration is displayed as a dotted line -- by eye, it could otherwise not be distinguished from the final result.\label{fig:fundamental solution}} 
\end{center}
\end{figure*}
\subsection{Subtraction constants, fundamental solutions}\label{sec:Subtraction}

For the phase shift parametrizations we are using, the integrands vanish above 1.7 GeV. 
Hence convergence is not an issue -- we could use unsubtracted dispersion 
integrals, i.e.~set $n_I=0$ in~\eqref{eq:DROmnes}. It is more convenient, however, to instead work with 
$n_0=2$, $n_1=1$, $n_2=2$, for two reasons: (i) Although the manifold of
solutions is exactly the same, for the solutions obtained with $n_I=0$, the 
dispersion integrals are quite sensitive to the behaviour of the phase 
shifts above 0.8 GeV, which is poorly known -- the sensitivity is compensated
by a corresponding sensitivity of the subtraction constants, but the correlation
leads to a clumsy error analysis. (ii) The choice is also more convenient for
comparison with earlier work where the dispersion integrals were written
in subtracted form. 

We now impose the constraints introduced in Sec.~\ref{sec:Polynomial}
to make the decomposition unique. Since $M_0(s)$ then grows only quadratically,
$\tilde{P}_0(s)$ is of the form $\alpha_0+\beta_0 s+\gamma _0 s^2+\delta_0 s^3$.
The linear growth of $M_1(s)$ leads to $\tilde{P}_1(s)=\alpha_1+\beta_1 s+\gamma_1 s^2$
and the condition $M_1(0)=0$ implies $\alpha_1=0$. Finally, the asymptotic behaviour
$M_2(s)\propto s$ implies $\tilde{P}_2(s)=\alpha_2+\beta_2 s$
and the condition $M_2(0)=M_2'(0)=0$ yields $\alpha_2=\beta_2=0$.
The dispersion relations thus take the following final form:
\begin{eqnarray}\label{eq:DROmega}
M_0(s)\al=\al\Omega_0(s)\left\{\rule[-0.5em]{0em}{2em}\alpha_0+\beta_0 s+\gamma_0 s^2+\delta_0s^3 \right. \\
 \al \al \hspace{7em}\left.+s^2\! \!\int_{4M_\pi^2}^\infty \hspace{-0.5em}d\mu_0(s')\frac{\hat M_0(s')}{s'-s-i\epsilon}\right\}\;,\nonumber
  \end{eqnarray}
\begin{eqnarray}
 M_1(s)\al=\al \Omega_1(s)\left\{\beta_1 s+\gamma_1 s^2+s\!\!\int_{4M_\pi^2}^\infty\hspace{-0.5em} d\mu_1(s')\frac{s'\hat M_1(s')}{s'-s-i\epsilon}\right\}\;,\nonumber\\
M_2(s)\al=\al \Omega_2(s)\,s^2\!\!\int_{4M_\pi^2}^\infty \hspace{-0.5em}d\mu_2(s')\frac{\hat M_2(s')}{s'-s-i\epsilon}\;,\nonumber
\end{eqnarray}
\noindent where the integration measure stands for
\begin{equation}\label{eq:dmu}d\mu_I(s')= \frac{ds'}{ \pi s'^2}\frac{\sin \delta_I(s')}{|\Omega_I(s')|}\;,\quad I=0,1,2\,.\end{equation}
The general solution of the constraints imposed by elastic unitarity and the asymptotic conditions thus
involves altogether six subtraction constants: $\alpha_0$, 
$\beta_0$, \ldots, $\gamma_1$. Note that these constraints are linear. The general solution
of our system of integral equations is a linear combination of six fundamental solutions:
 \begin{equation}M_I(s)=\alpha_0M_I^{\alpha_0}(s)+\beta_0 M_I^{\beta_0}(s)+\cdots +\gamma_1 M_I^{\gamma_1}(s)\;.\end{equation}

 The fundamental  solutions only depend on the $\pi\pi$ phase shifts, are uniquely determined by these and can
 be calculated once and for all. The first one, $M_I^{\alpha_0}(s)$, for instance, represents the solution of our
 integral equations for $\alpha_0=1$, $\beta_0= \ldots =\gamma_1=0$. It can be calculated iteratively.
As a starting point of the iteration, one may use the solution obtained if the phase shifts are set
equal to zero, so that the dispersion integrals in~\eqref{eq:DROmega}
vanish and $\Omega_I(s) = 1$. In the case of $M_I^{\alpha_0}(s)$, the starting point of the iteration 
is $M_0^{\alpha_0}(s) = 1$, $M_1^{\alpha_0}(s) =
M_2^{\alpha_0}(s) =0$. Inserting the corresponding angular averages in the
integrals in (\ref{eq:angularAverage}), the evaluation of~\eqref{eq:DROmega}
yields the result of the first iteration. The procedure can then be
repeated, using this result as a new start. From the second iteration
on, the complications in the evaluation of the angular averages discussed 
in Sec.~\ref{sec:Elastic unitarity} must be accounted for -- they do
affect the computing time, but the iteration only requires a 
few steps to converge.

Fig.~\ref{fig:fundamental solution} shows the result
for this particular fundamental solution. The comparison of the first and last panels shows that the neutral component
of the solution is dominated by the contribution from $M_0(s)$. 

\subsection{Taylor invariants}\label{sec:Taylor invariants}
The subtraction constants are closely related to the coefficients of the Taylor 
expansion of the functions $M_0(s)$, $M_1(s)$, $M_2(s)$ in powers of $s$:
\begin{equation}\label{eq:Taylor series} M_I(s)=A_I+ s\hspace{0.1em}B_I+ s^2 C_I+s^3 D_I+\cdots\end{equation}
In the form~\eqref{eq:DROmega} of the dispersion relations, the six coefficients
$A_0$, $B_0$, $C_0$, $D_0$, $B_1$, $C_1$ uniquely determine the
six subtraction constants $\alpha_0$, $\beta_0$, $\gamma_0$, $\delta_0$,
$\beta_1$, $\gamma_1$ and vice versa, but this only holds for the particular
choice made, where some of the subtraction constants 
are set equal to zero. 

The polynomial ambiguities in the isospin components amount to 
corresponding ambiguities in the Taylor coefficients. 
In the case of $M_1(s)$, for instance, the transformation law~\eqref{eq:gauge a} 
amounts to a linear transformation of the Taylor coefficients belonging to this component: 
$\tilde{A}_1=A_1+c$, $\tilde{B}_1=B_1+b$, $\tilde{C}_1=C_1 + 3a$. The sum over the
isospin components remains the same, provided the coefficients of
$M_0(s)$ and $M_2(s)$ are subject to corresponding transformations. 
The Taylor coefficients thus transform in a non-trivial manner under $G_5$,
but it is a simple matter to check that the six combinations
\begin{eqnarray}\label{eq:K}
K_0\al=\al A_0+\mbox{$\frac{4}{3}$}A_2+B_0s_0+\mbox{$\frac{4}{3}$} B_2s_0\;,\nonumber\\
K_1\al=\al A_1+\mbox{$\frac{1}{3}$}B_0-\mbox{$\frac{5}{9}$}B_2 -3\, C_1s_0^{\,2}-3\,C_2s_0\,\nonumber\\
K_2\al=\al C_0+\mbox{$\frac{4}{3}$}C_2\;, \\
K_3\al=\al B_1+C_2+9\,D_2 s_0\;,\nonumber\\
K_4\al=\al D_0+\mbox{$\frac{4}{3}$}D_2\;,\nonumber\\
K_5\al=\al C_1-3\,D_2\;,\nonumber
\end{eqnarray} 
are invariant. We refer to these quantities as {\em Taylor invariants}.
They fully characterize the representation in a manner that does not depend
on the choices made when decomposing $M_c(s,t,u)$ into the isospin components
$M_0(s)$, $M_1(s)$, $M_2(s)$: Knowledge of the invariants $K_0$, \ldots, $K_5$
determines the isospin components up to polynomials that are irrelevant because they
drop out in the sum. Instead of specifying the six subtraction constants, we can equally
well specify the six Taylor invariants. This will be useful when comparing the dispersive
solutions with the representations obtained from \chpt. 

For $K_0$, the expression in terms of the subtraction constants is particularly simple. In the form \eqref{eq:DROmega} used for the dispersion relations, the coefficients $A_2$ and $B_2$ vanish, so that this invariant is determined by the first two coefficients of the Taylor expansion of the function $M_0(s)$: $K_0= A_0+B_0\hspace{0.1em}s_0$. The dispersion relation for $M_0(s)$ shows that $A_0=\alpha_0$ and $B_0=\beta_0+ \omega_0\ \alpha_0$, where $\omega_0$ is the first derivative of the Omn\`{e}s factor $\Omega_0(s)$ at $s=0$. Hence $K_0$ is related to the subtraction constants by $K_0=(1+\omega_0\,s_0)\alpha_0+s_0\,\beta_0$. While $\alpha_0$ is dimensionless, $\beta_0$ is of dimension 1/Energy$^{2}$. Expressing  the value of $\beta_0$ in GeV units, the relation takes the form
\begin{equation}\label{eq:K0} K_0=1.368\,\alpha_0+0.1195\,\beta_0\;.\end{equation}
\subsection{Nonrelativistic expansion}\label{sec:NR}
The nonrelativistic region concerns the behaviour of the functions $M_0(s)$, $M_1(s)$, $M_2(s)$ in the vicinity of $s = 4 M_\pi^2$. The structure of the amplitude in that region is governed by the fact that the branch cut singularity generated by elastic final state interactions among two of the pions is of the square-root type: Below the inelastic thresholds, the amplitude has only two sheets -- the functions $M_0(s)$, $M_1(s)$, $M_2(s)$ are analytic in the variable $q=\sqrt{s/4M_\pi^2-1}$. They can be expanded in a Taylor series:
\begin{equation}\label{eq:M0q}M_0(s)=\sum_{k=0}^\infty m_0^k \,q^k \;,\quad s=4M_\pi^2(1+q^2)\;,\end{equation} 
and likewise for $M_1(s)$ and $M_2(s)$. The velocity of the two particles in their center-of-mass system is given by $v=q/\sqrt{1+q^2}$. Accordingly the series \eqref{eq:M0q} essentially amounts to an expansion in powers of the velocity.  

At a given value of $s$, the two sheets only differ in the sign of $q$. Hence the discontinuity is given by the contributions from the odd powers
\begin{equation}\label{M0disc}\mathrm{disc}\hspace{0.1em}M_0(s)=\frac{1}{i}\sum_{k=0}^\infty m_0^{2k+1}\, q^{2k+1}\;.\end{equation}
Our integral equations fully determine the amplitude as a linear combination of the subtraction constants and the coefficients of the nonrelativistic expansion inherit this property. This implies that only six of the coefficients are independent, $m_0^0$, $m_0^2$, $m_0^4$, $m_0^6$, $m_1^2$, $m_1^4$, for instance. All other coefficients of the nonrelativistic expansion can explicitly be expressed as linear combinations of these. In the nonrelativistic expansion, the integral equations thus boil down to an infinite set of linear relations among the expansion coefficients.  

The nonrelativistic effective theory \cite{Colangelo+2006a,Bissegger:2007yq,Bissegger:2008ff,Gullstrom:2008sy,Gasser:2011ju,Schneider+2011} represents an alternative framework for the analysis of the decay \mbox{$\eta\to3\pi$}. In the two-loop representation of the amplitude given in \cite{Bissegger:2007yq}, the $\pi\pi$ phase shifts only enter via the first few terms of the effective range expansion. Indeed, the values
\begin{eqnarray}\label{eq:effective range}a_0^0 \al=\al  0.22\,,\hspace{1.3em} a_2^0 = -0.0444\,,\hspace{1em} a_1^1 = 0.0379\,,\hspace{1em}\nonumber \\ b_0^0\al =\al 0.297\,,\hspace{1em} b_2^0= -0.0781\,, \\
 c_0^0 \al = \al -0.0466\,,\hspace{1em}c_2^0 = 0.00865\,,\nonumber\end{eqnarray}
do provide a rather accurate representation of the $\pi\pi$ scattering amplitude, throughout  the physical region of $\eta\to 3\pi$.  They determine the coefficients of the loop integrals occurring in the NREFT representation of the functions $M_0(s)$, $M_1(s)$, $M_2(s)$. The representation of Ref.~\cite{Bissegger:2007yq} does account for the mass difference between the charged and neutral pions, but otherwise neglects the electromagnetic interaction. It involves six low-energy-constants, denoted by $L_0$, $L_1$, $L_2$, $L_3$, $K_0$, $K_1$. 

To compare this framework with ours, we consider the isospin limit. In this limit, the pion mass difference disappears and only four of the LECs are independent:
\begin{eqnarray}\label{eq:KL}K_0  \al = \al - 3 L_0 - L_1 (M_\eta - 3 M_\pi) + L_3 (M_\eta - 3 M_\pi)^2\,, \nonumber\\
K_1 \al = \al -L_2 - 
  3 L_3\;.\end{eqnarray} 
In the isospin limit, the one-loop integrals of the nonrelativistic effective theory are described by the function $J(q)=i\hspace{0.2em} q/\!\sqrt{1+q^2}$, which only involves odd powers of $q$. At two loops, there are contributions proportional to the two-loop integral $F(q)$ as well as terms proportional to $J(q)^2$. The nonrelativistic expansion of $F(q)$  involves odd as well as even powers of $q$. Chopping the expansion off at $O(q^4)$ yields a very accurate representation of this function, throughout the physical region. If the loop contributions are dropped, $M_0(s)$ reduces to a quadratic polynomial in $s$, $M_1(s)$ becomes proportional to $s$, while $M_2(s)$ vanishes. 

The LECs $L_0, \ldots\,, L_3$ play a role analogous to the subtraction constants $\alpha_0,\ldots\,\gamma_1$ of the dispersive framework, but there is a qualitative difference: While the LECs are real, the subtraction constants can be complex. Note also that the decomposition of the amplitude into isospin components is unique only up to polynomials. When comparing the components of the NREFT representation with those of dispersion theory, the polynomial ambiguities must be taken into account. This can be done with the method used when matching the dispersive and chiral representations. The polynomial ambiguities only affect the coefficients of the even powers of $q$. There are analogs of the Taylor invariants -- suitable linear combinations of the coefficients $m_0^{2k}, m_1^{2k}, m_2^{2k}$ -- that do not depend on the choice made when decomposing the amplitude into isospin components. Four such invariants are within reach of the two-loop representation. Hence there is a unique dispersive solution with four subtraction constants that matches the generic two-loop representation in the isospin limit. Alternatively, one may compare the dispersive and nonrelativistic amplitudes in the physical region and minimize the difference between the two. We will carry this out for one particular nonrelativistic representation in  Sec.~\ref{sec:Comparison NREFT}.
\vspace{-0.2cm}
\section{Chiral perturbation theory}\label{sec:Chiral perturbation theory}
\subsection{Current algebra, Adler zero}\label{sec:Current algebra}

The leading term in the chiral expansion of the transition amplitude was 
worked out from current algebra, long before the formulation of
\chpt~\cite{Osborn+1970}. In the normalization~\eqref{eq:normalization}, it
exclusively involves $s$, $M_\pi$ and $M_\eta$:
\begin{equation}\label{eq:MLO}M^{\mathrm{LO}}_c(s,t,u)=T(s)\;,\quad T(s)\equiv\frac{3s-4M_\pi^2}{M_\eta^2-M_\pi^2}\;.\end{equation}
The formula exhibits an Adler zero at $s=\frac{4}{3}M_\pi^2$. The zero is outside the physical region, where $s$ is confined to $4M_\pi^2<s<(M_\eta-M_\pi)^2$. The rapid growth of the observed Dalitz plot distribution does show that the square of the amplitude grows with $s$, but the leading term represents a decent approximation to the full amplitude only at small values of $s$. Already at $s=4M_\pi^2$, the final state interaction generates a pronounced momentum dependence which in the chiral expansion starts showing up at NLO.   

\subsection{\chpt\ to one loop}\label{sec:One loop}

The chiral perturbation series of the transition amplitude was worked out to NLO in the framework of SU(3)$\times$SU(3) in~\cite{Gasser+1985a}. In this framework, the final state interaction manifests itself through one-loop graphs involving pions as well as kaons or $\eta$-mesons. The amplitude can be expressed in terms of the meson masses $M_\pi$, $M_K$, $M_\eta$, the decay constants $F_\pi$, $F_K$ and the low-energy constant $L_3$. We use the numerical values $F_\pi=92.28(9) \MeV$~\cite{Olive:2016xmw}, $F_K/F_\pi=1.193(3)$~\cite{Aoki:2016frl} and rely on the recently improved determination of $L_3$ from $K_{\ell 4}$ decay, $L_3=-2.63(46)\cdot 10^{-3}$~\cite{Colangelo:2015kha}, so that the one-loop representation does not contain any unknowns. 

While the dispersive representation yields an accurate description of the
momentum dependence in the entire range from $s=0$ to the physical region
and even beyond, the truncated chiral expansion is useful only at small
values of $s$, where it can be characterized by the lowest few coefficients
of the Taylor series~\eqref{eq:Taylor series}. The contributions from the loop graphs
are determined by the masses of the Nambu-Goldstone bosons and the pion decay constant. 
The tree graphs, on the other hand, yield polynomials of up to $O(p^4)$ in the momenta. The coefficients
of these polynomials are in one-to-one correspondence with the Taylor coefficients
$A_0$, $B_0$, $C_0$, $A_1$, $B_1$, $A_2$, $B_2$, $C_2$. Together with $F_\pi$,
these coefficients thus uniquely determine the one-loop representation. 

The polynomial ambiguities also show up in the decomposition of the
chiral representation. At one loop, the polynomial parts of $M_0(s)$, $M_2(s)$ are quadratic in $s$, 
while $M_1(s)$ is linear in $s$. The transformations~\eqref{eq:gauge a}, \eqref{eq:gauge b} retain
this property only if $a$ is set equal to zero. This shows that the polynomial ambiguities of the
one-loop representation form a four-dimensional subgroup $G_4$ 
of the general invariance group $G_5$ associated with the decomposition~\eqref{eq:RT}.
Only $8-4=4$ combinations of the eight Taylor coefficients listed above
are invariant under this group of transformations. We may identify these with what remains of the Taylor invariants $K_0$, $K_1$, $K_2$, $K_3$ if the coefficients $D_0$, $C_1$, $D_2$ are dropped:
\begin{eqnarray}\label{eq:H4}
H_0\al =\al A_0 + \mbox{$\frac{4}{3}$} A_2 + s_0 \left(B_0 + \mbox{$\frac{4}{3}$} B_2\right) \nonumber\\
H_1 \al=\al A_1 + \mbox{$\frac{1}{9}$}\left(3 B_0 -5 B_2\right)- 3 C_2 s_0\\
H_2\al =\al C_0 + \mbox{$\frac{4}{3}$} C_2\nonumber \\
H_3 \al= \al B_1 + C_2\;.\nonumber 
\end{eqnarray}
Since $K_0$ does not contain $D_0$, $C_1$ or $D_2$, the quantity $H_0$ is identical with it -- this combination is invariant under the full group $G_5$. For $H_1$, however, this is not the case: $K_1\equiv H_1-3\,C_1s_0^{\,2}$ involves the coefficient $C_1$, which is beyond reach at one loop, but is needed for $K_1$ to be invariant under the full group. The situation with $K_2$ and $K_3$ is similar: $K_2\equiv H_2$, $K_3\equiv H_3+9 \,D_2\,s_0$. The invariants $K_4$ and $K_5$ exclusively involve Taylor coefficients that are beyond reach of the one-loop representation.\footnote{In the letter version of the present paper, we shortened the presentation by working with a single set of invariants, completing the set \{$H_0$, $H_1$, $H_2$, $H_3$\} with $H_4\equiv K_4$ and $H_5\equiv K_5$.}  
This means that the quantities $H_1$ and $H_3$ are invariant only under the four-parameter subgroup $G_4$ formed by the elements of $G_5$ with $a=0$. Under the full group of polynomial ambiguities, $H_1$ and $H_3$ are invariant only up to terms of NNLO.

The constants $H_0,H_1,H_2,H_3$ contain the essence of the one-loop representation: If they are known, the transition amplitude is uniquely determined by unitarity, to NLO of the chiral expansion (an explicit proof of this statement can be found in 
Appendix~\ref{sec:DR NLO}). In this sense, the momentum dependence of the chiral representation is not of interest -- dispersion theory provides better control over that. The general principles that underly dispersion theory, however, do not determine the subtraction constants. That is where \chpt\ can offer useful information. 

In the following, we will make use of the remarkably accurate experimental determination of the Dalitz plot distribution~\cite{KLOE:2016qvh}, which subjects the Taylor invariants to strong constraints. More precisely, since the distribution is normalized to 1 at the center, these data concern their relative size rather than the constants themselves. We use the invariant $H_0$ to parametrize the 
normalization of the amplitude and describe the relative size
of the Taylor invariants  by means of the variables 
\begin{equation}h_i=\frac{H_i}{H_0}\;.\quad i=1,2,3\;.\end{equation}
While experiment yields strong constraints on $h_1,h_2,h_3$, it cannot shed
any light on the value of $H_0$, because this term fixes the normalization of
the amplitude $M_c(s,t,u)$ rather than $A_c(s,t,u)$, which is what can be
measured. We need to rely on \chpt\ to determine $H_0$.  

At leading order of the chiral expansion, the normalization~\eqref{eq:normalization} implies $H_0^\mathrm{LO} = 1$. Working out the Taylor coefficients of the one-loop representation, which is given explicitly  in 
Appendix~\ref{sec:DR NLO}, one readily verifies the representation 
\begin{eqnarray}\label{eq:H0}H_0 = \al \al1+ 
\frac{2(M_\eta^2-5M_\pi^2)}{3(M_\eta^2-M_\pi^2)}\Delta_\mathrm{GMO}+
\frac{8M_\pi^2}{3(M_\eta^2-M_\pi^2)}\Delta_\mathrm{F} \nonumber \\
\al \al + \mathrm{chilogs}+O(m_\mathrm{quark}^2)\;.\end{eqnarray}
The constants $\Delta_\mathrm{GMO}$ and $\Delta_\mathrm{F}$ stand for
\begin{equation}\label{eq:DGMODF} \Delta_\mathrm{GMO}\equiv \frac{4M_K^2-3M_\eta^2-M_\pi^2}{M_\eta^2-M_\pi^2}\;,\quad \Delta_\mathrm{F}\equiv\frac{F_K}{F_\pi}-1\;,\end{equation}
and the remainder contains the chiral logarithms typical of \chpt~-- in the present case, it involves contributions proportional to $M_\pi^2\ln (M_\pi^2/M_\eta^2)$ and to $\ln (M_K^2/M_\eta^2)$. The relation \eqref{eq:H0} amounts to a low energy theorem: Up to contributions of next-to-next-to-leading order, the invariant $H_0$ is determined by the masses and decay constants of the Nambu-Goldstone bosons. 

Remarkably, despite the fact that the  $\eta$ undergoes mixing with the $\eta^\prime$, the formula \eqref{eq:H0} only contains $M_\eta$, while $M_{\eta'}$ does not occur. The role played by the $\eta'$ in the low-energy structure of QCD is well understood. It can be studied in a systematic manner by invoking the large $N_c$ limit, where the $\eta^\prime$ becomes massless and can be treated on the same footing as the Nambu-Goldstone bosons \cite{Gasser+1985}.  This framework gives a good understanding of the size of the LEC $L_7$, which determines the deviation from the Gell-Mann-Okubo formula and enters the low-energy theorem via the term $\Delta_\mathrm{GMO}$.  Indeed, as shown in Ref.~\cite{Leutwyler:1996np}, the contribution from this term in the low energy theorem \eqref{eq:H0} fully accounts for the effects generated by $\eta$-$\eta'$-mixing at $O(m_\mathrm{quark})$ -- it would be wrong to supplement \chpt\ with an extra wheel to account for $\eta$-$\eta'$-mixing.

Note that the dependence on the decay constants is suppressed by a factor of $M_\pi^2$ -- if the two lightest quarks are taken massless, $H_0$ is fully determined by the masses of the Nambu-Goldstone bosons, up to NNLO contributions. At the physical values of the masses and decay constants, the term proportional to $\Delta_\mathrm{F}$ amounts to 0.036. The contribution from the chiral logarithms is also small: $\mathrm{chilogs}=0.037$. The dominating contribution stems from the term $\Delta_\mathrm{GMO}$ and amounts to $0.103$. The net result at one loop reads: $H_0^\mathrm{NLO}=1.176$. 

The change in the value of $H_0$ from tree level to one loop confirms a general experience with \chpt\ based on SU(3)$\times$SU(3): Unless the quantity of interest 
contains strong infrared singularities, subsequent terms in the chiral 
perturbation series are smaller by 20 to 30\,\%. The values\footnote{Throughout, numerical values of dimensionful quantities are given in GeV units.}  $h_1^\mathrm{LO}= 1/(\meta^2 - \mpi^2)=3.56$  and $h_1^\mathrm{NLO}=4.52$, are also consistent with this rule, but the correction is relatively large (27\,\%), 
because this quantity does contain a strong infrared singularity. In fact, $h_1$ explodes if $m_u$ and $m_d$ are sent to zero:
The expansion of $h_1$ in powers of $M_\pi$ starts
with a term that is inversely proportional to the square of $M_\pi$:
\begin{equation}\label{eq:h1lR}
h_1=\frac{M_\eta^2}{160\pi^2F_\pi^2M_\pi^2}+\cdots
\end{equation}
Numerically, the singular term dominates the difference between $h_1^\mathrm{NLO}$ and $h_1^\mathrm{LO}$.  

We conclude that it is meaningful to truncate the chiral expansion of the 
Taylor coefficients at NLO. The invariant $X$ is approximated with
the one-loop result $X^\mathrm{NLO}$ and the uncertainties
from the omitted higher orders are estimated at
$0.3\,|X^\mathrm{NLO}-X^\mathrm{LO}|$. This is on the conservative
side of the rule mentioned above and yields the following theoretical estimate 
for the four Taylor invariants:
\begin{eqnarray}\label{eq:HNLO} H_0 \al =\al1.176(53)\;,\quad h_1= 4.52(29)\;,\quad h_2= 16.4(4.9)\;, \nonumber\\
h_3 \al = \al 6.3(2.0)\;.\end{eqnarray}
The estimate used for $h_3$ in particular also covers the comparatively 
small uncertainty in the value of $L_3$. 
\subsection{\chpt\ to two loops}\label{sec:Two loops}

Bijnens and Ghorbani~\cite{Bijnens+2007} have worked out the chiral perturbation series of the transition amplitude to NNLO.
The amplitude retains the form~\eqref{eq:RT}, but the isospin components
$M_0(s)$, $M_1(s)$, $M_2(s)$ pick up additional contributions, which can be
expressed in terms of the meson masses and the LECs that occur in the
effective Lagrangian. As discussed above, elastic unitarity determines the
one-loop representation in terms of the tree graph amplitude up to a polynomial, which can be characterized by the four Taylor invariants $H_0,\ldots,H_3$. The situation at NNLO is analogous: Elastic unitarity determines the amplitude in terms of the one-loop 
representation up to a polynomial. Since the amplitude now includes terms of $O(p^6)$, the polynomial is of higher degree and now contains 
six independent terms rather than four: $p_0+p_1\,s+ p_2\, s^2+p_3\,\tau^2+p_4\,s^3+p_5\, s\,\tau^2$, with $\tau\equiv t-u$. Hence there are six combinations of Taylor coefficients that are independent of the choice of the decomposition. At two loops, all of the six Taylor invariants $K_0$, \ldots, $K_5$ are needed to characterize the representation.

The invariants $K_0$, \ldots, $K_5$ can also be used to characterize the solutions of our system of integral equations. The Taylor coefficients of the dispersive representation are given by linear combinations of the six subtraction constants and uniquely determined by these. Knowledge of the subtraction constants thus fixes the Taylor invariants $K_0$, \ldots, $K_5$ and vice versa: The degrees of freedom inherent in the two-loop representation are in one-to-one correspondence with the degrees of freedom occurring in our integral equations. 

The Taylor coefficients of the representation specified in~\cite{Bijnens+2007} can be worked out with the code provided by Bijnens and collaborators~\cite{BijnensCode}. For the numerical values of the corresponding invariants $K_0,\ldots,K_5$, we then obtain:  
\begin{eqnarray}\label{eq:KBG}
K_0^\mathrm{BG}\al=\al 1.27-0.0074\, i\;,\quad K_1^\mathrm{BG}=3.88+0.10\, i\;, \\
K_2^\mathrm{BG}\al=\al 37.2-0.22\,i\;,
\quad K_3^\mathrm{BG}= -6.2-2.8\, i\;, \nonumber\\
K_4^\mathrm{BG} \al = \al 113-2.0\,i\;,\hspace{1.8em} K_5^\mathrm{BG}=73+8.3\,i\;.\nonumber
\end{eqnarray}

The main problem with the two-loop representation is that it involves new low-energy constants. These arise from the effective Lagrangian of $O(p^6)$ and are not known to a precision comparable to the parameters that enter the one-loop representation. They show up in the real parts of $K_0,\ldots,K_5$. There is a parameter free prediction only for one of these: The invariant $K_4$ does not get a contribution from the low-energy constants of NNLO.\footnote{An analogous phenomenon occurs at one loop, where the invariant $H_3$ does not pick up any contribution from the effective Lagrangian of $O(p^4)$.}  Estimating the uncertainties in the prediction for $\Re K_4$ with the rule of Sec.~\ref{sec:One loop}, we obtain
\begin{equation}\label{eq:ReK4}\Re K_4=113(34)\;.\end{equation}
As we will see in Sec.~\ref{sec:Anatomy},  where we compare the representation of Bijnens and Ghorbani with the outcome of our dispersive analysis, this prediction is perfectly consistent with experiment.  

\subsection{Imaginary parts at two loops}\label{sec:Imaginary parts two loops}

The coefficients of the Taylor expansion of the Omn\`{e}s factors are real, but the expansion of the dispersion integrals in~\eqref{eq:DROmega} in powers of $s$ yields complex coefficients. Accordingly, the linear relations between the Taylor invariants and the subtraction constants involve complex coefficients. As the dispersion integrals arise from the discontinuities in the crossed channels, they are small: If the subtraction constants are real, the imaginary parts of the Taylor invariants are small. Indeed, in the chiral expansion, the Taylor invariants start picking up an imaginary part only at two loops. Unitarity implies that the leading terms in the chiral expansion of the imaginary parts only involve those low-energy constants that occur already in the one-loop representation of the transition amplitude, which are known: The imaginary parts of $K_0$, \ldots, $K_5$ represent parameter free predictions.  Applying the rule given in Sec.~\ref{sec:One loop} to estimate the uncertainties, we obtain
\begin{eqnarray}\label{eq:ImK}\Im K_0 \al=\al -0.0074(22),~\Im K_1 = 0.10(3),~\Im K_2 =-0.22(7),\nonumber\\
\Im K_3 \al =\al  -2.8(8),~\Im K_4 =-2.0(6),~\Im K_5 = 8.3(2.5). \
\end{eqnarray}
As they are small, the imaginary parts of the subtraction constants do not play an important role in our analysis. In the letter version of our work~\cite{Colangelo:2016jmc}, we shortened the presentation by simply setting the imaginary parts of the subtraction constants equal to zero and we stick to this approximation throughout the first part of the present paper. We will return to the issue in Sec.~\ref{sec:Imaginary parts} and determine the changes occurring if we do not take the subtraction constants real, but instead fix the imaginary parts of the Taylor invariants with Eq.~\eqref{eq:ImK}.  As we will see, the modification barely affects our results. 

\subsection{Matching the dispersive and one-loop representations}\label{sec:Matching}

At one loop, the Taylor invariants are known within rather small uncertainties. We now work out the dispersive representation that matches the one-loop representation in the sense that the behaviour of the functions $M_0(s)$, $M_1(s)$, $M_2(s)$ at small values of $s$ is the same: the dispersive solution that possesses the same Taylor invariants. More precisely, as we are working with real subtraction constants, we can match only the real parts of the Taylor invariants. 

Since only four of the invariants are within reach of the one-loop
representation, fixing these does not suffice to determine the solution
uniquely. We therefore consider a simplified setting by imposing stronger
asymptotic conditions on the dispersive representation: The amplitude
$M_c(s,t,u)$ is allowed to grow at most linearly when the Mandelstam
variables become large. The subtraction constants $\delta_0$ and $\gamma_1$
must then be set to zero because the fundamental solutions belonging to
them violate the stronger form of the asymptotic condition. We fix the
remaining four subtraction constants by requiring that the real parts of the four Taylor
invariants of the dispersive representation agree with those obtained at
one loop.  With the central values in~\eqref{eq:HNLO}, this gives (GeV
units) 
\begin{eqnarray}\label{eq:matching subtraction constants}\mathrm{fit}\chi_4:\quad
\al \al \alpha_0=-0.621\;,\quad\beta_0=16.9\;\quad\gamma_0=-29.5\;, \nonumber\\
\al \al \delta_0=0\;,\quad\beta_1=6.61\;,\quad\gamma_1=0\;.
\end{eqnarray}
We refer to this solution of our integral equations as the {\em matching solution}. Although it does not represent a fit to data, we denote it by fit$\chi_4$, to simplify the notation used when comparing the various solutions to be discussed below. The label $\chi$ indicates that this solution makes use of the constraints imposed by chiral symmetry and 4 is the number of subtraction constants used. 

In order to compare the isospin components of the matching solution with those of the one-loop representation, we need to fix the decomposition of the latter. This can be done in such a way that the two representations match not only in the real parts of the Taylor invariants within reach of the one-loop representation, but in the real parts of the Taylor coefficients themselves. With this choice of the decomposition,  the two representations for $\Re M_0(s)$,  $\Re M_1(s)$,  $\Re M_2(s)$ agree at small values of $s$.
\begin{figure*}[thb]
\begin{center}
\includegraphics[width=7.8cm]{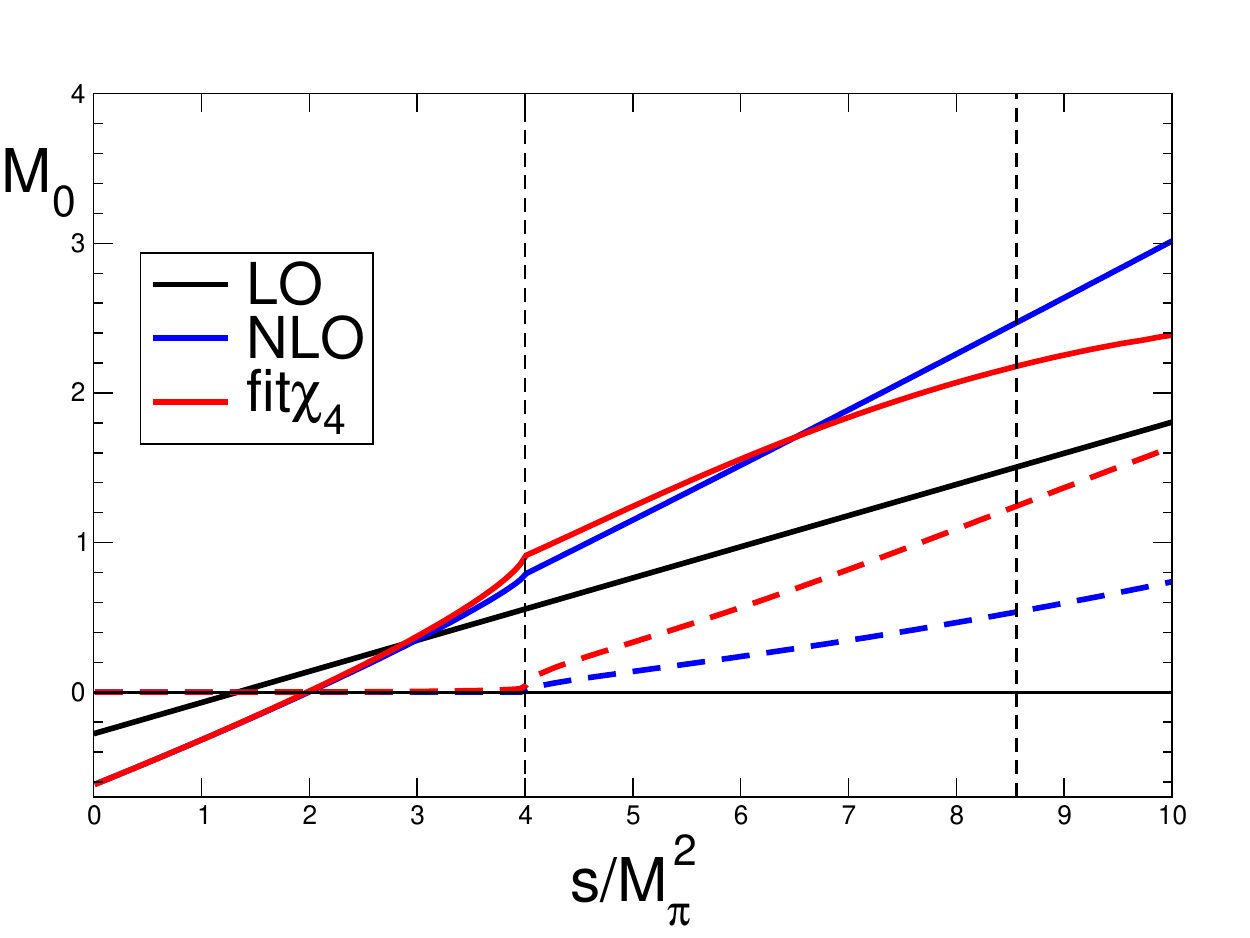}\hspace{1.cm} \includegraphics[width=7.8cm]{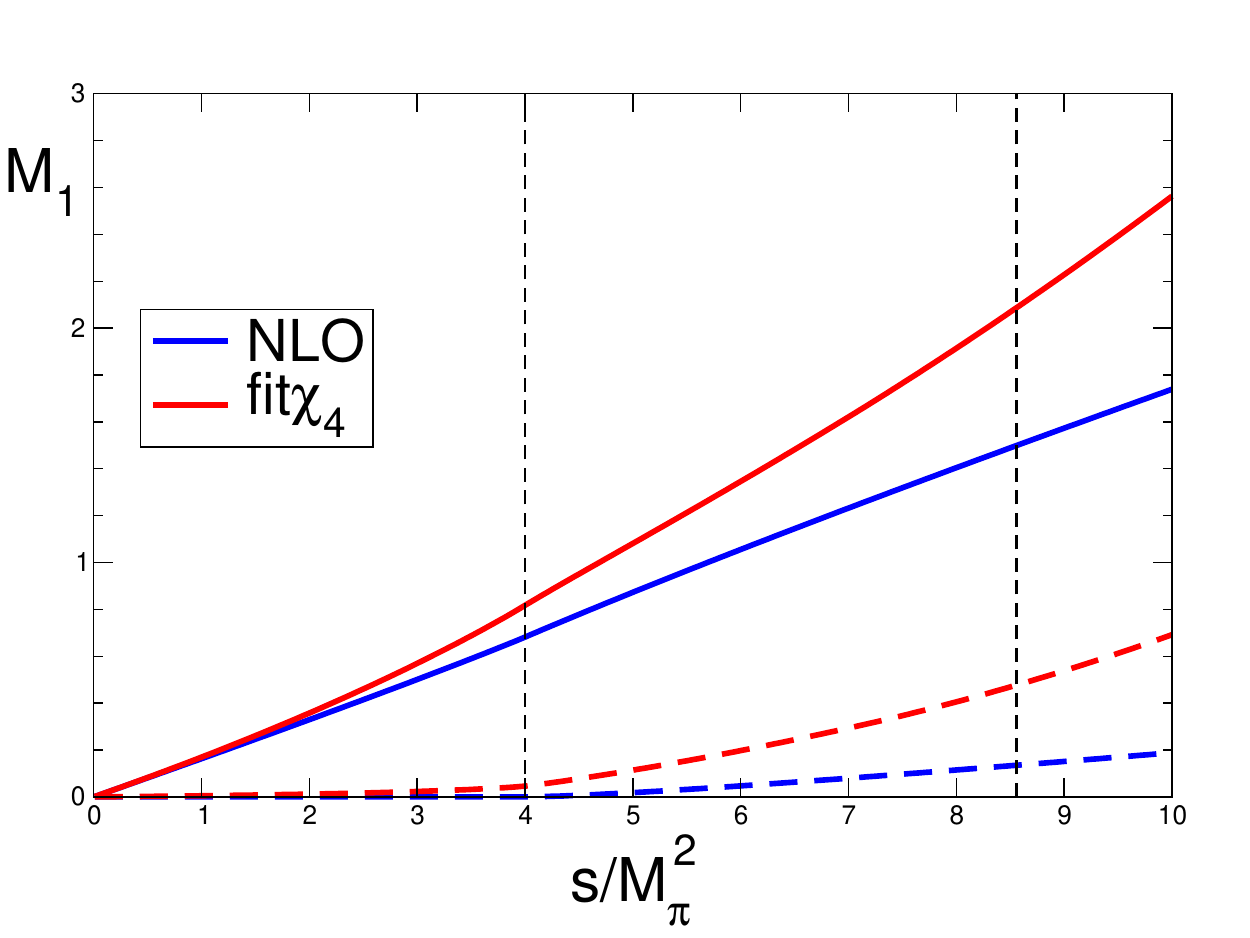}\\
\includegraphics[width=7.8cm]{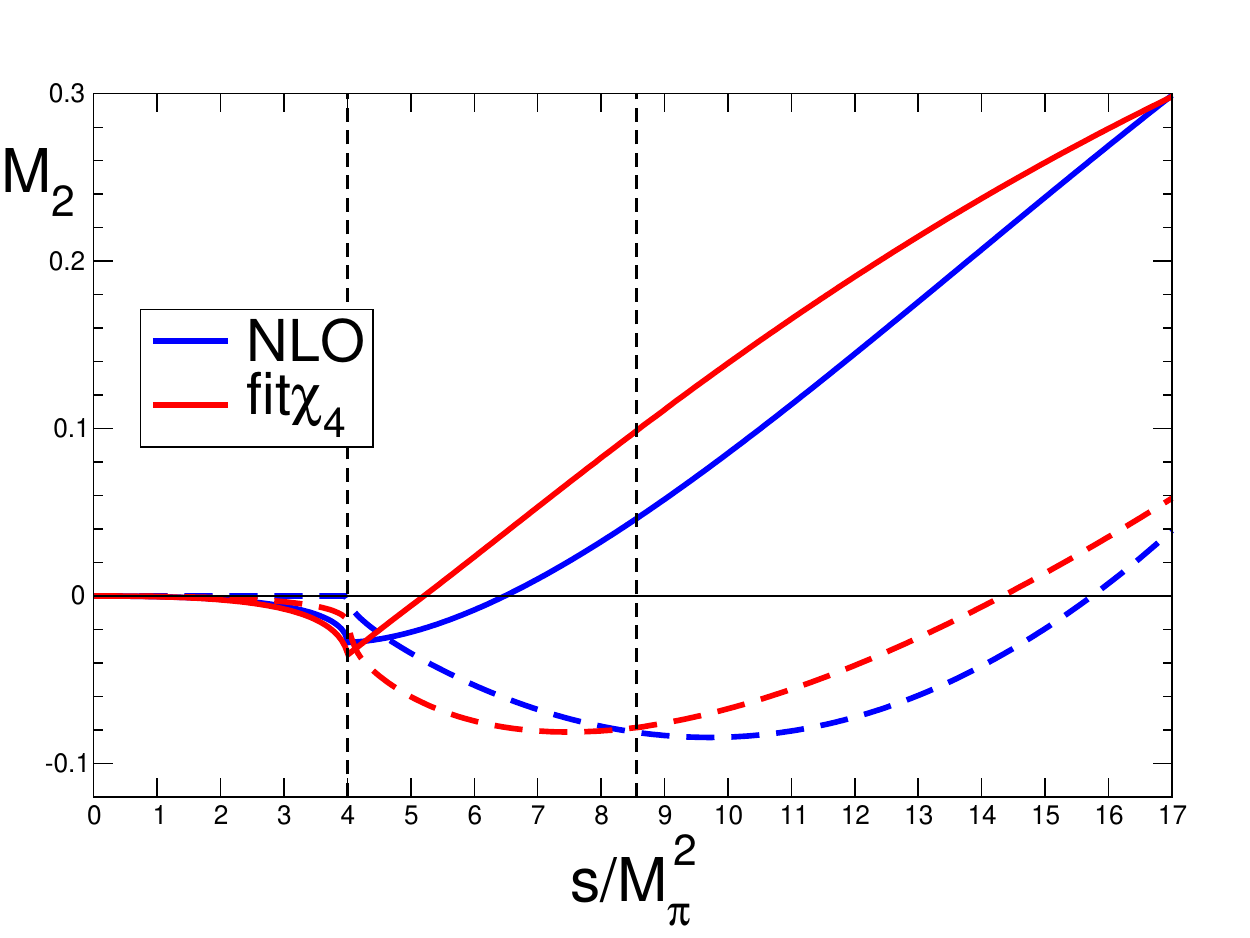}\hspace{1.cm} \includegraphics[width=7.8cm]{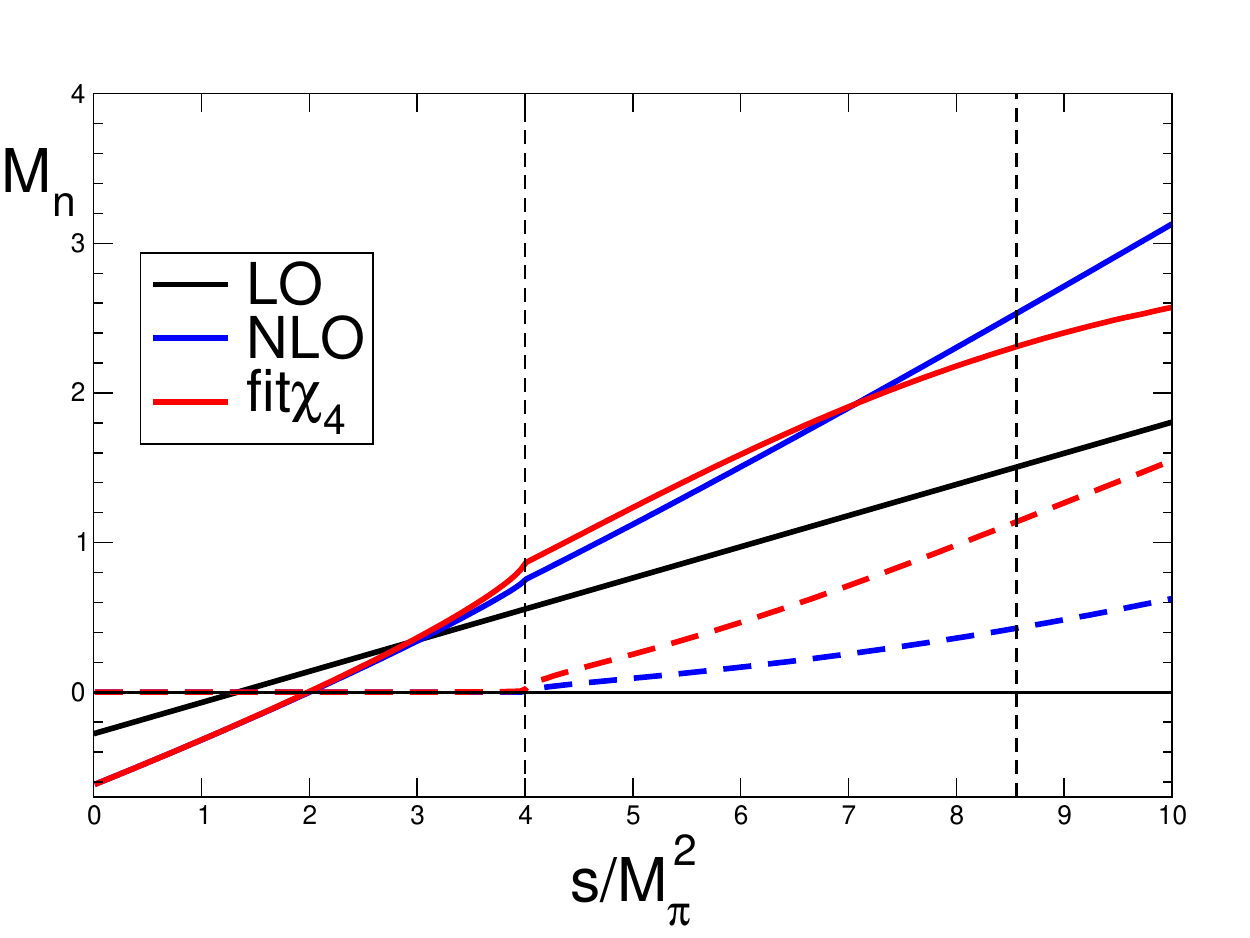}

\caption{Isospin components and neutral channel amplitude: comparison of the chiral representations  to leading and first non-leading order with the dispersive solution that matches the NLO representation at small values of $s$. Full and dashed lines show the real and imaginary parts, respectively. The dashed vertical lines indicate the lower and upper ends of the physical region of  the decay.\label{fig:Match}} 
\end{center}
\end{figure*}
Fig.~\ref{fig:Match} compares the matching solution with the chiral representation. By construction, the real parts of the two versions of the amplitude are very close at small values of $s$. The figure shows that, for the dominating contribution, $\Re M_0(s)$, the more precise treatment of the final state interaction only generates a rather modest change in the physical region. In the small components, $M_1(s)$, $M_2(s)$, the changes are more pronounced. The relative size of the corrections is larger because these components vanish altogether at LO, so that the one-loop representation only gives the leading term of the chiral series -- in $M_0(s)$, the one-loop representation is more accurate because it contains the leading as well as the first non-leading order of the series. 
\begin{figure*}[thb]\centering
\includegraphics[width=7.8cm]{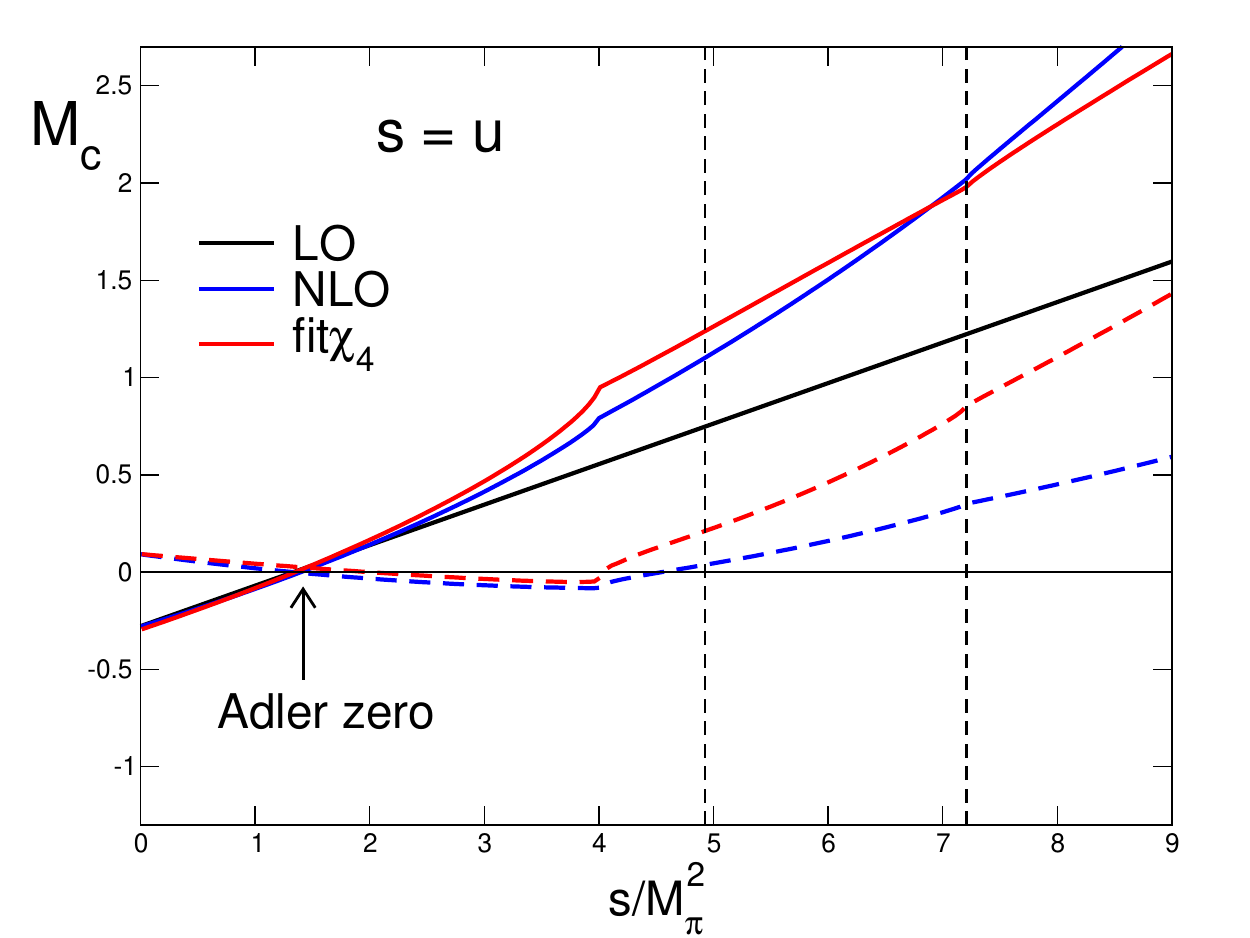}\hspace{1.cm}\includegraphics[width=7.8cm]{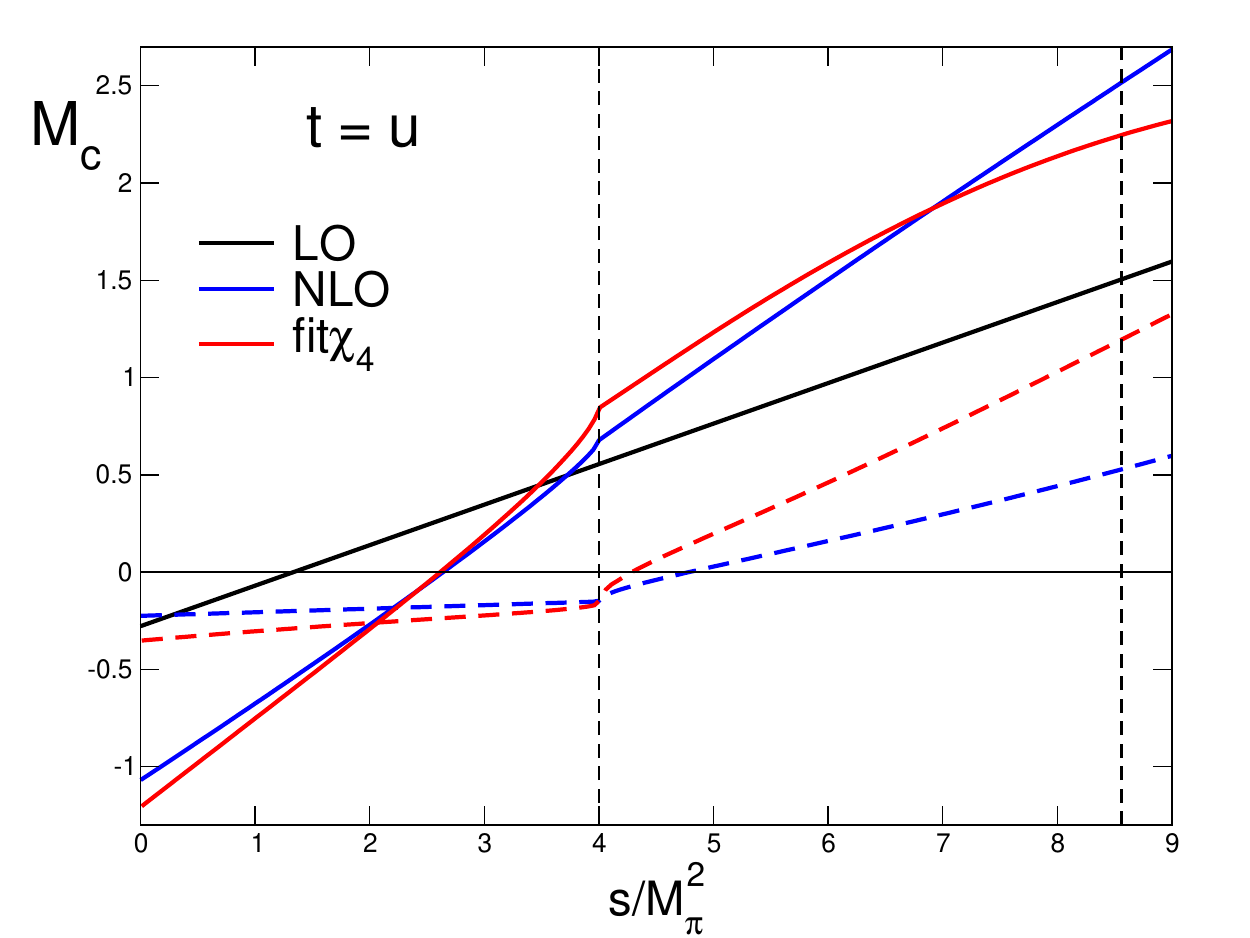}
\caption{Curvature generated by the final state interaction: comparison of the one loop representation with the dispersive solution
that matches it at low energies. Real parts (full lines) and imaginary parts (dashed lines) along the lines $s=u$ and $t=u$. The dashed vertical lines indicate the boundaries of the physical region.\label{fig:Mmatch}}
\end{figure*}
The imaginary parts of the chiral representation vanish for $s<4M_\pi^2$. Those of the dispersive representation are different from zero in that region, but are very small there because they exclusively arise from the crossed channels. Above threshold, however, the one-loop representation strongly underestimates the imaginary parts. It is not difficult to see why that is so: The dominating contribution to $\Im M_0$ is the one proportional to $\sin^2\!\delta_0$. At one-loop, the representation for the $\pi\pi$ phase shifts enters at LO, where the scattering length of the $I=0$ S-wave is given by Weinberg's current algebra result~\cite{Weinberg1966}: $a_0^\mathrm{LO}=0.16$ in pion mass units, below the prediction $a_0=0.220(5)$~\cite{Colangelo2001} by the factor 1.38. The one-loop representation underestimates the imaginary part of $M_0$ roughly by the square of this factor.

\subsection{Adler zero at one loop}\label{sec:Adler zero at one loop}

Fig.~\ref{fig:Mmatch} shows that the final state interaction generates curvature, but does not significantly affect the position of the Adler zero: At LO, it occurs at $s_A=\frac{4}{3}M_\pi^2$, while at one loop, the real part along the line $s=u$ vanishes at $s_A=1.40 M_\pi^2$. Note that the behaviour of the amplitude in the vicinity of the zero involves large values of $t$: for $s=u\simeq \frac{4}{3}M_\pi^2$, we get $t_A\simeq  15.7\, M_\pi^2$, i.e.~$\sqrt{t_A}\simeq 550\,\MeV$. As far as the isospin components $M_0(s)$ and $M_1(s)$ are concerned, only their behaviour at small arguments of order $s\simeq s_A$ matters, but $M_2(s)$ is needed for $s\simeq t_A$ as well as for $s\simeq s_A$. Adler's low-energy theorem thus concerns the behaviour of the amplitude not only at small values of $s$ and $u$, but also in the vicinity of $t=t_A$. In particular, the contributions from kaon loops to $M_2(t_A)$ are relevant. The fact that these do not move the 
position of the zero far away from the place where it occurs in current algebra shows that they do obey the constraints imposed by chiral symmetry.

For the matching solution, the Adler zero occurs in the same ball park: $s_A =1.36M_\pi^2$. By construction, the behaviour 
at small arguments is the same as for the one-loop representation, but Fig.~\ref{fig:Match} shows that the chiral and dispersive representations for $\Re M_2(s)$ differ significantly in the physical region. The graph for $\Re M_2$ in Fig.~\ref{fig:Match} is drawn on a sufficiently wide range to show that the two representations approach one another above the physical region and intersect at $s\simeq 16.8 M_\pi^2$ -- this ensures that the two solutions have the Adler zero at approximately the same place.
\begin{figure*}[thb]\centering
  \includegraphics[width=8.5cm]{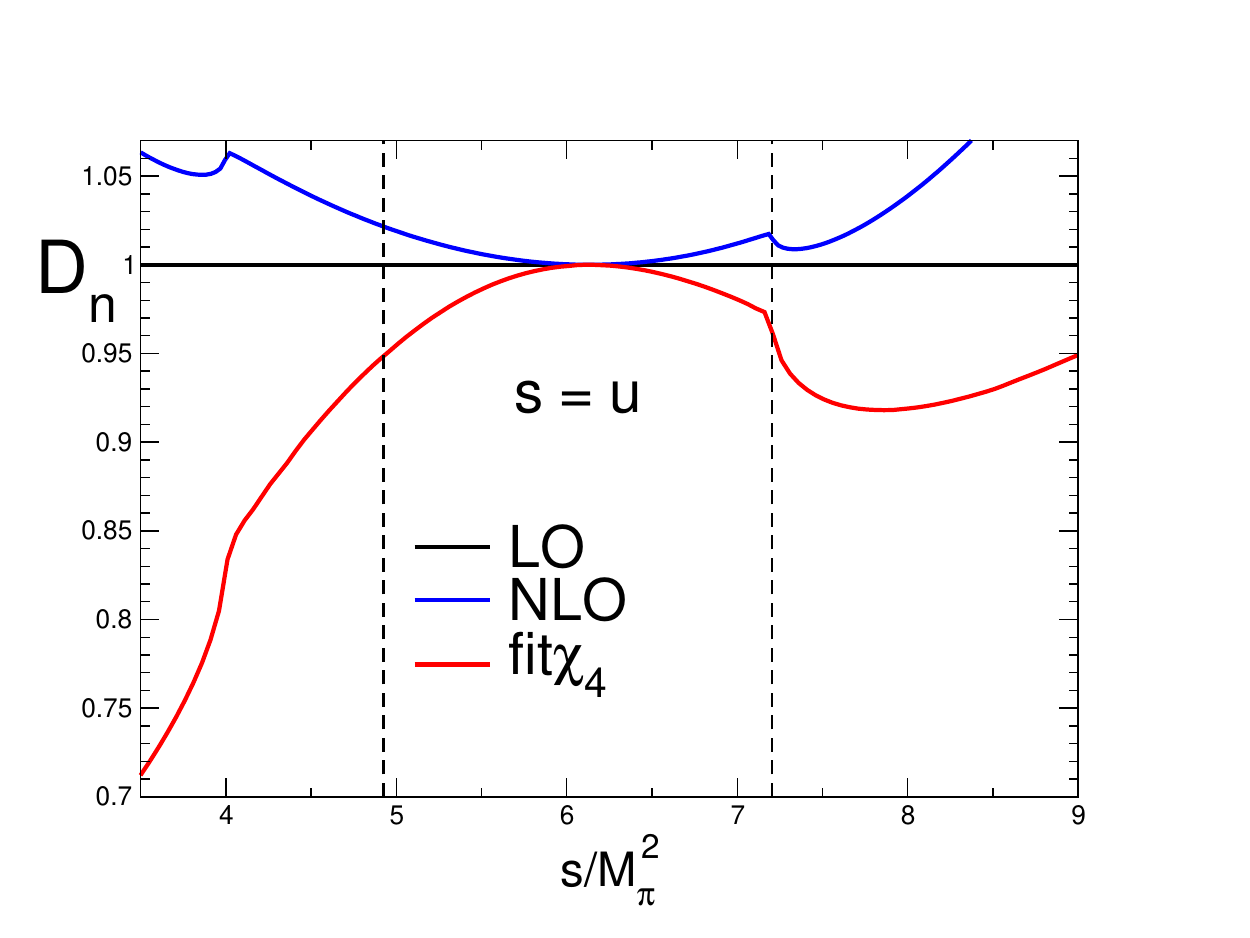}\hspace{0.3cm}\includegraphics[width=8.5cm]{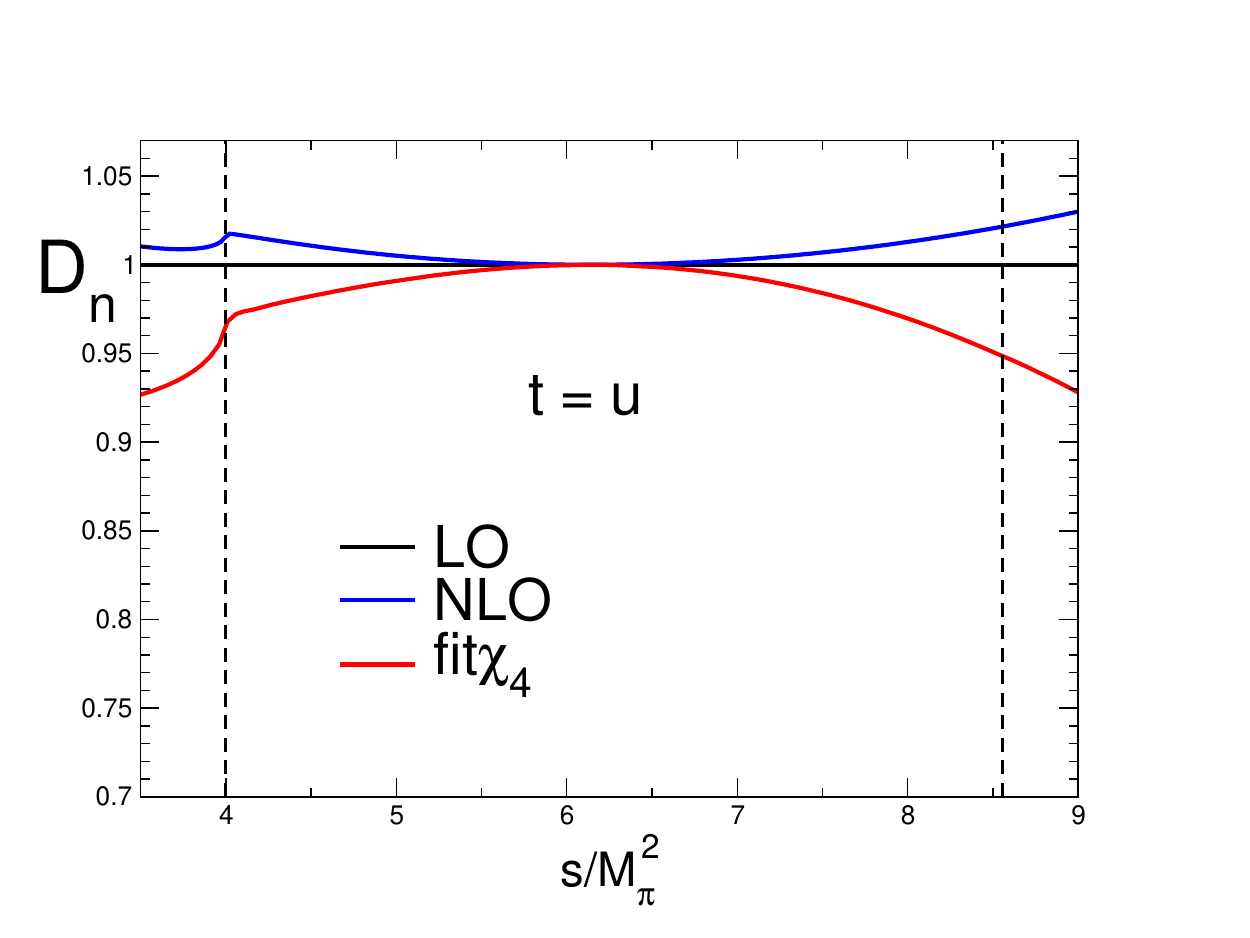}
\caption{Dalitz plot distribution (square of the amplitude normalized to 1 at the center) of the decay $\eta\to3\pi^0$, along the lines $s=u$ and $t=u$. The plots show that accounting properly for the final state interaction changes the sign of the curvature and hence the sign of the slope $\alpha$.\label{fig:Match Dalitzn}} 
\end{figure*}
%
\subsection{Neutral decay mode}\label{sec:Neutral decay mode}

The plot for the neutral isospin component $M_n(s)$ in Fig.~\ref{fig:Match} can again barely be distinguished from the one for $M_0(s)$, because the exotic component $M_2(s)$ is small (in particular, the final state interaction in the channel with $I=2$ is repulsive, so that the amplification seen in the channel with $I=0$ does not occur.) The picture gives the impression that, in the physical region, the one-loop and dispersive representations of the transition amplitude of the neutral mode are practically the same. This is not the case, however. Fig.~\ref{fig:Match Dalitzn} shows that the corresponding Dalitz plot distributions 
\begin{equation}\label{eq:Dn}D_n(s,t,u)=\rule[-1em]{0.04em}{2.5em}\frac{M_n(s,t,u)}{M_n(s_0,s_0,s_0)} 
\rule[-1em]{0.04em}{2.5em}^{\,2}\;, \end{equation}

are qualitatively different. At leading order, the Dalitz plot distribution of the neutral decay mode is flat, \mbox{$D_n^\mathrm{LO}(s,t,u)=1$}. At NLO, the distribution picks up a positive curvature: The parameter-free one-loop prediction for the slope of the $Z$-distribution~\cite{Kambor+1996} is positive and hence disagrees with experiment, even in sign (the definition and the properties of that distribution will be discussed in detail in Sec.~\ref{sec:Z distribution}). The more accurate account of the final state interaction provided by the matching solution (fit$\chi_4$) makes a qualitative difference here: The curvature of this solution is negative.  This points to a resolution of the puzzle mentioned in point 4. of the introduction. Indeed, as shown in~\cite{Colangelo:2016jmc} and discussed in detail in Sec.~\ref{sec:Slope},  the value of the slope predicted within our framework is in excellent agreement with experiment.  

Fig.~\ref{fig:Match} shows that at NLO, the neutral component $M_n(s)$ is quite close to the matching solution: In the physical region, the difference does not exceed 15\,\%. Fig.~\ref{fig:Match Dalitzn} shows, however, that in the corresponding Dalitz plot distributions, a difference of this size generates a qualitative change. To see why that is so, we expand the neutral component around the center of the Dalitz plot: 
\begin{equation}M_n(s)=M_n(s_0)\{1+a_n(s-s_0)+ b_n(s-s_0)^2+\cdots\}\end{equation}
In the total amplitude $M_n(s)+M_n(t)+M_n(u)$, the linear term drops out. For the Dalitz plot distribution, the expansion starts with the quadratic term: 
\begin{equation}D_n(s,t,u)=1+\mbox{$\frac{2}{3}$}\,\Re b_n(s^2+t^2+u^2-3s_0^2)+\cdots\
\label{eq:TaylorNeutral}\end{equation} 
The dimensionless quantity $\alpha=\frac{2}{9}M_\eta^2(M_\eta-3M_\pi)^2\,\Re b_n$ is referred to as the slope of the distribution.  In the one-loop approximation, the quadratic term is so small that it can barely be seen in Fig.~\ref{fig:Match}. In the matching solution, this term is more than twice as large and of opposite sign. 

As noted above, in connection with the imaginary parts, the chiral representation only offers a crude, semi-quantitative description of the final state interaction. The comparison of the LO and NLO representations for $M_n(s)$ shows that, at the center of the Dalitz plot, the effects generated by this interaction are large: The one-loop contributions modify the tree level amplitude by more than 50\,\%. We conclude that the truncated chiral series does not have the accuracy required to make a meaningful statement about the slope.

\section{Isospin breaking corrections}\label{sec:Isospin breaking corrections}

The decay $\eta\to 3\pi$ violates isospin conservation. As discussed in Sec.~\ref{sec:Isospin}, the dominating contribution to the transition amplitude can be represented in the form~\eqref{eq:normalization}, as a product of the factor $\DMKQCD$ which breaks isospin symmetry and the factor $M_c(s,t,u)$ which is invariant under isospin rotations.  The basic properties of the amplitude $M_c(s,t,u)$ were discussed in the preceding sections -- we now turn to the remainder, which is of order \mbox{$O[e^2,(m_u-m_d)^2]$}. While the effects due to $(m_u -m_d)^2$ are tiny, those from the electromagnetic interaction must properly be taken into account when comparing theory with experiment. In particular, the e.m.~self-energy of the charged pion generates a mass difference 
to the neutral pion which affects the phase space integrals quite significantly.  

In the literature, the corrections of  order \mbox{$O[e^2,(m_u-m_d)^2]$} have been calculated by several groups, to
different levels of accuracy -- {\em i.e.}~to different orders of the
expansion in the isospin breaking parameters.  In the present paper we will
rely on the work of Ditsche, Kubis and Mei{\ss}ner
(DKM)~\cite{Ditsche+2009}, who evaluated the transition amplitude
within the effective theory relevant for QCD+QED, to first non-leading order of the chiral expansion and to order $e^2$ in the electromagnetic interaction, with unequal up and down quark masses and in the presence of real as well as virtual photons. 
An earlier calculation by Baur, Kambor and Wyler
\cite{Baur+1996}, performed in the same framework, did not include effects
of order $e^2(m_u-m_d)$. These are of second order in isospin breaking and
were deemed to be negligible. Ditsche, Kubis and Mei{\ss}ner, however, correctly
observe that while terms of order $(m_u-m_d)^2$ are indeed negligible,
there are a number of effects which scale as $e^2(m_u-m_d)$ and should be
taken into account, like real and virtual photon corrections to the purely
strong amplitude, and also, and most importantly, effects related to the
pion mass difference, which are in particular responsible for the presence
of cusps in the Dalitz plot of $\eta \to 3 \pi^0$.

Isospin breaking also affects the phase shifts of $\pi\pi$ scattering. We take these from the solution of the Roy equations reported in~\cite{Colangelo2001}, which is done in the isospin limit. Our dispersive analysis is also carried out in that limit. In order to correct our results for isospin breaking effects, we make use of Chiral Perturbation Theory. We first study the effects of isospin breaking in this framework, comparing the representation of Ditsche, Kubis and Mei{\ss}ner~\cite{Ditsche+2009}, which does account for isospin breaking, with the one of Gasser and Leutwyler~\cite{Gasser+1985a}, which concerns the isospin limit. Our estimates for the size of the isospin breaking effects in the physical amplitudes rely on the assumption that these effects factorize, at least approximately. The branching ratio $B=\Gamma_{\eta\to 3\pi^0}/\Gamma_{\eta\to\pi^+\pi^-\pi^0}$ provides a strong test of the assumptions that underly our analysis.
\subsection{Kinematics}\label{sec:Kinematics}

The Mandelstam variables are not independent. We work with $s$ and $\tau\equiv t-u$.
The value of the sum $s+t+u$ depends on the masses of the particles occurring in the final state.
We reserve the symbols $s$, $t$, $u$ for the isospin symmetric world,
use the variables $s_c$, $t_c$, $u_c$ for the charged decay mode and $s_n$, $t_n$, $u_n$ for the neutral mode.
The constraints 
\begin{eqnarray}\label{eq:s+t+u}s+t+u\al=\al M_\eta^2+3M_\pi^2\;,\nonumber\\
s_c+t_c+u_c\al =\al M_\eta^2+2M_{\pi^+}^2+M_{\pi^0}^2\;,\\
 s_n+t_n+u_n\al=\al M_\eta^2+3M_{\pi^0}^2\;.\nonumber\end{eqnarray}
determine all of the Mandelstam variables in terms of ($s, \tau$), ($s_c,\tau_c$), ($s_n,\tau_n$).%
\begin{figure*}[thb]
\begin{center}
\includegraphics[width=7.6cm]{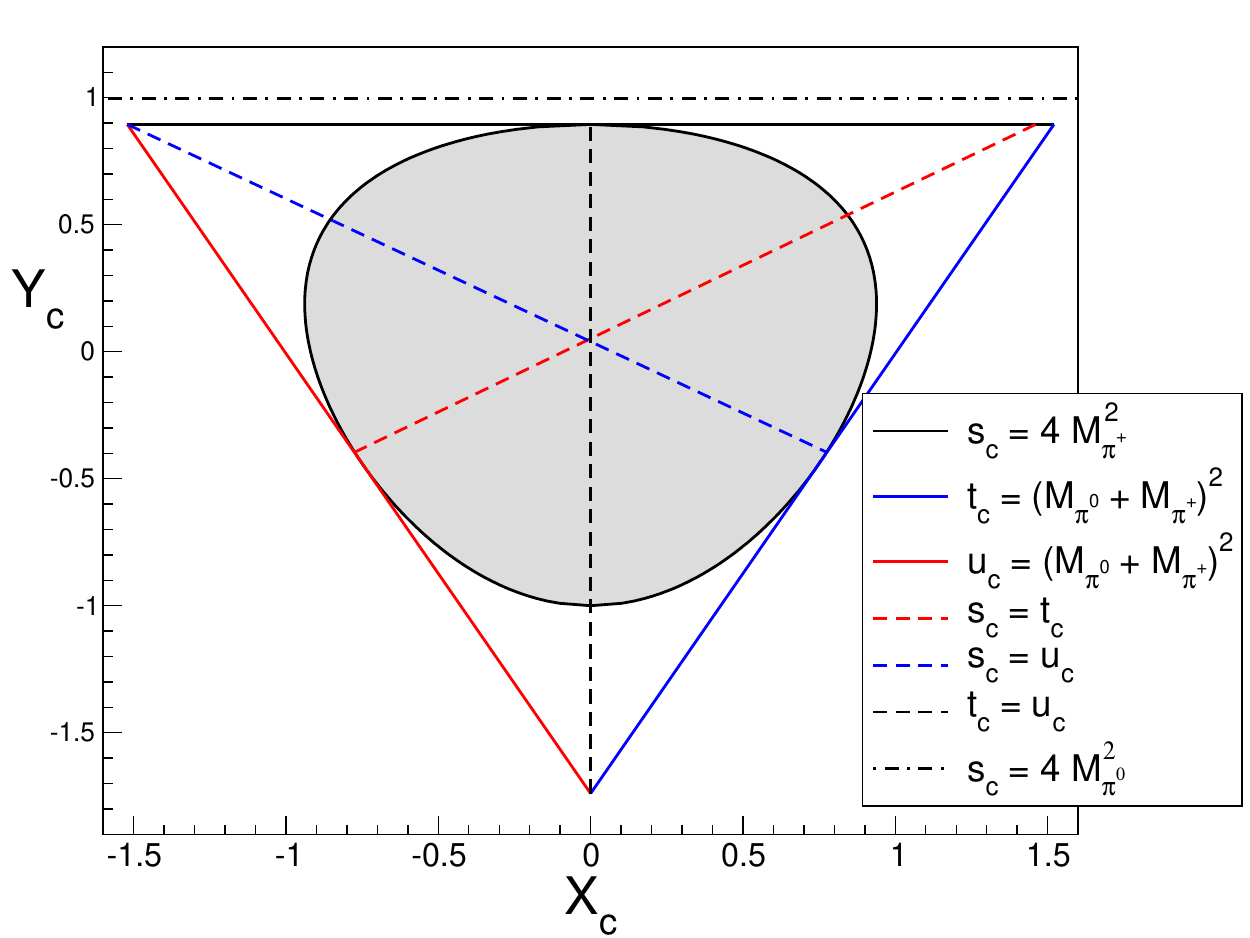}\hspace{1.5cm}\includegraphics[width=7.6cm]{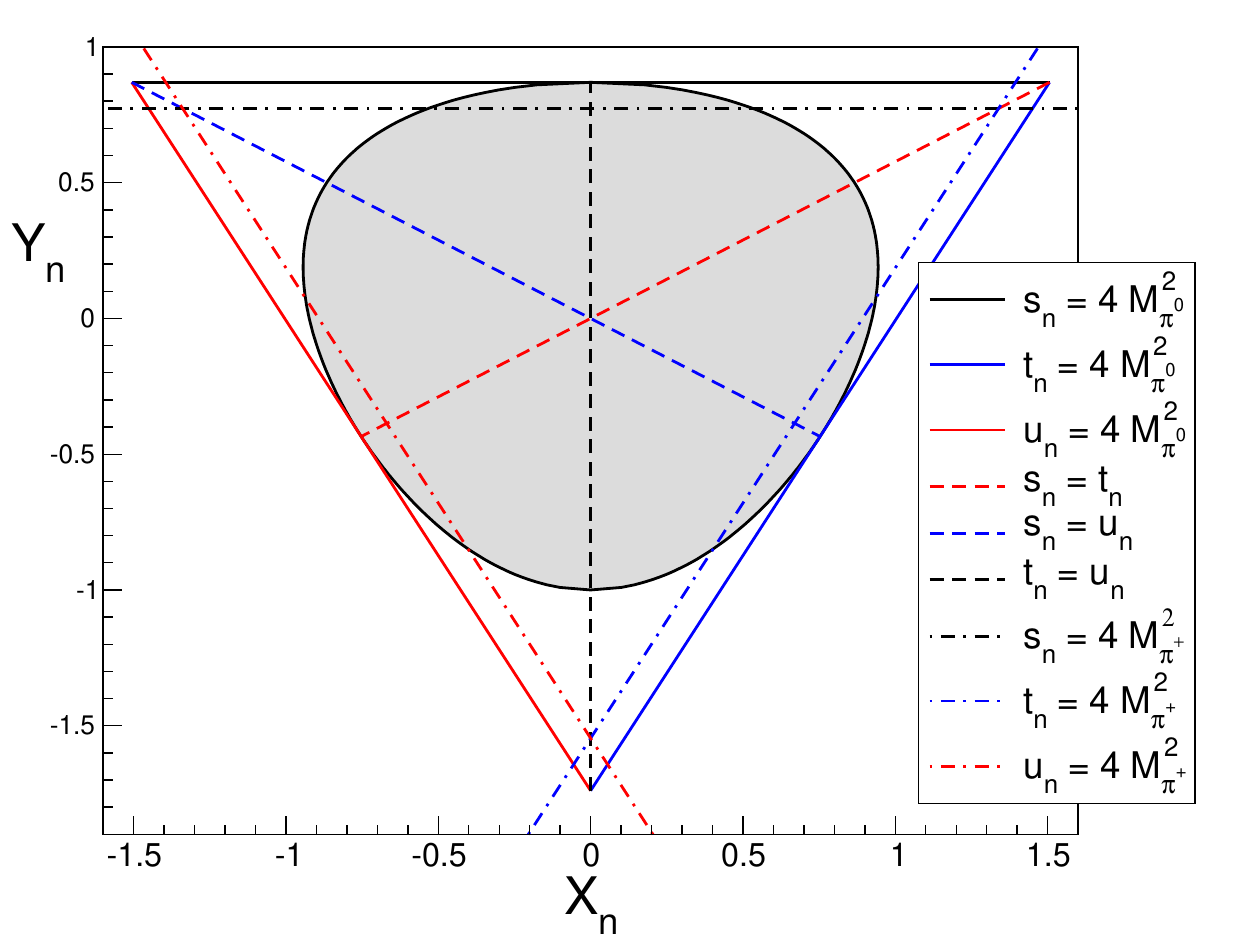}
\caption{The left panel shows the Dalitz plot geometry for the decay $\eta\to\pi^+\pi^-\pi^0$ in the plane of the two independent variables $X_c$, $Y_c$. The shaded area indicates the physical region, the full lines
that are tangent to this region represent singularities generated by the final state interaction. In addition to the branch cut at $s_c=4M_{\pi^+}^2$ (full), the $s$-channel contains a further such singularity outside the physical region, at $s_c=4M_{\pi^0}^2$ (dash-dotted). 
The right panel shows the kinematics of the decay $\eta\to 3\pi^0$. In this channel, Bose statistics implies that the amplitude is invariant under rotations by 120$^\circ$ as well as under reflections at the lines where $t_n=u_n$ or $s_n=u_n$ or $s_n=t_n$, which divide the physical region into six physically identical sextants -- the data points in one of these determine the entire distribution. The branch cut singularities where $s_n$, $t_n$ or $u_n$ are equal to $4M_{\pi^0}^2$ are tangent to the boundary while those at $4M_{\pi^+}^2$ are visible as cusps in the physical region. \label{fig:boundary}}
\end{center}
\end{figure*}

Note that, up to normalization, $\tau$ coincides with the standard Dalitz plot variable $X$, while $s$ is linear in $Y$. In the case of the charged decay mode, the relations read
\begin{eqnarray}\label{eq:XcYc}
s_c\al=\al -\frac{2}{3}M_\eta\,(M_\eta-2M_{\pi^+}-M_{\pi^0})\, Y_c \nonumber\\
\al \al + \frac{1}{3}\{M_\eta^2+3M_{\pi^0}^2+4M_\eta(M_{\pi^+}-M_{\pi^0})\}\;,\\
\tau_c\al=\al -\frac{2}{\sqrt{3}}M_\eta\,(M_\eta-2M_{\pi^+}-M_{\pi^0})\,X_c\;.\nonumber\end{eqnarray}
In these variables, the 
physical region is characterized by $4M_{\pi^+}^2\leq s_c\leq (M_\eta-M_{\pi^0})^2$ and $-\tau^\mathrm{max}_c(s_c)\leq \tau_c\leq \tau^\mathrm{max}_c(s_c)$. The maximal value of $\tau_c$ depends on $s_c$:
{\small 
\begin{equation}\label{eq:taucmax}
\tau^\mathrm{max}_c(s_c)=\sqrt{\frac{1-4M_{\pi^+}^2}{s_c}}\sqrt{(M_\eta+M_{\pi^0})^2-s_c}\sqrt{(M_\eta-M_{\pi^0})^2-s_c}~,
\end{equation}}
Since the masses of $\pi^0$ and $\pi^+$ differ, the final state interaction among the pions 
generates several different branch points. The left panel of Fig.~\ref{fig:boundary} shows the location 
of these singularities for the charged decay mode, in the plane spanned by $X_c$ and $Y_c$. 
They represent straight lines that touch the boundary of the physical region.
The $s$-channel contains two branch points, one at  $4M_{\pi^0}^2$, the other at $4M_{\pi^+}^2$.  
The straight line $s_c=4M_{\pi^+}^2$ also touches the boundary, while
the line $s_c=4M_{\pi^0}^2$ runs outside the physical region.  The singularities in the $t$- and $u$-channels occur
at \mbox{$t_c=(M_{\pi^0}+M_{\pi^+})^2$} and $u_c=(M_{\pi^0}+M_{\pi^+})^2$, respectively. 

The Adler zero discussed in Sec.~\ref{sec:Adler zero at one loop} occurs along the line $s_c=u_c$, which is indicated as a dashed line, but the relevant value of $s_c$ is around $\frac{4}{3}M_\pi^2$, which is outside the range shown in this figure. The symmetry with respect to $t\leftrightarrow u$ implies that an Adler zero also occurs along the line $s_c=t_c$, at the same value of $s_c$. 

The amplitude relevant for the decay into $3\pi^0$ is invariant under the exchange of the three Mandelstam variables also in the presence of
isospin breaking. Each of the three channels contains a pair of branch points at $4M_{\pi^0}^2$
and $4M_{\pi^+}^2$. The right panel of Fig.~\ref{fig:boundary} shows that the three straight lines with $s_n$, $t_n$ or $u_n$ equal to $4M_{\pi^0}^2$ touch the boundary of the physical region, while the other three branch cuts run across this region and manifest
themselves as cusps in the Dalitz plot distribution. The relations between $s_n$, $\tau_n$ and the variables $X_n,Y_n$ used in the figure 
are obtained from \eqref{eq:XcYc} by replacing $M_{\pi^+}$ with $M_{\pi^0}$, while those among the variables
$s$, $\tau$ and  $X$, $Y$ of the isospin symmetric world are reached with the substitutions $M_{\pi^+}\to M_\pi$,  $M_{\pi^0}\to M_\pi$. 
 \begin{figure*}[thb]\centering\includegraphics[width=8.2cm]{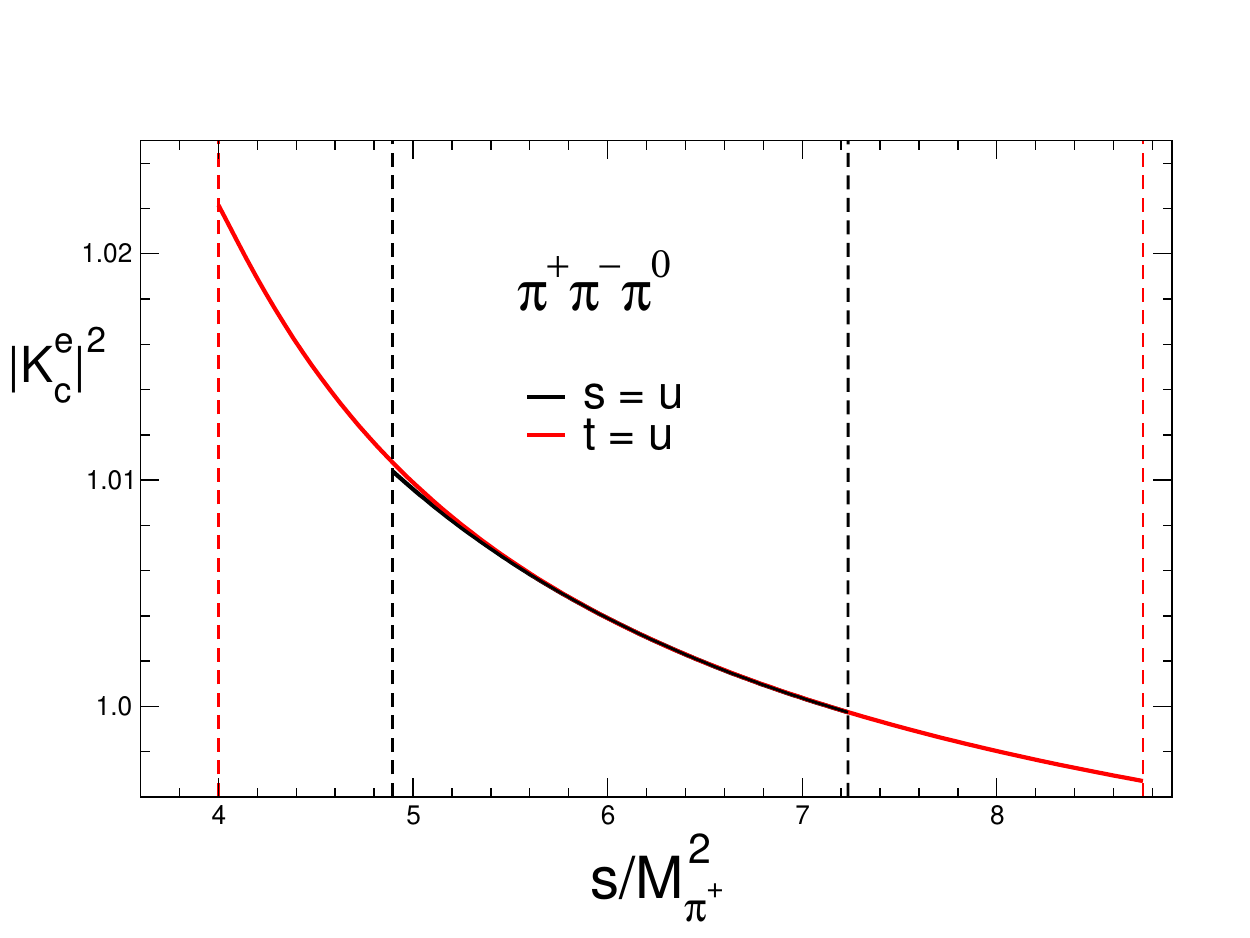}\hspace{0.3cm}\includegraphics[width=8.2cm]{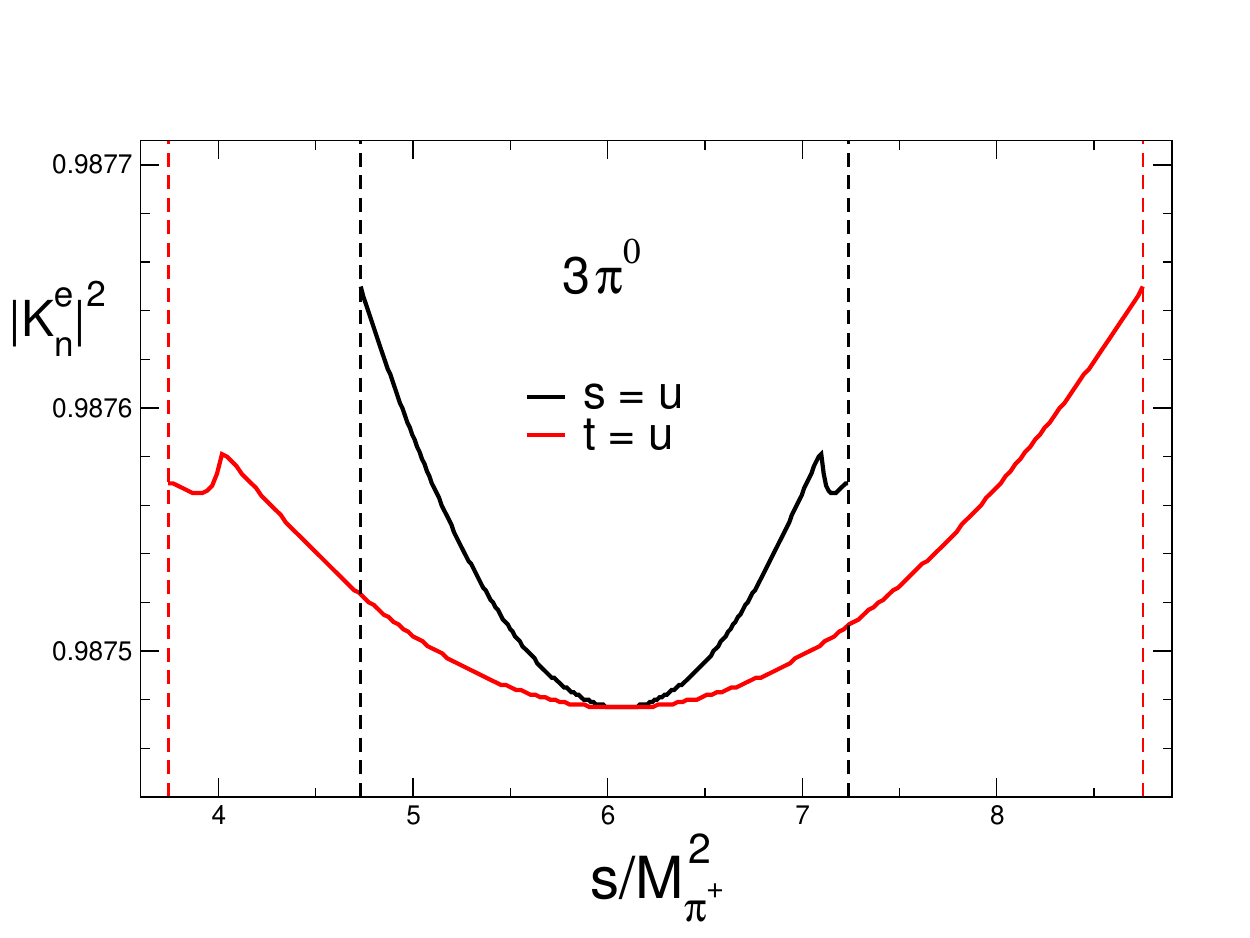}
 \caption{One-loop representation: electromagnetic effects that are not accounted for in the self-energies of the particles. The plots show the square of the ratio between the full amplitude and what remains if the meson masses are kept fixed at the physical values, while $e$ is set equal to zero. Note that the range of values seen in the right panel is 100 times smaller than the one on the left. \label{fig:Me}}
 \end{figure*} 
 %
\subsection{Isospin breaking at one loop}\label{sec:Isospin breaking one loop}

We denote the representations given in~\cite{Ditsche+2009} for the amplitudes of the decays $\eta\to \pi^+\pi^-\pi^0$ and $\eta\to3\pi^0$ 
 by $A^\mathrm{DKM}_c(s_c,t_c,u_c)$ and $A^{\mathrm{DKM}}_n(s_n,t_n,u_n)$, respectively. In addition to the constants $F_\pi$, $F_K$, $L_3$
that occur in the one-loop representation already in the isospin limit, the expressions involve the two isospin breaking parameters $\delta=m_d-m_u $ and $e$, the meson masses $M_{\pi^+}$, $M_{\pi^0}$, $M_{K^+}$, $M_{K^0}$, $M_\eta$, and a set of low-energy constants, $K_1, \ldots, K_{11}$, which stem from the effective Lagrangian for the electromagnetic interaction.  The infrared singularities occurring in loops that involve virtual photons are regularized by giving these a nonzero mass $m_\gamma$. We work in the normalization (the constant $N$ is specified in Eq.~\eqref{eq:normalization}): 
%
\begin{eqnarray}M^\mathrm{DKM}_c(s_c,t_c,u_c) \al \equiv\al -A^\mathrm{DKM}_c(s_c,t_c,u_c)/N\;,\\
M^\mathrm{DKM}_n(s_n,t_n,u_n) \al \equiv\al -A^\mathrm{DKM}_n(s_n,t_n,u_n)/N\;.\nonumber\end{eqnarray}
 We have checked that, in the limit $e\to 0$, $m_u\to m_d$, these quantities indeed
reduce to the isospin symmetric amplitudes $M^\mathrm{GL}_c(s,t,u)$, $M^\mathrm{GL}_n(s,t,u)$ of Gasser and Leutwyler~\cite{Gasser+1985a}. 

Photon exchange generates poles in $M^\mathrm{DKM}_c(s_c,t_c,u_c)$ at 
$s_c=0$.
Moreover, the exchange of a photon between the charged pions in the final state
gives rise to the so-called Coulomb pole, which in the one-loop representation is described
by a triangle graph. It only shows up in the amplitude for the charged decay mode in the 
form of a contribution to the $s$-channel discontinuity,
\begin{equation}\label{eq:Coulomb}M^\mathrm{Coulomb}_c(s_c,t_c,u_c)=\frac{e^2(1+\sigma^2)}{16\,\sigma}T(s_c)\;,~~ \sigma =\sqrt{1-\frac{4M_{\pi^+}^2}{s_c}}\;,\end{equation}
where $T(s_c)$ stands for the current algebra approximation to the transition amplitude specified in~\eqref{eq:MLO}.
This contribution diverges at the boundary of the Dalitz plot, where $s_c\to 4M_{\pi^+}^2$. 

Remarkably, despite these additional singularities, the one-loop representation obeys elastic unitarity also in the presence of photons: The amplitude $M^\mathrm{DKM}_c(s_c,t_c,u_c)$ can be expressed in terms of three functions of a single variable
according to~\eqref{eq:RT} and $M^\mathrm{DKM}_n(s_n,t_n,u_n)$
retains the form~\eqref{eq:RTn}. Only the explicit expressions for the components are
modified and the relation~\eqref{eq:Mn} between the components relevant for
the charged and neutral decay modes is lost. As it is the case without isospin
breaking, for the charged decay mode one function of a single variable is needed for the $s$-channel (S-wave)
and two functions (S-wave and P-wave) for the $t$-and $u$-channels. 
For the neutral decay mode, a single function $M_n^\mathrm{DKM}(s)$ again suffices (S-wave), but it now
differs from the combination $M_0^\mathrm{DKM}(s)+\frac{4}{3}M_2^\mathrm{DKM}(s)$ of amplitudes relevant for the charged mode.

The decay is necessarily accompanied by the emission of real photons and the comparison with the data must properly account for that. The main features of the phenomenon are universal and are thoroughly discussed in the literature~\cite{Isidori:2007zt}. Up to and including $O(e^2)$, the rate of the decay $\eta\to\pi^+\pi^-\pi^0$ contains two contributions, one from the square of the amplitude relevant for the decay without real photons in the final state, the other from the square of the amplitude for the emission of one real photon. It is well-known that both of these contributions are infrared divergent and that, in the sum of the two, the infinities cancel.  The only physical remnant of the infrared divergences is that the probability for generating a real photon depends logarithmically on the upper limit  set for the energy of the emitted photon. In the comparison with the data, the maximal photon energy in the rest frame of the $\eta$, which is denoted by $E_\mathrm{max}$, is determined by the 
experimental resolution.

The DKM-representation is regularized by giving the virtual photons a mass $m_\gamma$. The explicit expression for the amplitude $M_c^\mathrm{DKM}(s_c,t_c,u_c)$, which represents the transition without real photons, diverges logarithmically if $m_\gamma$ is sent to zero. To leading order in the chiral expansion, the divergent part is given by
\begin{eqnarray}\label{eq:IR divergence 1}M^\mathrm{DKM}_c(s_c,t_c,u_c) \al= \al -\frac{e^2}{8\pi^2}\ln \frac{m_\gamma^2}{M_\pi^2} 
 \left\{1-\frac{1+\sigma^2}{2\sigma} \right. \\
\al \al \left. \times \left (\ln\frac{1+\sigma}{1-\sigma}-i\hspace{0.05em} \pi\right )\right\}
T(s_c)+\mbox{finite}\;, \nonumber \end{eqnarray}
while the divergence of the soft-photon contribution is of the form
\begin{eqnarray}\label{eq:IR divergence 2}|M_{\pi^+\pi^-\pi^0\gamma}|^2 \al = \al \frac{e^2}{4\pi^2}\ln \frac{m_\gamma^2}{4E_\mathrm{max}^2} 
\left\{1-\frac{1+\sigma^2}{2\sigma}\ln\frac{1+\sigma}{1-\sigma} \right\}T(s_c)^2 \nonumber \\
\al \al +
\mbox{finite}\;,\end{eqnarray}
To leading order of the chiral expansion, where the finite part in~\eqref{eq:IR divergence 1} is given by $T(s_c)$, the divergences thus cancel as they should: In effect, adding the contribution from the production of real photons converts the divergent term $\ln (m_\gamma^2/M_\pi^2)$ into the finite expression $\ln (4E_\mathrm{max}^2/M_\pi^2)$. At leading order of the chiral expansion, the production of real photons with $E<E_\mathrm{max}$ can therefore be accounted for in a very simple manner: Stick to the amplitude relevant for the decay without emission of real photons, equip the virtual photons with a mass $m_\gamma$  and set  $m_\gamma=2E_\mathrm{max}$. This also provides us with an estimate of the sensitivity to $E_\mathrm{max}$: Replacing $m_\gamma$ by $2E_\mathrm{max}$ in the one-loop representation of~\cite{Ditsche+2009} and varying $E_\mathrm{max}$ in the range $M_\pi< 2E_\mathrm{max}<
M_\eta$, the quantity $|M_c^\mathrm{DKM}(s_c,t_c,u_c)|^2$ only changes by half a permille. We conclude that, at the present accuracy, the sensitivity to the experimental resolution is an academic problem and set $2E_\mathrm{max}=M_\pi$. Apart from that, we follow the prescriptions used by Ditsche, Kubis and Mei{\ss}ner~\cite{Ditsche+2009} to compare the calculated amplitudes with 
the experimental results (see the discussion in Sect.~3.2.6 therein). In particular, we assume that the Coulomb pole specified in~\eqref{eq:Coulomb} is accounted for in the data analysis and replace the amplitude of~\cite{Ditsche+2009} by $M^\mathrm{DKM}_c(s_c,t_c,u_c)-M^\mathrm{Coulomb}_c(s_c,t_c,u_c)$. 
Neither photon emission nor the Coulomb pole enter the amplitude
$M^\mathrm{DKM}_n(s_n,t_n,u_n)$, which we take over from Ref.~\cite{Ditsche+2009}
as it is. 

\subsection{Self-energy effects}\label{sec:Self-energy}

In the decay $\eta\to\pi^+\pi^-\pi^0$, the self-energy of the charged pion directly affects the
kinematics, as it is relevant for the size of the physical region
and for the value of $s_c+t_c+u_c$.  
The self-energy of the charged pion increases its mass and hence 
reduces the phase space available in the charged decay mode -- 
since phase space is small, this makes a significant difference, which must
be accounted for.  In early work on $\eta$-decay, this was done
only very crudely: In the calculation of the decay rate,
the square of the isospin symmetric amplitude was simply integrated
over the physical phase space rather than the isospin symmetric one.

The one-loop representation allows us to separate the self-energy effects from the remaining contributions generated by the electromagnetic interaction: The amplitude can be evaluated at the physical masses of the mesons even if $e$ is set equal to zero.  The left panel of Fig.~\ref{fig:Me} depicts the square of the ratio $K_c^e=M_c^\mathrm{DKM}(s,t,u)/M_c^\mathrm{DKM}(s,t,u)_{e=0}$, along the lines $s=u$ and $t=u$. It shows that the remaining electromagnetic contributions vary in the narrow range \mbox{$0.997< |K_c^e|^2< 1.022$}.  As seen in the right panel, the square of the correction factor $K_n^e=M_n^\mathrm{DKM}(s,t,u)/M_n^\mathrm{DKM}(s,t,u)_{e=0}$ relevant for the neutral channel is also of the order of 1\,\%, but nearly constant over the entire physical region:  $0.98757 < |K_n^e|^2<  0.98765$. This implies that in the Dalitz plot distribution of the decay $\eta\to3\pi^0$, the corrections generated by the electromagnetic interaction are totally dominated by the 
self-
energy effects.

\subsection{Kinematic map for $\eta\to\pi^+\pi^-\pi^0$}\label{sec:Kinematic map charged channel}

Any comparison of an isospin symmetric transition amplitude with experiment
requires that the values of $s$ and $\tau$ that correspond to a given point $s_c$ and $\tau_c$ 
of physical phase space are specified -- a map from the physical world into
the space spanned by the variables $s$ and $\tau$ is needed:
\begin{equation}\label{eq:map}s=s[s_c,\tau_c]\;,\quad\tau=\tau[s_c,\tau_c]\;.\end{equation} 
The map is all but unique, but not any choice
is acceptable. The simplest possible one, for instance, the trivial map $s=s_c$, $\tau=\tau_c$, 
fails because it generates fictitious singularities: The branch point $t=4M_\pi^2$ is mapped into a line of constant $t_c$, but
the value\footnote{Value obtained for the convention we are using, where $M_\pi=M_{\pi^+}$.} of the constant, $\frac{1}{2}M_{\pi^0}^2+\frac{7}{2}M_{\pi^+}^2$, is larger
than ($M_{\pi^0}+M_{\pi^+})^2$. Hence the image of the singularity crosses the 
physical region: The trivial map produces a fictitious cusp in the Dalitz plot distribution.

In current algebra approximation, the amplitude only depends on $s$ and the
one-loop representation shows that the variable $\tau$ does not play an important
role at NLO, either. The representation of Ditsche, Kubis and Mei{\ss}ner~\cite{Ditsche+2009}
indicates that this remains true even in the presence of isospin breaking:
The leading terms\footnote{Since the symmetry with respect to $\tau\leftrightarrow-\tau$ also holds
in the presence of isospin breaking, the first term in the Taylor series of
$\tau[s_c,\tau_c]$ with respect to $\tau_c$ vanishes.} of the Taylor series of the map~\eqref{eq:map} in powers of $\tau_c$,
\begin{equation}\label{eq:fg}s=f_c[s_c]\;,\quad\tau=g_c[s_c]\,\tau_c\;,\end{equation}
suffice to obtain a good understanding of the 
deformation of phase space generated by the electromagnetic interaction. The coefficients $f_c[s_c]$, $g_c[s_c]$ can
be chosen such that the map does not generate any fictitious singularities in the physical region:
It suffices to impose the condition that the boundary of physical phase space
is taken into the boundary of isospin symmetric phase space. We refer to such maps
as {\em boundary preserving}. Since the branch points of the isospin symmetric
amplitude relevant for the charged mode do not pass through the physical region, 
their image will automatically also have this property. The requirement amounts to the condition
 \begin{figure*}[thb]\centering\includegraphics[width=7.8cm]{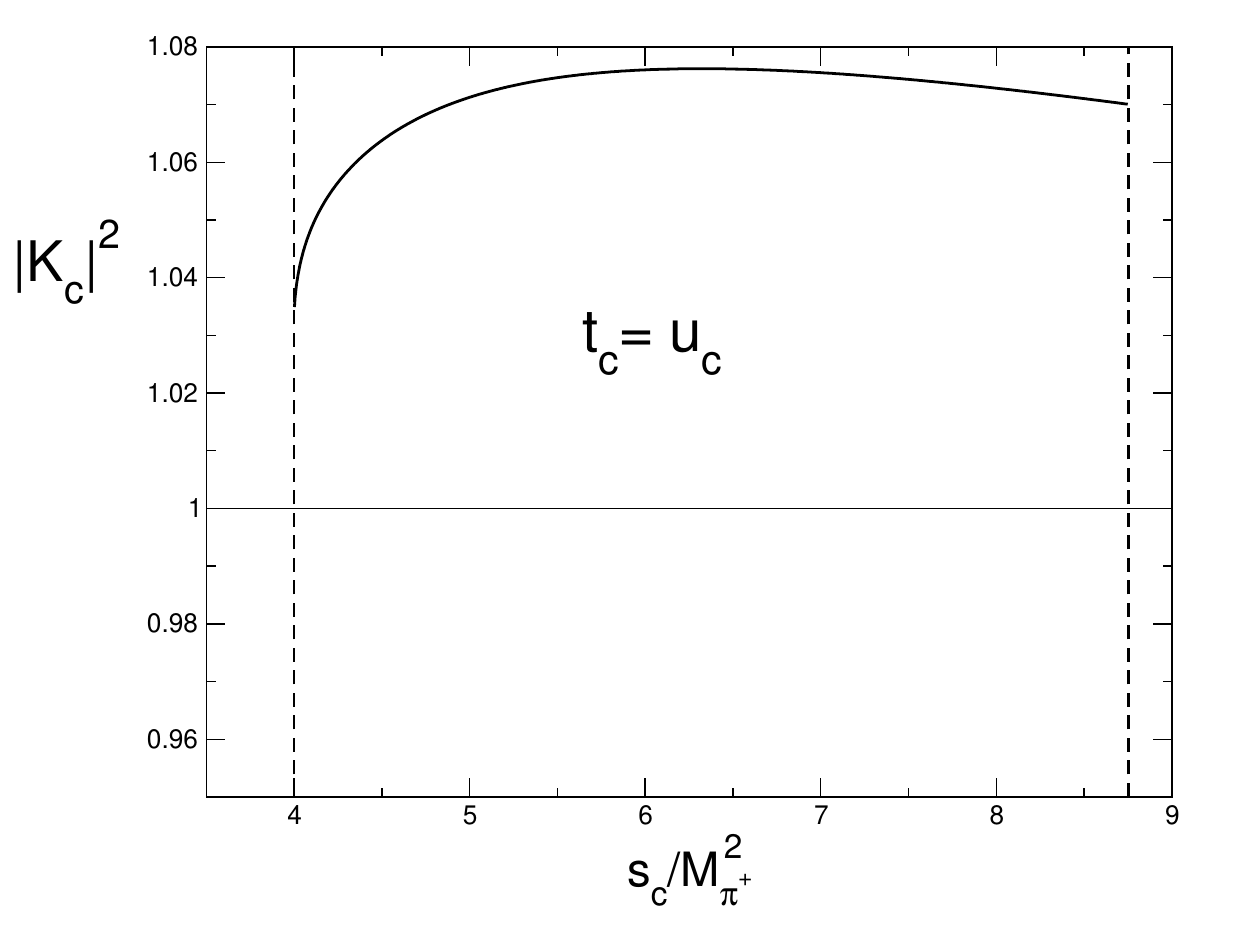}\includegraphics[width=7.8cm]{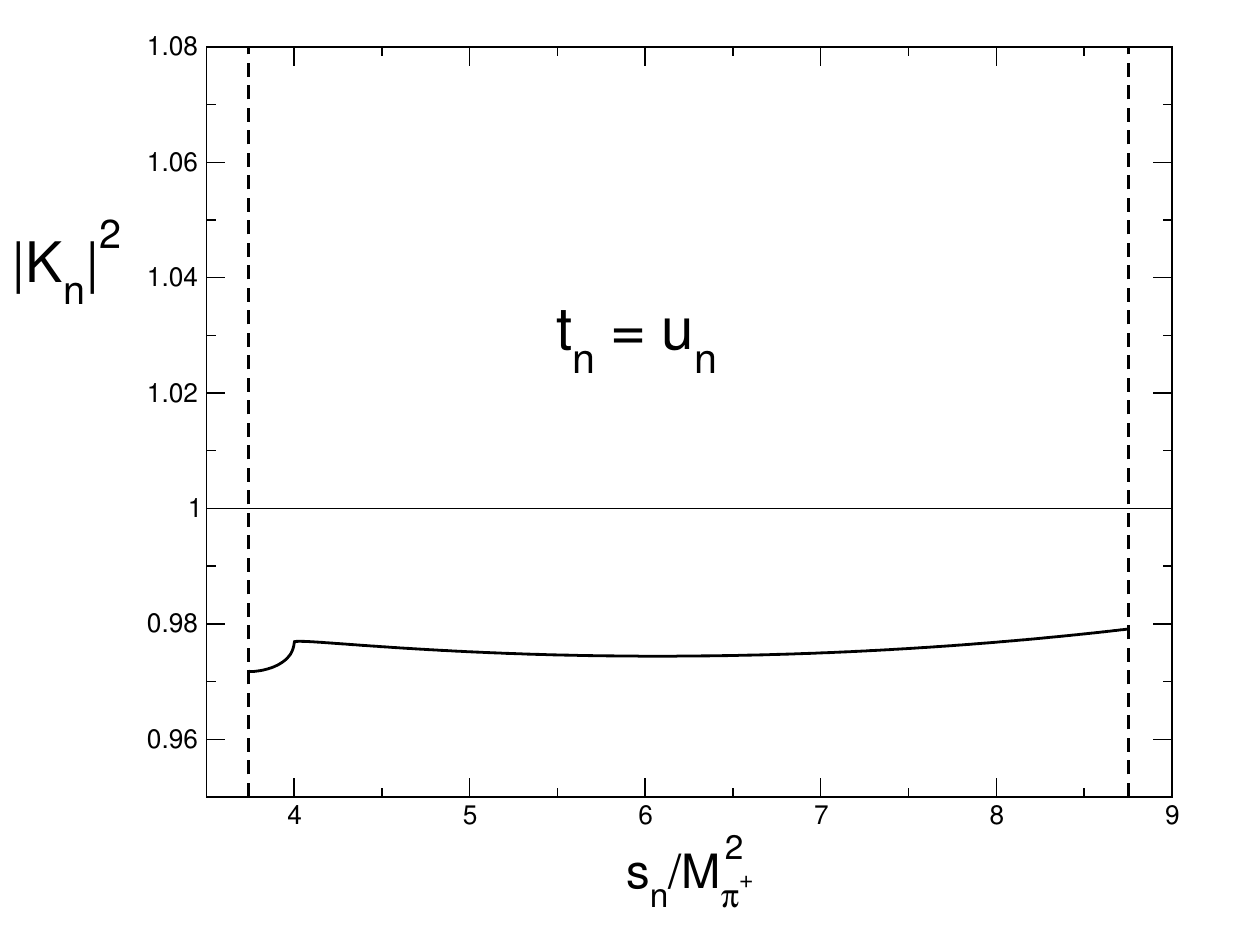}
 \caption{One-loop representation: residual corrections in physical region.\label{fig:K}}
 \end{figure*}
\begin{equation}\tau^\mathrm{max}(f_c[s_c])=g_c[s_c]\,\tau_c^\mathrm{max}(s_c)\;,\end{equation}
which fixes one of the coefficients of the map in terms of the other:
\begin{equation}\label{eq:g}g_c[s_c]=\frac{\tau^\mathrm{max}(f_c[s_c])}{\tau_c^\mathrm{max}(s_c)}\;.\end{equation}
The function $\tau_c^\mathrm{max}(s_c)$ is specified in~\eqref{eq:taucmax}, while $\tau^\mathrm{max}(s)$ is obtained
from  this one with $M_{\pi^0}\to M_\pi$, \mbox{$M_{\pi^+}\to M_\pi$}, $s_c\to s$. The function $f_c[s_c]$ remains free,
except for the boundary conditions $f_c[4M_{\pi^+}^2] =4M_\pi^2$ and \mbox{$f_c[(M_\eta-M_{\pi^0})^2] =(M_\eta-M_\pi)^2$}.
We choose a parabola that goes through these two points and, in addition, maps the center of the physical Dalitz plot into the center of the isospin symmetric one. We adopt the definition used in phenomenological analyses of the data, where the center is specified in terms of the standard Dalitz plot variables of Eq.~\eqref{eq:XcYc}, as the point with the coordinates $X_c=Y_c=0$. It sits at $s_c=\frac{1}{3}M_\eta^2+M_{\pi^0}^2+\frac{4}{3}M_\eta(M_{\pi^+}-M_{\pi^0})$, 
 slightly to the right of the place where $s_c=t_c=u_c$, i.e.~where the dashed lines in Fig.~\ref{fig:boundary} intersect.  
 The explicit expression for $f_c[s_c]$ involves $M_{\pi^+},M_{\pi^0}$ as well as $M_\pi,M_\eta$ and is rather clumsy. In the convention we are using, where the isospin limit is taken  such that $M_{\pi^+}$ stays put ($M_\pi= M_{\pi^+}$), it simplifies to
 \begin{eqnarray}\label{eq:pcqc}f_c[s_c]\al=\al s_c+p_c(s_c-4M_{\pi^+}^2) \nonumber\\
 \al \al +q_c(s_c-4M_{\pi^+}^2)(s_c-(M_\eta-M_{\pi^0})^2)\;,\nonumber\\
  p_c\al=\al -\frac{(M_{\pi^+}-M_{\pi^0})(2M_\eta-M_{\pi^+}-M_{\pi^0})}{(M_\eta-M_{\pi^0})^2-4M_{\pi^+}^2}\;,
\end{eqnarray}
{\small{   \begin{equation}
  q_c = \frac{3(M_{\pi^+}-M_{\pi^0}) (M_\eta-3M_{\pi^+})}{(M_\eta+6M_{\pi^+}-3M_{\pi^0})(M_\eta-2M_{\pi^+}-M_{\pi^0})^2(M_\eta+2M_{\pi^+}-M_{\pi^0})}\;.\nonumber\end{equation}}}
The deformation of the trivial map $s=s_c$ needed to preserve the boundary is measured by the coefficients $p_c$, $q_c$, which are proportional to $M_{\pi^+}-M_{\pi^0}$. This difference is dominated almost totally by the self-energy of the charged pion. Numerically, the deformation is small throughout the physical region: The difference between $s_c$ and $s$ reaches the maximum at the upper end of the range of interest and amounts to 2.2\,\% there, but this suffices to ensure that the lines $s=4M_\pi^2$, $t=4M_\pi^2$ and $u=4M_\pi^2$,  where the amplitude is singular, do not enter the physical region. Note that the map is fully specified by the meson masses -- in this sense, the deformation of phase space discussed in the present section represents a purely kinematic effect. As will be shown in the next section, the full modification brought about by isospin breaking at one loop includes a second, qualitatively different contribution that is approximately constant over phase space. Hence it affects the Dalitz 
plot distribution only little, but has an important effect on the rate of the decay. \\
\\
The extension to the decay $\eta\to 3\pi^0$ meets with a technical problem: The map obtained by applying the above construction to the corresponding transition amplitude does take the physical region of the neutral Dalitz plot onto the isospin symmetric one, but does not respect Bose statistics, because it does not treat $s$ on equal footing with $t$ and $u$. As shown in Appendix~\ref{sec:Kinematic map neutral channel}, this shortcoming is easily cured -- the kinematic map specified in~\eqref{eq:fgneutral}--\eqref{eq:Mntilde} does preserve the symmetry under exchange of $s$, $t$and $u$ as well as the boundary and the center of the physical region. In the following, we use this map to analyze isospin breaking effects in the neutral channel.
\vspace{-0.5cm}
\subsection{Applying the kinematic map to the one-loop representation}
\label{sec:Kinematic map at one loop}

We now apply the map constructed in the preceding section 
to the one-loop representation. At that level, the isospin symmetric amplitude
is given by $M^\mathrm{GL}_c(s,t,u)$. The boundary preserving map defined
in~\eqref{eq:fg}, \eqref{eq:g}, \eqref{eq:pcqc} expresses the variables
$s$ and $\tau=t-u$ in terms of those relevant for the physical phase space
of the charged decay mode. With the constraint~\eqref{eq:s+t+u} for
$s+t+u$, the variables $t$ and $u$ can also be expressed in terms of $s$ and $t-u$.
We denote the resulting expressions for $s,t,u$ by $\tilde{s}_c,\tilde{t}_c,\tilde{u}_c$:
\begin{eqnarray}\label{eq:sctcuctilde}\tilde{s}_c\al=\al f_c[s_c]\;,\nonumber\\
\tilde{t}_c\al=\al \mbox{$\frac{1}{2}$}\{3s_0-f_c[s_c]+(t_c-u_c) g_c[s_c]\}\;,\\
\tilde{u}_c\al=\al \mbox{$\frac{1}{2}$}\{3s_0-f_c[s_c]-(t_c-u_c) g_c[s_c]\}\;,\nonumber\end{eqnarray}
with $s_0=\frac{1}{3}M_\eta^2+M_\pi^2$.  The amplitude 
\begin{equation}\label{eq:Mctilde}\Mtilde^\mathrm{GL}_c(s_c,t_c,u_c)\equiv M^\mathrm{GL}_c(\tilde{s}_c,\tilde{t}_c,\tilde{u}_c)\end{equation}
then lives on physical phase space and has the three branch points that occur at the boundary of the physical region, $s_c=4M_{\pi^+}^2$, $t_c=(M_{\pi^0}+M_{\pi^+})^2$, $u_c=(M_{\pi^0}+M_{\pi^+})^2$, at the proper place. The only qualitative
 difference with the full one-loop amplitude $M^\mathrm{DKM}_c(s_c,t_c,u_c)$ is that the branch
 cut due to $\pi^+\pi^-\to\pi^0\pi^0\to \pi^+\pi^-$, which occurs outside the physical region at
 $s_c=4M_{\pi^0}^2$, is missing. We use the ratio
\begin{equation}\label{eq:Kc}K_c(s_c,t_c,u_c)\equiv
\frac{M^\mathrm{DKM}_c(s_c,t_c,u_c)}{\Mtilde^\mathrm{GL}_c(s_c,t_c,u_c)}\;\end{equation}
 to account for the difference between the full amplitude and the one obtained from the isospin symmetric representation with a purely kinematic map. The left panel of Fig.~\ref{fig:K} shows that, in the
physical region and along the line $t_c=u_c$, this ratio is roughly constant at one loop. The same is true along the
line $s_c=u_c$. Indeed, in the entire physical region, the factor $|K_c(s_c,t_c,u_c)]^2$  only varies in the range 
 $1.031 < |K_c|^2 < 1.078$.

The right panel of Fig.~\ref{fig:K} shows the square of the analogous factor relevant in the neutral channel,
 \begin{equation}\label{eq:Kn}K_n(s_n,t_n,u_n)\equiv
\frac{M^\mathrm{DKM}_n(s_n,t_n,u_n)}{\Mtilde^\mathrm{GL}_n(s_n,t_n,u_n)}\;.\end{equation}
It describes  those effects in the one-loop representation of the decay $\eta\to3\pi^0$ that are not already accounted for by the kinematic map (the explicit expression for $\Mtilde^\mathrm{GL}_n$ is given in Appendix~\ref{sec:Kinematic map neutral channel}). Visibly, in the neutral decay mode, the residual corrections are even smaller than in the charged mode: Their square only varies in the range  $0.972<|K_n|^2< 0.978$. The Dalitz plot distribution of the decay $\eta\to 3\pi^0$ is affected by less than half a percent. For $t_n=u_n$, the physical region is characterized by $4M_{\pi^0}^2\leq s_n\leq(M_\eta-M_{\pi^0})^2$. The small cusp generated by the virtual transition $\pi^0\pi^0\to\pi^+\pi^-\to\pi^0\pi^0$ occurs within that range, at $s_n=4M_{\pi^+}^2$. In the right panel of Fig.~\ref{fig:K}, it shows up near the vertical line that marks the lower end of the physical region. 

\subsection{Correcting the dispersive solutions for isospin breaking effects}\label{sec:Correcting}

In order to clearly distinguish the isospin symmetric dispersive representations $M_c(s,t,u)$, $M_n(s_n,t_n,u_n)$ from those that include isospin breaking effects, we denote the physical amplitudes by $M_c^\mathrm{phys}(s,t,u)$,  $M_n^\mathrm{phys}(s,t,u)$ and work in the normalization
\begin{eqnarray}\label{eq:Ac in terms of Mphysc}A_c(s,t,u)\al=\al -N M_c^\mathrm{phys}(s,t,u)\;,\\
A_n(s,t,u)\al=\al -N M_n^\mathrm{phys}(s,t,u)\;,\nonumber\end{eqnarray}
The approximation we are using to account for isospin breaking applies two steps:

(i) We first apply the kinematic map, replacing
the solutions $M_c$, $M_n$ of our integral equations by the amplitudes $\Mtilde_c$, $\Mtilde_n$. In the charged channel, the explicit expression reads $\Mtilde_c(s_c,t_c,u_s)\equiv M_c(\tilde{s}_c,\tilde{t}_c,\tilde{u}_c)$, where $\tilde{s}_c$, $\tilde{t}_c$, $\tilde{u}_c$ are specified in~\eqref{eq:sctcuctilde}.  Since this operation takes the constraint \mbox{$s_c+t_c+u_c=M_\eta^2+2M_{\pi^+}^2+M_{\pi^0}^2$} into \mbox{$\tilde{s}_c+\tilde{t}_c+\tilde{u}_c=M_\eta^2+3M_\pi^2$}, it ensures that the solutions $M_c(s,t,u)$ are used only for values of the Mandelstam variables that obey $s+t+u=M_\eta^2+3M_\pi^2$ -- this is where they are uniquely defined. Moreover, the map takes center and boundary of the physical Dalitz plot into center and boundary of the isospin symmetric phase space. Analogous statements hold for the neutral channel -- the kinematic map relevant in that case is specified in Appendix~\ref{sec:Kinematic map neutral channel}.

(ii) We assume that the remaining isospin breaking effects can be estimated with the one-loop representation and approximate the physical amplitude with
\begin{eqnarray}\label{eq:Mphys}M^\mathrm{phys}_c(s,t,u)\al=\al K_c(s,t,u)\Mtilde_c(s,t,u)\;,\\
M^\mathrm{phys}_n(s,t,u)\al=\al K_n(s,t,u)\Mtilde_n(s,t,u)\;.\nonumber\end{eqnarray}
Note that we are treating the residual corrections multiplicatively. We expect this prescription to provide a decent estimate
even in the physical region: While Fig.~\ref{fig:Mmatch} shows that the one-loop representation as such has a pronounced momentum dependence and reproduces the curvature of the dispersive solution only semi-quantitatively, the ratios $K_c$, $K_n$ vary comparatively slowly and stay close to unity throughout the physical region. 

The main difference between the two decay modes is that, for $\eta\to\pi^+\pi^-\pi^0$, the residual corrections increase the square of the amplitude at the center by 7.6\,\% and hence increase the decay rate, while for $\eta\to 3\pi^0$, the opposite is the case: At the center, the square of the amplitude is reduced by 2.6\,\%.  As will be discussed in Sec.~\ref{sec:Branching ratio}, the comparison of the results obtained for the branching ratio $B=\Gamma_{\eta\to3\pi^0}/\Gamma_{\eta\to\pi^+\pi^-\pi^0}$ with the experimental results offers a strong test of the approximations used to account for isospin breaking.
\begin{table*}[t]
\begin{center}
\begin{tabular}{l | l l l l l}
\textbf{Experiment} & \multicolumn{1}{c}{\hspace{-4em}$-a$} &
\multicolumn{1}{c}{\hspace{-3em}$b\cdot 10$} & \multicolumn{1}{c}{\hspace{-3em}$d \cdot 10^2$} & 
\multicolumn{1}{c}{\hspace{-3em}$f \cdot 10$} & \multicolumn{1}{c}{\hspace{-1em}$g \cdot 10^2$} \\
\hline
Gormley(1970)\hfill\cite{Gormley+1970} & $1.17(2)$ & $2.1(3)$ & $6(4)$ & $-$ & $-$\\
Layter(1973)\hfill\cite{Layter+1973} & $1.080(14)$ & $0.34(27)$ & $4.6(3.1)$ & $-$ & $-$\\
CBarrel(1998)\hfill\cite{Abele:1998yj} & $1.22(7)$ & $2.2(1.1)$ & $6$(fixed) & $-$&$-$\\
KLOE(2008)\hfill\cite{Ambrosino:2008ht} & $1.090(5)(^{+8}_{-19})$ &
$1.24(6)(10)$ & $5.7(6)(^{+7}_{-16})$ & 1.4(1)(2)&$-$ \\ 
WASA(2014)\hfill\cite{Adlarson:2014aks} & 1.144(18) & 2.19(19)(37)  &
8.6(1.8)(1.8)& 1.15(37)&$-$\\ 
BESIII(2015)\hfill\cite{Ablikim:2015cmz} &  1.128(15)(8) & 1.53(17)(4)
&8.5(1.6)(9) & 1.73(28)(21)&$-$\\ 
KLOE$^a$(2016)  \hfill\cite{KLOE:2016qvh}  &$1.104(3)$ &$ 1.420(29)$ & $7.26(27)$ & $1.54(6)$  & $ 0$ \\ 
KLOE$^b$(2016)  \hfill\cite{KLOE:2016qvh}  & $1.095(3)$ &$ 1.454(30)$ & $8.11(33)$ & $1.41(7)$  & \hspace{-0.8em}$-4.4(9)$  \\ 
\hline
\end{tabular}
\caption{Experimental values of the Dalitz plot parameters of $\eta \to
  \pi^+ \pi^- \pi^0$. The two entries for KLOE(2016) correspond to their fits with 4 and 5 free coefficients, respectively  (fit\#3 and fit\#4).\label{tab:Dalitz-exp}}
  \end{center}
\end{table*} 

While in the neutral channel, the residual corrections affect the Dalitz plot distribution only very little, the momentum dependence of the amplitude relevant for the charged decay mode is not properly accounted for by the kinematic map. The contribution from the triangle graph is singular at $s=4M_{\pi^+}^2$, but we have removed that singularity by subtracting the Coulomb pole specified in~\eqref{eq:Coulomb}.  As shown in Appendix~\ref{sec:Dalitz plot distribution}, the spike occurring there does not arise from the triangle graph, but from the interference between the contributions generated by the branch cuts in the $s$-channel (final state interaction among the pairs $\pi^+\pi^-$ and $\pi^0\pi^0$) with those in the $t$- and $u$-channels due to $\pi^\pm\pi^0$ pairs. We assume that the one-loop approximation does provide a decent estimate for the distortion of the discontinuities generated by the electromagnetic interaction and expect that multiplying the amplitudes of the charged and neutral decay modes 
with the ratios $K_c=M_c^\mathrm{DKM}/M_c^\mathrm{GL}$ and $K_n=M_n^\mathrm{DKM}/M_n^\mathrm{GL}$ yields a good approximation of the physical distribution. This implies, in particular, that we are accounting for the cusps that run through the physical region of the decay $\eta\to3\pi^0$ only in one-loop approximation. We will compare the resulting parameter free prediction for the Dalitz plot distribution of the decay $\eta\to3\pi^0$ with experiment in Sec.~\ref{sec:etato3pi0} -- this comparison offers another good check on the internal consistency of our framework.

\section{Dalitz plot distribution for $\eta\rightarrow\pi^+\pi^-\pi^0$}
\label{sec:Dalitz plot}
\subsection{Experiment}
\label{sec:Experiment}

The most precise measurement of the Dalitz plot of $\eta \to \pi^+ \pi^-
\pi^0$ and the one on which our analysis has been based is the recent one
by KLOE~\cite{KLOE:2016qvh}, but the experimental measurements of this
decay in the charged and neutral channel have a long history, which we are
going to briefly review here. 
The first measurements of the Dalitz plot of $\eta \to \pi^+ \pi^- \pi^0$
have been performed already in the 
seventies~\cite{Gormley+1970,Layter+1973} and led to a rough 
determination of the leading coefficients  
occurring in the standard parametrization of the distribution,\footnote{The original notation allowed for additional terms ($c,e$) with odd powers of $X_c$. Since crossing symmetry implies that the amplitude is even under $X_c\rightarrow -X_c$, we are omitting these.}\newpage
\begin{equation}D_c(X_c,Y_c)=1+a\,Y_c+ b \,Y_c^2+ d\,X_c^2 + f\, Y_c^3+gX_c^2Y_c+\ldots\;,\label{eq:Dcpoly}\end{equation}
as quoted in Table~\ref{tab:Dalitz-exp}. The
same measurement was performed by Crystal Barrel at LEAR in
1998~\cite{Abele:1998yj}, with less precise (because of the low statistics)
but compatible results. 

Only more recently has the interest in such a measurement been revived
again and thanks to the existence of experimental facilities, like 
DA$\Phi$NE, MAMI or COSY, and detectors like KLOE and WASA, a new series of
more precise measurements has been performed. 
KLOE made a first measurement
in 2008~\cite{Ambrosino:2008ht}, with a much more precise determination of
the three parameters $a$, $b$ and $c$ and for the first time of the
parameter $f$. This measurement has been repeated by the WASA-at-COSY
collaboration~\cite{Adlarson:2014aks} and more recently by the BESIII
collaboration~\cite{Ablikim:2015cmz}. The latest measurement is again due
to KLOE~\cite{KLOE:2016qvh}, and is based on the largest statistic sample
of about 5 million decays (for comparison, WASA has 30 and BESIII 60 times
less events). The values of the individual Dalitz plot parameters, all
shown in Table~\ref{tab:Dalitz-exp}, seem to differ somewhat among these
recent measurements but it is difficult to draw conclusions about a
possible discrepancy by just looking at central values and errors, because
there are strong correlations among the parameters. 
A more effective way to judge the compatibility of the different
measurements is to fit them with the same parametrization and
calculate the $\chi^2$ for each of the data sets. Unfortunately this is
only possible for the latest KLOE data~\cite{KLOE:2016qvh} and for those of
WASA~\cite{Adlarson:2014aks}, because only these have published 
unfolded data in the form of a bidimensional bin distribution.
For these two data sets, we find:
\begin{itemize}
\item
In view of the much larger statistics, KLOE data dominate any common fit; the inclusion of 
the WASA data barely shifts the parameters and any outcome of the fit.
\item
The compatibility among the two data sets is marginal: A common fit (with
six subtraction constants, i.e. five fit parameters) gives
$\chi^2_\mathrm{K}=371$ for 371 data points and
$\chi^2_\mathrm{W}=84$ for 59 data points. 
\item
Fitting WASA data by themselves gives a much better $\chi^2$:
$\chi^2_\mathrm{W}=49$, but this would be totally incompatible with
KLOE, as the corresponding $\chi^2$ is huge.
\end{itemize}
\begin{figure*}[thb]\centering \includegraphics[width=7.8cm]{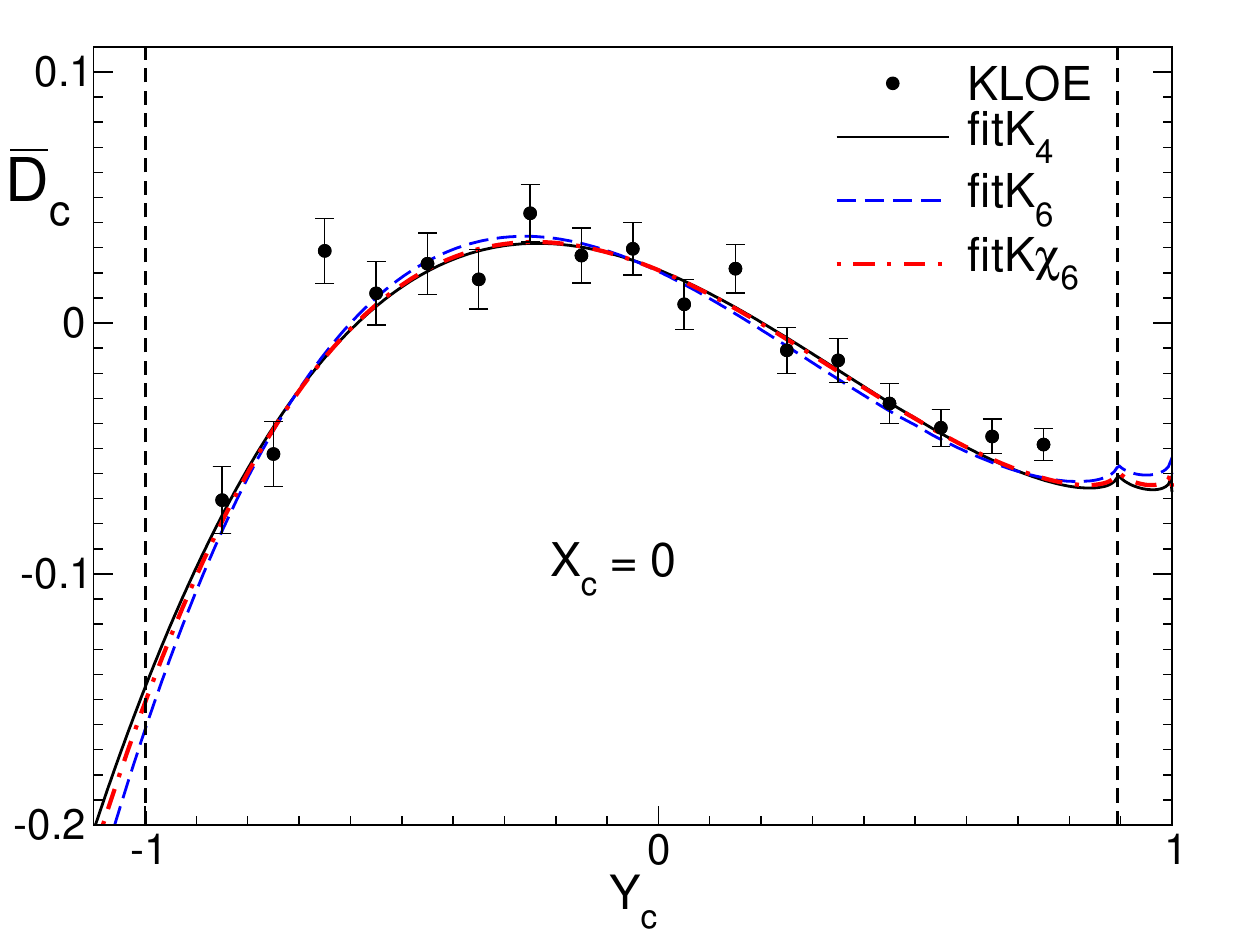}\includegraphics[width=7.8cm]{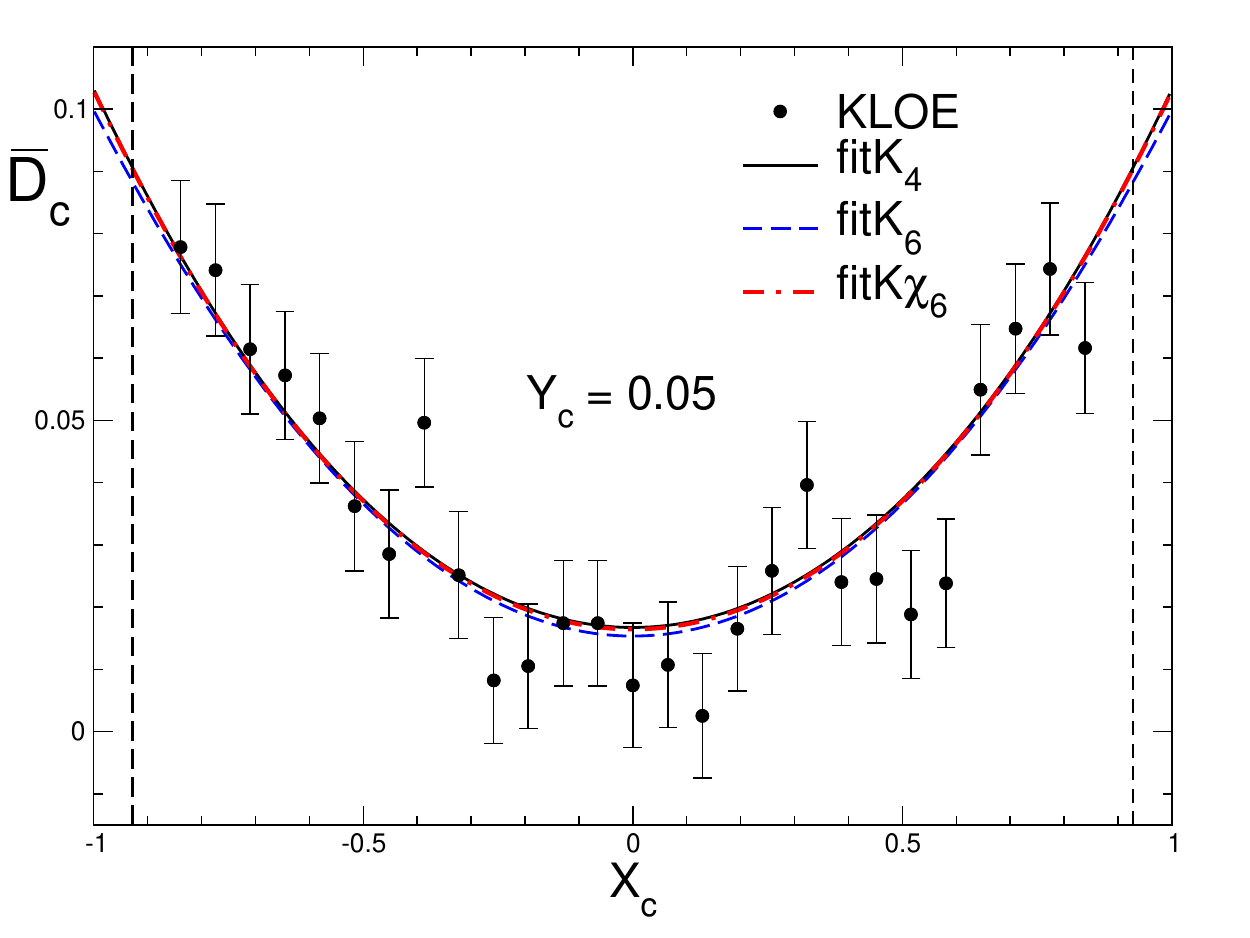}
\caption{Fits to the KLOE data on the Dalitz plot distribution of $\eta\to\pi^+\pi^-\pi^0$. To make the different entries visible, the distribution obtained from current algebra is subtracted.\label{fig:Dcbar}}
\end{figure*}
%
\subsection{Fitting the KLOE distribution for $\mathbf\eta\to\pi^+\pi^-\pi^0$}\label{sec:Fits to KLOE}

In our analysis, the recent KLOE data~\cite{KLOE:2016qvh,CaldeiraBalkestahl:2016trc} play the central role. In this experiment, the Dalitz plot distribution of the decay 
$\eta \to \pi^+ \pi^- \pi^0$ is determined to high accuracy, splitting phase space into altogether 371 bins.
The binning is based on the Dalitz plot variables $X_c,Y_c$ specified in Eq.~\eqref{eq:XcYc}. We denote the values of $X_c,Y_c$ at the center of bin \#$i$ by $X_c^i,Y_c^i$ and use the symbols $D_c^\mathrm{i}$, $\Delta D_c^\mathrm{i}$ for the experimental central values and errors in that bin. These values are to be compared with the Dalitz plot distribution that belongs to the amplitude $M_c^\mathrm{phys}(X_c,Y_c)$ obtained from the one defined in~\eqref{eq:Mphys} by expressing the variables $s_c,t_c,u_c$ in terms of $X_c,Y_c$ according to~\eqref{eq:XcYc}:
\begin{equation}\label{eq:Dcphys}D^\mathrm{phys}_c(X_c,Y_c)=\rule[-1em]{0.04em}{2.5em}\frac{M^\mathrm{phys}_c(X_c,Y_c)}{M^\mathrm{phys}_c(0,0)}\rule[-1em]{0.04em}{2.5em}^{\,2}\;.\end{equation}
When comparing with the data, we let the normalization of the observed distribution
float and define the discrepancy function by
\begin{equation}\label{eq:discrepancy KLOE}\chi^2_\mathrm{K}=\sum_i\left( \frac{D_c^\mathrm{phys}(X_c^i,Y_c^i)-  \Lambda_\mathrm{K}\,D_c^\mathrm{i}}{\Lambda_\mathrm{K}\,\Delta D_c^\mathrm{i}}\right)^2\;,\end{equation}
where the sum extends over the 371 bins of the KLOE data.

Since the normalization of the amplitude drops out in the Dalitz plot distribution, the value of $H_0$ is irrelevant -- the discrepancy function is independent thereof.  We fix it at the central value obtained at one loop, $H_0=1.176$.  The relation \eqref{eq:K0} between $H_0\equiv K_0$ and the subtraction constants  thus ties $\alpha_0$ to $\beta_0$ according to $\alpha_0=0.8594-0.08736\,\beta_0$, so that  $\chi_\mathrm{K}^2$ contains six independent real parameters: $\beta_0$, $\gamma_0$, $\delta_0$, $ \beta_1$, $\gamma_1$, $\Lambda_\mathrm{K}$. 

\subsection{Dispersive fits to the KLOE data without theoretical constraints}\label{sec:without theoretical constraints}

In Sec.~\ref{sec:Matching}, we determined the dispersive solution that matches the one-loop representation at low energies, allowing for only four subtraction constants. We now consider the opposite: Ignore the information obtained from \chpt\ and exclusively make use of the data  on the Dalitz plot distribution. Again, we only allow for four subtraction constants, setting $\delta_0=\gamma_1=0$. The minimum occurs at 
\begin{eqnarray}\mathrm{fitK_4:}\label{eq:alphaK4}\quad \al\al\beta_0=17.6\;,\quad \gamma_0=-35.2\;,\quad \delta_0=0\;,\quad\\
\al\al \beta_1=5.9\;,\quad \gamma_1=0\;,\quad  \ \Lambda_\mathrm{K}=0.938\;,\quad \chi^2_\mathrm{K}=390\;.\nonumber\end{eqnarray}
We refer to this fit to KLOE with 4 subtraction constants as $\mathrm{fitK_4}$.  It is of remarkably good quality: $\chi^2_\mathrm{K}=390$ for 371 data points and 4 free parameters. 

Fig.~\ref{fig:Dcbar} compares various fits with the KLOE data. Since the value of $\Lambda_\mathrm{K}$ depends on the fit, we leave the data as they are and divide the dispersive representations by this factor -- instead of showing the normalized observed distribution. Moreover, for better visibility, the leading term of the chiral expansion, $D_c^\mathrm{LO}=(3s-4M_\pi^2)^2/(M_\eta^2-M_\pi^2)^2$, is subtracted. The data points in the left panel of Fig.~\ref{fig:Dcbar} represent the remainder, $ D_c^i-D_c^\mathrm{LO}$, for the  bins centered at $X_c=0$. The full line shows the value of  $\,\overline{\hspace{-0.2em}D}_c=D_c^\mathrm{phys}\!/\Lambda_\mathrm{K} -D_c^\mathrm{LO}$, where $D_c^\mathrm{phys}$ is the isospin corrected Dalitz plot distribution belonging to  $\mathrm{fitK_4}$.  The right panel shows the analogous picture for the bins centered at $Y_c=0.05$ (the significance of the other two fits shown in this figure is discussed in the next section).

The left panel of Fig.~\ref{fig:Dcbar} corresponds to the one on the left of Fig.~\ref{fig:K}: $X_c=0$ implies $t_c=u_c$. While Fig.~\ref{fig:K} concerns the correction factor $|K_c|^2$ used to account for some of the isospin breaking effects, 
 we are now considering the Dalitz plot distribution of the full amplitude. The comparison shows that the spike occurring in $|K_c|^2$ near $s_c=4M_{\pi^+}^2$ also manifests itself in the Dalitz plot distribution near $Y_c=0.895$, but in rather modest form. For the reasons given in Sec.~\ref{sec:Self-energy}, the spikes in $|K_c|^2$ and in $D_c$ are of opposite sign. A dedicated experimental study is required to resolve the structure in the vicinity of $s_c=4 M_{\pi^+}^2$.  

The most important aspect of the solution obtained by fitting the measured
Dalitz plot distribution concerns the comparison with the matching solution
discussed earlier. The two solutions exclusively differ in the values of
the subtraction constants: While those relevant for the matching solution
are given in Eq.~\eqref{eq:matching subtraction constants}, the fit to the
KLOE data is characterized by Eq.~\eqref{eq:alphaK4}.  In order to compare
$\mathrm{fitK_4}$ with the estimates obtained from \chpt, 
we work out the real parts of the Taylor
invariants belonging to this fit. The result reads: 
\vspace{-0.11cm}
\begin{equation}
\Re h_1^\mathrm{K_4}=4.6\;, \qquad \Re h_2^\mathrm{K_4}= 12.8\;,
\qquad \Re h_3^\mathrm{K_4}=6.0 \; .
\end{equation} 
Remarkably,  these numbers are within the range estimated in~\eqref{eq:HNLO}: Although chiral symmetry was not made use of in the derivation of $\mathrm{fitK_4}$, the resulting transition amplitude is consistent with the estimates based on the low-energy theorems that follow from it. This neatly confirms that the uncertainty estimates we are attaching to the Taylor invariants are on the conservative side. Moreover, the solution fitK$_4$ does contain an Adler zero along the line $s=u$, at $s_A^\mathrm{K_4}=1.50\, M_\pi^2$, not far from the point $s_A=\frac{4}{3}M_\pi^2$, where it was predicted long ago, on the basis of current algebra~\cite{Osborn+1970}. This provides a good check on the internal consistency of our framework.  
\begin{figure*}[thb]\centering
  \includegraphics[width=7.5cm]{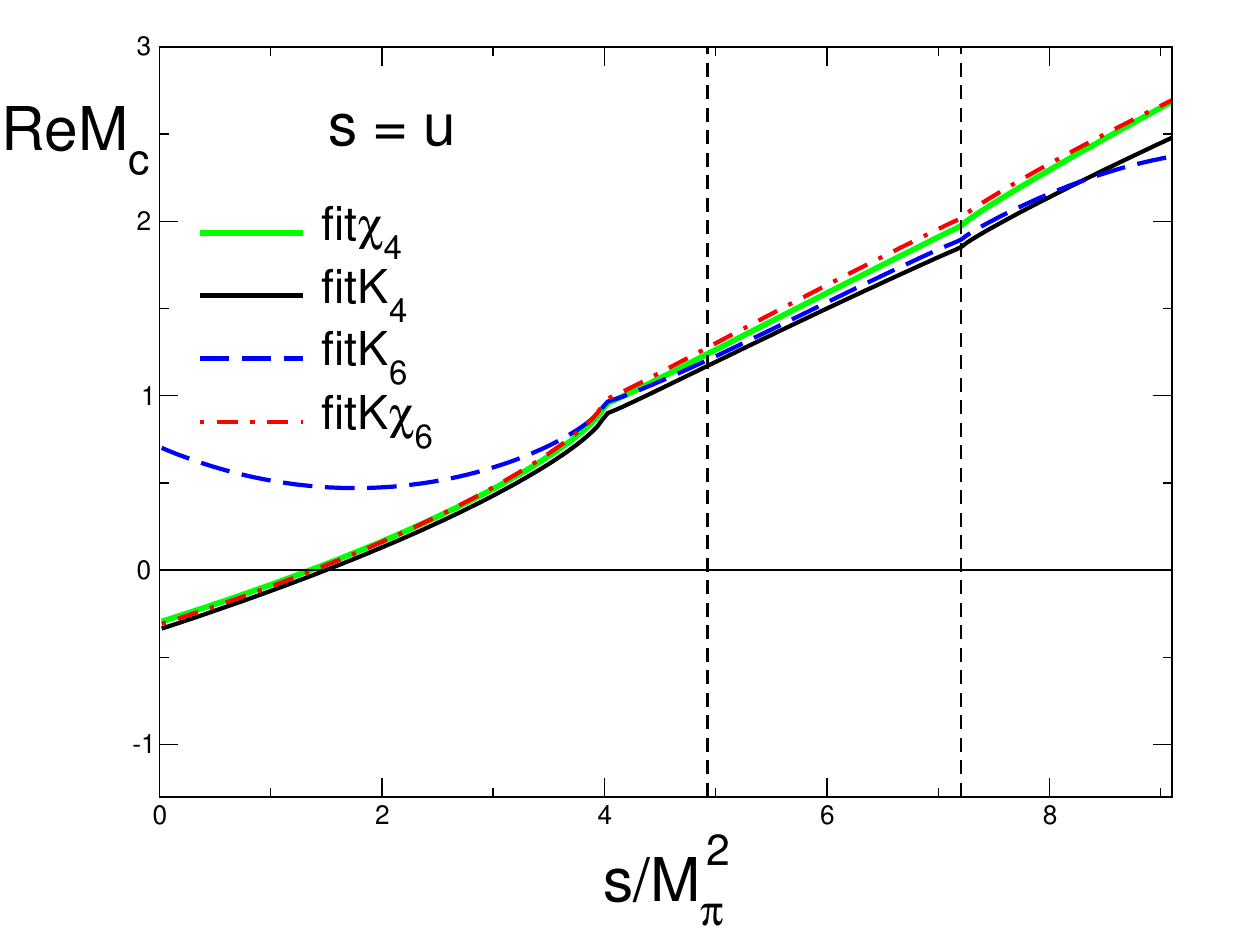}\hspace{0.7em}
  \includegraphics[width=7.5cm]{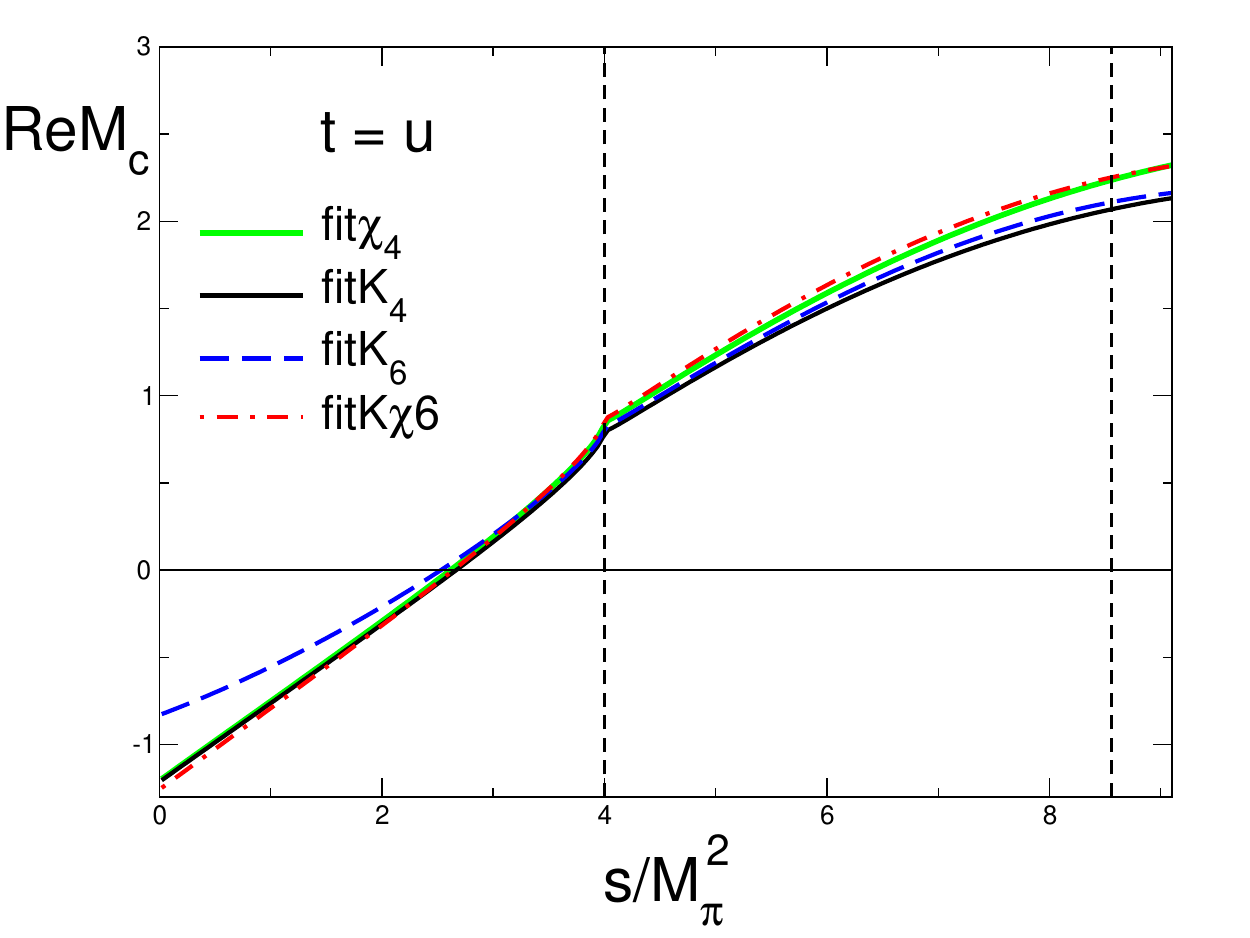} 
\caption{Real parts of various dispersive solutions along the lines $s=u$ and $t=u$.\label{fig:ReMdisp}}
\end{figure*}
\subsection{Theoretical constraints}\label{sec:Theoretical constraints}

Since the experimental and theoretical sources of information are consistent with one another, it is meaningful to combine them. We do this by introducing a discrepancy function that measures the deviation from the theoretical estimates: 
\begin{equation}\label{eq:chith}
\chi^2_\mathrm{th}=\frac{(H_0-H_0^\mathrm{NLO})^2}{\Delta H_0^2}+\sum_{i=1}^3\frac{(\Re h_i-h_i^\mathrm{NLO})^2}{\Delta h_i^2} \;.\end{equation}
The quantities $H_0^\mathrm{NLO}$, $h_i^\mathrm{NLO}$ represent the central values listed in~\eqref{eq:HNLO} and $\Delta H_0^\mathrm{NLO}$, $\Delta h_i^\mathrm{NLO}$ denote the uncertainties quoted there.  We identify the central solution of our integral equations with the minimum of the sum of the two discrepancy functions:
\begin{equation}\label{eq:chitot}\chi^2_\mathrm{tot}=\chi^2_\mathrm{K}+\chi^2_\mathrm{th}\;.\end{equation}

Let us first treat all six subtraction constants as well as the normalization $\Lambda_\mathrm{K}$ of the Dalitz plot distribution as free parameters. We use the symbol $\mathrm{fitK\chi_6}$ for this fit, to indicate that it relies both on the KLOE data and on the theoretical constraints obtained from \chpt\ and involves 6 subtraction constants. The fit represents a compromise between the minima of the experimental and theoretical discrepancies: 
\begin{eqnarray}\mathrm{fitK\chi_6:}\quad\beta_0\al =\al  16.2\;,\hspace{0.2em}\gamma_0=-20.8\;, \hspace{0.2em} \delta_0= -37.8\;,\hspace{0.2em} \beta_1 = 8.5\;,\label{eq:fitKchi6} \nonumber\\
 \gamma_1\al = \al-3.8\;, \hspace{0.2em}\Lambda_\mathrm{K} = 0.938\;,\hspace{0.4em}\chi^2_\mathrm{K}=384\;,\hspace{0.4em}\chi^2_\mathrm{th}=1.47.\nonumber \\\end{eqnarray}
The quality of the fit to the data is slightly better than in the case of $\mathrm{fitK_4}$ -- not a surprise: We are allowing for six rather than only four subtraction constants. The price to pay is that the theoretical discrepancy increases. By construction $\chi_\mathrm{th}^2$ vanishes for $\mathrm{fit\chi_4}$,  takes the value $\chi_\mathrm{th}^2=0.67$ for $\mathrm{fitK_4}$ and reaches $\chi_\mathrm{th}^2=1.47$ for $\mathrm{fitK\chi_6}$. 

Fig.~\ref{fig:ReMdisp} displays the behaviour of the real parts belonging to the various dispersive solutions all the way down to $s=0$ (while the curves for the Dalitz plot distribution shown in Fig.~\ref{fig:Dcbar} account for the corrections due to isospin breaking, those for Re$M$ represent the isospin symmetric solutions as they are). Remarkably, in the entire range shown, $\mathrm{fitK\chi_6}$ runs close to fit$\chi_4$, the matching solution specified in Sec.~\ref{sec:Matching}. 

In addition to the representations fit$\chi_4$, $\mathrm{fitK_4}$ and $\mathrm{fitK\chi_6}$ we discussed above, Fig.~\ref{fig:ReMdisp} shows a fourth solution, $\mathrm{fitK_6}$. The only difference between this solution and $\mathrm{fitK_4}$ is that $\delta_0$ and $\gamma_1$ are not set equal to 0, but are treated as free parameters. Accordingly, this fit follows the data even more closely: $\chi_\mathrm{K}^2=371$ for 371 data points and 6 free parameters. Fig.~\ref{fig:Dcbar} shows that, in the physical region, the Dalitz plot distributions belonging to $\mathrm{fitK_4}$ and $\mathrm{fitK_6}$ are nearly the same. Outside the physical region, however, $\mathrm{fitK_6}$ goes astray: This solution of our system of integral equations is not acceptable, because it does not have an Adler zero at all. The clash with chiral symmetry also manifests itself in  the Taylor invariants: $\mathrm{fitK_6}$ yields $\Re h_3^\mathrm{K_5}=59.8$, for instance, which differs from the 
theoretical estimate $h_3=6.3(2.0)$ in~\eqref{eq:HNLO} by 28 $\sigma$. This  indicates that -- with six subtraction constants -- there is too much freedom in the space of solutions for the experimental information about the Dalitz plot distribution to control the behaviour of the transition amplitude outside the physical region. 

The fact that $\mathrm{fitK\chi_6}$ does have an Adler zero at $s_A=1.39\,M_\pi^2$ shows that the theoretical constraints do provide the missing information: The only difference between $\mathrm{fitK_6}$ and $\mathrm{fitK\chi_6}$ is that the latter accounts for these while the former does not. The theoretical constraints barely matter in the physical region, but play an important role in the extrapolation to small values of $s$. The properties of the amplitude at small values of $s$ are essential, because theory is needed to determine the normalization of the amplitude. Since the relevant Taylor invariant, $H_0$, represents a linear combination of the subtraction constants $\alpha_0$ and $\beta_0$, it concerns the value and the first derivative of the component $M_0(s)$ at $s=0$. 

\begin{table*}[thb]\centering
\begin{tabular}{lccccccc}
&$\beta_0$&$\gamma_0$&$\delta_0$&$\beta_1$&$\gamma_1$ &$\chi^2_\mathrm{K}$&$\chi^2_\mathrm{th}$ \\
\hline
$\mathrm{fit\chi_4}$ &16.9(1.7) &-29.5(10.6)  & \hspace{1.2em}\rule[0.3em]{1em}{0.05em} & \hspace{0.15em} 6.6(2.3) & \hspace{0.1em}\rule[0.3em]{1em}{0.05em} & (801) &0 \\
$\mathrm{fitK_4}$ &17.6(7) &-35.2(7.2)  & \hspace{1.2em}\rule[0.3em]{1em}{0.05em}&  \hspace{0.15em} 5.9(8) & \hspace{0.1em}\rule[0.3em]{1em}{0.05em}&  390&(0.67) \\
$\mathrm{fitK\chi_4}$ &17.5(6) &-35.0(7.2)  & \hspace{1.2em}\rule[0.3em]{1em}{0.05em}& \hspace{0.15em} 6.0(7) & \hspace{0.1em}\rule[0.3em]{1em}{0.05em}&  390 &0.59\\
\hline
$\mathrm{fitK_5}$ &13.3(2.2)&\hspace{0.3em}23.8(26.9) & -147(66)  & 13.4(3.6)&  \hspace{0.1em}\rule[0.3em]{1em}{0.05em}&379&(46)\\
$\mathrm{fitK\chi_5}$ &16.6(8)&-20.1(9.1) & \hspace{0.2em} -38(17)  & \hspace{0.15em} 7.8(1.1) &  \hspace{0.1em}\rule[0.3em]{1em}{0.05em}&   384&1.43\\
\hline
$\mathrm{fitK_6}$ &\hspace{-0.35em}-20.0(10.2)& -35.6(89.3) &  \hspace{0.2em} -75(91) &  77(19)  & -308(88) &     371&(1005) \\
$\mathrm{fitK\chi_6}$ &16.2(1.2)&-20.8(10.1) &  \hspace{0.2em} -38(17) &  \hspace{0.15em} 8.5(2.2)  & -3.8(10.7) &     384&1.47 \\
\hline
\end{tabular}
\caption{Comparison of the matching solution $\mathrm{fit\chi_4}$ with fits to the KLOE Dalitz plot distribution for $\eta\to\pi^+\pi^-\pi^0$. The presence or absence of the label $\chi$ indicates whether or not the theoretical discrepancy~\eqref{eq:chith} is included in the minimization procedure and the index specifies whether four, five, or six subtraction constants are taken different from zero (in the chosen normalization, $\alpha_0$ is tied to $\beta_0$ according to $\alpha_0=0.8594-0.08736\,\beta_0$). For fits obtained by dropping either the experimental or the theoretical part of the discrepancy function, the values of $\chi^2_\mathrm{K}$ or $\chi^2_\mathrm{\chi}$ are put in brackets. \label{table:fitK}}
\end{table*}

\subsection{Error analysis}\label{sec:Error analysis}

The uncertainties in our results are dominated by the statistical errors. These are determined by the behaviour of the discrepancy function in the vicinity of the minimum. In connection with the fits to the measured Dalitz plot distribution of the charged decay mode, the normalization constant $H_0$ is irrelevant -- we keep it fixed at the value found at one loop. Also, since none of the observables of interest in the present context depends on $\Lambda_\mathrm{K}$, we fix this parameter at the minimum, which is nearly the same  for all fits: $\Lambda_\mathrm{K}\simeq 0.938$. The discrepancy function $\chi^2_\mathrm{tot}$ then depends on five independent real variables, which can, for instance, be identified with $\beta_0$, $\gamma_0$, $\delta_0$, $\beta_1$, $\gamma_1$. We rely on the Gaussian approximation, which exploits the fact that, in the vicinity of the minimum, the discrepancy function can be approximated by the truncated Taylor series in all five variables. 
The calculation is described in detail in Appendix~\ref{sec:Gaussian errors}. 

The uncertainties inherent in the input used for the $\pi\pi$ phase shifts must also be accounted for. These were discussed in Sec.~\ref{sec:Phase shifts}. We have worked out the response of the dispersive representation to variations in the Roy solutions of~\cite{Colangelo2001}, not only below 800 MeV where the uncertainties are small, but also at higher 
energies where dispersion theory does not provide strong constraints -- for details see 
Appendix~\ref{sec:Sensitivity to phase shifts}. The resulting uncertainties in the subtraction constants are small compared to the Gaussian errors discussed above, except for $\gamma_0$: This term is relatively sensitive to the high energy tail of the dispersion integrals -- the corresponding uncertainty is comparable to the Gaussian error.   

The kinematic map we are using to embed 
the isospin symmetric dispersive representation in the physical world accounts for the effects due to the mass difference between the charged and neutral pions only rather crudely.  We rely on the one-loop approximation of Ditsche, Kubis and 
Mei{\ss}ner~\cite{Ditsche+2009} to correct for all other effects that (i) are generated by the e.m.~interaction and (ii) are not taken care of when applying radiative corrections to the data. We consider the difference between our results and those obtained by neglecting the isospin breaking effects altogether and estimate the uncertainty of our treatment of these effects at 30\,\% of that difference. 

The errors listed in Table~\ref{table:fitK}  are obtained by adding the Gaussian errors, those from the $\pi\pi$ phase shifts and those related to isospin breaking in quadrature,
\vspace{-0.3cm}
\subsection{Number of subtraction constants, significance of theoretical constraints}
\label{sec:Number of subtraction constants}
The number of subtraction constants occurring in the dispersive form of the chiral representation increases with the order: four subtraction constants at NLO, six at NNLO, etc. We impose theoretical constraints based on the NLO representation of $\chi$PT -- four subtraction constants are a suitable choice in this context, but our framework does leave room for two further subtractions. In the present section, we compare the solutions of our integral equations obtained with four, five or six subtraction constants and discuss the role of the theoretical constraints. 

The approach in \cite{Kampf+2011} differs from ours as it relies on the NNLO representation of $\chi$PT \cite{Bijnens+2007}. Six subtraction constants are used ab initio to impose the theoretical constraints. In particular, the representation obtained in this way invokes the estimates for the LECs obtained from resonance saturation in the scalar channel -- our analysis avoids the use of such estimates. For a comparison of their results with ours, we refer to Sec.~\ref{sec:Comparison}. 

The first two lines in Table~\ref{table:fitK} represent two extremes: While fit$\chi_4$ only relies on theory, fitK$_4$ only relies on experiment.  For a detailed comparison of these two solutions, we refer to the end of Sec.~\ref{sec:without theoretical constraints}. Table~\ref{table:fitK} shows  that the central values of all of the subtraction constants of fitK$_4$ are within the uncertainty range of fit$\chi_4$ and vice versa. In other words, the fit to the data automatically satisfies the theoretical constraints. This can also be seen in the value $\chi^2_\mathrm{th}=0.67$ obtained with fitK$_4$: The central values of $h_1$, $h_2$, $h_3$ obtained from the KLOE data are all in the predicted range. 

The entries for $\chi_\mathrm{K}^2$, on the other hand, show that fit$\chi_4$ differs strongly from fitK$_4$: While the latter represents an excellent fit of the 371 data points with $\chi_\mathrm{K}^2=390$, the former yields a value of $\chi^2_\mathrm{K}$ that is more than twice as large. Superficially, this may give the impression that the matching solution is ruled out by experiment, but this is by no means the case. In view of the uncertainties attached to the predictions for $h_1$, $h_2$, $h_3$, the matching procedure leads to an entire family of solutions -- fit$\chi_4$ merely represents the central one of these. The very fact that fitK$_4$ is a member of this family shows that the KLOE data on the Dalitz plot distribution of $\eta\to\pi^+\pi^-\pi^0$ confirm the theoretical estimates based on the assumption that the strong interaction possesses a hidden approximate symmetry.

In the derivation of fitK$\chi_4$, both the KLOE data and the theoretical constraints are made use of. The comparison with fitK$_4$  shows, however,  that this barely makes any difference. In particular, the values of $\chi_\mathrm{th}^2$ and $\chi_\mathrm{K}^2$ obtained with these two fits are nearly the same.  

The solution fitK$_5$ differs from fitK$_4$ in that the subtraction constant $\delta_0$ is not set equal to zero, but is treated as a free parameter. Table~\ref{table:fitK} shows that the solution then changes quite drastically: (1) the minimum occurs at a value of $\delta_0$  that differs from zero by about two standard deviations, (2) the quantities $\beta_0$, $\gamma_0$ and $\beta_1$ are also pushed outside the range found with fitK$_4$ or fitK$\chi_4$ and (3) the value of $\chi_\mathrm{th}^2$ becomes very large. This shows that fitK$_5$ very strongly violates the theoretical constraints. The situation is similar to the one encountered with 
fitK$_6$ in Sec.~\ref{sec:Theoretical constraints}: The data are not accurate enough to pin down more than four parameters. Both fitK$_5$ and fitK$_6$ must be discarded -- they represent unphysical solutions of our integral equations. 

The theoretical constraints domesticate the manifold of solutions if more than four subtraction constants are treated as free parameters. In fact, it does then not make much of a difference whether five or six subtraction constants are treated as free parameters. In either case, the solution is consistent with the theoretical constraints and the common subtraction constants  agree within errors. Moreover, fitK$\chi_6$, which treats $\gamma_1$ as a free parameter, yields a result with a broad uncertainty range -- the value $\gamma_1=0$ that corresponds to fitK$\chi_5$ is within that range. The discrepancy function $\chi_\mathrm{th}^2$ punishes strong deviations from the values of the Taylor invariants obtained at one loop. The fit yields $ \Re h_1^\mathrm{K\chi_6}=4.52(14) $, $\Re h_2^\mathrm{K\chi_6}=21.7(4.3) $, \mbox{$\Re h_3^\mathrm{K\chi_6}= 7.3(1.7)$}. The comparison with~\eqref{eq:HNLO} shows that, within errors, these numbers are consistent with the estimates based on \chpt.

The shape of the Dalitz plot distribution is tightly constrained by experiment. Indeed, 
Fig.~\ref{fig:Dcbar} shows that for the behaviour in the physical region, it barely makes a difference whether four or six subtraction constants are treated as free parameters. The numbers for $\chi_\mathrm{K}^2$ in Table~\ref{table:fitK} confirm this: The fits fitK$\chi_4$, fitK$\chi_5$ and fitK$\chi_6$ all describe the data very well. We conclude that, as far as the momentum dependence in the physical region is concerned, the description of the observed behaviour does not require more than four subtraction constants. 

In order to establish contact with QCD and with the quark mass ratio $Q$, however, we need to be able to calculate the decay rate. In this connection, the normalization of the amplitude plays a key role -- it is not accessible experimentally because it drops out in the Dalitz plot distribution. As discussed above, we specify the normalization of the dispersive representation with the Taylor invariant $H_0$, which only concerns the behaviour of the component $M_0(s)$  at small values of $s$. For the rate, the value of the amplitude instead counts at the center of the Dalitz plot. We need to understand the relation between the two. For this purpose, we consider the quantity
\begin{equation}N_1=\rule[-0.7em]{0.05em}{2.2em}\hspace{0.1em}\frac{M_c(0,0)}{H_0}\hspace{0.1em}\rule[-0.7em]{0.05em}{2.2em}\;,\end{equation}
which compares the value of the dispersive representation at the center of the Dalitz plot ($X_c=Y_c=0$) with the Taylor invariant $H_0$. Qualitatively, $N_1$ represents the amplification generated by the final state interaction at the center of the physical region. At tree level, the final state interaction is ignored: $N_1=1$. The one loop representation yields $N_1=1.33$. For those fits to the KLOE data that are physically meaningful, the value of $N_1$ is listed in Table~\ref{table:N}.
The result shows that the number of subtraction constants matters: The amplification factor obtained if five or six subtraction constants  are used differs significantly from what is obtained if $\delta_0$ and $\gamma_1$ are set equal to zero.
\begin{table}[thb]\centering
\begin{tabular}{lcccc}
&fitK$_4$&fitK$\chi_4$&fitK$\chi_5$&fitK$\chi_6$\\
\hline
$N_1$&1.371(22)&1.372(22)&1.499(64)&1.494(66)\\
\hline
\end{tabular}
\caption{Value of the amplitude at the center of the Dalitz plot: sensitivity to the number of subtraction constants.\label{table:N}}
\end{table}

\vspace{-3em}
\begin{figure}[thb]\centering \includegraphics[width=8.7cm]{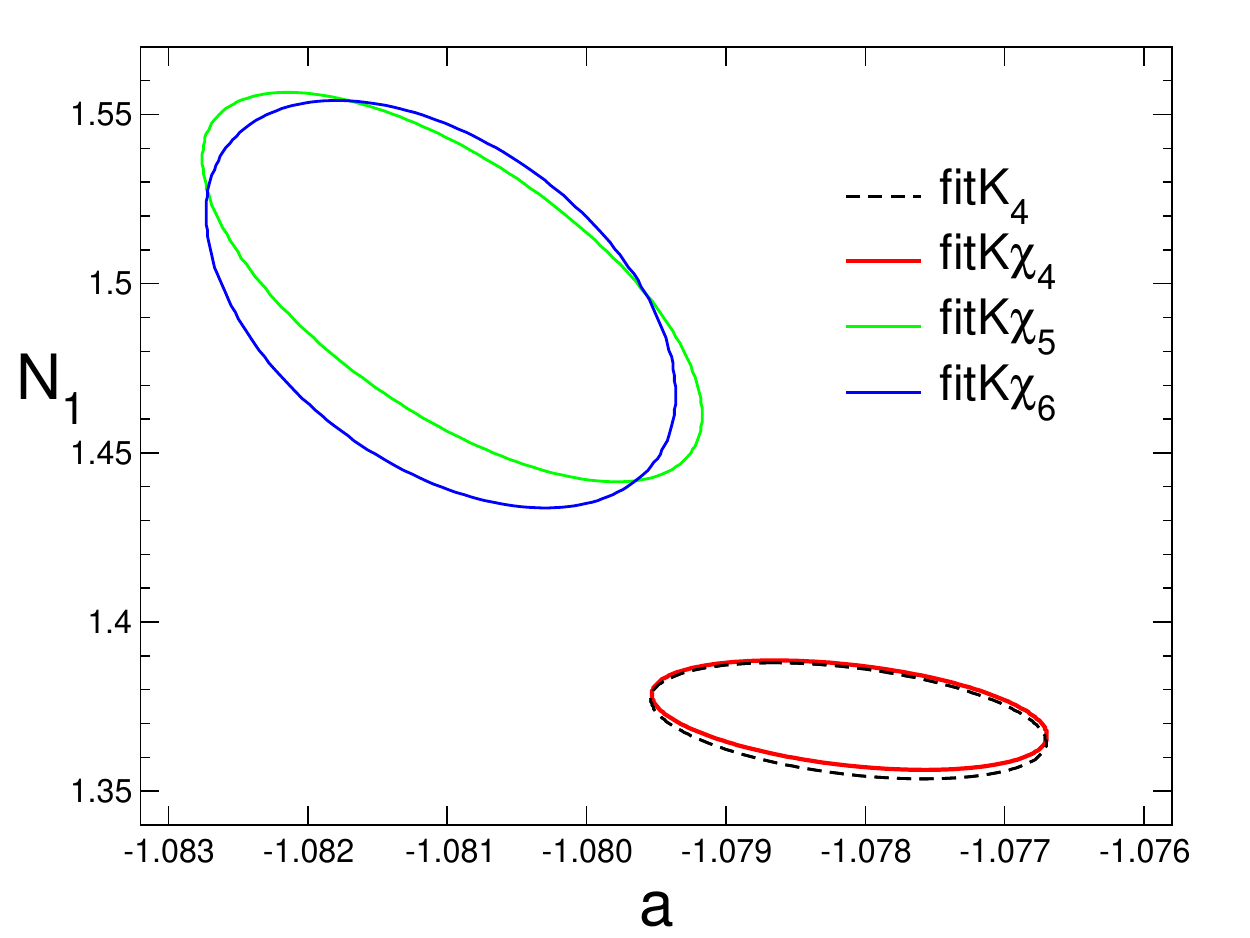} 
\caption{Value of the amplitude at the center versus slope of the Dalitz plot distribution in the charged channel: sensitivity to the number of subtraction constants.\label{fig:Nversusa}}
\end{figure}
\begin{table*}[thb]\centering
\begin{tabular}{lccccccc}
&$\Re \beta_0$&$\Re \gamma_0$&$\Re \delta_0$&$\Re \beta_1$&$\Re \gamma_1$&$\chi^2_\mathrm{K}$&$\chi^2_\mathrm{th}$\\
\hline
$\mathrm{fitK\chi_6}$&16.2(1.2) &-20.8(10.1)  & -38(17)  & 8.5(2.2) &   -3.8(10.7)&   384.1&1.47 \\
$\mathrm{FitK\chi_6}$&16.2(1.2) &-21.0(10.0) &  -38(17) &  8.6(2.2)  & -4.8(10.7)  &    384.8&  1.58\\
\hline
\end{tabular}
\caption{Central values and errors for two versions of the central solution: While for $\mathrm{fitK\chi_6}$, the subtraction constants are taken real, in the case of $\mathrm{FitK\chi_6}$, they are instead calculated from the two-loop prediction for the imaginary parts of the Taylor coefficients. \label{table:comparison of fits}}
\end{table*}

\vspace{3em}
\begin{table*}\centering
\begin{tabular}{lcccccc}
&$\Re K_1$&$\Re K_2$&$\Re K_3$&$\Re K_4$&$\Re K_5$&$s_A$\rule{0.8em}{0em}\\
\hline
$\mathrm{fitK\chi_6}$&4.51(25) &  25.6(5.4)&   -3.7(1.8)&   90.2(5.0)&   52.9(7.0)&   1.39(11)$M_\pi^2$  \\
\hline
$\mathrm{FitK\chi_6}$&4.55(24) &  25.8(5.2)&   -3.6(1.9)&   90.8(5.2)&   52.3(6.9)&   1.38(11)$M_\pi^2$  \\
\hline
\end{tabular}
\caption{Taylor invariants and position of the Adler zero for the two variants of the central solution.\label{table:ReK}}
\end{table*}

\vspace{1em}
 To discuss the implications of this result, we consider the correlation between $N_1$ and the slope $a$ of the Dalitz plot distribution at the center, that is, the term linear in $Y_c$ in~\eqref{eq:Dcpoly}. 
Fig.~\ref{fig:Nversusa} shows that it makes a significant difference whether the subtraction constant $\delta_0$ is set equal to zero (fitK$_4$, fitK$\chi_4$) or treated as a free parameter (fitK$\chi_5$, fitK$\chi_6$). If $\delta_0$ is set equal to zero then $N_1$ is determined very sharply. In fact, the solution then becomes so stiff that the result for $N_1$ is outside the range obtained if $\delta_0$ is allowed to float. In somewhat milder form, the problem also manifests itself in Table~\ref{table:fitK}: The value $\delta_0=0$ is about two standard deviations away from the results obtained with fitK$\chi_5$ or fitK$\chi_6$.  This shows that setting $\delta_0=0$ amounts to introducing a systematic theoretical error, which pulls the amplitude down by about 9 percent. 
 
Four subtraction constants do suffice to properly describe the momentum dependence in the physical region of the decay, but to cope with the theoretical constraints that follow from the fact that the particles involved in this decay are Nambu-Goldstone bosons of a hidden approximate symmetry, an extrapolation from the physical region all the way down to the Adler zero is required. We conclude that with only four subtractions, the dispersive representation does not provide a controlled extrapolation: $\delta_0$ cannot simply be set equal to zero, but needs to be determined by experiment.

For $\gamma_1$, the situation is different: Since the value $\gamma_1=0$ is  close to the center of the range obtained if this parameter is allowed to float, it does not make much of a difference whether or not we keep it fixed at zero. The advantage of using six subtractions rather than five is that the uncertainties associated with the contributions from the high energy tails of the dispersion integrals are then reduced. For this reason, we identify our central solution with fitK$\chi_6$.

\subsection{Imaginary parts of the subtraction constants}\label{sec:Imaginary parts}
As discussed in Sec.~\ref{sec:Imaginary parts two loops}, the subtraction constants pick up an imaginary part at NNLO of the chiral expansion. In fact, at two loops, the imaginary part is fully determined by the one-loop representation and does therefore not involve any unknowns. The imaginary parts of the Taylor coefficients depend on the choice of the decomposition, but those of the invariants $K_0$, \ldots , $K_5$ are unambiguous. In the present section, we investigate the changes occurring in our central solution if instead of taking the subtraction constants to be real, the values of Im$K_0$, \ldots, Im$K_5$ are taken from the two-loop representation of Bijnens and Ghorbani~\cite{Bijnens+2007}, which are listed in Eq.~\eqref{eq:ImK}. We denote this version of the central solution by $\mathrm{FitK\chi_6}$, to distinguish it from the solution $\mathrm{fitK\chi_6}$ considered above, for which the subtraction constants are real.  
For the Dalitz plot distribution, the normalization of the amplitude is irrelevant. We fix it by using the one-loop result for the real part of $K_0\equiv H_0$. 

Table~\ref{table:comparison of fits} compares the real parts of the subtraction constants belonging to FitK$\chi_6$ with those of fitK$\chi_6$, which are real by construction. It shows that the differences between the two versions of our central solution are negligibly small compared to the uncertainties therein.

Table~\ref{table:ReK} shows that the same conclusion is reached if instead of the real
 parts of the subtraction constants we compare the real parts of the Taylor invariants $\Re K_1$, \ldots, $\Re K_5$ or the position of the Adler zero for the two variants of our central solution. The Adler zero is determined to an accuracy of about 8\,\% and occurs in the immediate vicinity of the current algebra prediction, $s_A=4/3\,M_\pi^2$. 

Since the difference between the two versions of the central solution is in the noise of our calculation, we do not pursue it further.  In Sec.~\ref{sec:Anatomy}, where we discuss the difference between the two-loop representation of \chpt\ and the dispersive representation that matches it at low energies, we consider the version FitK$\chi_6$, because it matches the imaginary parts as well as the real parts.  Throughout the remainder of the paper, however, where we draw the conclusions from our analysis, we stick to real subtraction constants and work with the version fitK$\chi_6$ of the central solution. 

\subsection{Dalitz plot coefficients of our central solution}\label{sec:coefficients}
To complete this discussion of the dispersive representation in the charged channel, we approximate
our central solution with a polynomial of the form~\eqref{eq:Dcpoly}. The result reads
\begin{eqnarray}\label{eq:Dalitz fitKchi6}a \al =\al-1.081(2)\;,\; b=0.144(4)\;,\;  d=0.081(3)\;, \nonumber\\
f \al = \al 0.118(4)\;,\; g=-0.069(4)\;.
\end{eqnarray}

It is not surprising  that these numbers are close to those obtained by KLOE (last row in Table~\ref{tab:Dalitz-exp}) -- the two representations of the Dalitz plot distribution differ by less than 1.2\,\%, in the entire physical region. The difference arises because we are imposing theoretical constraints. Indeed, dropping these, i.e.~replacing our central solution by fitK$_6$, the coefficients of the polynomial approximation reproduce those obtained by KLOE within errors. This shows that (i) with 6 subtraction constants, the dispersive framework is flexible enough to describe the KLOE data well and (ii)  the available experimental information is consistent with the theoretical constraints. 

The parametrization~\eqref{eq:Dcpoly} amounts to a polynomial in the  Mandelstam variables $s, t, u$. Unitarity generates branch points at the boundary of the physical region (the corresponding cusps in the real part of the amplitude can be seen e.g.~in Fig.~\ref{fig:ReMdisp}). Outside the physical region, a polynomial parametrization of the Dalitz plot distribution cannot provide a reliable improvement of the current algebra formula, $D_c^\mathrm{LO}=(3\,s-4M_\pi^2)^2/(M_\eta^2-M_\pi^2)^2$. The dispersive framework we are using does account for the singularities required by unitarity, but as discussed in Sec.~\ref{sec:Number of subtraction constants}, a fit to the KLOE distribution that simply treats the subtraction constants as free parameters leads to solutions that violate chiral symmetry. We are exploiting the fact that this symmetry imposes strong conditions on the amplitude at small values of $s$, in particular also near the Adler zero.  Although these conditions do not significantly constrain the 
amplitude in the physical region, they are essential for the interpretation of the experimental results in the framework of the Standard Model. 

\subsection{Comparison with the nonrelativistic effective theory}\label{sec:Comparison NREFT}
As discussed above, the Dalitz plot distribution is well described by the dispersive representation with four real subtraction constants. The fit to the KLOE data obtained in that framework, fitK$_4$, does have an Adler zero in the vicinity of the current algebra prediction and also yields values for the Taylor invariants $h_1$, $h_2$, $h_3$ that are consistent with the theo\-retical constraints. We now compare the dispersive solutions with the two-loop representation of the nonrelativistic effective theory for the transition $\eta\to 3\pi$ set up in Ref.~\cite{Bissegger:2007yq}. As this representation does not account for the electromagnetic interaction, we consider the isospin limit, setting $M_{\pi^0}=M_{\pi^\pm}$ and fixing the low-energy constants $K_0$, $K_1$ with \eqref{eq:KL}.  Since the Dalitz plot distribution does not fix the normalization of the amplitude, we set $L_0=1$. The fit to the KLOE data then yields the following values in GeV units:
\begin{eqnarray}\label{eq:NRK}L_0 \al = \al 1\,\hspace{1em}L_1 =-3.91\;,\hspace{1em} L_2 = -48.2\;,\hspace{1em} L_3 = 4.92\,,
\nonumber\\
\Lambda_\mathrm{K}\al=\al0.9383\;.\end{eqnarray}
With $\chi^2_\mathrm{K}=370.3$ for 371 data points, the fit is of excellent quality, even better than fitK$_4$. 

Next, we look for a solution of our integral equations that matches the nonrelativistic representation. Instead of matching the coefficients of the nonrelativistic expansion as discussed in Sec.~\ref{sec:NR}, we minimize the difference between the nonrelativistic and relativistic representations of the amplitude in the physical region. To do this, we allow for four subtraction constants and treat these as complex free parameters. The minimum occurs at
\begin{eqnarray}\label{eq:NRdisp}\mathrm{fitNRK}_4:\hspace{2em}\alpha_0\al=\al -0.235 - i\,0.252\,,\hspace{1em}\beta_0= 7.20 + i\,3.48\,,\nonumber \\
\gamma_0\al=\al -14.1 - i\,11.6\;,\hspace{2em} \beta_1= 3.69 - i\,1.50\;.\nonumber\\\end{eqnarray}
We denote this solution of our integral equations by fitNRK$_4$. It may be viewed as a relativistic extension of the NR representation: In contrast to the latter, it is meaningful also at small values of $s$. Indeed, fitNRK$_4$ does have an Adler zero at $s_A= 1.36\,M_\pi^2$. Moreover, the real parts of the Taylor invariants $h_1$, $h_2$, $h_3$ are given by $ 4.4$, $ 12.3$, $7.1$, respectively -- these values are consistent with the theoretical constraints.

\begin{figure*}[thb] \centering
\includegraphics[width=7cm]{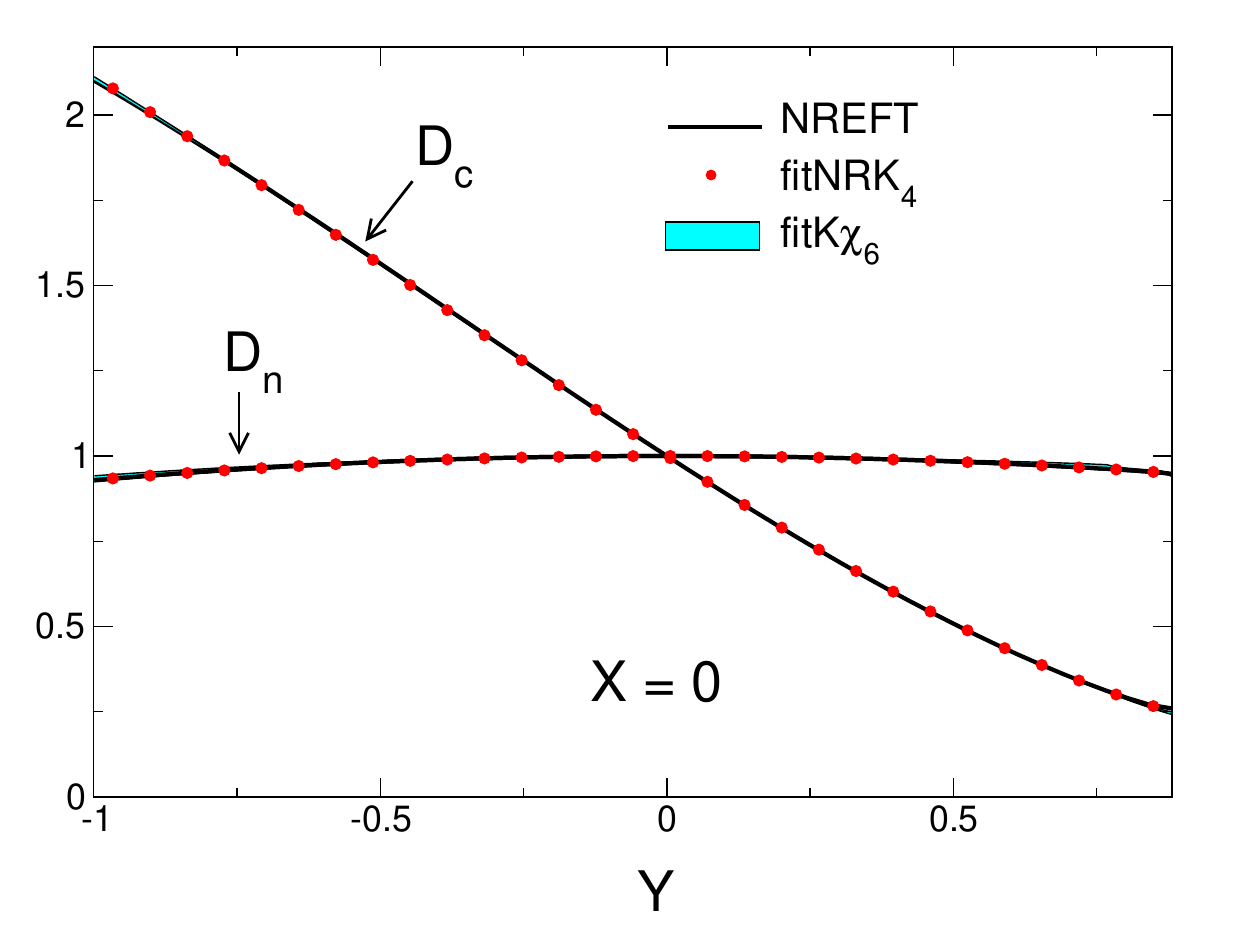}\hspace{0.7em} \includegraphics[width=7cm]{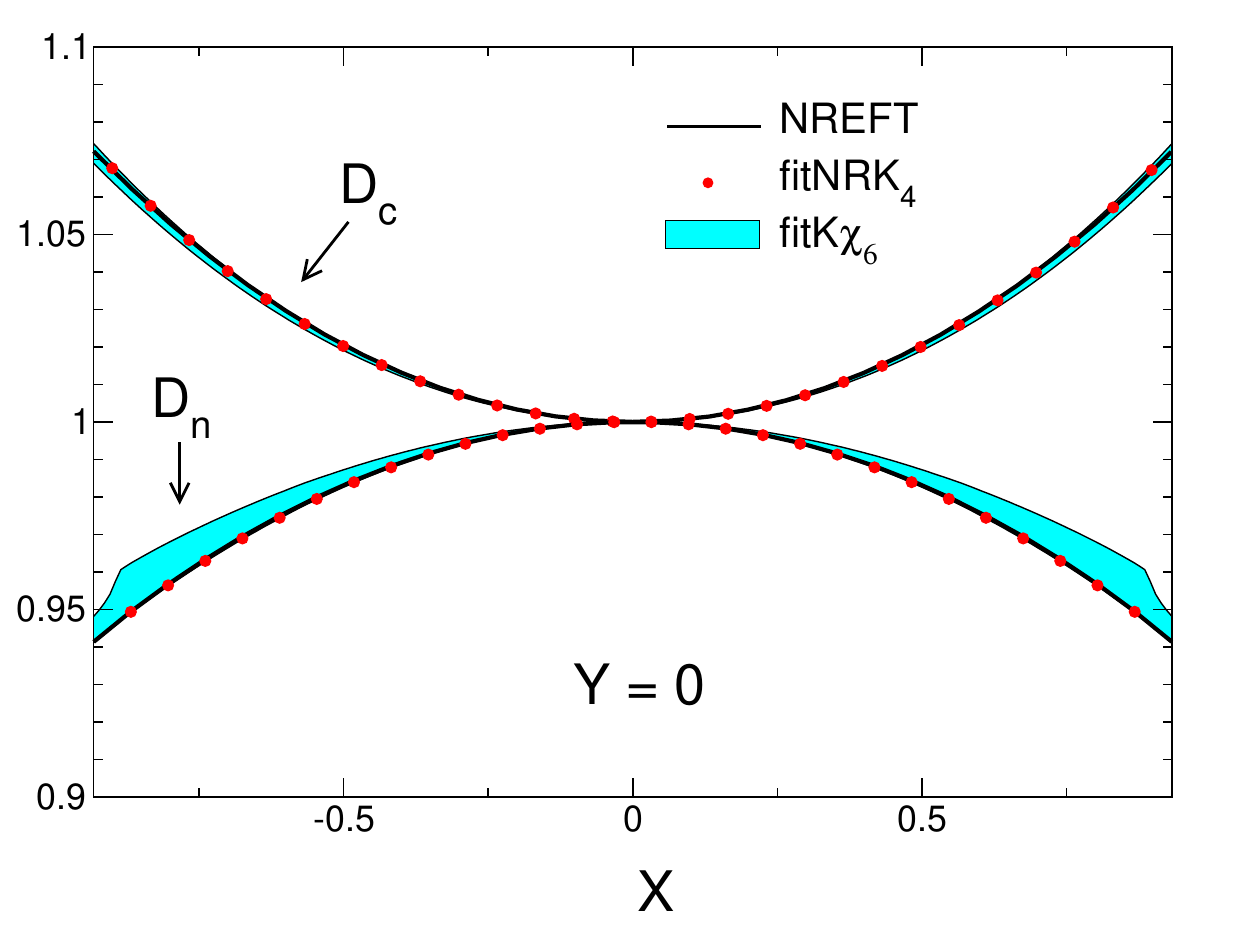} 
\caption{Comparison of the nonrelativistic two-loop representation (black dots) with the dispersive solution that matches it (red dots): Dalitz plot distributions for the charged and neutral channels in the isospin limit. The uncertainty band belongs to our central solution, fitK$\chi_6$, which does account for isospin breaking effects. The left and right panels indicate the behaviour along the lines $t=u$ and $s=\frac{1}{3}M_\eta^2+M_\pi^2$, respectively.  \label{fig:DNR}} \end{figure*}

We conclude that the two-loop representation of NREFT yields a decent approximation of the momentum dependence also for $\eta$-decay. In the case of kaon-decay, the contributions due to the electromagnetic interaction were worked out in the framework of NREFT and the cusps generated by the transition $\pi^0\pi^0\to\pi^+\pi^-\to\pi^0\pi^0$ were studied in detail. The two-loop representation of Ref.~\cite{Bissegger:2007yq} does properly account for the mass difference between the charged and neutral pions -- an evident advantage compared to our analysis, which takes care of the mass difference only in a purely kinematic way. For those electromagnetic effects that do not show up in the self-energies of the pions, we are relying on the relativistic one-loop representation \cite{Ditsche+2009}. The work done in the framework of NREFT \cite{Bissegger:2008ff,Gasser:2011ju} would provide the basis for a more thorough analysis of the contributions generated by the electromagnetic interaction, but we must leave this for future work.

The numerical values found for the subtraction constants of fitNRK$_4$ are very different from those of the dispersive solutions listed in Table \ref{table:fitK}. One of the reasons is that the normalization differs: While the nonrelativistic two-loop representation is normalized by setting $L_0=1$, the solutions in Table \ref{table:fitK} are normalized by fixing the Taylor invariant $H_0$ at the value found at one loop. The Taylor invariants are outside the reach of the nonrelativistic effective theory. We can instead fix the normalization such that the magnitude of the amplitude at the center of the Dalitz plot is the same as for our central solution, fitK$\chi_6$. This is achieved by simply stretching all of the LECs: $L_n\to \lambda L_n$, with $\lambda=2.353$. The subtraction constants of fitNRK$_4$ must be stretched by the same factor. 

There is a further difference: For the dispersive solution to match the NR representation, the subtraction constants must be allowed to have an imaginary part -- those of the solutions listed in Table \ref{table:fitK} are real.  We investigated the sensitivity of our results to the imaginary parts of the subtraction constants in Sec.~\ref{sec:Imaginary parts}. There, we observed that, in the chiral expansion, the Taylor invariants become complex at NNLO. We worked out the dispersive solution obtained  if the imaginary part of the Taylor invariants are taken from the two-loop representation of the relativistic effective theory and found that the imaginary parts do not significantly affect our results. Matching with the NR effective theory at two loops confirms this experience: Although the subtraction constants of fitNRK$_4$ have sizeable imaginary parts while those of the solutions listed in Table \ref{table:fitK} are real, the results obtained for quantities of physical interest are in the same ballpark.  As we are not in a position to properly account for isospin breaking effects, we do not continue the comparison with the nonrelativistic framework further, but will briefly return to related work in Sec.~\ref{sec:Nonrelativistic}.  

Figure~\ref{fig:DNR} shows that the Dalitz plot distributions of the two representations can barely be distinguished, in the entire physical region and for $\eta\to\pi^+\pi^-\pi^0$ as well as for $\eta\to3\pi^0$. Note the difference in the scale used in the two panels. In the left panel, the difference between the nonrelativistic fit to KLOE and our central solution  can barely be seen, but it does show up in the right panel: The cusps generated by the final state interaction represent an isospin breaking effect, which is clearly seen in the band belonging to fitK$\chi_6$,  but is absent in the other Dalitz plot distributions, because these are shown in the isospin limit. Visibly,  $D_n=1+2\alpha(X_n^2+Y_n^2)+\ldots$ stays close to 1, with a negative value of the slope parameter $\alpha$. 
\section{Anatomy of the two-loop representation}\label{sec:Anatomy}

As discussed in Sec.~\ref{sec:Two loops}, elastic unitarity determines the NNLO representation of \chpt\ in terms of the one valid at NLO, up to a polynomial. The non-polynomial part does not contain any unknowns, but the polynomial does, in the form of the low-energy constants that occur in the effective Lagrangian at $O(p^6)$ -- for some of these, only crude theoretical estimates are available. Note that the two-loop representation is unique up to a {\it real} polynomial. To consistently compare the dispersive and chiral representations at $O(p^6)$ of the chiral expansion, the subtraction constants must be given the proper imaginary part. In particular, for the central solution, we need to consider the version $\mathrm{FitK\chi_6}$, so that the imaginary parts of the Taylor invariants do agree with those of the two-loop representation. 

\subsection{Final state interaction at two loops}\label{sec:Final state interaction}
We first investigate the non-polynomial part: How well does the two-loop
representation account for the final state interaction? To answer this
question, we construct the two-loop representation that
matches our central solution for the functions $M_0(s)$, $M_1(s)$,
$M_2(s)$ at low values of $s$ -- the only difference between the two
representations then arises from the fact that the dispersive one describes
the final state interaction effects more accurately. Finally, we will
compare the chiral representation obtained in this way with the one of
Bijnens and Ghorbani~\cite{Bijnens+2007} -- these two only differ in the
LECs of $O(p^6)$. 
\begin{figure*}[thb]\centering
\includegraphics[width=7.8cm]{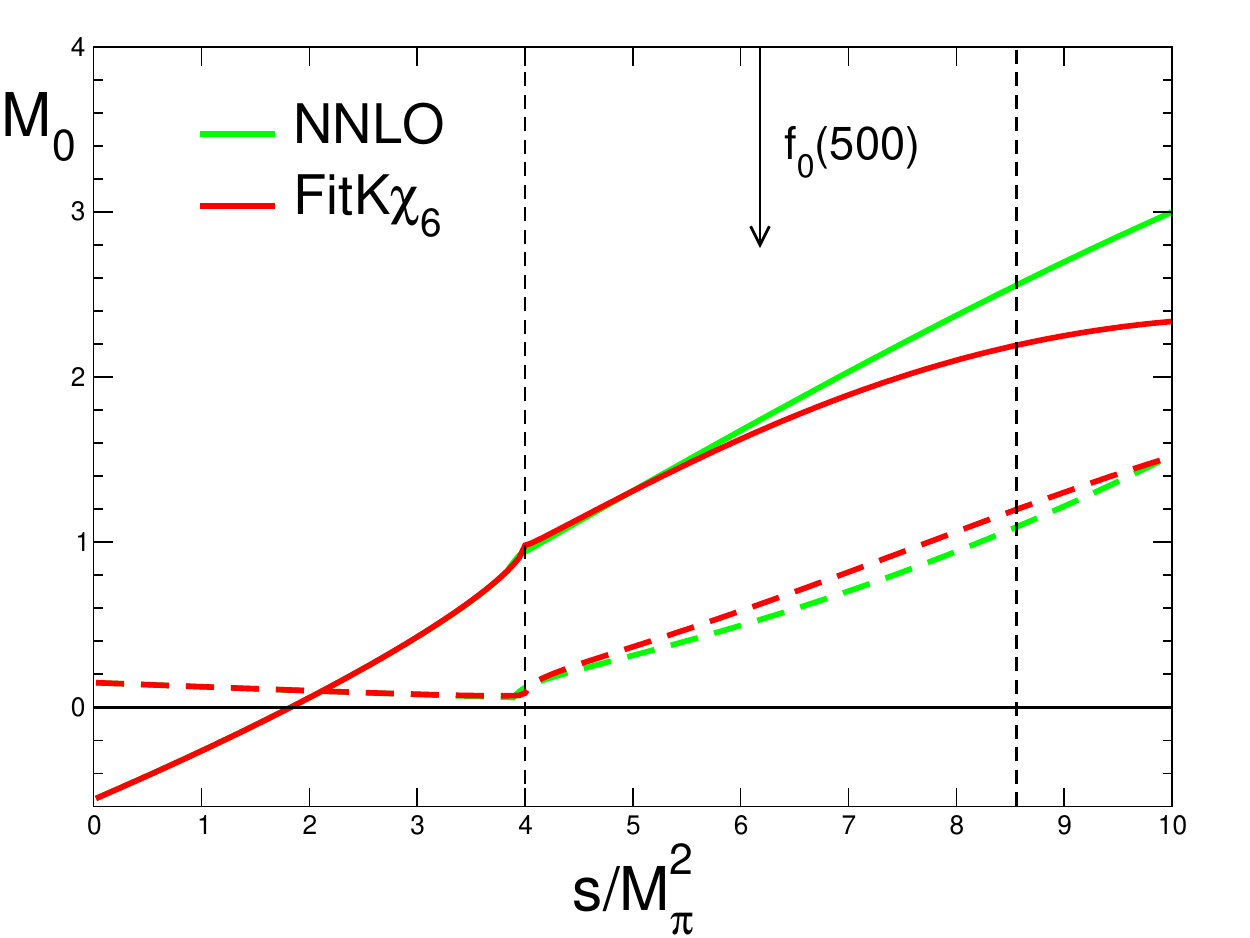}\hspace{0.4cm}\includegraphics[width=7.8cm]{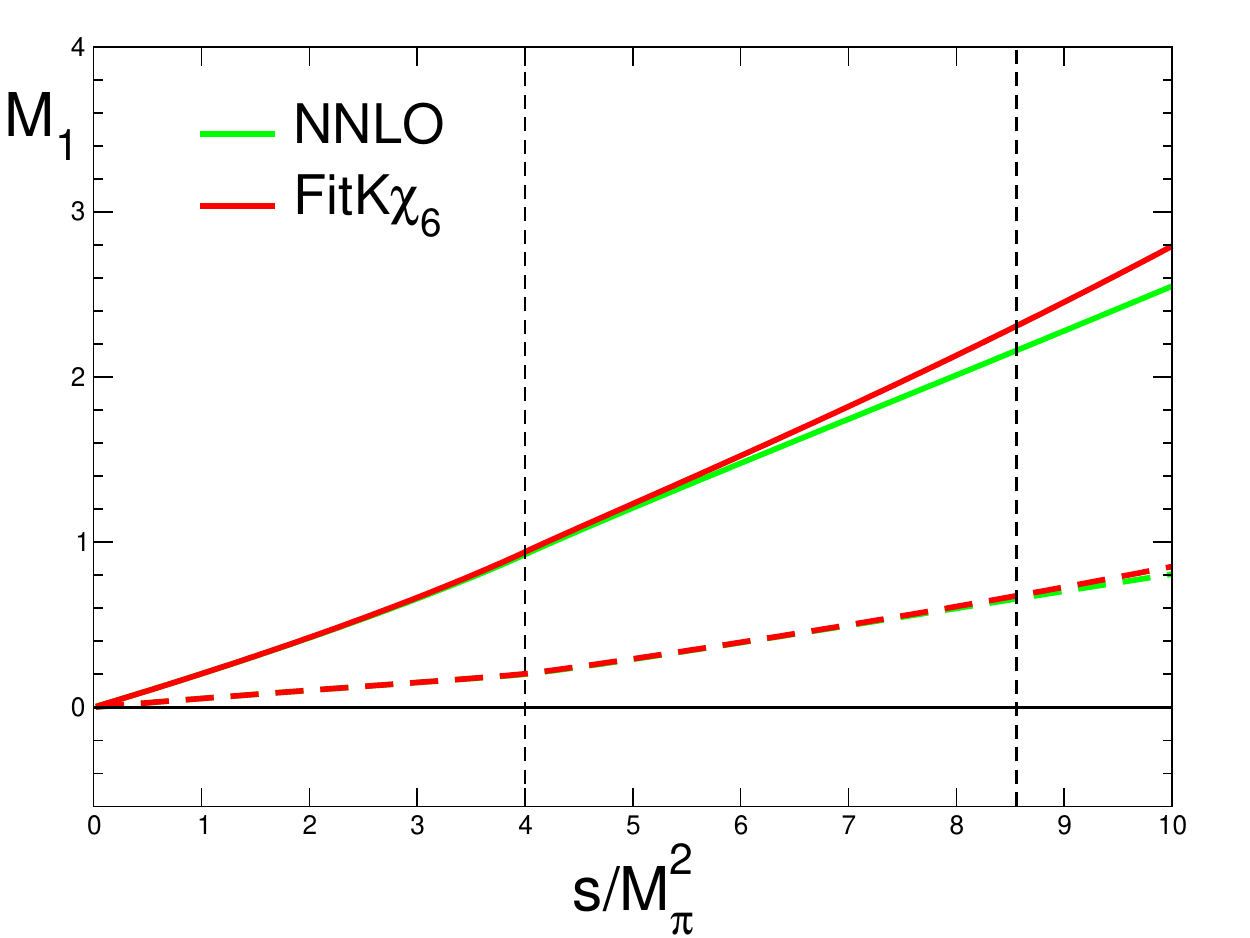}\\
\includegraphics[width=7.8cm]{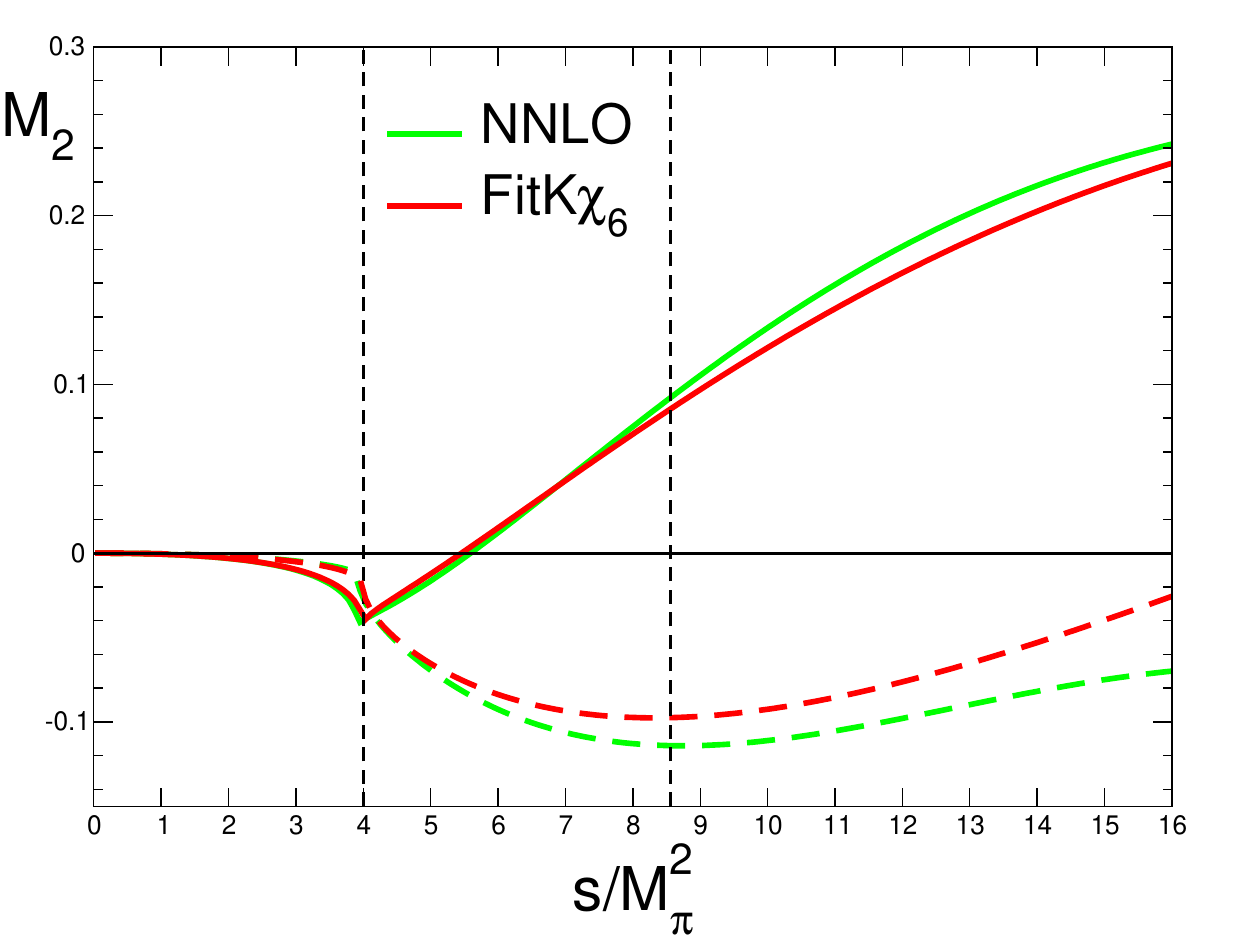}\hspace{0.4cm}\includegraphics[width=7.8cm]{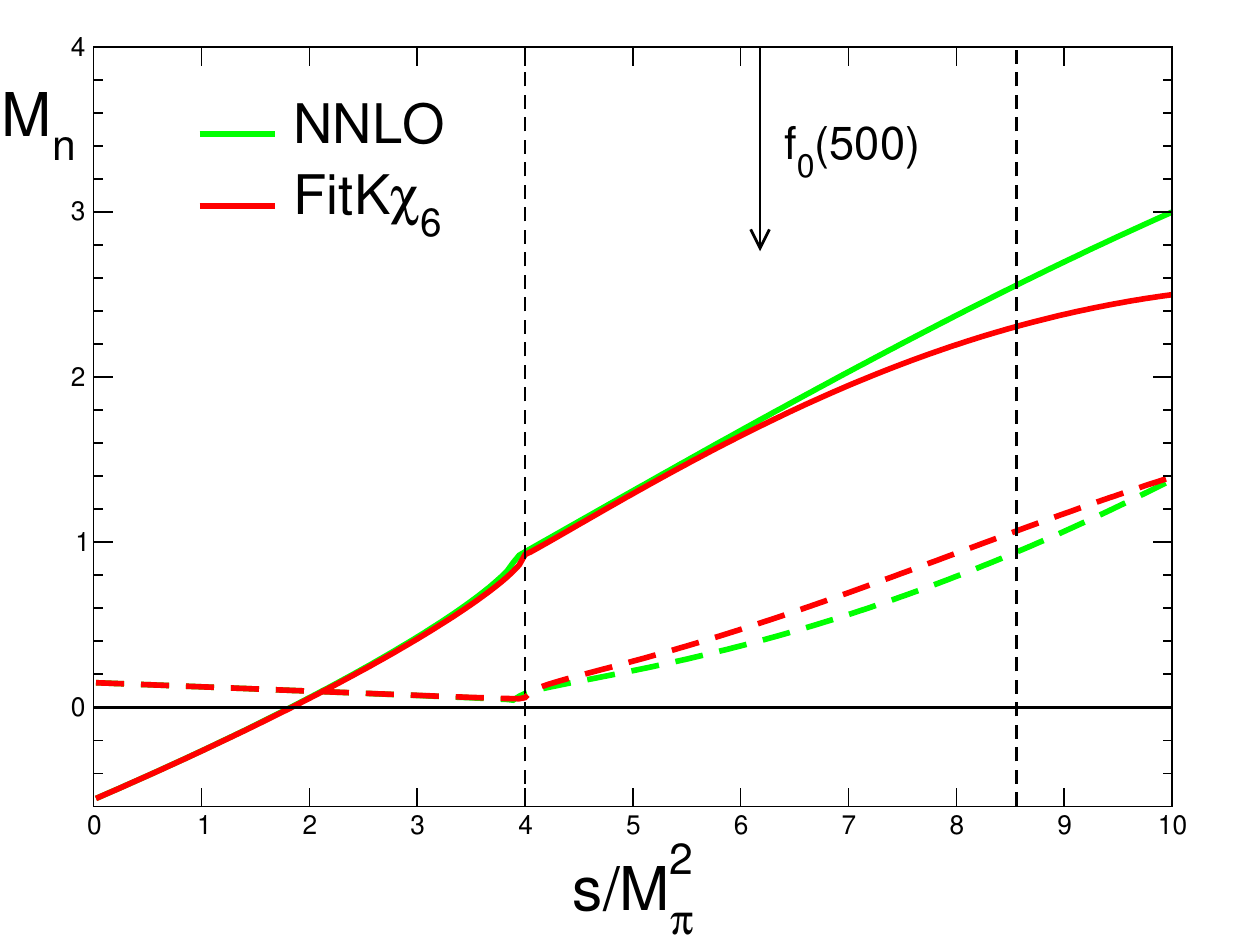}
\caption{Comparison of the central solution with the two-loop representation that matches it at low energies.\label{fig:MNNLOmatch}}\end{figure*}
In Sec.~\ref{sec:Matching}, we determined the solution of our integral equations which matches the one-loop representation of \chpt\ at low energies: fit$\chi_4$. We now extend this to the two-loop level, exploiting the fact that the contributions from the loop graphs are determined by the one-loop representation and do not involve any unknowns. For the explicit numerical evaluation of these contributions, we rely on the work of Bijnens and Ghorbani, more precisely on the code provided by these authors~\cite{BijnensCode}. Concerning the tree graph contributions, we make use of the fact that these are polynomials in the momenta. Instead of calculating the coefficients of the polynomials with the effective Lagrangian and then inserting the available estimates for the LECs contained therein, we determine the polynomial part in such a way that the amplitude matches our central solution at low energies. 
In the sum over the isospin components, the polynomial part contains six independent coefficients, which are in one-to-one correspondence with the Taylor invariants $K_0$, \ldots , $K_5$. In order to construct the two-loop representation that matches FitK$\chi_6$, we simply need to match these invariants. 

In contrast to the one-loop representation, where the Taylor coefficients are real, those of the two-loop representation have an imaginary part, which can only be matched if we allow the subtraction constants of the dispersive representation to be complex. Indeed, in the construction of the solution FitK$\chi_6$, we pinned the imaginary parts of the subtraction constants down with the requirement that the imaginary parts of the Taylor invariants agree with those obtained from the code~\cite{BijnensCode}, which are listed in Eq.~\eqref{eq:ImK}.  The two-loop representations of the functions $M_0(s)$, $M_1(s)$, $M_2(s)$ that match the solution FitK$\chi_6$ differ from those of Ref.~\cite{Bijnens+2007} only by a polynomial:
\begin{eqnarray}M_0^\mathrm{NNLO}(s)\al=\al M_0^\mathrm{BG}(s)+dA_0+dB_0\,s+ dC_0\,s^2+dD_0\,s^3\;,\nonumber\\
M_1^\mathrm{NNLO}(s)\al=\al M_1^\mathrm{BG}(s)+dA_1+dB_1\,s+ dC_1\, s^2\;,\\
M_2^\mathrm{NNLO}(s)\al=\al M_2^\mathrm{BG}(s)+dA_2+dB_2\,s+ dC_2\,s^2+dD_2\,s^3\;.\nonumber
\end{eqnarray}
The coefficients of the polynomial are given by the difference between the Taylor coefficients of the two representations, for instance:
\begin{equation}dA_0= A_0^\mathrm{K\chi_6}- A_0^\mathrm{BG}\end{equation}
and likewise for the remaining coefficients. Note that the differences are complex -- only for the Taylor invariants, the imaginary parts are the same. This property ensures that the quantity of physical interest, $M_c^\mathrm{NNLO}(s,t,u)$, which is given by the sum over the components, differs from $M_c^\mathrm{BG}(s,t,u)$ only by a real polynomial in the Mandelstam variables. The polynomial reflects the fact that the LECs of $O(p^6)$ are not the same for the two versions of the two-loop representation -- the contributions from these constants are real.

Fig.~\ref{fig:MNNLOmatch}  compares the isospin components of the two-loop representation with those of $\mathrm{FitK\chi_6}$.
Below threshold, the two representations can barely be distinguished from one another. The components with $I=1$ and $I=2$ of the two-loop representation closely follow those of the central solution even for $s>4M_\pi^2$ (note that the range shown for $M_2(s)$ is substantially wider than for the other components, because this is of interest in connection with the position of the Adler zero -- see below). In $M_0(s)$, however, a significant difference can be seen in the physical region. It implies that  the real part of the isospin combination relevant for the transition $\eta\rightarrow3\pi^0$, $M_n^\mathrm{NNLO}(s)=M_0^\mathrm{NNLO}(s)+\frac{4}{3}M_2^\mathrm{NNLO}(s)$ nearly follows a straight line. This answers the question raised above: The two-loop representation accounts sufficiently well for the final state interaction only for $s\lsim 5 M_\pi^2$. Above that energy, the lowest resonance of QCD, the $f_0(500)$, manifests itself.
The corresponding pole occurs on the second sheet, in the vicinity of $s_\mathrm{pole}\simeq (441-i \,272\, \MeV)^2\simeq 6.2-i \,12.3 \,M_\pi^2$~\cite{Caprini:2005zr,Pelaez:2015qba} (the arrows in Fig.~\ref{fig:MNNLOmatch} indicate the real part of the pole position). Although the resonance is very broad -- the pole is far away from the real axis -- the truncated expansion in powers of momentum cannot properly cope with it above $5M_\pi^2$, not even at NNLO.
\begin{table*}[thb]\centering
\begin{tabular}{lccccccc}
&$\Re K_0$&$\Re K_1$&$\Re K_2$&$\Re K_3$&$\Re K_4$&$\Re K_5$&$s_A$\rule{0.8em}{0em}\\
\hline
NNLO&1.176(53)&4.55(24) &  25.8(5.2)&   -3.6(1.9)&   90.8(5.2)&   52.3(6.9)&   1.33(14)$M_\pi^2$ \\
BG &1.27&3.88&37.2&-6.2&113(34)&73&1.17$M_\pi^2$\\
\hline
\end{tabular}
\caption{Comparison of the Taylor invariants belonging to the two-loop representation constructed in Sec.~\ref{sec:Final state interaction} with those of the two-loop representation of Bijnens and Ghorbani~\cite{Bijnens+2007}.\label{table:ReKBG}}
\end{table*}
 
As discussed in Sec.~\ref{sec:Neutral decay mode}, the curvature of the
function $M_n(s)$ determines the slope parameter $\alpha$ of the neutral
decay mode. Since the curvature of $M^\mathrm{NNLO}_n(s)$ nearly vanishes, 
the slope of this representation is very small -- numerically, we obtain
$\alpha^\mathrm{NNLO}= + 0.002$. In the neutral channel, the NNLO representation
of the Dalitz plot distribution can thus barely be distinguished from the horizontal line
in Fig.~\ref{fig:Match Dalitzn}, which indicates the tree level
result. This is lower than the value $\alpha= +0.011$ that belongs to the NLO curve, which is also shown
in Fig.~\ref{fig:Match Dalitzn}, or the two-loop estimate $\alpha=+0.013(32)$ given in
\cite{Bijnens+2007}, but the 
discrepancy with the experimental value $\alpha=-0.0318(15)$
\cite{Olive:2016xmw} is not removed. We conclude that a substantial part of
the discrepancy is due to the fact that the two-loop result does not fully
account for the enhancement of the final state interaction generated by the
resonance $f_0(500)$. Closely related aspects of the same problem were
discussed already earlier, by Schneider, Kubis and Ditsche 
(see in particular Sec.~4.3 of Ref.~\cite{Schneider+2011}).  

The Adler zero of $\Re M_n^\mathrm{NNLO}(s,t,u)$ occurs at $s_A=1.35(11)\,M_\pi^2$, remarkably close to the value $s_A=1.37(11)$  where the real part of FitK$\chi_6$ has its zero. By construction, the isospin components belonging to the two-loop approximation  $M^\mathrm{NNLO}(s,t,u)$ agree with those of the dispersive representation at small values of $s=u$, but as discussed in Sec.~\ref{sec:Adler zero at one loop}, the behaviour of the sum over the isospin components at small values of $s=u$ is not controlled exclusively by their behaviour in that region, but also depends on the properties of the comparatively small  component $\Re M_2(s)$ in the vicinity of $s = 16 M_\pi^2$. Fig.~\ref{fig:MNNLOmatch} shows that even there, the two-loop approximation follows the dispersive representation for $M_2(s)$ rather well. This explains why that approximation is rather accurate also in the vicinity of the Adler zero. 

The differences between the curves labeled Fit$\chi_6$ and NNLO in Fig.~\ref{fig:MNNLOmatch} yield an estimate for the size of those uncertainties of the two-loop representation that arise solely from the fact that it describes the final state interaction very well only at low energies. In particular, the two-loop representation for the dominating contribution, $M_0(s)$, represents an accurate approximation only in part of the physical region -- the Dalitz plot distribution is not reproduced well, neither in the charged channel, nor in the neutral one.

\subsection{Contribution from the low-energy constants at NNLO}\label{sec:LEC}

Finally, we compare the polynomial part of the amplitude of Bijnens and Ghorbani~\cite{Bijnens+2007} with the two-loop representation constructed in the preceding section. The numbers in the row NNLO of Table~\ref{table:ReKBG} represent central values and uncertainties of the Taylor invariants belonging to that representation -- by construction, these coincide with the invariants of the dispersive solution FitK$\chi_6$. The values in the row BG are obtained with the code~\cite{BijnensCode} mentioned earlier.

\begin{figure}[thb]\hspace{-1.2em}\includegraphics[width=8.8cm]{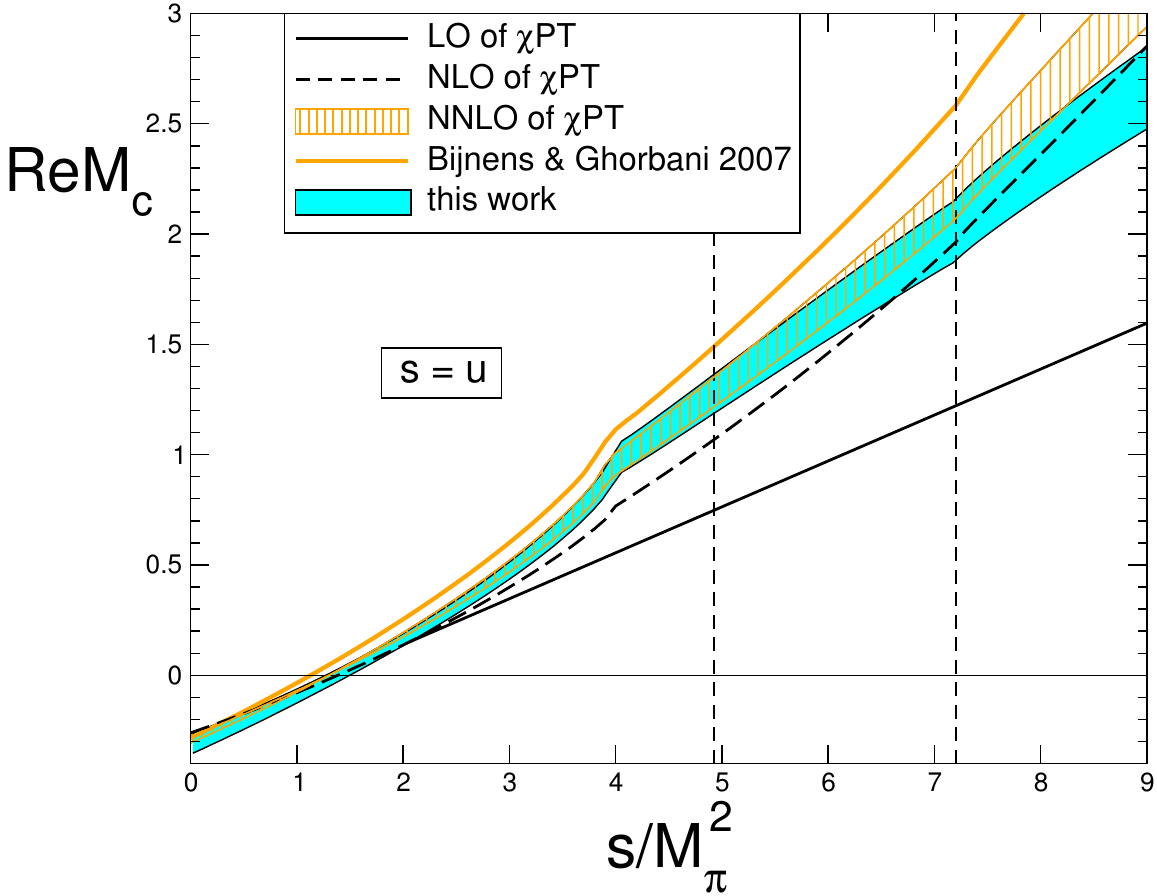}\caption{Comparison of our result with the representations based on \chpt\ at LO, NLO and NNLO (real part of the amplitude along the line $s=u$). While the first two orders of the chiral perturbation series are parameter free, the NNLO representation does involve a set of low-energy constants that are not determined by the symmetry properties of the theory. The band labeled NNLO is obtained by determining these experimentally as outlined in Sec.~\ref{sec:LEC}.  \label{fig:ReMsu}} 
\end{figure}

We recall that the experimental information about the Dalitz plot distribution exclusively concerns the relative size of the invariants, not the invariants themselves. The value quoted for $\Re K_0$ relies on theory, more precisely on the expansion of $K_0$  in powers of the masses of the three lightest quarks. This expansion starts with $K_0=1+O(m_\mathrm{quark})$.  As discussed in Sec.~\ref{sec:One loop}, the coefficient of the next-to-leading term of the 
expansion can be worked out from the one-loop representation of the transition amplitude, which does not involve any unknowns. Numerically, the correction is of typical size: $K_0=1+0.176+O(m_\mathrm{quark}^2)$. The error quoted in Table~\ref{table:ReKBG} is based on the estimate of the higher order contributions described in Sec.~\ref{sec:One loop}.
The table shows that the value obtained for $\Re K_0$ from the estimates used for the LECs in~\cite{Bijnens+2007} is outside our range (disregarding the uncertainty in the number 1.27, the difference amounts to $1.7\hspace{0.05em}\sigma$). Since $K_0$ is not plagued by infrared singularities -- in particular, this invariant remains finite in the limit $M_\pi\to 0$ -- we see no reason why it should pick up unusually large corrections from higher orders and stick to the value quoted in the table. 
\\
\indent The value of $K_0$ is important for the determination of the kaon mass difference and of the quark mass ratio $Q$, to be discussed in Sec.~\ref{sec:MKQ}, but in the present section, we compare the chiral and dispersive representations for the Dalitz plot distribution of the charged channel, the slope $\alpha$ of the $Z$-distribution in the neutral channel and the position of the Adler zero with our central solution -- these quantities only involve the ratios $K_1/K_0,\ldots,K_5/K_0$. We set $\Re K_0=1.176$ and fix the imaginary parts with the two-loop representation of Bijnens and Ghorbani~\cite{Bijnens+2007}. \\
\indent As pointed out in Sec.~\ref{sec:Two loops}, the Taylor invariant $K_4$ does not get any contribution from the LECs of $O(p^6)$. 
The corresponding entry for $\Re K_4$ in the table includes our uncertainty estimate from Eq.~\eqref{eq:ReK4}. The value obtained with our central solution is indeed within the range of this prediction (the imaginary parts are identical by construction). $\Re K_3$  also agrees within the uncertainties attached to our central solution, but for $\Re K_1$, $\Re K_2$ and $\Re K_5$, the two results differ by up to $2\hspace{0.05em}\sigma$. We conclude that the values of some of the LECs used in~\cite{Bijnens+2007} are not consistent with the experimental information on $\eta\to3\pi$ available today.\\
\indent As discussed in Sec.~\ref{sec:Final state interaction}, a direct comparison of the two-loop representation with the data in the physical region is not meaningful -- the $f_0(500)$ is the stumbling block. Dispersion theory is needed to establish a controlled connection between the region that is accessible to experiment and the domain $s\lsim 5M_\pi^2$, where the two-loop approximation for $M_0(s)$ is sufficiently accurate. \\
\indent The Taylor invariants provide the bridge. The dispersive representation reliably determines the behaviour of the amplitude in the physical region in terms of these. Their imaginary parts are known to NNLO of the chiral expansion. Using this, and keeping $\Re K_0$ fixed at the central value, the KLOE data on the Dalitz plot distribution of $\eta\to\pi^+\pi^-\pi^0$ imply that the real parts of the remaining five invariants are in the range indicated in the row NNLO of Table~\ref{table:ReKBG}. \\
\indent As already mentioned, unitarity fixes the two-loop representation for $M_c(s,t,u)$ in terms of known quantities up to a real polynomial. The polynomial contains six independent coefficients that are in one-to-one correspondence with the real parts of the Taylor invariants $K_0,\ldots,K_5$. In the representation of the amplitude obtained with \chpt, the Taylor invariants represent linear combinations of some of the LECs of $O(p^6)$. In particular, those relevant for the scalar channel with $I=0$ contribute, which are notoriously difficult to estimate because the contribution from the $f_0(500)$ to the corresponding spectral functions is not easily accounted for. The experimental information about the Taylor invariants and their correlations obtained from our analysis should make it possible to reliably determine these particular couplings, which also enter in many other applications of \chpt.  An update of the LECs of $\chi$PT (for a recent 
review, see~\cite{Bijnens:2014lea}) that accounts for this information would be of considerable interest, but is beyond the scope of the present work.\\
\indent Fig.~\ref{fig:ReMsu} compares our central solution,  fitK$\chi_6$, with the results obtained on the basis of \chpt\ (real part, along the line $s=u$ and in the isospin limit: $m_u=m_d$, $e = 0$). The error band attached to the NNLO representation is obtained with the calculation described in Sec.~\ref{sec:Final state interaction}, which relies on the KLOE data. It concerns the two-loop representation as such -- the contributions from higher orders, which grow with the energy, are not accounted for. The orange solid line corresponds to the amplitude of Bijnens and Ghorbani~\cite{Bijnens+2007}, which exclusively differs in the values of the LECs. 
\section{Consequences for $\mathbf\eta\to3\pi^0$}\label{sec:etato3pi0}
\subsection{Branching ratio}\label{sec:Branching ratio}
The rates $\Gamma_{\eta\to\pi^+\pi^-\pi^0}$ and $\Gamma_{\eta\to 3 \pi^0}$ involve the overall normalization factor $N$, as well as the constant $K_0$ that normalizes the amplitudes $M_c(s,t,u)$ and $M_n(s,t,u)$, but in the branching ratio,
\begin{equation}\label{eq:B definition}B=\frac{\Gamma_{\eta\to 3 \pi^0}}{\Gamma_{\eta\to\pi^+\pi^-\pi^0}}\;,\end{equation}
these quantities drop out. Hence we obtain a parameter free prediction for $B$.

In the branching ratio, the uncertainties of the dispersive representation also cancel out 
almost completely -- not only the errors occurring in the determination of the 
subtraction constants, but also those generated by the uncertainties in the phase shifts.
The main source of error in $B$ arises from isospin breaking. In particular,
the mass difference between the charged and neutral pions generates a substantial
difference in shape and size of the region over which the square of the amplitude
must be integrated to calculate the rate. As the corrections for the charged and neutral decay modes are of opposite sign, 
the branching ratio is affected quite strongly -- they dominate our estimate of the
error:
\vspace{-0.11cm}
 \begin{equation}\label{eq:B value}B=1.44(4)\;.\end{equation} 
The experimental values given by the
 Particle Data Group are $B = 1.426(26)$ [`our fit'] and $B = 1.48(5)$ [`our
 average']~\cite{Olive:2016xmw}. The comparison with our result in~\eqref{eq:B value}
 shows that the value predicted for the decay rate of the
 neutral mode (on the basis of Dalitz plot distribution and decay rate of
 the charged mode) is in good agreement with experiment. This provides a very strong test of
 the approximations used to account for isospin breaking.   
\subsection{Dispersive representation of the Dalitz plot distribution}
Equation~\eqref{eq:Mn(s,t,u)} shows that, in the isospin limit, the amplitude for the neutral decay mode is determined by the one for the charged mode. With the approximate formulae~\eqref{eq:Mphys}, this statement remains true even in the presence of isospin breaking. The physical amplitude $M_n^\mathrm{phys}(s_n,t_n,u_n)$ is expressed as the product of a factor $K_n(s_n,t_n,u_n)$ that stems from the one-loop representation and a factor $\Mtilde_n(s_n,t_n,u_n)$, that represents the isospin symmetric dispersive amplitude, evaluated with the kinematic map. In this approximation, the Dalitz plot distribution of the neutral mode is  given by  
\begin{equation}\label{eq:Dnphys}D^\mathrm{phys}_n(X_n,Y_n)=\rule[-1em]{0.04em}{2.5em}\frac{M^\mathrm{phys}_n(X_n,Y_n)}{M^\mathrm{phys}_n(0,0)}\rule[-1em]{0.04em}{2.5em}^{\,2}\;, \end{equation}
where $M^\mathrm{phys}_n(X_n,Y_n)$ is obtained from $M_n^\mathrm{phys}(s_n,t_n,u_n)$ by expressing the independent Mandelstam variables $s_n$ and $\tau_n=t_n-u_n$ in terms of the Dalitz variables $X_n$ and $Y_n$:
\begin{eqnarray}\label{eq:XnYn}s_n\al=\al -\frac{2}{3}M_\eta\,(M_\eta-3M_{\pi^0})\, Y_n+ \frac{1}{3}(M_\eta^2+3M_{\pi^0}^2 )\\
\tau_n\al=\al -\frac{2}{\sqrt{3}}M_\eta\,(M_\eta-3M_{\pi^0})\,X_n\;.\nonumber\end{eqnarray}
This implies that the central solution fitK$\chi_6$, which we constructed in Sec.~\ref{sec:Dalitz plot},  yields a parameter free prediction for the Dalitz plot distribution of the decay $\eta\to 3\pi^0$, together with an estimate of the uncertainties to be attached to this prediction.
 
The main difference compared to the charged channel is that the Dalitz plot distribution is nearly flat: The experimental values differ from the current algebra prediction, $D_n=1$, only by a few percent. This limits the precision not only of the experimental determination, but also of the theoretical prediction for the parameters that describe the deviation from unity. A further difference compared to the charged channel arises from the fact that a single physical decay into three neutral pions is mapped into six distinct points of the physical region, so that the values of $D_n$ on a sextant of phase space fully determine the distribution (compare Sec.~\ref{sec:Kinematics}). Accordingly,  the Dalitz plot distribution of the decay $\eta\to3\pi^0$ is invariant under $120^\circ$ rotations around the center of  the $(X_n,Y_n)$ plane as well as under reflections at the $Y_n$-axis. Expressed in terms of radial coordinates,
\begin{equation}\label{eq:Z phi}X_n=\sqrt{Z}\cos\varphi\;,\quad Y_n=\sqrt{Z}\sin\varphi  \;,\quad Z\equiv X_n^2+Y_n^2\;,\end{equation}
the transition amplitude is periodic in $\varphi$ with period  $2\pi/3$ and even under $\varphi\to \pi\!-\!\varphi$. 
\subsection{Slope}\label{sec:Slope}

As discussed in Sec.~\ref{sec:Neutral decay mode}, the symmetry of the transition amplitude with respect to interchange of the Mandelstam variables implies that the expansion around the center of the physical region starts with a quadratic term. Expressed in the variables $X_n$ and $Y_n$, this term is proportional to $X_n^2+Y_n^2=Z$:
\begin{equation}M_n(X_n,Y_n)=M_n(0,0)\{1+\overline{\alpha}\,Z+\ldots\}\;.\end{equation}
Only the real part of the coefficient, $\alpha=\Re \overline{\alpha}$,  shows up in the Dalitz plot distribution: 
\begin{equation}D_n(X_n,Y_n)=1+2\,\alpha\, Z+\ldots\end{equation} 
For our central solution (fitK$\chi_6$), we obtain
\begin{equation}\label{eq:alpha}\alpha=-0.0303(12)\;.\end{equation}
The uncertainty is dominated by the Gaussian error,
but includes our estimates for the noise generated by all sources that play a role in our analysis. The result is consistent with the experimental value $\alpha=-0.0318(15)$ quoted by the Particle Data Group~\cite{Olive:2016xmw}. This solves a long-standing puzzle: Our dispersive framework  not only yields the proper sign of the slope, but predicts a value that is consistent with experiment. 

Since $\alpha$ is very small, details of the evaluation matter. In particular, as demonstrated in Sec.~\ref{sec:Neutral decay mode},  $\alpha$ is very sensitive to the final state interaction. As an example, consider isospin breaking. Although the isospin breaking effects in the decay $\eta\to 3\pi^0$ are small, dropping them in the calculation of the slope changes the central value of the prediction from $-0.0303$ to $-0.0327$.  Details of the evaluation also matter in the analysis of the data: The number quoted in~\eqref{eq:alpha} is the derivative of the $Z$-distribution at $Z=0$. In the past, the experimental determination of the slope was instead determined by fitting the data with the linear formula $1+2\hspace{0.3mm}\alpha Z$ on a finite range of $Z$-values. The sensitivity of the result to this range and to the fact that -- at the accuracy reached -- the curvature of the distribution cannot be neglected will be discussed in Sec.~\ref{sec:polynomial}. 
\vspace{-0.5cm}
\subsection{Experiment\label{sec:experiment neutral channel}}
The  experimental determination of the slope $\alpha$
has an even longer recent history than that of the measurement of the
Dalitz plot in the charged channel: A list of all the measurements and the
references can be found in Table~\ref{tab:alpha}.

  \begin{table}[thb]
	\begin{tabular}{llr} & \hspace{2em} $\alpha$	& \\ \hline
 		GAMS-2000 (1984)	\rule{0cm}{1.1em}				&$-$0.022(23)					&\cite{Alde+1984} \\ 
 		Crystal Barrel@LEAR (1998)			&$-$0.052(20)					&\cite{Abele+1998} \\ 
 		Crystal Ball@BNL (2001)				&$-$0.031(4)					&\cite{Tippens+2001} \\ 
 		SND (2001)						&$-$0.010(23)					&\cite{Achasov+2001} \\ 
 		WASA@CELSIUS (2007)				&$-$0.026(14)					&\cite{Bashkanov+2007} \\ 
 		WASA@COSY (2008)				&$-$0.027(9)					&\cite{Adolph+2009} \\ 
 		Crystal Ball@MAMI-B (2009)			&$-$0.032(3)					&\cite{Unverzagt+2009} \\ 
 		Crystal Ball@MAMI-C (2009)			&$-$0.032(3)					&\cite{Prakhov+2009} \\ 
		KLOE (2010)						&$-$0.0301($_{-49}^{+41}$)		&\cite{Ambrosino+2010} \\ 
		BESIII (2015)						&$-$0.055(15)					&\cite{Ablikim:2015cmz} \\ 
		PDG average						&$-$0.0318(15)					&\cite{Olive:2016xmw} \\ 
		Crystal Ball@MAMI-A2 (2018)			&$-$0.0265(10)(9)				&\cite{Prakhov+2018}\\
		\hline
 		Kambor et al.\ (1996) 				&$-0.007$&\cite{Kambor+1996} \\ 
		Bijnens \& Gasser (2002)				&$-$0.007					&\cite{Bijnens+2002} \\ 
		Bijnens \& Ghorbani (2007)			&\hspace{0.8em}0.013(32)	&\cite{Bijnens+2007}\\
 		Schneider et al.\ (2011)				&$-$0.025(5)				&\cite{Schneider+2011} \\ 
		Kampf et al.\ (2011)					&$-$0.044(4)				&\cite{Kampf+2011} \\ 
 		JPAC (2016)					&$-$0.025(4)				&\cite{Guo:2016wsi} \\ 
 		Albaladejo \& Moussallam (2017)	 	&$-$0.0337(12)				&\cite{Albaladejo+2017} \\ 
		this work 							&$-$0.0303(12)				&\\
		\hline
	\end{tabular}  
	\caption{Various experimental and theoretical results for the slope parameter $\alpha$. We have added systematic
and statistical uncertainties in quadrature. The PDG average is based on the experimental results listed here. For comparison, the above numbers are visualized in Fig.~\ref{fig:alpha}.}
	\label{tab:alpha}
\end{table}

\begin{figure}[thb]\suppressfloats\centering\includegraphics[width=8.2cm]{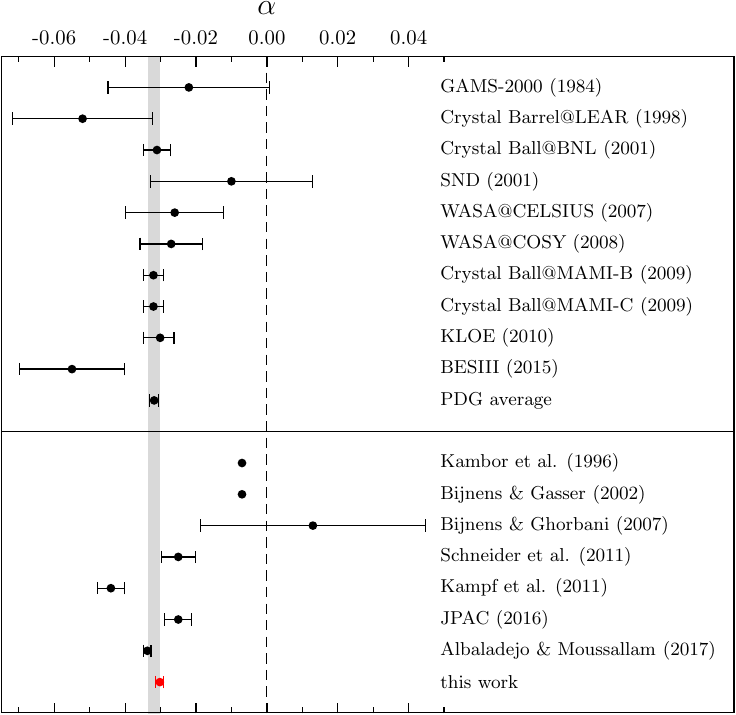}\caption{Comparison of experimental and theoretical results for the slope $\alpha$ of $\eta\to 3\pi^0$. \label{fig:alpha}}\end{figure}
\begin{figure}[thb]\centering\includegraphics[width=8cm]{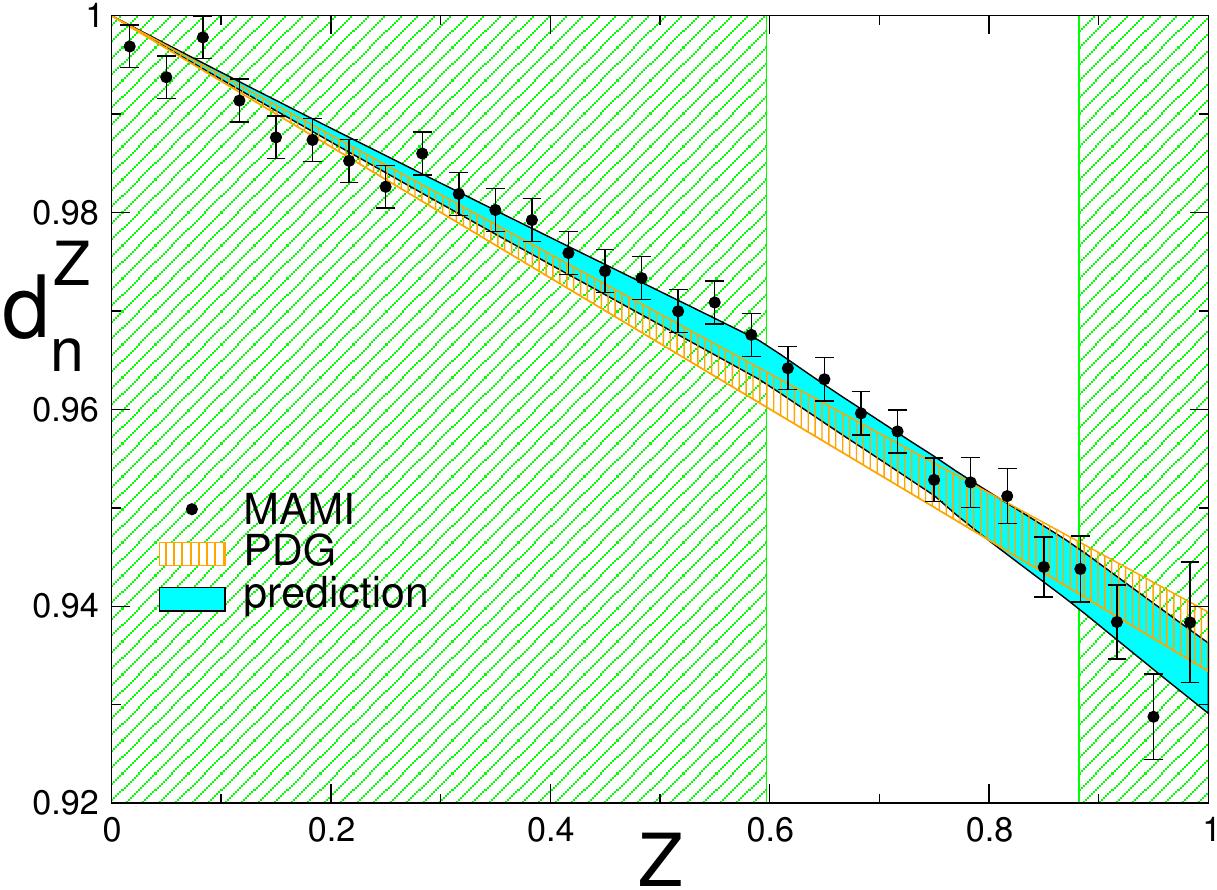}\caption{Prediction obtained from 
the KLOE measurements of $\eta\to\pi^+\pi^-\pi^0$~\cite{KLOE:2016qvh} for the $Z$-distribution of
the decay for $\eta\to 3\pi^0$ compared with the most recent MAMI results \cite{Prakhov+2018}. The shaded areas indicate the region where the cusps generated by the final state interaction do not show up.\label{fig:z-distribution}}\end{figure}

The most precise determination of the 
Dalitz plot distribution and its slope parameter $\alpha$ is based on the
data collected at the Mainz Microtron: $1.8$ million events were
analyzed at MAMI-B~\cite{Unverzagt+2009}, another three million $\eta\to 3\pi^0$
decays were collected at  MAMI-C~\cite{Prakhov+2009} and, very recently,
the A2 Collaboration came up with an update based on altogether  7 million 
events \cite{Prakhov+2018}. KLOE has performed such a measurement
too~\cite{Ambrosino+2010}, on the basis of about half a million 
events. The  PDG average $\alpha=-0.0318(15)$~\cite{Olive:2016xmw}
is largely dominated by the MAMI measurements.  As discussed in the preceding
section, the result for $\alpha$ is sensitive to the range over which
the data are approximated with the linear formula $1+2\alpha Z$. 
 A more controlled determination that does not rely on this approximation
became possible only very recently \cite{Prakhov+2018}. 
We will discuss it in detail in Sec.~\ref{sec:Fits to MAMI}.

\subsection{$Z$-distribution}\label{sec:Z distribution}

The $Z$-distribution is obtained by averaging the Dalitz plot distribution over the angle $\varphi$. As mentioned above, the events collected in one sextant of phase space fully determine the distribution. We consider the sextant with $30^\circ<\varphi<90^\circ$, i.e.~the upper one of the two sectors  between the lines $s=t$ and $t=u$ (these are shown as dashed red and black lines in the right panel of Fig.~\ref{fig:boundary}). If $Z$ is below the value 
\begin{equation}\label{eq:Zcrit} Z^\mathrm{crit}=(M_\eta+3M_{\pi^0})^2/4M_\eta^2 \simeq 0.756\;,\end{equation}
the circle $Z= \mathrm{constant}$ runs inside the physical region, so that the average is given by 
\begin{equation}d_n^Z(Z)=\frac{1}{(\varphi_2-\varphi_1})\int_{\varphi_1}^{\varphi_2}\hspace{-0.3em} d\varphi \, D^\mathrm{phys}_n(\sqrt{Z}\cos\varphi,\sqrt{Z}\sin \varphi)\;,\end{equation}
with $\varphi_1=\frac{1}{6}\pi$ and $\varphi_2=\frac{1}{2}\pi$. For $Z>Z^\mathrm{crit}$, the interval relevant for the average shrinks.
The lower end stays at  $\varphi_1=\frac{1}{6}\pi$ , but the upper end is lowered to the value of $\varphi$, where the circle $Z= \mbox{constant}$ intersects the boundary of the physical region, which is determined by 
\begin{eqnarray}\sin(3\,\varphi_2)=\frac{3\,Z(M_\eta^2+3M_{\pi^0}^2)-(M_\eta+3M_{\pi^0})^2}{2\,Z^\frac{3}{2}M_\eta(M_\eta-3M_{\pi^0})}\;,\nonumber \\
\mbox{$\frac{1}{6}$}\pi\leq\varphi_2\leq\mbox{$\frac{1}{2}$}\pi\;.
\end{eqnarray}
\begin{figure}[thb]\centering\includegraphics[width=8cm]{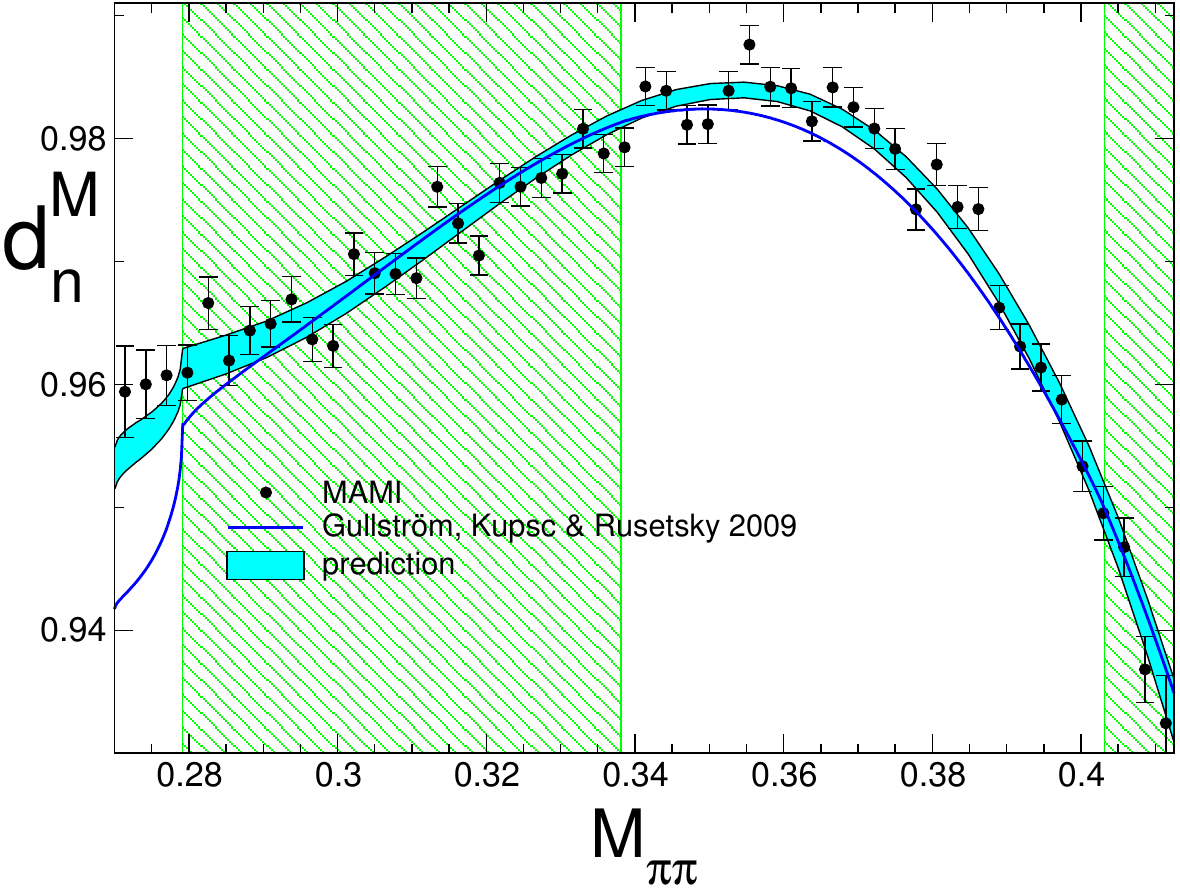}
\caption{Distribution in the variable $\mathrm{M}_{\pi\pi}=\sqrt{s}$ (GeV units). Prediction obtained from 
the KLOE measurements of $\eta\to\pi^+\pi^-\pi^0$~\cite{KLOE:2016qvh} compared with the MAMI results on $\eta\to 3\pi^0$~
\cite{Prakhov+2018}.  The shaded areas indicate the cusp-free regions.\label{fig:s-distribution}}
\end{figure}

The band in Fig.~\ref{fig:z-distribution} shows the result obtained for the $Z$-distribution
from our central solution, $\mathrm{fitK\chi_6}$. The width of the band represents the uncertainties in $d_n^Z$,
which are worked out as described in Sec.~\ref{sec:Error analysis}. The data points represent the $Z$-distribution obtained by the A2 collaboration at MAMI \cite{Prakhov+2018}. In earlier accounts of the data collected at MAMI, the normalization of the $Z$-distribution was fixed by fitting the data with the linear approximation, $d_n^Z=1+2\alpha Z$, but at the accuracy reached, this is not legitimate any more, because the curvature cannot be neglected. In Ref. \cite{Prakhov+2018}, the normalization of the $Z$-distribution is left open.  When comparing these data with our prediction,  we multiply the observed distribution by the factor $\Lambda $, which is treated as a free parameter. Visibly, the resulting normalized distribution, $\Lambda \,d_n^\mathrm{Z\; exp}$,  is in excellent agreement with the prediction. Quantitatively, we obtain $\Lambda =0.974$, $\chi^2=24.9$  for 30 data points and one free parameter.   

\subsection{$M$-distribution}
Fig.~\ref{fig:s-distribution} shows  the distribution over the center-of-mass energy of one of the pion pairs in the final state, which we denote by $M_{\pi\pi}$. It is given by the mean value of $D_n^\mathrm{phys}(X_n,Y_n)$ over the variable $X_n$ at the fixed value of $Y_n$ that belongs to $M_{\pi\pi}=\sqrt{s}$: 
\begin{equation}\label{eq:dM}d_n^M (M_{\pi\pi})=\frac{1}{X_n^\mathrm{max}}\int_0^{X_n^\mathrm{max}}\hspace{-1.5em}dX_n\,D_n^\mathrm{phys}(X_n,Y_n)\;.\end{equation}
We refer to $d_n^M$ as the $M$-distribution. The data points represent the MAMI results (Runs I and II combined)~\cite{Prakhov+2018}, while the band indicates the prediction obtained on the basis of the KLOE data for the decay $\eta\rightarrow\pi^+\pi^-\pi^0$.  In contrast to the distribution in the variable $Z$, which barely shows any structure at all, the prediction for the $M$-distribution  clearly exhibits a cusp at $M_{\pi\pi}=2M_{\pi^+}$. The data, however, do not show any sign of such a cusp. We return to this discrepancy in Sec.~\ref{sec:Fits to MAMI}, where we discuss various fits to the MAMI data. The figure also indicates the $M$-distribution obtained in Ref.~\cite{Gullstrom:2008sy} on the basis of the nonrelativistic effective theory. For a brief discussion of this approach, we refer to Sec.~\ref{sec:Nonrelativistic}. 
 
\subsection{Polynomial approximation\label{sec:polynomial}}
Bose  statistics interrelates the coefficients of the expansion in powers of $X_n$ and $Y_n$: Up to and including quartic terms, the expansion takes the form\footnote{We stick to the notation introduced by Schneider et al.~\cite{Schneider+2011}}
\begin{eqnarray}\label{eq:DnTaylor}D_n^\mathrm{poly}(X_n,Y_n)\al=\al 1+2\hspace{0.3mm}\alpha (X_n^2+Y_n^2)+2\hspace{0.3mm}\beta\,(3X_n^2Y_n-Y_n^3) \nonumber\\
\al \al + 2\hspace{0.3mm}\gamma\, (X_n^2+Y_n^2)^2 \\
\al=\al 1+2\hspace{0.3mm}\alpha\,Z+2\hspace{0.3mm}\beta\,Z^\frac{3}{2}\sin(3\,\varphi)+2\hspace{0.3mm}\gamma\,Z^2 \;.\nonumber \end{eqnarray}
The analogous approximation relevant for the charged decay mode was discussed in Sec.~\ref{sec:Experiment}. There is a significant difference between the two channels: Instead of the 5 independent coefficients $a$, $b$, $d$, $f$, $g$ needed if all terms up to third order are retained in the charged channel, the two coefficients $\alpha$, $\beta$ suffice in the neutral channel. At the next order of the expansion, $D_c$ contains the three independent terms $X_c^4$, $X_c^2Y_c^2$, $Y_c^4$, while the symmetry under exchange of the three particles only allows a single contribution in $D_n$:  $ \gamma\,(X_n^2+Y_n^2)^2$.    
 
In the neutral channel, the presence of cusps in the physical region implies that a para\-metrization of the Dalitz plot distribution in terms of a polynomial in the variables $X_n,Y_n$ is limited to values of $Z$ below 
\begin{equation}\label{eq:Zcusp}Z^\mathrm{cusp}=\left(\frac{M_\eta^2-12M_{\pi^+}^2+3M_{\pi^0}^2}{2M_\eta(M_\eta-3M_{\pi^0})}\right)^2\simeq 0.597\;.\end{equation}
For $Z>Z^\mathrm{cusp}$, the square root singularities generated by the virtual transition  $\eta\to\pi^+\pi^-\pi^0\to 3\pi^0$ need to be accounted for, but below this value of $Z$, only  the coefficients $\alpha$ and $ \gamma$ contribute to the $Z$-distribution -- the angular average of the term proportional to $ \beta \sin(3\,\varphi)$ vanishes below $Z^\mathrm{cusp}$: 
\begin{equation}\label{eq:dnZ}d_n^Z(Z) = 1 +2\hspace{0.3mm}\alpha Z+ 2\hspace{0.3mm}\gamma Z^2\;,\quad Z < Z^\mathrm{cusp}\;.\end{equation}
In Fig.~\ref{fig:z-distribution}, the left shaded region corresponds to the range $0<Z<Z^\mathrm{cusp}$. In this region, the $Z$-distribution is very well described by a straight line: Evidently, the coefficient $\gamma$, which measures the curvature, is very small. The same figure also shows that the slope changes at $Z=Z^\mathrm{cusp}\simeq 0.597$, on account of the contributions from the cusps. In the $Z$-distribution, the term proportional to $\beta$ only manifests itself for $Z>Z^\mathrm{crit}\simeq 0.756$, but it does affect the $M$-distribution, even in the region above the cusp, \mbox{$2M_{\pi^+}<M_{\pi\pi}<0.338\, \GeV$}. 

Minimizing the square of the difference between the polynomial \eqref{eq:DnTaylor}  and the Dalitz plot distribution of our central solution on the disk $Z<Z^\mathrm{cusp}$, we obtain the following polynomial approximation:
\begin{eqnarray}\label{eq:polyKchi6}\mathrm{fitK}\chi_6:\quad \alpha \al = \al -0.0307(17)\;,\quad \beta=-0.0052(5)\,, \nonumber \\
\gamma \al = \al 0.0019(3)\;.\end{eqnarray}
where the errors  cover all sources of uncertainty encountered in the dispersive analysis.
The  polynomial approximation represents our result remarkably well: In the region $Z<Z^\mathrm{cusp}$, the difference between $D_n^\mathrm{poly}$ and the Dalitz plot distribution obtained from our central solution of the dispersion relations (corrected for isospin breaking effects) is below 0.2 permille. Within errors, the result for $\alpha$ agrees with the one obtained for the quadratic term of the Taylor series in the variables $X_n$, $Y_n$ in \eqref{eq:alpha}. This demonstrates that the slope of the $Z$-distribution at $Z=0$ can accurately be measured by fitting the observed Dalitz plot distribution on the disk $Z\leq Z^\mathrm{cusp}$ with the formula \eqref{eq:DnTaylor}.   

\subsection{Strength of the cusps\label{sec:cusp}}
The polynomial approximation \eqref{eq:DnTaylor} is adequate only in the singularity-free part of the physical region. We now turn to the remainder, $Z>Z^\mathrm{cusp}$, where the cusps do manifest themselves. 
The pioneering work of  Budini, Fonda and Cabibbo~\cite{Budini+1961,Cabibbo:2004gq} on the physics of the cusps occurring in the decays $K^+\rightarrow \pi^+\pi^0\pi^0$ and $K_L\rightarrow 3\pi^0$ and the subsequent thorough analysis in~\cite{Cabibbo:2005ez,Gamiz:2006km,Colangelo+2006a,Bissegger:2007yq,Bissegger:2008ff,Gasser:2011ju} led to a very satisfactory understanding of the phenomenon.  As shown in~\cite{Colangelo+2006a,Bissegger:2007yq,Bissegger:2008ff,Gasser:2011ju}, it can be analyzed by means of nonrelativistic effective theory. Indeed, the precision of the data on kaon decays even allows a determination of $\pi\pi$ scattering lengths~\cite{Colangelo+2006a,Bissegger:2007yq,Bissegger:2008ff,Gasser:2011ju,Cabibbo:2004gq,Cabibbo:2005ez,Batley:2000zz}.  The situation for $\eta\to 3\pi^0$ is essentially the same as for $K_L\to 3 \pi^0$, but the knowledge is much more limited, both experimentally and theoretically. The work reported 
in two theoretical investigations~\cite{Gullstrom:2008sy,Schneider+2011} will briefly be discussed in Sec.~\ref{sec:Nonrelativistic}.

The branch cut required by unitarity is of the square-root type: The expansion of the function $M_n(s)$ around the point 
$s=4M_{\pi^+}^2$ contains a term proportional to  $\sqrt{4M_{\pi^+}^2-s}$, which changes from real to imaginary when $s$ passes through this point. In the $M$-distribution, this term is responsible for the discontinuity in the derivative at $M_{\pi\pi}=2M_{\pi^+}$, as well as for the rapid fall-off below this point seen in Fig.~\ref{fig:s-distribution}. In the Dalitz plot distribution, the leading term generated by the branch cut in the $s$-channel only shows up in the narrow strip between the line  $s=4M_{\pi^+}^2$ and the boundary of the physical region.  We approximate the contributions from the cusps with the leading term:
\begin{eqnarray}\label{eq:Dncusp} \al \al D_n^\mathrm{cusp}(s,t,u) = 2\hspace{0.3mm}\delta\, \{\rho(s)+\rho(t)+\rho(u)\}\;,\;\; \nonumber\\
 \al \al \rho(s) \equiv \theta(4M_{\pi^+}^2- s)\sqrt{1-s/4M_{\pi^+}^2}\;.\end{eqnarray}
The parameter $\delta$ measures the strength of the cusps;  $\theta(x)$  is the Heaviside step function. For the background underneath the cusps, we simply extrapolate the terms of the Taylor series listed in Eq.~\eqref{eq:DnTaylor} and use the approximation  
\begin{eqnarray}\label{eq:Dnapprox}D_n(X_n,Y_n) \al \simeq \al1+2\hspace{0.3mm}\alpha\,Z+2\hspace{0.3mm}\beta\,Z^\frac{3}{2}\sin(3\,\varphi)+2\hspace{0.3mm}\gamma\,Z^2 \nonumber \\
\al \al +D_n^\mathrm{cusp}(s,t,u)  \;. \end{eqnarray}
on the entire phase space. Although the formula now involves square roots as well as powers of the Mandelstam variables, we continue using the term 'polynomial approximation'. 

While this approximation is very accurate on the disk $Z<Z^\mathrm{cusp}$, where the Taylor expansion converges and $D_n^\mathrm{cusp}$ vanishes, it describes the contributions from the cusps comparatively crudely. For this reason, we do not simply minimize the difference between this approximation and our dispersive representation over the entire physical region, but fix the coefficients $\alpha$, $\beta$, $\gamma$ at the values listed in Eq.~\eqref{eq:polyKchi6} and determine $\delta$ by minimizing the discrepancy over the remainder of the physical region, $Z>Z^\mathrm{cusp}$. The minimum occurs at 
\begin{equation} \label{eq:deltaKchi6}\mathrm{fitK}\chi_6:\quad \delta=-0.017(4)\;.\end{equation}
%
\begin{table*}\centering
\begin{tabular}{llllllll}
&\hspace{2em}$\alpha$& \hspace{2em}$\beta$&\hspace{1.9em}$ \gamma$&\hspace{1.5em}$\delta$&\raisebox{0.2em}{$\chi^2_\mathrm{M}$}&\raisebox{0.2em}{$\chi^2_\mathrm{K}$}&\raisebox{0.2em}{$\chi^2_\mathrm{th} $} \\
\hline
fitMZ&$-$0.0265(59)&&$+$0.0017(96) &&10.2&&\\ 
fitMZ$_1$&$-$0.0267(15)&&$+$0.0019$^*$ &&10.2&&\\ 
\hline
fitMD  &$-$0.0301(64) &$-$0.0069(18)&$+$0.0087(110)&$-$0.027(14) &343&& \\  
fit\#9 \hspace{0.45em}\cite{Prakhov+2018}&$-$0.0265(10)&$-$0.0073(10)& \hspace{0.8em}$0^\star$&$-$0.017(7)&408&& \\
fit\#10 \cite{Prakhov+2018}&$-$0.0247(30)&$-$0.0070(12)&$-$0.0023(40)&$-$0.015(7) &363&&\\
\hline
fit$\chi_4$ &$-$0.0222(117)&$-$0.0039(7)&$+$0.0015(8)&$-$0.0169(4) &352&&0\\
fitK$_4$&$-$0.0310(17)&$-$0.0043(3)&$+$0.0021(3)&$-$0.017(4) &354&390&0.67\\
fitKM$_4$&$-$0.0303(13)&$-$0.0042(4)&$+$0.0020(2)&$-$0.017(4)&352&391&0.46\\
fitK$\chi_6$ & $-$0.0307(17)&$-$0.0052(5)&$+$0.0019(3)&$-$0.017(4) &352&384&1.47 \\
fitKM$\chi_6$&$-$0.0296(12)&$-$0.0055(4)&$+$0.0018(3)&$-$0.017(4) &387&383&5.12\\
\hline
\end{tabular}
\caption{Polynomial representations for the decay $\eta\to3\pi^0$. The parametrization is specified in Eq.~\eqref{eq:Dnapprox}. The first two lines represent fits to the MAMI data for the $Z$-distribution. The next three lines show polynomial fits to the MAMI data on the Dalitz plot distribution -- two of these stem from Table I of Ref.~\cite{Prakhov+2018}. The lower half of the table contains polynomial approximations to various dispersive representations obtained within our framework. The coefficients $\alpha$, $\beta$ and $\gamma$ are determined with a fit in the region $Z<Z\mathrm{cusp}\approx 0.597$, where $\delta$ does not contribute (18 bins of the $Z$-distribution and 266 bins of the Dalitz plot distribution are in this region -- the values quoted for $\chi^2_\mathrm{M}$ give the contributions to the discrepancy function from these bins). The values of $\delta$ are obtained by fitting the remaining 140 bins of the Dalitz plot distribution, varying $\alpha$, $\beta$, $\gamma$ in the range found in the first step. The asterisks mark values used as input. \label{tab:poly}}\end{table*}
With the values of the coefficients in \eqref{eq:polyKchi6}, \eqref{eq:deltaKchi6}, the parametrization \eqref{eq:Dnapprox} reproduces our dispersive representation of the Dalitz plot distribution within 0.6 permille, throughout the physical region. It does not quite reach the remarkable precision of the polynomial representation on the disk $Z<Z^\mathrm{cusp}$, presumably because the extrapolation of the first few terms of the Taylor series does not describe the background underneath the cusps very accurately -- the presence of the resonance f$_0$(500) may accurately be accounted for only in the dispersive representation.\\
\indent The error in the result for $\delta$ reflects the uncertainties of the dispersive representation. These subject the coefficients  $\alpha$, $\beta$, $\gamma$ to the errors listed in \eqref{eq:polyKchi6} and also lead to correlations among them. When minimizing the discrepancy in the region $Z>Z^\mathrm{cusp}$, the errors then propagate into $\delta$. The evaluation shows that the strength of the cusps is rather sensitive to the uncertainties in the isospin breaking corrections -- the corresponding contribution to the error budget is even slightly larger than the Gaussian error, while the one from the noise in the phase shifts is negligible.   \\
\indent The prediction for the slope mainly relies on the experimental information concerning the Dalitz plot distribution of $\eta\to\pi^+\pi^-\pi^0$ -- the theoretical constraints are not important in this connection. This can be seen by comparing  the polynomial approximations for the two dispersive solutions obtained if either the data on this decay or the theoretical constraints are ignored: fit$\chi_4$ versus fitK$_4$ -- the first represents the matching solution, which exclusively relies on theory, while the second is instead based on the KLOE data alone. The coefficients of the corresponding polynomial approximations  are listed in Table \ref{tab:poly}.
The  comparison shows that the two representations of the Dalitz plot distribution in the neutral channel are consistent with one another. Concerning $\delta$, the results are even the same and for $\beta$, there is not much of a difference, either. For fit$\chi_4$,  however, the uncertainties in $\alpha$ and $\gamma$  are much larger than for fitK$_4$:  In this regard, the theoretical constraints are much weaker than the experimental ones.

\section{Fits to the MAMI data\label{sec:Fits to MAMI}}
\subsection{$Z$-distribution \label{sec:Z-distribution MAMI}}
Next, we compare the experimental information with the polynomial parametrization in the region where the Taylor series converges, $Z<Z^\mathrm{cusp}$. The simplest way to determine the slope experimentally is to measure the $Z$-distribution. In the singularity-free region, only the coefficients $\alpha$ and $\gamma$ of the polynomial approximation show up in this distribution -- $\alpha$ specifies the slope, while $\gamma$ measures the curvature. In the recent update of the MAMI data (Runs I and II combined)~\cite{Prakhov+2018}, the $Z$-distribution is not normalized. Allowing for a free normalization factor $\Lambda_\mathrm{M}$ and fitting the data with the polynomial representation \eqref{eq:dnZ}, we obtain a fit of excellent quality, which we denote by fitMZ:  $\Lambda_\mathrm{M}=0.9762(15)$, $\chi^2=10.2$ for 18 data points and 3 parameters. The corresponding values for $\alpha$ and $\gamma$ are listed in Table \ref{tab:poly}. The allowed range is represented by the green ellipse in the left panel of Fig,~\ref{fig:AlphaVersusGamma}. The central value of $\alpha$ is somewhat smaller than our prediction, fitK$\chi_6$,  which is based on the KLOE data for $\eta\to\pi^+\pi^-\pi^0$, while the result for $\gamma$ is close to what we obtain on this basis. The uncertainties are large, however -- the data on the $Z$-distribution do not provide an accurate determination of $\alpha$ or $\gamma$, but impose a strong correlation between these two coefficients. If $\gamma$ is not treated as a free parameter, but is held fixed at the value in fitK$\chi_6$, we obtain fitMZ$_1$. The quality remains excellent: $\chi^2=10.2$, and the central value of $\alpha$ nearly stays the same, but the uncertainty drops by a factor of four. If we extend the range and fit the data on the entire physical region, $0<Z<1$, the coefficients $\beta$ and $\delta$ do show up, but the $Z$-distribution does not determine them well and the result for $\alpha$ and $\gamma$ barely changes.
\subsection{Dalitz plot distribution on the disk $Z<Z^\mathrm{cusp}$\label{sec:Dalitz plot MAMI}}
\begin{figure*}[thb]
\centering
\includegraphics[height = 5.5cm]{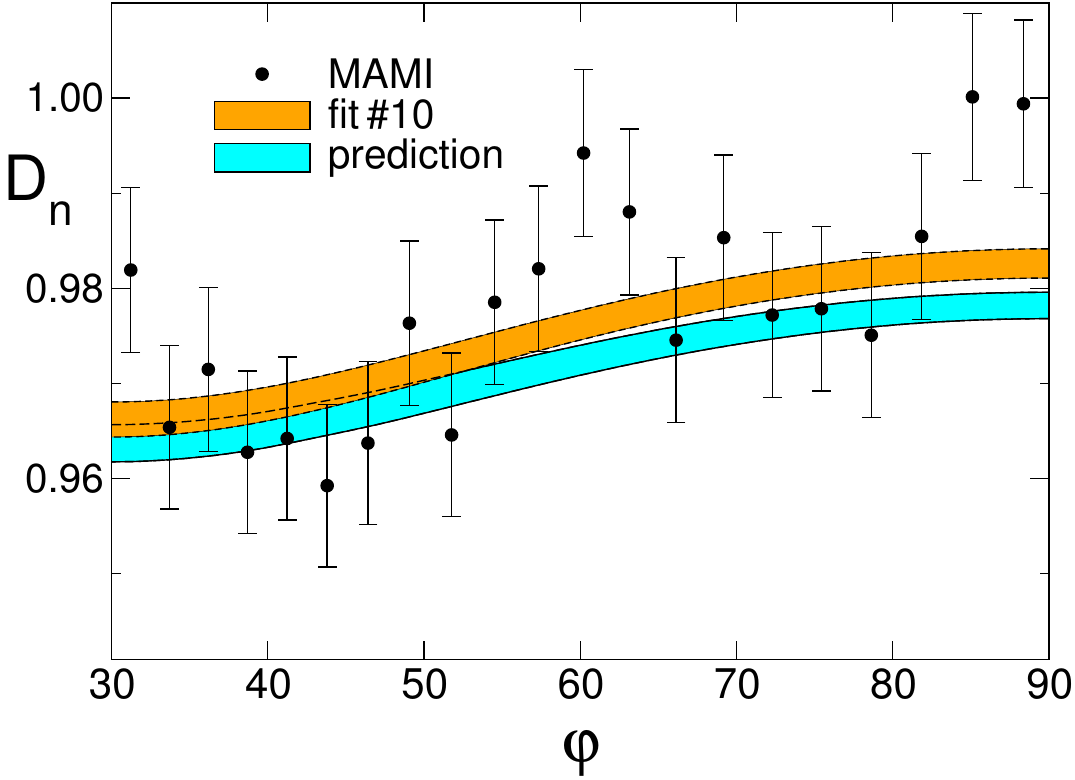}\hspace{0.7cm}\includegraphics[height= 5.5cm]{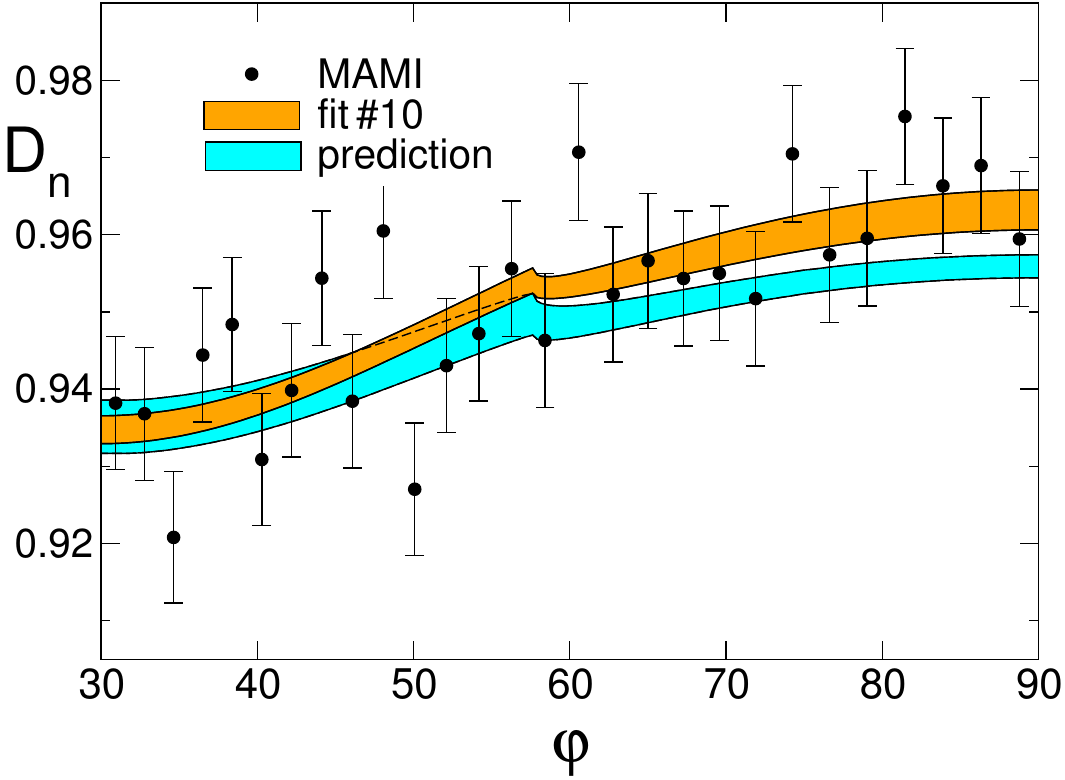}\caption{Angular dependence of the Dalitz plot distribution in the neutral channel. The left panel compares our prediction with the MAMI data contained in band \#21 ($0.719<\lambda <0.754$). For comparison, we also show the polynomial fit\#10 of Ref.~\cite{Prakhov+2018}. The right panel concerns band \#28 ($0.955<\lambda < 1$), which is located at the boundary of the physical region. \label{fig:TM28phase}}
\end{figure*}
Next, we consider the MAMI data on the Dalitz plot distribution.  As noted above, each event is represented by 6 different points in the physical region. The binning in the variables $X_n$, $Y_n$  does preserve the symmetry under $X_n\to-X_n$, but  not  the one under reflections at the lines $\varphi=\pm\, 30^\circ$. Accordingly, a subset of bins that contains each event exactly once does not exist.

This problem is readily solved by sampling the data in the radial coordinates  $Z,\varphi$ defined in Eq.~\eqref{eq:Z phi} rather than in $X_n$, $Y_n$: The sextant $30^\circ<\varphi< 90^\circ$ contains each event exactly once. At the boundary of the physical region, however, the pair $Z$, $\varphi$ is no better than $X_n$, $Y_n$, because the boundary value of $Z$ depends on the angle: $Z=Z_b(\varphi)$. We propose to instead use the coordinates $\lambda$, $\varphi$, where $\lambda$ stands for
\begin{equation}\label{eq:lambda phi}\lambda =\sqrt{\frac{Z}{Z_b(\varphi)}}\;.\end{equation}
In these variables, each event gives rise exactly to one point in the sextant $0 < \lambda<1$, $30^\circ<\varphi < 90^\circ$, so that the binning is easy to implement, not only at the boundaries of the sextant, but also at the boundary of the physical region -- for a detailed account of the procedure, we refer to Appendix~\ref{sec:Binning}.  We thank Sergey Prakhov for providing us with the corresponding sampling of the MAMI data \cite{Prakhov2018}. All of the fits to the Dalitz plot distribution discussed in the following are based on this data set (Runs I and II combined). Fig.~\ref{fig:TM28phase} compares the angular dependence of two subsets of these data with our prediction (fitK$\chi_6$). The difference between the prediction and the polynomial approximation to it is too small to be visible in this figure.

A polynomial fit to the MAMI data on the Dalitz plot distribution that does not invoke dispersion theory at all is listed in the entry fitMD of Table \ref{tab:poly}: The coefficients $\alpha$, $\beta$, $\gamma$ are determined with a fit to the data in those bins that are contained in the disk $Z<Z^\mathrm{cusp}$, where the Taylor series converges and where $\delta$ does not contribute. Treating the overall normalization of the experimental distribution as a free parameter, the fit returns the central values for $\alpha$, $\beta$, $\gamma$ listed in the table, together with $\Lambda_\mathrm{M}=0.976$ and $\chi^2=343.3$ for 266 data points and 4 parameters. The errors are obtained in the same way as for the subtraction constants of the dispersive representation, except that the discrepancy function now contains an additional parameter, $\Lambda_\mathrm{M}$. The result for $\alpha$ and $\gamma$ confirms what we found when fitting the $Z$-distribution: fitMD and fitMZ agree within errors. The uncertainties are large, but the values are strongly correlated. In contrast to fitMZ,  however,  the likelihood of fitMD is not satisfactory: $\chi^2/\mathrm{dof}=1.31$. Since the polynomial approximation of the dispersive representation is very accurate in the disk $Z<Z^\mathrm{cusp}$, we consider it very unlikely that the problem originates in the lack of flexibility of the parametrization. 

\subsection{Cusps\label{sec:Cusps MAMI}}
Next we study the behaviour of the data in the remainder of the physical region,  where the final state interaction generates cusps. The problem encountered at the boundary of the disk $X_n^2+Y_n^2=Z^\mathrm{cusp} $ repeats itself at the boundary of the physical region. We have checked, however, that restricting the fit to those bins that are entirely contained in the physical region does not significantly modify the result. In the following, we determine the strength of the cusp with a fit to all of the  bins for which $D_n^\mathrm{cusp}$ contributes.
\begin{figure*}[thb]\centering\includegraphics[height = 5cm]{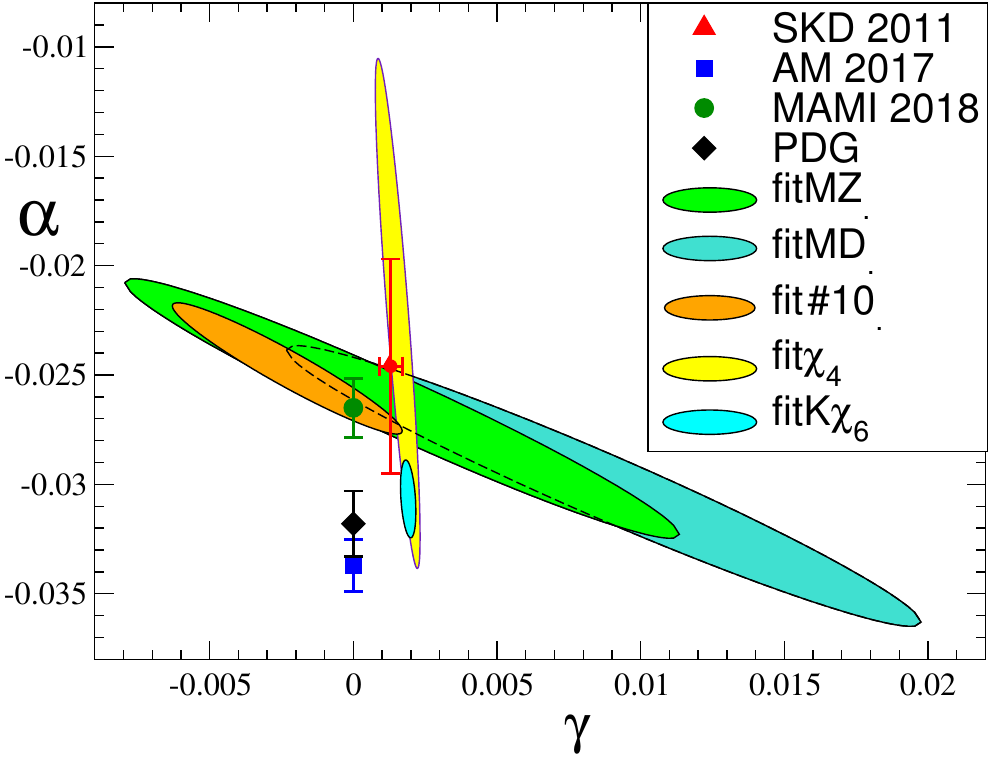}\hspace{0.3cm}\includegraphics[height= 5cm]{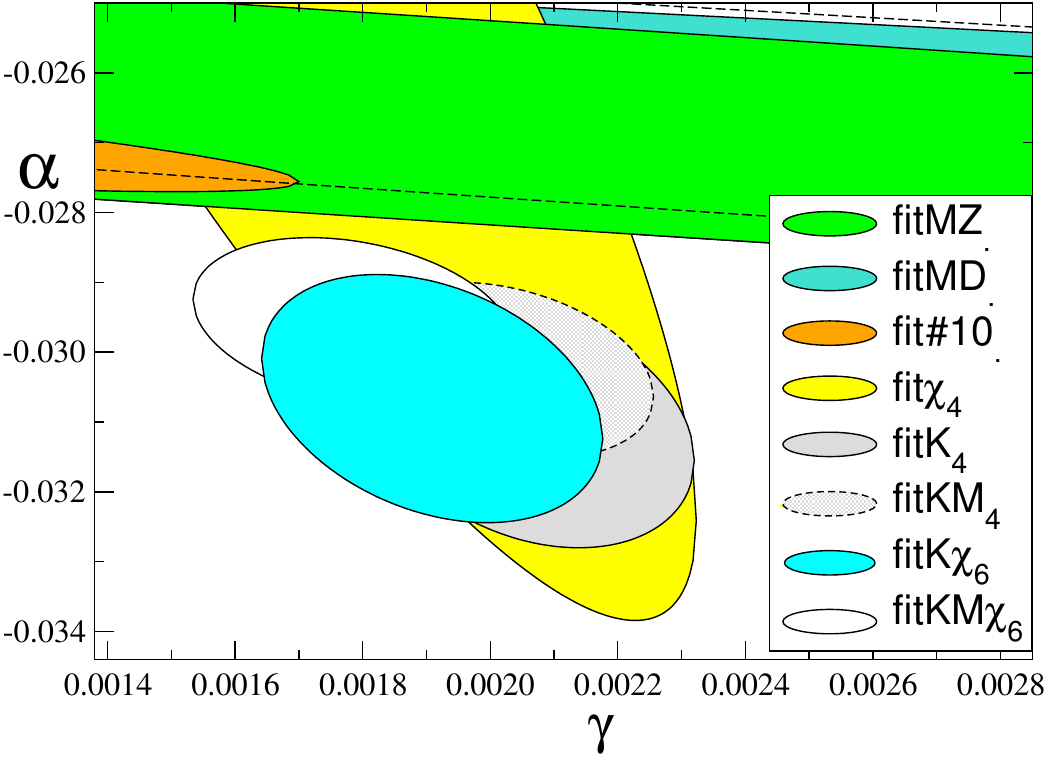}\caption{Correlation between slope and curvature. The polynomial fits to the MAMI data for the decay $\eta\to3\pi^0$  correspond to the large, slightly tilted ellipses in the left panel. They are compared with the results of Schneider, Kubis and Ditsche \cite{Schneider+2011}, Albaladejo  and Moussalam \cite{Albaladejo+2017}, the A2 collaboration at MAMI \cite{Prakhov+2018} and the Particle Data Group \cite{Olive:2016xmw}. The latter three neglect the curvature and are shown at $\gamma=0$. The matching solution fit$\chi_4$, which exclusively relies on theory, is indicated by the large yellow ellipse. All other representations obtained within our dispersive framework cluster around the comparatively small cyan ellipse, which represents our prediction, fitK$\chi_6$.  The right panel focuses on these and compares the dispersive representations fitK$_4$ and fitK$\chi_6$ based on the KLOE data for the decay $\eta\to\pi^+\pi^-\pi^0$ alone with the common fits to the KLOE and MAMI data, denoted by fitKM$_4$ and fitKM$\chi_6$, respectively. \label{fig:AlphaVersusGamma}}
\end{figure*}
To evaluate the strength of the cusps for fitMD, we use the same procedure as in the construction of an approximate representation for our central dispersive solution: Keep the values of $\alpha$, $\beta$ and $\gamma$ fixed at fitMD, vary $\delta$ and minimize the difference between the parametrization \eqref{eq:Dnapprox} and the data in the  region $Z>Z^\mathrm{cusp}$.  The quality of the fit is worse than for the bins contained in the disk $Z<Z^\mathrm{cusp}$: $\chi^2=233$ for 140 data points and 1 free parameter, $\chi^2/\mathrm{dof}=1.68$. The error calculation follows the same steps: First determine $\delta$ for prescribed values of $\alpha$, $\beta$, $\gamma$, $\Lambda_\mathrm{M}$, then vary these within the range obtained when minimizing the discrepancy in the disk $Z<Z^\mathrm{cusp}$, accounting for the correlations among them. Finally, the additional uncertainty arising from the statistical fluctuations in the region $Z>Z^\mathrm{cusp}$ is added in quadrature.  For $\delta$, the error is dominated by the contribution from the uncertainties and correlations encountered in the first step. 
Table \ref{tab:poly} shows that the result for fitMD is consistent with our prediction, also concerning $\delta$.  Although the cusps do not stick out from the fluctuations visible in Fig.~\ref{fig:s-distribution}, the quantitative analysis on the basis of formula \eqref{eq:Dnapprox} does confirm their presence. 

For the dispersive representation of the amplitude, it does not make much of a difference whether the slope is determined with a fit in the disk $Z<Z^\mathrm{cusp}$ or in the entire physical region. Fitting the parametrization \eqref{eq:Dnapprox} to our central solution fitK$\chi_6$ in the entire physical region, we obtain  $\alpha=-0.0307(18)$, $\beta=-0.0049(5)$, $\gamma=0.0018(3)$, $\delta=-0.016(4)$. These numbers barely differ from those quoted in Table \ref{tab:poly} for the polynomial approximation to fitK$\chi_6$. This shows that the dispersive representation provides a stable extrapolation from the region below $Z^\mathrm{cusp}$ to the region where the cusps occur. 

When fitting data with the polynomial approximation, the situation is very different,  because the correlation between the behaviour at small values of $Z$ and in the region where the cusps manifest themselves is then absent. This is illustrated with two fits taken from Table I of Ref.~\cite{Prakhov+2018},  which are also based on the combined data of Runs I and II, but use all three sextants with  $X_n>0$.  Apart from that, the analysis differs from ours only in one respect:  While we determine the coefficients $\alpha$, $\beta$, $\gamma$ with a fit to the data in the disk $Z<Z^\mathrm{cusp}$ and make use of those in the remaining bins exclusively to estimate the strength of the cusps, fit\#9 and fit\#10  treat all coefficients on the same footing (except that in the case of fit\#9 $\gamma$ is set to zero). The comparison of the two illustrates the strong correlation between $\alpha$ and $\gamma$: The uncertainty in the result for the slope becomes much smaller if $\gamma$ can be taken as known. Note that for all of the entries in Table \ref{tab:poly}, the values quoted for $\chi^2_\mathrm{M}$ refer to the 266 independent bins in the disk $Z<Z^\mathrm{cusp}$.

The three polynomial representations 
fitMD, fit\#9 and fit\#10 agree within uncertainties, but the latter two have substantially smaller errors. The left panel of Fig.~\ref{fig:AlphaVersusGamma} illustrates the difference, which arises because the polynomial terms grow with $Z$; extending the region over which the approximation is fit to the data leads to smaller errors in the coefficients. While fitMD is consistent with our prediction \eqref{eq:polyKchi6}, \eqref{eq:deltaKchi6}, the values obtained for $\alpha$ and$\beta$ with fits \#9 and \#10 are not. In fact, the entries for $\chi^2_\mathrm{M}$ show that, in the region $Z<Z^\mathrm{cusp}$, the polynomial approximation to our prediction follows the data more closely than these two fits. Concerning the parameter $\delta$, which measures the strength of the cusps, however, they are in very good agreement with our prediction. 
\\
\indent The main problem we are facing here is that one is dealing with small effects. In current algebra approximation, the Dalitz plot distribution is flat, $D_n^\mathrm{LO}(X_n,Y_n)=1$. The MAMI data do allow an accurate measurement of  the slope $\alpha$ of the distribution, but what remains  is tiny: For our prediction, the difference $D_n^\mathrm{phys}(X_n,Y_n)-1- 2\,\alpha\, Z$ stays below 7 permille, throughout the region $Z<Z^\mathrm{cusp}$, where the Taylor series converges. Although the set we are analyzing is based on more than 7 million events, the statistical errors in the mean value of the Dalitz plot distribution for a given bin are of order 8 permille and the systematic ones must be small compared to this for the measurement to be sound. Isospin breaking effects are by no means negligible at this level of accuracy. In the approximation we are using, they yield a positive contribution to the slope: $\delta \alpha = + 0.0024(7)$. At $Z=Z^\mathrm{cusp}$, it affects  the value of the Dalitz plot distribution by about 3 permille. Note also that the cusps are visible in the physical region only because the physical masses of the charged and neutral pions differ -- isospin breaking is crucial for an accurate analysis of the Dalitz plot distribution in the region $Z > Z^\mathrm{cusp}$.  The fact that the result obtained for the branching ratio agrees with experiment gives us confidence that our estimates for the effects due to  isospin breaking in the integrals over the square of the amplitude are adequate, but resolving the Dalitz plot distribution at the level of accuracy needed to reliably determine small quanitites like $\beta$ and $\gamma$ and to measure the strength of the cusps is a different matter.  

\subsection{Dispersive analysis of the MAMI data \label{sec:dipersive fit MAMI}}

The errors attached to the values of $\gamma$ listed in the lower half of Table \ref{tab:poly} are much smaller than those in the upper half: Dispersion theory fixes the curvature term much more accurately than the data on the Dalitz plot distribution in the neutral channel -- even the theoretical constraints alone (fit$\chi_4$) yield a rather sharp value for this coefficient. We now investigate the impact of the MAMI data on the dispersive analysis. The discrepancy function relevant for these data is of the same form as the one for the KLOE data in Eq.~\eqref{eq:discrepancy KLOE}:
\begin{equation}\label{eq:discrepancy MAMI}\chi^2_\mathrm{M}=\frac{1}{3}\sum_i\left( \frac{D_n^\mathrm{phys}(X_n^i,Y_n^i)-  \Lambda_\mathrm{M}\,D_n^\mathrm{i}}{\Lambda_\mathrm{M}\,\Delta D_n^\mathrm{i}}\right)^2\;.\end{equation}

Taken by themselves, the data on the neutral channel do not suffice to pin down the subtraction constants. In particular, as evidenced by the current algebra approximation, the neutral channel does not contain information about the slope of the amplitude in the charged channel or about the position of the Adler zero. We combine the experimental information available in the charged and neutral channels, first ignore the theoretical constraints and look for the minimum of $\chi^2_\mathrm{K}+\chi^2_\mathrm{M}$. The normalization of the dispersive representation plays no role here -- we again fix it with $H_0=H_0^\mathrm{NLO}$ and restrict the fits to the data contained in the disk $Z<Z^\mathrm{cusp}$. As noted above, the correlations present in the dispersive representation imply that the results are essentially the same if that restriction is dropped.  

We first allow  for only four subtraction constants, set $\delta_0=\gamma_1=0$ and denote the simultaneous fit to the KLOE and MAMI data by fitKM4. Table \ref{tab:poly} shows that  the inclusion of the MAMI data lowers the value of the slope $\alpha$ from $-0.0310(17)$ (fitK$_4$) to $-0.0303(13)$ (fitKM$_4$), while the coefficients $\beta$, $\gamma$, $\delta$ nearly stay put. The ratio 
\mbox{$\chi^2_\mathrm{M}/\mathrm{dof}= 1.34$} shows that the quality of the fit is not satisfactory, even slightly worse than for the polynomial representation fitMD, where $\chi^2_\mathrm{M}/\mathrm{dof}= 1.31$. On the other hand, the value $\chi^2_\mathrm{th}=0.46$ indicates that, although the theoretical constraints that follow from the presence of a hidden approximate symmetry are not made use of in the derivation of fitKM$_4$, the MAMI data for $\eta\to 3\pi^0$ are consistent with these, as well as with the KLOE data for $\eta\to\pi^+\pi^-\pi^0$.
 
If more than four subtraction constants are treated as free parameters, the minimization again goes astray. When analyzing the KLOE data we found that simply adding the term $\chi_\mathrm{th}^2$ to the discrepancy function suffices to ensure that the theoretical constraints are respected. In the present case, this is not the case, however: The contributions from the 371 and 406 data points of KLOE and MAMI, respectively, overwhelm the one from the theoretical part of the discrepancy function. The minimum occurs at $\chi_\mathrm{th}^2=5.12$, indicating that the constraints are still violated -- fitKM$\chi_6$ does not represent a physically acceptable solution of our integral equations. For the determination of $Q$, the extrapolation below threshold is needed and the theoretical constraints do play an essential role in this connection. 

As far as the behaviour in the physical region is concerned, however, fitKM$\chi_6$ does represent an acceptable parametrization of the amplitude. The violation of the theoretical constraints can be cured without significantly changing the behaviour of the amplitude there. It suffices, for instance, to give the theoretical discrepancy in $\chi_\mathrm{tot}^2=\chi_\mathrm{K}^2+\chi_\mathrm{M}^2+\chi_\mathrm{th}^2$ more weight. If we multiply that term by 3, the value of $\chi_\mathrm{th}^2$ falls to 1.20 while $\alpha$, $\beta$, $\gamma$, $\delta$ nearly stay put at the values obtained for fitKM$\chi_6$  listed in Table \ref{tab:poly}. The white ellipse in the right panel of Fig.~\ref{fig:AlphaVersusGamma} illustrates the result. The comparison shows that fitKM$\chi_6$ is close to fitKM$_4$, consistent with fitMZ and fitMD (MAMI data alone) as well as with our prediction, fitK$\chi_6$ (KLOE data plus theoretical constraints). 
The result for $\beta$, $\gamma$ and $\delta$ can barely be distinguished from the prediction. The inclusion of the MAMI data reduces the value of the slope, irrespective of whether four or six subtraction constants are allowed.  As emphasized in Ref.~\cite{Prakhov+2018}, these data imply a smaller value than the average $\alpha=-0.0318(15)$ quoted by the Particle Data Group \cite{Olive:2016xmw}.  

\section{Kaon mass difference and quark mass ratios}\label{sec:MKQ}
\subsection{Mass difference between charged and neutral kaons}\label{sec:MK}

According to Eqs.~\eqref{eq:Gammac} and \eqref{eq:Gamman}, the rates of the charged and neutral decay modes are proportional to integrals over the square of the transition amplitude, denoted by $J_c$ and $J_n$, respectively. Solving for $\hat{M}_{K^0}^2-\hat{M}_{K^+}^2$, the relations can be rewritten in the form: 
 \begin{equation}\label{eq:DMKJ}\hat{M}_{K^0}^2-\hat{M}_{K^+}^2=\left\{\mathrm{\Large\begin{array}{l}\left(\frac{N_a\,\Gamma_c\rule[-0.2em]{0em}{0em}}{J_c}\right)^\frac{1}{2}\\
\left(\frac{N_a\,\Gamma_n\rule[-0.2em]{0em}{0em}}{J_n}\right)^\frac{1}{2} \end{array}}\right.\hspace{2em} N_a=6912\,\pi^3 F_\pi^4 M_\eta^3\;. \end{equation}

 with $\Gamma_c\equiv\Gamma_{\eta\to\pi^+\pi^-\pi^0}$ and $\Gamma_n\equiv \Gamma_{\eta\to 3\pi^0}$. The constant $N_a$ does not involve any unknowns. 
The phase space integrals are quadratic in the subtraction constants
 $\{k_1,\ldots,k_6\}=\{\alpha_0,\beta_0,\gamma_0,\delta_0,\beta_1,\gamma_1\}$:
 \begin{equation}\label{eq:J} J_r=\sum_{a,b=1}^6 J_r^{ab}k_a\bar{k}_b\;\hspace{1em}r=c,n\;.\end{equation}
 The coefficients $J_c^{ab}$ and $J_n^{ab}$ represent  integrals over our fundamental 
 solutions, which only depend on the input used for the phase shifts.  They can 
 be worked out once and for all, but to evaluate the uncertainties due to the
 noise in the phase shifts, the calculation needs to be done separately for the eight different 
 phase shift configurations specified in Appendix~\ref{sec:Sensitivity to phase shifts}.
 
 For our central solution, fitK$\chi_6$, we obtain 
 \begin{equation}\label{eq:JGeV}J_c=1.96(24)\times 10^{-2}\,\mathrm{GeV}^4\;,\quad J_n=2.82(32)\times 10^{-2}\,\mathrm{GeV}^4\;.\end{equation}
 Note that, in contrast to the Dalitz plot distribution and the branching ratio, where the normalization of the
 amplitude drops out, the integrals $J_c$ and $J_n$  do depend on it. While the relative size 
 of the subtraction constants is strongly constrained by experiment, the overall normalization is not.
 We fix it with the theoretical estimate $H_0=1.176(53)$ derived in Sec.~\ref{sec:One loop}. The uncertainty therein and the Gaussian errors
  contribute about equally to the uncertainties in the integrals $J_c$, $J_n$, while those associated with the phase shifts and with the 
  estimates used for isospin breaking barely affect the result (for more details concerning the error budget, we refer to  Sec.~\ref{sec:Q}). 
    
With  the experimental values   
 $\Gamma_c=299(11)$ eV and $\Gamma_n= 427(15)$ eV~\cite{Olive:2016xmw}, the relations \eqref{eq:DMKJ} lead to two independent
 determinations  of the kaon mass difference in QCD:\footnote{The numerical values differ slightly from those given in 
 Ref.~\cite{Colangelo:2016jmc}, partly because  the experimental results for the decay rates quoted by the Particle Data Group have changed  in the meantime, partly because we improved the accuracy of the numerical representation of the fundamental solutions. As the shift in the central values amounts to less than a tenth of the quoted uncertainties, it is without significance.}
 \begin{equation}\label{eq:DMKQCDcn}\hat{M}_{K^0}^2-\hat{M}_{K^+}^2=\left\{ \begin{array}{ll}
6.25(41) \times 10^{-3}\, \mathrm{GeV}^{2} & \eta\to\pi^+\pi^-\pi^0 \\&\\
6.23(37)\times 10^{-3}\, \mathrm{GeV}^{2} & \eta\to3\pi^0
\end{array}
\right.
\end{equation}
Since our prediction  for the branching ratio agrees with experiment, the two results are nearly
 the same, but they are statistically independent only with regard to the
 uncertainties in the experimental values of the rates, which are responsible for only a small
 fraction of the error. Combining the two, we can determine the mass
 difference to an accuracy of 6\,\%:
\begin{equation}\label{eq:DMKQCD}
\hat{M}_{K^0}^2-\hat{M}_{K^+}^2 =  6.24(38)\times 10^{-3}\,\mbox{GeV}^2\,.
\end{equation} 

As discussed in the introduction, $\eta \to 3 \pi$ is uniquely sensitive to
isospin breaking due to the quark masses. This is thanks to Sutherland's theorem
which proves the suppression of electromagnetic isospin breaking in this
decay. In most other quantities which are sensitive to isospin breaking
there is a competition of effects of strong and electromagnetic origin and
it is difficult to disentangle the two. It is for this reason that lattice
calculations, which in principle would be ideally suited to determine the
size of the light quark mass difference, only recently have become able to
determine this quantity: This task had to wait for simulations of QCD and QED
close to the physical point, which have become possible only in the current
decade. A detailed understanding of the systematic effects related to the
inclusion of QED in the lattice action is still ongoing, but the latest 
results on strong isospin breaking from the lattice are already of
significant precision. A comparison with our results is therefore highly
relevant.

There are two recent lattice calculations which have evaluated the kaon
mass difference in QCD in a simulation where both QCD and QED were
included: one by the BMW collaboration~\cite{Fodor:2016bgu} and one by the
RM123 collaboration~\cite{Giusti:2017dmp}. The details of the calculations
differ, of course, but the outcomes are in very good agreement, not only with one another:

\begin{equation}\label{eq:DMKlattice}
\hat{M}_{K^0}^2-\hat{M}_{K^+}^2 =\left\{ \begin{array}{ll}
6.088(26)(68)(219) \times 10^{-3}\, \mathrm{GeV}^{2} & \mbox{\protect \cite{Fodor:2016bgu}} \\
& \\
5.950(150) \times 10^{-3}\, \mathrm{GeV}^{2} & \mbox{\protect \cite{Giusti:2017dmp}}
\end{array}
\right.
\end{equation}
but also with our determination from $\eta$-decay in Eq.~(\ref{eq:DMKQCD}).
\subsection{Electromagnetic contributions to the meson masses, Dashen theorem}\label{sec:Dashen}

Theoretical determinations of the meson self-energies started in the sixties of the last century~\cite{Socolow:1965zz,Das:1967it,Dashen1969}. The difference between $M_{\pi^+}$ and $M_{\pi^0}$ is well understood and is due almost exclusively to the 
electromagnetic self-energy of the $\pi^+$. Estimating the small contribution
proportional to $(m_u-m_d)^2$ with \chpt\ yields $\hat{M}_{\pi^+}-\hat{M}_{\pi^0} =0.17(3)\, \MeV$~\cite{Gasser+1985}. We denote the electromagnetic contribution to the square of the mass of a particle by $\Delta_P^\gamma\equiv M_P^2-\hat{M}_P^2$~\cite{Aoki:2016frl}. Together with  the observed mass difference, the above estimate for the mass difference in QCD implies  
\begin{equation}\label{eq:MpiQED}\Delta_{\pi^+}^\gamma-\Delta_{\pi^0}^\gamma=1.21(1)  10^{-3}\,\mbox{GeV}^2\,.\end{equation}

Dashen's theorem~\cite{Dashen1969} states that, at leading order of \chpt, the electromagnetic self-energies
of the neutral pions and kaons vanish, while the contributions to $M_{\pi^+}^2$ and $M_{K^+}^2$  are the same. 
The comparison of our result~\eqref{eq:DMKQCD} with the observed mass difference yields a result that is about twice as large: 
\begin{equation}\label{eq:DMKQED}
\Delta_{K^+}^\gamma-\Delta_{K^0}^\gamma=2.33(38) 10^{-3}\,\mbox{GeV}^2\;.\end{equation}
Indeed, Langacker and Pagels had pointed out  that the chiral perturbation series of the meson self-energies contains
unusually large logarithmic infrared singularities~\cite{Langacker:1974nm}. The numerical estimates based on the $1/N_c$-expansion~\cite{Bijnens1993} or on the Cottingham formula~\cite{Donoghue+1993} indicated that the Dashen theorem is strongly violated. The effective Lagrangian relevant for the evaluation of the contributions generated by
virtual photons was set up~\cite{Urech1995,Neufeld:1995mu}, but the evaluation of the self-energies on that basis~\cite{Baur:1995ig} did not confirm the picture -- the numerical estimates used for the LECs of order $e^2 p^2$ led to corrections of rather modest size. 

The corrections to the Dashen theorem from higher orders of the chiral expansion can  be characterized with the dimensionless parameter $\epsilon$, which is defined by~\cite{Aoki:2016frl} 
\begin{equation}\label{eq:epsilon definition}\Delta_{K^+}^\gamma-\Delta_{K^0}^\gamma =\Delta_{\pi^+}^\gamma-\Delta_{\pi^0}^\gamma+\epsilon\,
(M_{\pi^+}^2-M_{\pi^0}^2)\;.\end{equation}
In this notation, our results for the electromagnetic self-energy differences amount to 
\begin{equation}\label{eq:epsilon result}\epsilon=0.9(3)\;.\end{equation}
We emphasize that our calculation of the difference  $\Delta_{K^+}^\gamma-\Delta_{K^0}^\gamma$ does not face the problem with the strong infrared singularities encountered in direct evaluations of the self-energies and conclude that the Dashen theorem does receive large corrections from higher orders of the chiral expansion. 

The lattice results in Eq.~\eqref{eq:DMKlattice} lead to the same conclusion. For comparison we include other recent determinations as well as the value quoted in the FLAG review\footnote{Whenever
  three errors are given they are in the order: statistical, systematic,
  and systematic related to QED (quenching and finite volume).}: 
\begin{equation}
\epsilon=\left\{ \begin{array}{ll}
0.7(3) & \mbox{FLAG~\cite{Aoki:2016frl}} \\
0.50(6) & \mbox{QCDSF~\cite{Horsley:2015vla} } \\
0.73(3)(13)(5) & \mbox{MILC 2016~\cite{Basak:2016jnn}  } \\
0.73(2)(5)(17) & \mbox{BMW~\cite{Fodor:2016bgu}} \\
0.801(48)(25)(96) & \mbox{RM123~\cite{Giusti:2017dmp}}\\
0.78(1)(\rule{0em}{0em}^{+\,8}_{-11}) & \mbox{MILC 2018~\cite{Basak:2018zyd}}\\
\end{array}
\right. .
\end{equation}
Except for the marginal disagreement with QCDSF, where the quoted error is statistical only, all of these values are consistent with our result in Eq.~(\ref{eq:epsilon result}).

\subsection{Determination of the quark mass ratio $Q$}\label{sec:Q}

Finally, we invoke the low-energy theorem that relates the quark mass ratio $Q$
\begin{equation}\label{eq:Q}Q^2\equiv\frac{m_s^2-m_{ud}^2}{m_d^2-m_u^2}\;,\quad m_{ud}\equiv\mbox{$\frac{1}{2}$}(m_u+m_d)\;,\end{equation}
to a ratio  of meson masses~\cite{Gasser+1985a}:  
 \begin{equation}\label{eq:DMQ}
\frac{M_K^2\, (M_K^2-M_\pi^2)}{
 M_\pi^2 (\hat{M}_{K^0}^2-\hat{M}_{K^+}^2)}=Q^2(1+\Delta_Q) \,.\end{equation}
($\hat{M}_{K^0}$, $\hat{M}_{K^+}$ denote the mass of the neutral and charged kaons in QCD, while $M_\pi$, $M_K$ represent the mass of the pions and kaons in the isospin limit, respectively.) The low-energy theorem states that the chiral expansion of the left hand side in powers of $m_u$, $m_d$, $m_s$ starts with $Q^2$ and does not contain terms of next-to-leading order: 
\begin{equation}\label{eq:DQ}\Delta_Q=O(m_\mathrm{quark}^2)\,.\end{equation}
The expansion of the meson masses in powers of the quark masses with $m_u\neq m_d$ was worked out to NNLO in \cite{Amoros+2001}. The formulae involve the low-energy-constants of $\chi$PT, in particular also those arising from the effective Lagrangian at next-to-next-to-leading order. As the algebraic formulae are very lengthy, the authors only quote numerical results obtained by inserting numerical estimates for these constants. The estimates rely on the saturation of sum rules by resonances. In connection with the meson masses, the scalar channel plays the key role, where the resonance $f_0(500)$ is notoriously difficult to cope with in the framework of the chiral expansion -- in our opinion, the estimates for the LECs do not have the accuracy required to make a significant statement about the size of $\Delta_Q$. As discussed below, an evaluation of this quantity on the lattice would be of high interest.

The low-energy-theorem \eqref{eq:DQ} implies that, instead of normalizing the amplitude with the kaon mass difference in QCD, we can equally well normalize it with the quark mass ratio $Q$. The analog of the formula \eqref{eq:DMKJ} for $\hat{M}_{K^0}^2-\hat{M}_{K^+}^2$ reads
 \begin{equation}\label{eq:QJ}Q=\left\{\mathrm{\Large\begin{array}{l}\left(\frac{N_b\,J_c\rule[-0.2em]{0em}{0em}}{\Gamma_c}\right)^\frac{1}{4}\\
\left(\frac{N_b\,J_n\rule[-0.2em]{0em}{0em}}{\Gamma_n}\right)^\frac{1}{4} \end{array}}\right.\hspace{2em} 
N_b=\frac{M_K^4(M_K^2-M_\pi^2)^2}{6912\,\pi^3F_\pi^4M_\pi^4M_\eta^3}\;. \end{equation}
In either case, the relations only hold modulo corrections of next-to-next-to-leading order  in the chiral expansion. Apart from the phase space integrals $J_c$, $J_n$ and the decay rates,  they only contain the isospin limit of the meson masses and the pion decay constant.   

Concerning $M_\pi$, we rely on the estimates given
 in section 3.1.1 of the FLAG review~\cite{Aoki:2016frl}, which lead to %
\begin{equation}\label{eq:Mpibarnum}M_\pi=134.8(3)\, \mathrm{MeV}\;.\end{equation}
The result $M_K=494.2(3)$, on the other hand, must be reexamined, because it is based on the FLAG estimate $\epsilon=0.7(3)$ for the violation of the Dashen theorem. The change occurring if we instead use our own determination of $\epsilon$ in Eq.~\eqref{eq:epsilon result} is tiny:  The value of $M_K$ is lowered to 
\begin{equation}\label{eq:MKbarnum}
M_K=494.1(3)\,\mathrm{MeV}\;.\end{equation}

Using our central solution, fitK$\chi_6$,  the experimental values of the two decay rates then yield
\begin{equation}\label{eq:Qcn}Q=\left\{ \begin{array}{ll}
22.04(72)  &\hspace{1em} \eta\to\pi^+\pi^-\pi^0 \\&\\
22.08(66)  &\hspace{1em} \eta\to3\pi^0
\end{array}
\right.
\end{equation}
The  uncertainty in the theoretical estimate for $H_0$ contributes $\delta_1 Q=0.49$ to the error in the result for Q. The Gaussian error in the fit to the data is of similar size: $\delta_2 Q=0.44$ (this includes the uncertainties used for the theoretical part of the discrepancy function). The  noise in the representation used for the phase shifts only generates an uncertainty of $\delta_3 Q=0.05$. While the error arising from our treatment of the isospin breaking effects in the charged channel is more important, $\delta_4 Q_c=0.12$, the corresponding uncertainty in the neutral channel  is even smaller: $\delta_4 Q_n=0.04$. Finally, the experimental uncertainties in the decay rates of the charged and neutral channels yield an error of $\delta_5 Q_c=0.20$ and $\delta_5 Q_n =0.19$, respectively. The errors quoted in~\eqref{eq:Qcn} are obtained by adding these contributions up in quadrature. Combining the results obtained in the two channels, we obtain 
\begin{equation}\label{eq:Qnum} Q = 22.1(7)\,.
\end{equation} 
Note  that the value of the amplitude at the center of the Dalitz plot plays an important role here. As discussed in Sec.~\ref{sec:Number of subtraction constants}, this value is sensitive to the number of subtractions made. The   systematic theoretical error introduced by setting $\gamma_1=\delta_0=0$ reduces the value of the amplitude at the center of the Dalitz plot by the factor 1.483/1.366, so that $Q$ is lowered by almost one unit. 

\begin{table}[thb]
	\renewcommand{\arraystretch}{1.1}
	 { \small
		\begin{center}\begin{tabular}{llr}	
									& \hspace{0.3em} $Q$						& \\ \hline
		Gasser \& Leutwyler (1975)		& 30.2				& \cite{Gasser+1974} \\
		Weinberg (1977)				& 24.1				& \cite{Weinberg1977} \\  
		Gasser \& Leutwyler (1985)		& 23.2(1.8)			& \cite{Gasser+1985a} \\
		Donoghue et al.	(1993)		& 21.8				& \cite{Donoghue+1993} \\  
		Kambor et al.\ (1996)				& 22.4(9)				& \cite{Kambor+1996} \\  
		Anisovich \& Leutwyler (1996)	  	& 22.7(8)				& \cite{Anisovich+1996} \\
		Walker (1998)					& 22.8(8)				& \cite{Walker1998} \\  
		Amoros et al.\ (2001)                         & 21.3				& \cite{Amoros+2001} \\
		Martemyanov \& Sopov (2005)		& 22.8(4)				& \cite{Martemyanov+2005} \\  
		Bijnens \& Ghorbani	 (2007) 		& 23.2				& \cite{Bijnens+2007} \\   
		Kastner \& Neufeld (2008)			& 20.7(1.2)			& \cite{Kastner+2008} \\  
		Kampf et al.\ (2011)				& 23.1(7)				& \cite{Kampf+2011} \\  
		Lanz (2011)                                        & 21.31($\rule{0em}{0em}^{+59}_{-50}$)              & \cite{Lanz-PhD}\\
		FLAG ($N_f = 2 + 1$) (2016)		& 22.5(8)				& \cite{Aoki:2016frl} \\  
		FLAG ($N_f = 2 + 1 + 1$) (2016) 	& 22.2(1.6)			& \cite{Aoki:2016frl} \\  
		BMW ($N_f = 2 + 1$) (2016)		& 23.4(6)				& \cite{Fodor:2016bgu} \\  
		JPAC (2017)					& 21.6(1.1)			& \cite{Guo:2016wsi} \\  
		Albaladejo \& Moussallam (2017) 	& 21.5(1.0) 			& \cite{Albaladejo+2017} \\  
		RM123 ($N_f = 2 + 1 + 1$) (2017)	& 23.8(1.1)			& \cite{Giusti:2017dmp} \\ 
		this work						& 22.1(7)				&\\
		\hline
	\end{tabular} \end{center} }
	\caption{Theoretical results for the quark mass ratio $Q$ (statistical and systematic uncertainties added in quadrature).}  
	\label{tab:resultsQ}
\end{table}
Table~\ref{tab:resultsQ} compares our value of $Q$ with results found in the literature. The numbers listed are either given in the quoted papers or are calculated from the estimates for the quark masses or mass ratios given therein.  The first crude estimate for the masses of the three lightest quarks within QCD, $m_u\simeq 4$ MeV, $m_d\simeq 6$ MeV, $m_s\simeq135$ MeV~\cite{Gasser+1974} appeared in 1975 -- the entry in the first line is calculated from these numbers.  The value given in the second line is obtained from the current algebra formulae for $M_{\pi^+}^2$, $M_{K^+}^2$ and $M_{K^0}^2$, corrected for electromagnetic self-energies with Dashen's theorem~\cite{Weinberg1977} (tree approximation of \chpt). The significance of the quark mass ratio $Q$ for the chiral expansion of the meson masses was noticed only in 1985~\cite{Gasser+1985}.  The third line represents the result of a \chpt\ calculation to one loop~\cite{Gasser+1985a}, where the quantity $\kappa\equiv1/Q^2$ was determined from the experimental decay rate. Note that, at that time, the rate was still subject to substantial uncertainties -- since then, the value of $\Gamma_{\eta\to\pi^+\pi^-\pi^0}$ quoted by the Particle Data Group increased by more than three standard deviations: from 197(29) eV to 299(11) eV.  As the result for $Q$ is inversely proportional to the fourth root of the rate, the one-loop result 23.3(1.8) quoted in Ref.~\cite{Gasser+1985a} drops to $Q=20.9(1.6)$ if the erroneous input used for the width is corrected.

\subsection{Chiral expansion of the meson masses}\label{sec:Other}
As mentioned above, the correction term $\Delta _Q$ is beyond the accuracy of our calculation. Our result relies on the assumption that this term is too small to matter at the precision reached. This assumption concerns the properties of the strong interaction and could be examined with the same methods that are used in lattice determinations of the quark mass ratio 
\begin{equation}\label{eq:Salg}S\equiv\frac{m_s}{m_{ud}}\;.\end{equation}
The lattice results for this quantity have reached remarkable precision~\cite{Aoki:2016frl}. In particular, it has been shown that the result is not sensitive to the heavy quarks. FLAG quotes the values $27.34(31)$ and $27.30(34)$ for simulations of QCD with three and four dynamical flavours, respectively.  Since the most recent lattice results on the light quark masses are obtained with four dynamical flavours, we work with the second number,
\begin{equation}\label{eq:Snum} S=27.30(34)\;.\end{equation}

The quark mass ratio $S$ also represents the leading term in the chiral expansion of a ratio of meson masses. The formula analogous to the low-energy theorem~\eqref{eq:DMQ} reads~\cite{Gasser+1985}\footnote{In the notation used in that reference, $\Delta_S$ stands for $\Delta_M$.} 
\begin{equation}\label{eq:DMalg} \frac{2M_K^2}{M_\pi^2}= (S+1)(1+\Delta_S)\;,\end{equation}
but there is an important difference. While $\Delta_Q$ is of second order in the breaking of chiral symmetry,  $\Delta_S$ is of first 
order and involves the low-energy constants $L_5$ and $L_8$ of \chpt:
\begin{equation}\label{eq:DMalg1}
\Delta_S=O(m_\mathrm{quark})\;.\end{equation}
The lattice result in~\eqref{eq:Snum} implies that the correction $\Delta_S$ is rather small: 
\begin{equation}\label{eq:DMnum}\Delta_S=-0.051(12)\;.\end{equation}

The situation with the quark mass ratio 
\begin{equation}\label{eq:Ralg}R\equiv\frac{m_s-m_{ud}}{m_d-m_u}\end{equation}
is very similar. It compares the breaking of SU(3)-symmetry with the breaking of isospin symmetry; in current algebra approximation, $R$ is given by the ratio of the mass differences $M_K^2-M_\pi^2$ and $\hat{M}_{K^0}^2-\hat{M}_{K^+}^2$. The correction
\begin{equation}\label{eq:DRalg}\frac{M_K^2-M_\pi^2}{\hat{M}_{K^0}^2-\hat{M}_{K^+}^2}=R(1+\Delta_R) \end{equation}
is of the same order as in the case of $S$: $\Delta_R=O(m_\mathrm{quark})$.  

To evaluate $R$ numerically, we make  use of the fact that only two of the three ratios $Q$, $R$ and $S$ are algebraically independent: 
\begin{equation}\label{eq:QRS}2\,Q^2\equiv R(S+1)\;.\end{equation}
With our result~\eqref{eq:DMQ} for $Q$ and the lattice determination for $S$ in~\eqref{eq:Snum}, we obtain
\begin{equation}\label{eq:Rnum}R=34.4(2.1)\;.\end{equation}
The correction in the low-energy theorem~\eqref{eq:DRalg} is of about the same size as for $S$, but of opposite sign:
\begin{equation}\label{eq:DRnum} \Delta_R=0.053(14)\;.\end{equation}
It is not difficult to understand why that is so. The above formulae show that the higher order contributions in $Q$, $S$ and $R$ are related by 
\begin{equation}\label{eq:DQRS}(1+\Delta_Q)=(1+\Delta_S)(1+\Delta_R)\;.\end{equation}%
For the  first order contributions on the right hand side of this relation to cancel one another, the corrections $\Delta_R$ and $\Delta_S$ must be of opposite sign and comparable in size. There is no reason for this cancellation to be complete, but we expect $\Delta_Q$ to be too small to significantly affect our result for $Q$. 

We conclude that, together with the lattice value of $S$, our result for $Q$ leads to a coherent picture for the chiral expansion of the meson masses. The corrections of first order in the breaking of chiral symmetry are small. The well-known fact that the Gell-Mann-Okubo formula holds to good accuracy corroborates this picture further. The formula predicts the value of $M_K$ in terms of $M_\eta$ and $M_\pi$:\footnote{In the notation of Ref.~\cite{Gasser+1985}, $\Delta_{M_K}$ stands for $(M_\eta^2+M_\pi^2)/(3M_\eta^2+M_\pi^2)\Delta_\mathrm{GMO}$ and involves the LECs $L_5$, $L_6$ and $L_7$.}
\begin{equation}M_K^2=(\mbox{$\frac{3}{4}$}M_\eta^2+\mbox{$\frac{1}{4}$}M_\pi^2)(1+\Delta_{M_K})\;.\end{equation}
The correction $\Delta_{M_K}$ is comparable with those in $S$ and $R$, algebraically, $\Delta_{M_K}=O(m_\mathrm{quark})$, as well as numerically, $\Delta_{M_K}=0.063(1)$. 

Since the ratio $m_u/m_d$ is also determined by $S$ and $Q$, our framework leads to an estimate for the relative size of $m_u$ and $m_d$ as well. Neglecting $\Delta_Q$ also here, we obtain
\begin{equation}\label{eq:muovermd}\frac{m_u}{m_d}=0.45(3)\;.\end{equation}
For a while, the theoretical possibility of a massless $u$-quark
was taken seriously as a solution of the strong CP-problem~\cite{Kaplan+1986,Choi:1988sy},
but as pointed out long ago~\cite{Leutwyler:1989pn}, that idea is not consistent with the
observed pattern of chiral symmetry breaking. Our calculation fully confirms this, as it excludes
 the value $m_u=0$ by about 16 standard deviations. 
 
The upshot of the above discussion is that, in QCD, the chiral expansion of the squares of the Nambu-Goldstone masses is dominated by the leading terms. At the physical values of $m_u$, $m_d$, $m_s$, the corrections $\Delta_S$, $\Delta_R$, $\Delta_{M_K}$ from the higher order terms were found to be remarkably small and the low-energy theorem \eqref{eq:DQ} suggests that $\Delta_Q$ is even smaller.
We emphasize that these statements concern the dependence of the meson masses on the masses of the quarks and do not apply to the expansion in powers of the momenta. The example of $\pi\pi$ scattering shows that even within SU(2)$\times$SU(2), the expansion in powers of the momenta picks up sizeable contributions from the final state interaction already at threshold. It is essential that our analysis relies on dispersion theory for the momentum dependence -- as discussed in detail in Sec.~\ref{sec:Anatomy},  \chpt\ does not describe the momentum dependence of the transition amplitude sufficiently well in the physical region of the decay, even if the contributions arising at NNLO of the chiral perturbation series are taken into account.

\subsection{Comparison with the lattice results for $Q$}\label{sec:lattice Q}

Finally, we compare our results for $Q$ with the most recent determinations on the lattice. Table~\ref{tab:resultsQ} shows that, while the results reviewed in the FLAG report~\cite{Aoki:2016frl} for simulations with 3 or 4 flavours are quite consistent with ours, the most recent determinations, BMW ($N_f=2+1$)~\cite{Fodor:2016bgu} and RM123 ($N_f=2+1+1)$~\cite{Giusti:2017dmp} are higher than our 
value~(\ref{eq:Q}) by $1.5$ and $1.4$  standard
deviations, respectively. As mentioned in Sec.~\ref{sec:MK}, the results obtained 
in these references for the kaon mass difference are consistent with ours. Also,
the uncertainties in the values of the isospin limits $M_\pi$ and $M_K$ are
much too small to explain the discrepancy. Hence the difference must arise from the correction 
term $\Delta_Q$ in the low-energy theorem~\eqref{eq:DMQ}, which is beyond the accuracy of our calculation.

To identify the core of the problem, we stick to the central values for $M_\pi$ and $M_K$ in 
\eqref{eq:Mpibarnum}, \eqref{eq:MKbarnum}. Also, in order to respect the identity~\eqref{eq:QRS}, 
we fix the value of $S$  with those for $R$ and $Q$ given in the two references. Using the values 
for the mass difference $\hat{M}_{K^0}^2-\hat{M}_{K^+}^2$ listed in Eq.~\eqref{eq:DMKlattice}, 
the relations~\eqref{eq:DMalg}, \eqref{eq:DRalg} and \eqref{eq:DMQ} can then be solved
for  $\Delta_S$, $\Delta_R$ and $\Delta_Q$, respectively. The results are listed in Table~\ref{tab:DSDRDQ}.
\begin{table}[thb]\centering
\begin{tabular}{lllll}
&\hspace{0.7em}$Q$&$\hspace{0.9em}\Delta_S$&\hspace{0.9em}$\Delta_R$&\hspace{0.9em}$\Delta_Q$\\
\hline
BMW  \cite{Fodor:2016bgu}&23.4(6)&$-$0.063&$-$0.028&$-$0.089\\
\hline
RM123 \cite{Giusti:2017dmp}& 23.8(1.1)&$-$0.042&$-$0.060&$-$0.099\\
\hline
this work&22.1(7)&$-$0.051(12)&$+$0.053(14)&\hspace{0.9em}0\\
\hline\end{tabular}\caption{Corrections to the current algebra results for the quark mass ratios $S$, $R$ and $Q$.\label{tab:DSDRDQ}}
\end{table}
We only list the central values -- since the quantities $\hat{M}_{K^0}^2-\hat{M}_{K^+}^2$,
$R$ and $Q$ are strongly correlated, a meaningful error estimate requires knowledge
of the correlations and is thus beyond our reach.  The outcome for $\Delta_S$ and $\Delta_R$ 
confirms that the first order corrections are small, but $\Delta_R$ is of the same sign as $\Delta_S$:
on the right hand side of~\eqref{eq:DQRS}, the two contributions cannot possibly cancel. Hence
the result for $\Delta_Q$ is in conflict with the expectation that effects of second order are smaller 
than those of first order. 

The lattice approach is ideally suited to resolve this conundrum. At least in principle, it should be possible to determine $\Delta_Q$ with the same accuracy as $m_s/m_{ud}$ -- the issue concerns QCD and is not plagued by the long range contributions from QED, which are difficult to account for at finite volume. The calculation requires the simulation of QCD with three (or more) quark flavours of unequal mass. More precisely, one needs to calculate the meson masses  $M_{\pi^+}$,  $M_{K^+}$, $M_{K^0}$ in this theory as a function of the quark masses $m_u$, $m_d$, $m_s$. The scale $\Lambda_\mathrm{QCD}$ can be pinned down with the pion decay constant, for instance, and if the simulation includes charmed quarks, the corresponding mass can be fixed with $M_{D^+}$. The quantities of interest are the following combinations of meson and quark masses : 
\begin{eqnarray}\label{eq:DeltaSRQ}\Delta_S \al = \al \frac{2M_K^2}{M_\pi^2(S+1)}-1\;,\hspace{1em}\Delta_R= \frac{M_K^2-M_\pi^2}{(M_{K^0}^2-M_{K^+}^2)R}-1\;,\nonumber\\
\Delta_Q\al=\al\Delta_S+\Delta_R+\Delta_S\Delta_R\;,\end{eqnarray}
with $M_\pi^2\equiv \frac{1}{2}(M_{\pi^0}^2+M_{\pi^+}^2)$ and $M_K^2\equiv \frac{1}{2}(M_{K^0}^2+M_{K^+}^2)$. If the pion decay constant as well as the relative size of the quark masses are held fixed, $\Delta_S$ and $\Delta_R$ grow in proportion to $m_s$ while $\Delta_Q$ is proportional to $m_s^2$. For sufficiently small quark masses, chiral symmetry guarantees that $\Delta_Q$ is small compared to $\Delta_S$ and $\Delta_R$, but if the breaking of chiral symmetry becomes comparable to the scale of the theory, there is no reason for this to be so. Table \ref{tab:DSDRDQ} indicates that, for quark masses in the vicinity of the physical values, $\Delta_S$ amounts to about 0.05.  What is the size of $\Delta_Q$ there? 

While completing the present work, the Fermilab Lattice, MILC \& TUMQCD collaborations came up with a new lattice determination of the quark masses
~\cite{Bazavov+2018}. Unfortunately, the paper does not contain a result for the ratio $Q$, but neglecting correlations and adding errors in quadrature, the mass ratios which are given therein, $S= 27.182(46)(56)(1)$ and $m_u/m_d=0.4517(55)(101)$, imply $Q=22.1(3)$ and $R=34.7(1.0)$. The central values are very close to our numbers in Eqs.~\eqref{eq:Q} and~\eqref{eq:Rnum}. Accordingly, the outcome of this calculation appears to be consistent with a coherent chiral expansion of the meson masses and to confirm that the corrections to the current algebra formulae are small. Although the paper focuses on the determination of the masses of the heavy quarks, the ratios $m_u/m_d$ and $m_s/m_{ud}$ are given to remarkable accuracy. In particular, the precision claimed for $S$ is breathtaking -- the quoted uncertainty is about four times smaller than for the FLAG value~\eqref{eq:Snum} we are relying on and the uncertainty in the outcome for $Q$ is smaller than ours by  more than a factor of two. 
Concerning the comparison with~\cite{Fodor:2016bgu,Giusti:2017dmp}, the main difference is that the calculation is done within QCD rather than QCD + QED. The outcome for the masses $m_u$, $m_d$ and $m_s$ is corrected for e.m.~effects, but for details of the procedure used, the reader is referred to a  forthcoming paper by the MILC collaboration.  

\section{Comparison with other work}\label{sec:Comparison}
\subsection{Dispersive approaches}\label{sec:Dispersion theory}
Early papers on $\eta \to 3 \pi$ which have followed a similar approach to
the one presented here are~\cite{Anisovich+1996,Kambor+1996}. Indeed, in
spirit, the calculations are very similar, but there are significant
differences which make a detailed comparison of the results difficult:
\begin{itemize}
\item
The phase shifts adopted in~\cite{Kambor+1996,Anisovich+1996} were taken
from~\cite{Schenk1991}, whereas we are now able to use solutions of Roy
equations matched to \chpt~\cite{Ananthanarayan+2001,Colangelo2001}.
\item
At that time, accurate data on the Dalitz plot in the charged channel were 
not available yet, so that the best one could do to fix the subtraction
constants was to match them to \chpt.
\item
The available \chpt\ calculation was at one loop, and therefore there
was no possibility to go beyond four subtraction constants.
\item
The treatment of isospin breaking corrections available at that time
\cite{Baur+1996} was not yet as complete as the one provided in
\cite{Ditsche+2009}.
\end{itemize}
The  result
 $Q=22.4(9)$ obtained by Kambor, Wiesendanger and Wyler~\cite{Kambor+1996} 
and the value $Q=22.7(8)$ of Anisovich and Leutwyler~\cite{Anisovich+1996}
are slightly higher than ours, but the difference is mainly due to the fact that, in the 
meantime, the experimental value of the decay rate quoted by the Particle Data Group
increased (updating the calculation of~\cite{Anisovich+1996} with  
Ref.~\cite{Walker1998}, the result is lowered to
$Q=22.3(8)$~\cite{Leutwyler:2009jg}).  

The formulae derived by Kambor et al.\ have been used later to fit
KLOE data by Martemyanov and Sopov~\cite{Martemyanov+2005}. 
The paper is very short and does not give any detail about the calculation
-- other than a formula of Kambor et al., on which the authors based their 
analysis~\cite{Kambor+1996}. All the differences pointed
out above between the present analysis and the one by Kambor et al.\ apply
also to this calculation -- in particular that isospin breaking effects
have not been accounted for. For completeness we nonetheless quote the
value of $Q$ they obtained: $Q=22.8(4)$. The central value is the same
as the one quoted by Walker~\cite{Walker1998} and therefore higher than the one obtained by
Kambor et al., but the error much reduced. It is difficult to
understand why the effect of the KLOE data is to increase the value
obtained for $Q$, with respect to what Kambor et al.\ obtained by doing a
matching to \chpt\ to one loop. In his PhD thesis~\cite{Lanz-PhD} one of the
authors of the present paper (S.L.) showed that if one applies the same
formulae and simply replaces \chpt\ with data to fix the subtraction
constants, the value obtained for $Q$ decreases (see also
\cite{Lanz:2013ku}).  
\begin{figure}[thb]\hspace{-0.5em}\includegraphics[width = 8.5cm]{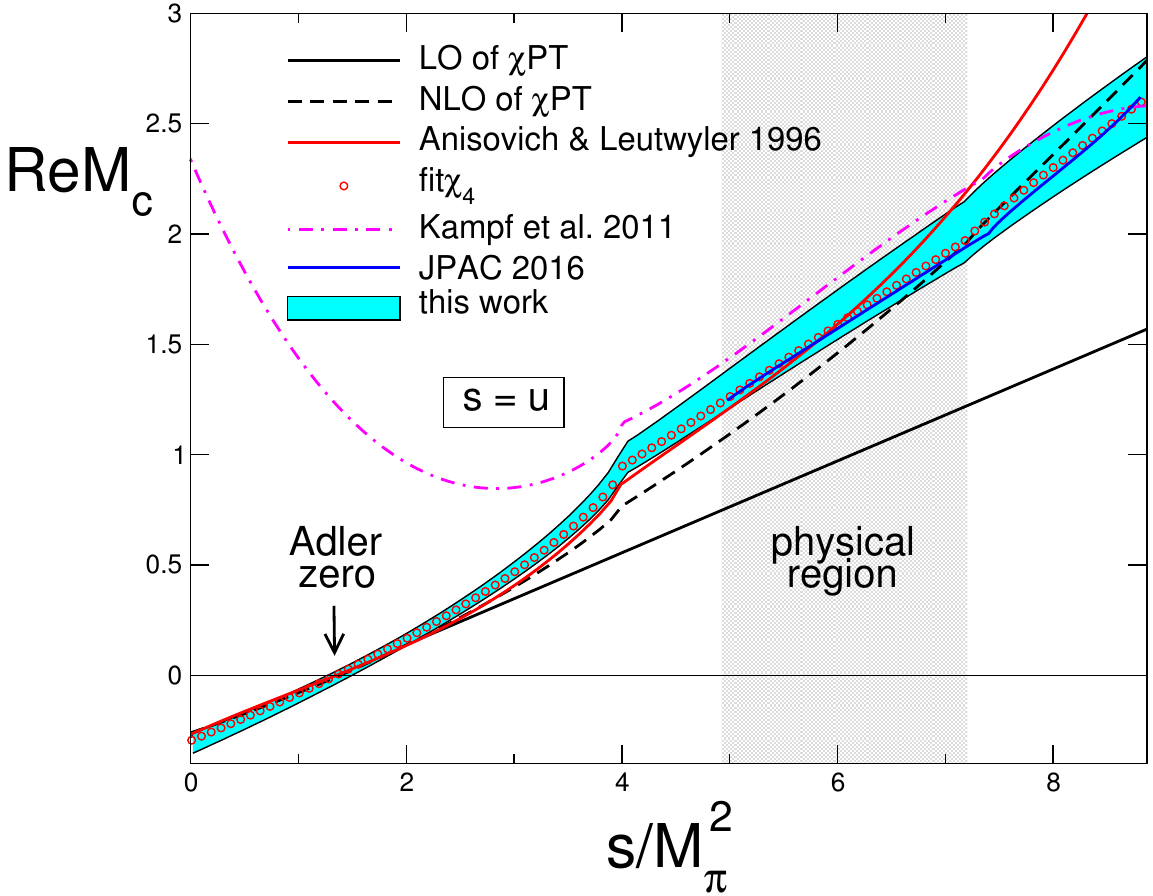}\caption{Real part of the amplitude along the line $s=u$.\label{fig:ReMsuA}}
\end{figure} 

Fig.~\ref{fig:ReMsuA} amounts to an update of a picture drawn by Anisovich and 
Leutwyler, more than twenty years ago, in order to illustrate the effects generated 
by the final state interaction~\cite{Anisovich+1996}. The framework underlying that 
paper is essentially the same as the one used in the construction of the matching
 solution fit$\chi_4$ in Sec.~\ref{sec:Matching}: a dispersive analysis with four subtraction constants, 
 which are determined by imposing theoretical constraints derived from \chpt. The 
 figure concerns the behaviour of the real part of the amplitude $M_c(s,t,u)$ along the 
 line $s=u$, in the isospin limit. 
 
 In the present work, the convention used for the value of 
 the pion mass in the isospin limit is irrelevant, because we account for isospin breaking 
 when comparing our calculation with experiment. In Fig.~\ref{fig:ReMsuA}, however, it 
 does matter: The straight line that shows the behaviour at leading order (LO), for instance, 
 depends on it. We identify the isospin limit of the pion mass with the mass of the charged pion, 
 while in~\cite{Anisovich+1996}, the mass of the neutral pion was used. If isospin 
 breaking corrections are not applied, that choice is preferable because isospin breaking
 in the masses of the pions is dominated by electromagnetism, which barely affects the
 mass of the neutral pion. We correct for the difference in the same way as for 
 the isospin breaking corrections, using \chpt. At LO, the transformation of the amplitude from 
 one convention to the other amounts to a mere rescaling of the vertical axis, by the factor  
 $M_{\pi^+}^2/M_{\pi^0}^2\,(M_\eta^2-M_{\pi^0}^2)/(M_\eta^2-M_{\pi^+}^2)\simeq 1.074$. 
At one-loop, the isospin limit of the chiral representation is given by $M_c^\mathrm{GL}(s,t,u)$ 
and the real parts are readily worked out for $M_\pi=M_{\pi^0}$ as well as for $M_\pi=M_{\pi^+}$.
The ratio of the real parts remains roughly constant, but at a slightly larger value. We expect this
to be the case for the dispersive representation as well -- the red curve in Fig.~\ref{fig:ReMsuA}  
is obtained from the one shown in the old figure  by stretching the values with the one-loop
result for the ratio of the real parts.  

For comparison, the open circles in Fig.~\ref{fig:ReMsuA} show the real part of the amplitude
belonging to the matching solution,  fit$\chi_4$. The main difference between this representation 
and  the one obtained in Ref.~\cite{Anisovich+1996} is that the $\pi\pi$ phase shifts are now known 
much more precisely. The figure shows that the old calculation underestimates the amplification 
of the amplitude by the final state interaction at threshold, but overestimates its growth with the energy. 

The figure also shows the outcome of two  more recent calculations~\cite{Kampf+2011,Guo:2016wsi}. 
Kampf, Knecht, Novotn\'{y} and Zdr\'{a}dhal~\cite{Kampf+2011} have adopted a
dispersive approach as well, but instead of solving the dispersion relations
numerically, they have solved them analytically by iterations, stopping
at the second iteration. This corresponds to a two-loop \chpt\
representation from the analytic point of view, but the subtraction
constants are not exactly related to the LEC of \chpt,  as the authors 
explain in their paper. In this connection, we refer to the 
detailed comparison of the 
dispersive approach with the two-loop representation of \chpt\
given above (Sec.~\ref{sec:Anatomy}). Their approach also differs
from ours in the way the normalization of the amplitude is fixed
from theory: While we use the value of the Taylor
invariant $K_0$, they use the imaginary part of the amplitude along the
line $t=u$. 

Fig.~\ref{fig:ReMsuA} compares their result for the real  
part of the amplitude along the line $s=u$ 
with the outcome of the present work. By construction, both representations
reproduce the Dalitz plot distribution of KLOE -- in the physical region of the decay,
they are nearly the same up to normalization. Below threshold, however, the difference
is very clearly visible: At small values of $s$, where current algebra
predicts the occurrence of an Adler zero at $s=\frac{4}{3}M_\pi^2$,
the amplitude of Kampf et al.~goes astray. We encountered a similar
phenomenon in Sec.~\ref{sec:Theoretical constraints}: Fig.~\ref{fig:ReMdisp}
shows that our calculation also goes astray if we allow for 6 subtraction
constants and fit the data on the Dalitz plot distribution by treating these
as free parameters. According to Martin Zdr\'{a}hal~\cite{Zdrahal-PC}, this deficiency can
be repaired without affecting significantly the rest of the calculation and
in particular the fit to data, but detailed results for this improved
analysis within their approach have not been published. Note also
that their work does not account for isospin breaking corrections. The published value
$Q=23.3(8)$ is significantly higher than ours, but in view of the shortcomings
of the underlying analysis, this does not come as a surprise. 

More recently, the JPAC collaboration~\cite{Guo:2014vya,Guo:2015zqa,Guo:2016wsi} has also analyzed
$\eta \to 3 \pi$ decays, and in particular KLOE data, with a dispersive
approach and the aim to determine the value of $Q$. The spirit is similar
to the one adopted here, but the way in which the dispersion relations for
this process are solved differs significantly from ours and
isospin breaking corrections are not applied. The authors
make an approximate treatment of the left-hand cut for the partial wave
amplitudes, and assume that it can be well described by a polynomial. As we
have demonstrated here (following~\cite{Anisovich+1996}), the iterative
procedure for deriving solutions of the dispersion relation converges fast
and takes into account crossed channels (responsible for the left-hand cut)
exactly. It is possible that the polynomial approximation adopted
in~\cite{Guo:2015zqa,Guo:2016wsi} works reasonably well, but having the
exact solution available, this becomes an academic question. We are indebted
to Igor Danilkin for providing us with the numerical values shown in 
Fig.~\ref{fig:ReMsuA}. In the physical region of the decay, their results
are consistent with ours and the same holds for the value obtained for the
quark mass ratio, $Q=21.6(1.1)$.  Unfortunately, the method used does not 
work below the physical region, so that the behaviour in the vicinity of the 
Adler zero cannot be compared.  
 
In Refs.~\cite{Kolesar:2016iyz,Kolesar:2016jwe,Kolesar:2017xrl,Kolesar:2017jdx} 
Koles\'{a}r and Novotn\'{y} take a very different
point of view from the one adopted here -- namely that the reason for the
bad convergence of \chpt\ for this decay is understood and has to do with
large final-state rescattering effects -- and try to identify the reasons
for the bad convergence within the framework of the so-called resummed
Chiral Perturbation Theory (rChPT)
\cite{DescotesGenon:2003cg,DescotesGenon:2007ta}. In this approach, vacuum
fluctuations of $\bar{s} s$ pairs are treated in a special way and their
effect resummed. Their size is left unconstrained, which implies that both
the SU(3) condensate and decay constant are treated as free parameters,
having possibly a very different value than their SU(2) counterparts. The
idea is very intriguing and if one could find a way to rigorously determine
the size of these SU(3) parameters, this would be a very interesting
result.

The present work shows that rescattering effects can be accounted for in
a systematic, nonperturbative manner. Causality and unitarity determine the momentum 
dependence of the transition amplitude up to a set of subtraction
constants -- \chpt\ is used exclusively to work out the constraints on these constants
arising from chiral symmetry. Our analysis, in particular, does 
not rely on the chiral expansion for quantities that contain strong infrared 
singularities and are notoriously difficult to deal with in \chpt.

Very recently, Albaladejo and Moussallam~\cite{Albaladejo+2015} have shown
how to extend the dispersive formalism we have used in the present work to
include the effect of inelastic two-body effects, like $\bar{K}K$ and $\eta
\pi$. This remarkable and very useful technical advance allowed them to
explicitly take into account effects related to narrow resonances in the
one-GeV region, like the $a_0(980)$ and the $f_0(980)$. From their
numerical analysis, they conclude that the effect on the determination of
$Q$ are of the order of $0.2$ units, and therefore much smaller than the
error. They also invoke the KLOE data on the Dalitz plot distribution
in the charged channel to constrain their representation and
to predict the coefficients of the distribution in the neutral channel. 
Setting $\gamma=0$, they obtain $\alpha=-0.0337(12)$, $\beta=-00054(1)$, 
to be compared with our result \eqref{eq:polyKchi6}. While our value 
for $\alpha$ is smaller than theirs by about 2 $\sigma$, we do confirm
their value of $\beta$. The difference may
in part arise because their analysis does not account for isospin breaking 
corrections, in part because the terms proportional to $\alpha$ and $\beta$
in the Taylor series \eqref{eq:DnTaylor} provide a decent approximation
only in the immediate vicinity of $Z=0$. As discussed in Sec.~\ref{sec:etato3pi0},
the curvature term $\gamma$ affects the behaviour away from the center of the
physical region -- setting it to zero distorts the result for $\alpha$. At any rate,
we consider it very unlikely that the difference has to do with the presence of inelastic
channels. The plots shown in~\cite{Albaladejo+2015} indicate that -- in the physical 
region of the decay -- the effects generated by these are well described by a polynomial.
In our calculation, such contributions are absorbed in the subtraction constants.
We do therefore not expect that explicitly accounting for inelastic channels
would lead to a significant change in our results.   
\subsection{Nonrelativistic effective field theory}
\label{sec:Nonrelativistic}

A different approach which has been applied to $\eta \to 3 \pi$ decays is
the one relying on a nonrelativistic Lagrangian. This has been very
successful in describing $K \to 3 \pi$ decays and in particular the cusp
structure at the opening of the $ \pi^+ \pi^-$ channel in the $2 \pi^0$
spectrum of the $K^\pm \to \pi^\pm 2 \pi^0$ decay
\cite{Colangelo+2006a,Bissegger:2007yq,Bissegger:2008ff,Gasser:2011ju}.
In this framework one makes a nonrelativistic expansion both at the level
of the Lagrangian as well as in the calculation of rescattering 
effects.
The importance of the latter is controlled by the scattering lengths, which
happen to be small (as a consequence of the Nambu-Goldstone-boson nature of the
pions): Technically, the NREFT also relies on an expansion in the
scattering lengths. From the calculation point of view, rescattering
effects are taken care of automatically by the loop expansion of quantum
field theory.  A significant advantage of this approach is that one does
not rely on an expansion in the quark masses: The tree-level decay
amplitude near to threshold is expanded in the spatial momentum squared,
and the coefficients of this expansion are treated as free parameters. Which
means that in this approach one does not have to worry about the slow
convergence of \chpt\ for the scattering lengths, for example, because
these are by definition the physical values.  The only question that
matters in this case is whether one is close enough to threshold that the
nonrelativistic expansion works.

The nonrelativistic approach is applied to the decay $\eta\to 3\pi$ in  
Refs.~\cite{Gullstrom:2008sy,Schneider+2011}. 
The mass difference between the charged and neutral pions
is accounted for and the cusp due
to the opening of the $\pi^+ \pi^-$ channel in the $\pi^0 \pi^0$ spectrum
of the decay $\eta \to 3 \pi^0$ is analyzed in detail. Moreover, fitting the
free parameters in the nonrelativistic representation of the transition
amplitude to the KLOE data available at the time, the authors of
Ref.~\cite{Gullstrom:2008sy} did obtain a negative value
for the slope $\alpha$ in the neutral channel, as observed. A comparison of the
predicted Dalitz plot in the neutral channel with the data by MAMI-C shows
that the calculation is in reasonable agreement with the data: In
particular that, as one moves from tree-level to one and then to two loops
(in the NR expansion), the curves obtained move towards the data and show a
good convergent behaviour. 

It is worth emphasizing here the difference
between our approach and the NR expansion: While in a dispersive treatment
rescattering effects (in the $S$ and $P$ waves) are treated exactly, the NR
expansion applies a perturbative scheme to account for these. However, the
treatment of isospin breaking effects can be done in a theoretically much
cleaner way within the NR approach. We have relied on one-loop \chpt\ and a
factorization hypothesis, which can only be approximately correct. To
exemplify the difference between the two approaches it is useful to compare
the Dalitz plot in the neutral channel: In the NR approach the strength of
the cusp effect is exactly described in terms of the $S$-wave scattering
lengths, according to a venerable low-energy theorem~\cite{Budini+1961}. 
If these were taken from experiment, 
then the strength of the cusp would be correct by definition. 

In Ref.~\cite{Schneider+2011} this approach has been further refined and
extended to include isospin breaking corrections beyond the $\pi^+-\pi^0$
mass difference, and a complete set of formulae describing these decays in
the NR expansion have been provided. In this paper the question whether
fitting the Dalitz plot data in the charged channel correctly reproduces
the Dalitz plot in the neutral channel has been addressed thoroughly. The
conclusion is similar to the one obtained by Gullstr\"om et
al.~\cite{Gullstrom:2008sy}, namely that the agreement with the data in the
neutral channel is marginal. In particular, only at the two loop level does
the value of $\alpha$ become negative, and only after a partial resummation
of rescattering effects does it get close to the measured value.
For the coefficients of the Dalitz plot distribution in the neutral channel, Schneider et
al.~\cite{Schneider+2011} obtain $\alpha=-0.0246(49)$, $\beta = -0.0042(7)$, 
$ \gamma=0.0013(4)$, based on matching to \chpt\ and resummation of 
bubble graphs. Although the ingredients of this calculation are quite different
from ours, the comparison with the numbers in  \eqref{eq:DnTaylor} shows that 
the qualitative properties of the prediction
for the Dalitz plot distribution in the neutral channel are the same. 
 
Ref.~\cite{Schneider+2011} also proposes a different approach to the
determination of $\alpha$ within the NREFT formalism: The authors derive an exact
relation (in the isospin limit) between the Dalitz plot parameters in the
charged channel and the slope $\alpha$ in the neutral channel and show that
if one inputs the parameters measured by KLOE and estimates the imaginary
part of a combination of Dalitz plot parameters (defined as
$\Im \,\bar{a}$) within the NR expansion, one obtains a value for
$\alpha$ which is only in marginal agreement with the measured value. This
remains true even after calculating isospin breaking corrections. We have
analyzed this apparent clash in some detail and came to the conclusion that
the estimate of the parameter $\Im \,\bar{a}$ within the NR
expansion does not seem to be reliable. The reasoning is as follows: If we
fit the KLOE data and calculate the slope at $Z=0$ with our dispersive
representation we get $\alpha=-0.0302(13)$, in agreement with the
PDG value. This evaluation accounts for isospin breaking effects. 
As discussed in Sec.~\ref{sec:coefficients}, the polynomial approximation to 
our central solution agrees well with the experimental determination by KLOE. 
If we now insert these numbers in Eq.~(6.9) of Ref.~\cite{Schneider+2011} and rely on their estimate of 
 $\mathrm{Im}\,\bar{a}$ we get $\alpha = -0.0474$, in substantial disagreement with our
own direct determination. Since Eq.~(6.2) of Ref.~\cite{Schneider+2011} is
algebraically exact, and the estimate of the isospin breaking effects
(leading to Eq.~(6.9)) only gives a small correction, the problematic step
must be in the estimate of $\Im  \,\bar{a}$. 

An even better test of the NREFT approach would be to analyze the data along the lines 
of Sec.~\ref{sec:Comparison NREFT}
\section{Summary and Conclusions}\label{sec:Conclusions}
\hspace{0.9em}1.~The essential properties of the framework we are using to analyze the transition amplitude of the decay $\eta\to 3\pi$ were derived long ago~\cite{Gribov:1958ez,Khuri+1960,Anisovich+1966}.  The decay violates the conservation of isospin. Since chiral symmetry suppresses the electromagnetic interaction in this transition~\cite{Bell+1968}, the dominating contribution arises from QCD and is proportional to the difference $m_d-m_u$ of quark masses.  
It is convenient to normalize the amplitude with  
\begin{equation} A_{\eta\to\pi^+\pi^-\pi^0}=-\frac{\hat{M}_{K^0}^2-\hat{M}_{K_+}^2}{3\sqrt{3} F_\pi^2}\,M_c(s,t,u)
\end{equation}
where $\hat{M}_{K^0}$ and $\hat{M}_{K^+}$ denote the kaon masses in QCD. 

2.~The first part of the present paper reviews the dispersion theory of the amplitude $M_c(s,t,u)$ in the isospin limit ($e\to 0$, $m_u\to m_d$),  where this function  also determines the amplitude relevant for the transition $\eta\to 3\pi^0$.
We follow the dispersive analysis set up in~\cite{Anisovich+1996}, which exploits the fact that, at low energies, the angular momentum barrier suppresses the imaginary parts of the D- and higher partial waves. Neglecting these, the amplitude can be decomposed into three isospin components, which only depend on a single variable:  $M_0(s)$, $M_1(s)$, $M_2(s)$ -- see Eq.~\eqref{eq:RT}. 

3.~Elastic unitarity determines the discontinuities of the isospin components across the branch cuts associated with collisions among pairs of pions, in terms of the S- and P-wave $\pi\pi$ phase shifts. We write the corresponding dispersion relations in the form~\eqref{eq:DROmega}, allowing for six subtraction constants: $\alpha_0$, $\beta_0$, $\gamma_0$, $\delta_0$, $\beta_1$, $\gamma_1$. These relations represent a set of integral equations that uniquely determine the amplitude in terms of the subtraction constants. Moreover, since the equations are linear in the subtraction constants, the general solution is given by a linear combination of six fundamental solutions that can be determined once and for all. 

4.~At the experimental accuracy reached, the electromagnetic interaction cannot be ignored. In particular, the e.m.~self-energy of the charged pion modifies the amplitude obtained from QCD quite significantly. We rely on the representation of Ditsche, Kubis and Mei{\ss}ner~\cite{Ditsche+2009}, who evaluated the transition amplitude within the effective theory of QCD+QED, to first non-leading order of the chiral expansion and to order $e^2$ in the electromagnetic interaction. Their analysis in particular also accounts for the emission of the soft photons that necessarily accompany the decay as well as for the Coulomb pole generated by the attraction among the charged pions in the final state. We assume that the data are radiatively corrected in accordance with their analysis.

5.~A substantial part of the e.m.~interaction can be accounted for with a purely kinematic map that takes the physical phase space of the decay $\eta\to\pi^+\pi^-\pi^0$ onto the phase space of the isospin symmetric world. Applying this map and removing the Coulomb pole, the isospin breaking corrections reduce to an approximately constant numerical factor, except near $s=4M_{\pi^+}^2$, where a visible structure due to the interference of the branch cuts from $\pi^+\pi^-$ and $\pi^0\pi^0$ intermediate states remains (left panel of Fig.~\ref{fig:K}). Isospin breaking in the decay $\eta\to 3\pi^0$ can be treated analogously. In that case, a Coulomb pole does not occur. Instead there is a small cusp due to the virtual transition \mbox{$\pi^0\pi^0\to\pi^+\pi^-\to \pi^0\pi^0$}  (right panel of Fig.~\ref{fig:K}). Those isospin breaking effects that are not taken care of by the kinematic map are accounted for only in one-loop approximation. 

6.~The theoretical constraints that follow from the fact that the pions are Nambu-Goldstone bosons of a hidden approximate symmetry can be worked out by means of Chiral Perturbation Theory. The representation of the amplitude obtained on this basis does have the structure of Eq.~\eqref{eq:RT}, up to and including NNLO. The only qualitative difference compared to the dispersive framework we are using is that the chiral representation corresponds to an extended version of elastic unitarity, which also accounts for the discontinuities generated by $K\bar{K}$, $\eta\eta$ and $\pi\eta$ intermediate states. In the region relevant for $\eta$ decay, the contributions generated by these singularities are very small and well described by their Taylor expansion in powers of $s$. As we are working with sufficiently many subtractions, they can be absorbed in the subtraction constants. 

7.~At leading order of the chiral expansion (current algebra), the transition amplitude of the decay $\eta\to\pi^+\pi^-\pi^0$ is independent of $t$ and $u$, grows linearly with $s$ and has an Adler zero at $s=\frac{4}{3}M_\pi^2$: $M_c(s,t,u)=(3s-4M_\pi^2)/(M_\eta^2-M_\pi^2)$.  Although the zero occurs outside the physical region, the data on the Dalitz plot distribution beautifully confirm its presence: Ignoring the theoretical constraints altogether and allowing only four subtraction constants, the dispersive representation yields a very good fit of the data (Sec.~\ref{sec:without theoretical constraints}, fitK$_4$). Along the line $s=u$, the real part of this representation indeed passes through zero at $s=1.43M_\pi^2$, close to the place where current algebra predicts this to happen. 

8.~The information provided by \chpt\ is essential, because the Dalitz plot distribution leaves the normalization of the amplitude open. To establish contact between the dispersive and chiral representations, we consider the region where the uncertainties in the latter are smallest, i.e.~focus on small values of $s$ in $M_0(s)$, $M_1(s)$, $M_2(s)$ and compare Taylor coefficients. The requirement that the one-loop representation, which does not involve any unknowns, yields an acceptable approximation at low energies allows us to consistently combine the two. In particular, we normalize the dispersive representation with the one-loop value of the coefficient $H_0$, accounting for the higher order contributions merely by attaching an uncertainty estimate to this value.

9.~There is an alternative to fitK$_4$, which we denote by fit$\chi_4$: A dispersive representation that also uses only four subtraction constants, but incorporates the theoretical information instead of the one obtained at KLOE. It is uniquely determined by the requirement that the isospin components of the dispersive representation match those of the one-loop representation at small values of $s$. Fig.~\ref{fig:Match} shows that the one-loop approximation accurately follows the dispersive representation only below threshold -- in the physical region, it underestimates the strength of the final state interaction. This manifests itself particularly clearly in the Dalitz plot distribution of the neutral decay mode: Fig.~\ref{fig:Match Dalitzn} shows that the curvature of the two representations differs even in sign. 

10.~The same deficiency also shows up at two loops: The lowest resonance of QCD, the $f_0(500)$, is not described well enough even at NNLO of the chiral expansion. This implies that the two-loop representation does not have the necessary accuracy in the physical region -- a meaningful comparison of theory and experiment is possible only in the framework of dispersion theory.  The problem is illustrated in Fig.~\ref{fig:MNNLOmatch}, which compares our central solution with the two-loop representation that matches it at low energies.  

11.~We emphasize that the analysis reported here became possible only very recently, with the accurate measurement of the Dalitz plot 
distribution for the decay $\eta\to\pi^+\pi^-\pi^0$ at KLOE~\cite{KLOE:2016qvh}. For the central solution of our system of equations, the errors arising from the experimental and theoretical uncertainties are of comparable size -- $\eta$-decay is a showcase for a fruitful interplay between theory and experiment. 

12.~As discussed in detail in Sec.~\ref{sec:Error analysis}, the simpler framework obtained by dropping the subtraction constants $\delta_0$ and $\gamma_1$ is too stiff -- doing this amounts to imposing constraints that distort the transition amplitude.  The need for the term $\delta_0\hspace{0.05em}s^3$ in the subtraction polynomial of $M_0(s)$ also shows up in connection with the polynomial approximation of the kaon loops: The contributions from the $K\bar{K}$ cuts to $M_0(s)$ are not accounted for sufficiently well by a quadratic polynomial, but a cubic one does suffice. Moreover, working with six subtraction constants has the advantage that -- in the region of interest -- the solutions are then not sensitive to the high energy tails of the dispersion integrals, where elastic unitarity does not represent a good approximation. In the error analysis, the uncertainties associated with the high energy tails are booked together with those in the phase shifts at low energies, where the Roy equations provide very good control -- with six subtraction constants,  the net uncertainty from 
these sources is very small. 

13.~The decomposition of the amplitude $M_c(s,t,u)$ into its isospin components $M_0(s)$, $M_1(s)$, $M_2(s)$ is unique only up to polynomials [see Eqs.~\eqref{eq:gauge a}, \eqref{eq:gauge b}]. For the dispersive representation, the ambiguity is disposed of when bringing the dispersion relations to the form~\eqref{eq:DROmega}. Alternatively, the solutions can be characterized by invariant combinations of Taylor coefficients: Two solutions yield the same representation $M_c(s,t,u)$ if and only if these invariants are the same. This allows us to unambiguously characterize the two-loop representation that matches our central solution at low energies (see Sec.~\ref{sec:LEC}). A corresponding update of the low-energy constants occurring in the effective Lagrangian at $O(p^6)$ would be of considerable interest but is beyond the scope of the present work. 

14.~Isospin symmetry leads to a prediction for the branching ratio of the neutral and charged decay modes, $B=\Gamma_{\eta\to3\pi^0}/\Gamma_{\eta\to\pi^+\pi^-\pi^0}$. The result of our calculation, $B=1.44(4)$ is in good agreement with the values $B=1.426(26)$ and $B=1.48(5)$ quoted by the Particle Data Group~\cite{Olive:2016xmw}.
\begin{figure*}[thb] \centering
\includegraphics[width=7.5cm]{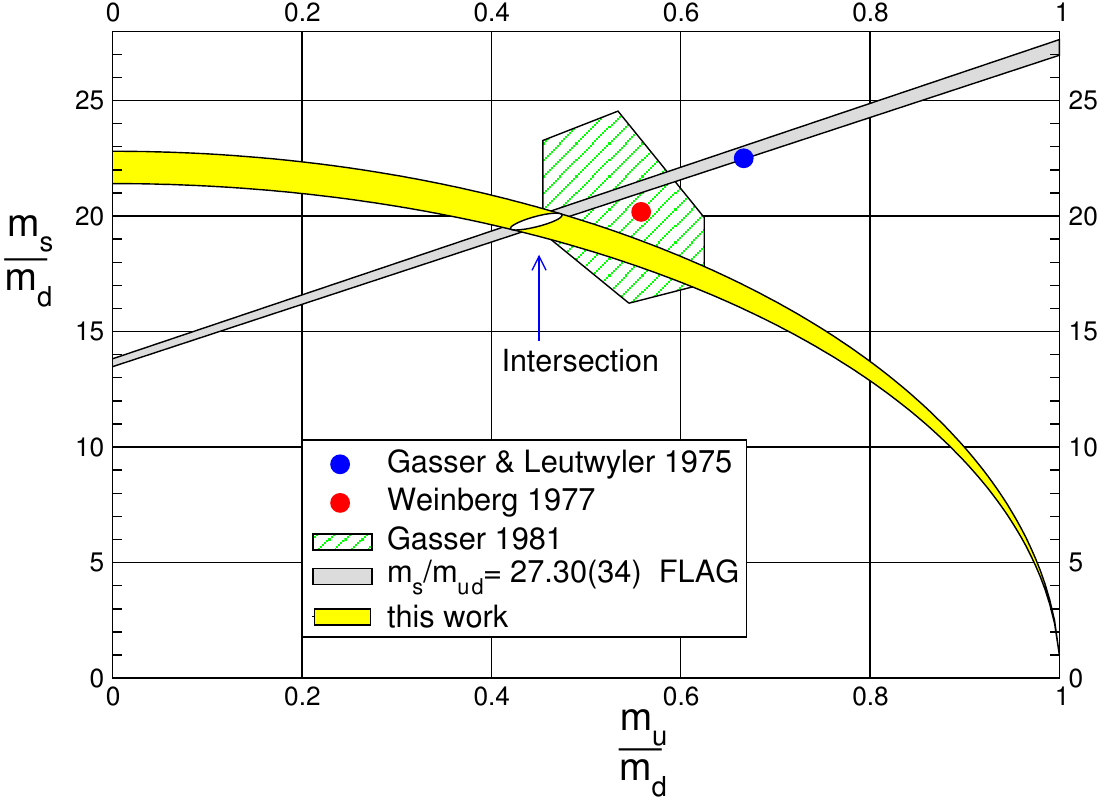}\hspace{2.em} \raisebox{0.1em}{\includegraphics[width=7.8cm]{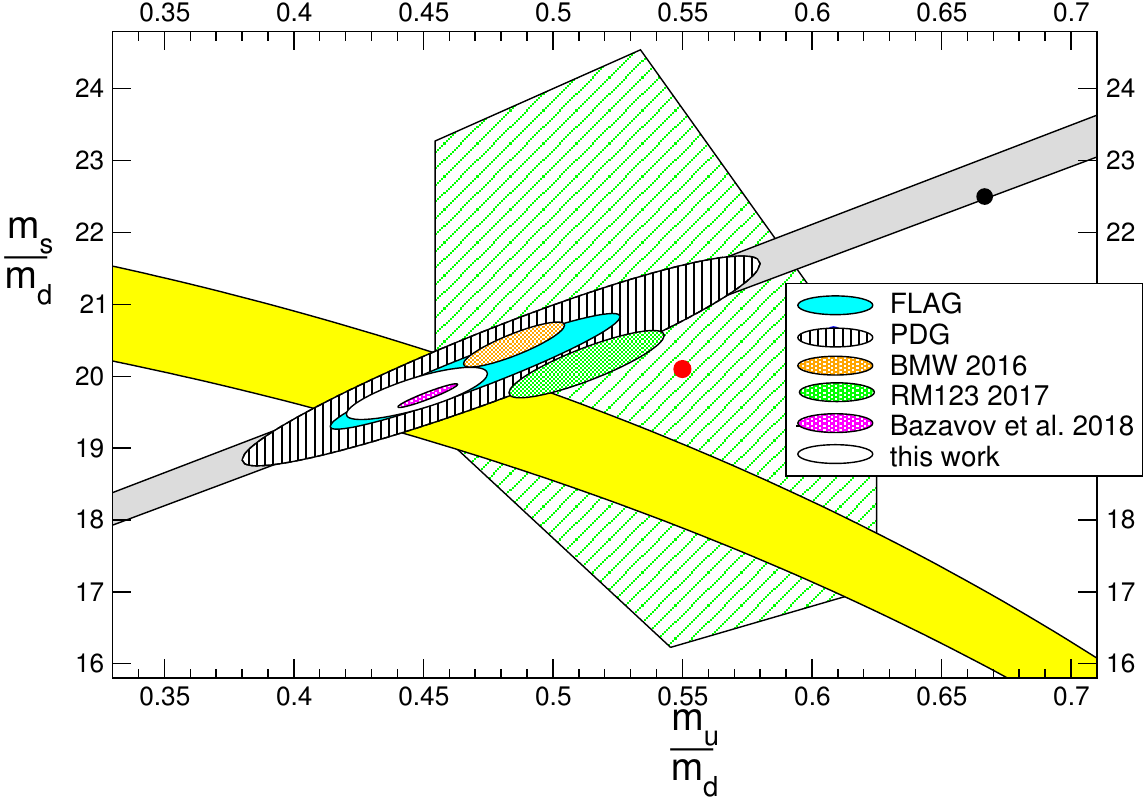}}\caption{Quark mass ratios (FLAG shown for $N_f=4$).\label{fig:Quark mass ratios}}
\end{figure*}
15.~The Dalitz plot distribution of the decay $\eta\to 3\pi^0$ can be expanded in powers of the variables $X_n$, $Y_n$. In the region where the series converges, $X_n^2+Y_n^2<0.6$, our prediction is remarkably well approximated by the  polynomial~\eqref{eq:DnTaylor} -- the coefficients are specified in \eqref{eq:polyKchi6}. In the remainder of the physical region, the singularities generated by the final state interaction manifest themselves as cusps. The dominating contribution from these  is described by the formula \eqref{eq:Dncusp}.  Although they are too weak to stick out from the fluctuations in the data, the quantitative analysis does confirm their presence at the strength required by dispersion theory. 

16.~The MAMI data on the decay $\eta\to 3\pi^0$~\cite{Unverzagt+2009,Prakhov+2009,Prakhov+2018} allow a strong test of our calculation. Isospin symmetry implies that the amplitude of this transition is described by the combination $M_n(s)\equiv M_0(s)+\frac{4}{3}M_2(s)$ of the isospin components relevant for the charged channel -- the KLOE data thus lead to a parameter free prediction for this decay. Fig.~\ref{fig:z-distribution} shows that the calculated distribution is in excellent agreement with the MAMI results.

17.~The recent update provided by the A2 collaboration \cite{Prakhov+2018} now allows an analysis of the Dalitz plot distribution that goes beyond the linear approximation. The data in the neutral channel do not by themselves determine the slope very accurately, but impose a strong correlation between the slope $\alpha$ and the curvature $\gamma$. Dispersion theory provides the missing element as it determines the curvature within narrow limits. Our analysis, which relies on the KLOE data for $\eta\to\pi^+\pi^-\pi^0$ and on the theoretical constraints that follow from the presence of a hidden approximate symmetry, predicts both the slope and the curvature rather precisely: $\alpha= -0.0303(12)$,  $\gamma=0.0019(3)$. The slope is somewhat smaller than the average $\alpha=-0.0318(15)$ quoted by the Particle Data Group~\cite{Olive:2016xmw}. Including the MAMI data \cite{Prakhov+2018} in the dispersive analysis, we obtain a result that is even a little smaller: $\alpha = -0.0294(10)$. Unfortunately, the likelihood of the fits to the MAMI results is not satisfactory: $\chi^2_\mathrm{M}/\mathrm{dof}=1.25$ for the polynomial fit to these data alone and $\chi^2_\mathrm{M}/\mathrm{dof}=1.27$ for the dispersive fit, which combines them with the data from KLOE. 

18.~Our result $\hat{M}_{K^0}^2-\hat{M}_{K^+}^2 =  6.3(4) 10^{-3}\,\mbox{GeV}^2$ for the kaon mass difference in QCD agrees with recent determinations of the electromagnetic self-energies on the lattice~\cite{Fodor:2016bgu,Giusti:2017dmp}. We thus confirm that the strong infrared singularities occurring in the chiral expansion of the kaon self-energies subject the Dashen theorem to a large correction from higher orders.  For the parameter which measures the size of this correction, we find $\epsilon=0.9(3)$. 

19.~Finally, we invoke the low-energy theorem which relates the kaon mass difference to the ratio $Q^2\equiv(m_s^2-m_{ud}^2)/(m_d^2-m_u^2)$ of quark masses \cite{Gasser+1985}. The theorem can be compared with the Gell-Mann-Okubo formula, but there is an important difference: While that formula only holds at leading order of the chiral expansion and picks up corrections of first non-leading order, the relation relevant for $Q$ receives corrections only at next-to-next-to-leading order. This implies that, instead of expressing the decay rate in terms of the kaon mass difference, we can just as well express it in terms of the quark mass ratio $Q$. Conversely, the measured decay rates in the charged and neutral channels yield two independent determinations of this mass ratio. The two results agree very well with one another -- combining them, we obtain $Q =22.1(7)$, where the error includes all sources of uncertainty encountered in the calculation, including an estimate for the neglected higher order contributions in the chiral series.  

20.~The ratio $S\equiv m_s/m_{ud}$ is now known remarkably well from lattice calculations.  With the value $S=27.30(34)$ quoted by FLAG for simulations  with four quark flavours~\cite{Aoki:2016frl}, our result for $Q$ leads to $R\equiv (m_s-m_{ud})/(m_d-m_u)= 34.2(2.2)$ and $m_u/m_d=0.44(3)$. These numbers indicate that, within QCD, the chiral expansion of the square of the Nambu-Goldstone masses is dominated by the leading terms, i.e.~by the linear formulae of current algebra. At the physical values of $m_u$, $m_d$, $m_s$, the higher order contributions amount to remarkably small corrections.

21.~While the outcome of our calculation for the kaon mass difference in QCD agrees with the lattice results within errors, the values obtained for the isospin breaking quantities $Q$, $R$ and $m_u/m_d$  in two of the three most recent lattice calculations~\cite{Fodor:2016bgu,Giusti:2017dmp} do not. We point out that the discrepancy concerns the size of the corrections arising in the low-energy theorems for the corresponding ratios of meson masses. While the pattern obtained with our result for $Q$ leads to a coherent picture, these lattice results imply that the corrections in $R$ and $S$, which are of first order in chiral symmetry breaking are smaller than those in $Q$, despite the fact that the latter represent contributions of second order. In Sec.~\ref{sec:lattice Q}, we indicate a way to resolve this conundrum by means of a lattice simulation within QCD. 

22.~In  the plane of the quark mass ratios $m_u/m_d$ and $m_s/m_d$, a given value of $Q$ corresponds to an ellipse, while a given value of $S$ corresponds to a straight line.  The yellow band in the left panel of Fig.~\ref{fig:Quark mass ratios} represents the region allowed by our result for $Q$, while the grey band represents the region allowed by the lattice result for $S$ quoted by FLAG. For comparison, the figure also indicates the first estimates of the three lightest quark masses~\cite{Gasser+1974,Weinberg1977}, which appeared shortly after the discovery of QCD. The hexagon represents the rough estimates for the range in the variables $S$, $R$ and $m_u/m_d$ where the chiral expansion yields a coherent picture, obtained many years ago~\cite{Gasser1981}.

The right panel focuses on the region of physical interest and includes recent results obtained on the lattice. In particular, it compares the outcome of our work with the region allowed by the lattice results according to FLAG~\cite{Aoki:2016frl} and to the Particle Data Group~\cite{Olive:2016xmw}. The outcome of the three most recent lattice calculations (BMW \cite{Fodor:2016bgu}, RM123 \cite{Giusti:2017dmp}, Bazavov et al.~\cite{Bazavov+2018}) is also indicated -- the regions shown are obtained by treating the values obtained for $S$ and $m_u/m_d$ as statistically independent.\footnote{Ref.~\cite{Giusti:2017dmp}, which is about isospin breaking, does not explicitly quote a value of $S$. The relevant one is $S=26.66(32)$, as given in~\cite{Carrasco:2014cwa}. }

23.~In Sec.~\ref{sec:Comparison}, our analysis is compared with related work. There are two significant  improvements compared to the early dispersive analyses in Refs.~\cite{Kambor+1996,Anisovich+1996}: The experimental information about $\eta$-decay improved very substantially and the phase shifts of $\pi\pi$ scattering are now under much better control. Concerning the properties of the Dalitz plot distribution, the various investigations are now in reasonable agreement. 
In order to establish contact with QCD and to extract information about the quark masses from $\eta$-decay, however, the theoretical constraints that follow from the fact that the pions and the $\eta$-meson are Nambu-Goldstone bosons of a hidden approximate symmetry play a crucial role. These constraints can be analyzed in a controlled manner in the framework of \chpt, but care must be taken not to leave the region where the first few terms of the chiral perturbation series provide a decent approximation. Some of the analyses found in the literature, for instance, rely on matching the dispersive and chiral representations directly in the physical region of the decay. Since the first few terms of the chiral perturbation series do not represent a good approximation there, this leads to incorrect conclusions.  

24.~The nonrelativistic effective theory provides a representation of the transition amplitude for the decay $K\to 3\pi$ that works very well~\cite{Colangelo+2006a,Bissegger:2007yq,Bissegger:2008ff,Gasser:2011ju}. The method even leads to a coherent analysis of the contributions from the electromagnetic interaction. Since $M_\eta$ is not much larger than $M_K$, this approach can be expected to work for $\eta\to 3\pi$ as well. We have verified that the amplitude of Ref.~\cite{Bissegger:2007yq} indeed fits the KLOE data perfectly well. Moreover, in the isospin limit and in the physical region, the NR framework yields an excellent approximation of our solutions. The subtraction constants of the dispersive solutions that match the NR amplitude have a sizeable imaginary part, but, throughout the physical region, the difference between the two representations is very small, for the imaginary part as well as for the real part. This demonstrates that the NR effective theory provides a suitable framework for the analysis of $\eta$-decay.  

25.~It is not a straightforward matter to establish contact between the nonrelativistic effective theory and the quark masses which occur in the QCD Lagrangian. Our approach relies on the assumption that, in the vicinity of the Adler zero, the one-loop representation of \chpt\ provides a good approximation. The Adler zero is outside the region where the truncated expansion of the nonrelativistic effective theory represents a good approximation, but the link can be established by matching the dispersive and nonrelativistic representations in the isospin limit: (i) Determine the Dalitz plot distributions in the charged and neutral channels within the nonrelativistic framework. (ii) Take the isospin limit of the transition amplitude and expand it in powers of the spatial momenta of the three pions in the rest frame of the $\eta$. (iii) Match the coefficients of this expansion -- the analogues of the scattering lengths -- to those of the generic dispersive representation. It would be most interesting to carry this out, but we leave this for the future.  

\section*{Acknowledgments}
We are indebted to J\"urg~Gasser and Akaki Rusetsky for letting us use the solutions of the integral equations, which they obtained with an entirely new method that yields significantly more accurate results than the one we used ourselves. Moreover, we warmly thank them as well as Bastian Kubis for carefully reading the manuscript, in particular for comments on the NREFT approach and on the analytic properties of the amplitude. Very useful information about the MAMI results on $\eta\to 3\pi^0$ -- in particular also a sampling of the data that accounts for the indistinguishability of the three pions in the final state -- was provided by Sergey Prakhov and is gratefully acknowledged. We also thank
P.~Adlarson, J.~Bijnens, L.~Caldeira Balkest\r{a}hl, I.~Danilkin, A.~Fuhrer, K.~Kampf, A.~Kup\'s\'c, B.~Moussallam,  S.~Simula and P.~Stoffer for useful information.  This work is supported in part by Schweizerischer Nationalfonds and the U.S.~Department of Energy (contract DE-AC05-06OR23177) and National Science Foundation (PHY-1714253).

\begin{appendices}
\section{Angular averages}\label{sec:Angular averages}
\subsection{Analytic continuation in $M_\eta$}
\begin{figure*}\centering\includegraphics[width=4.5cm]{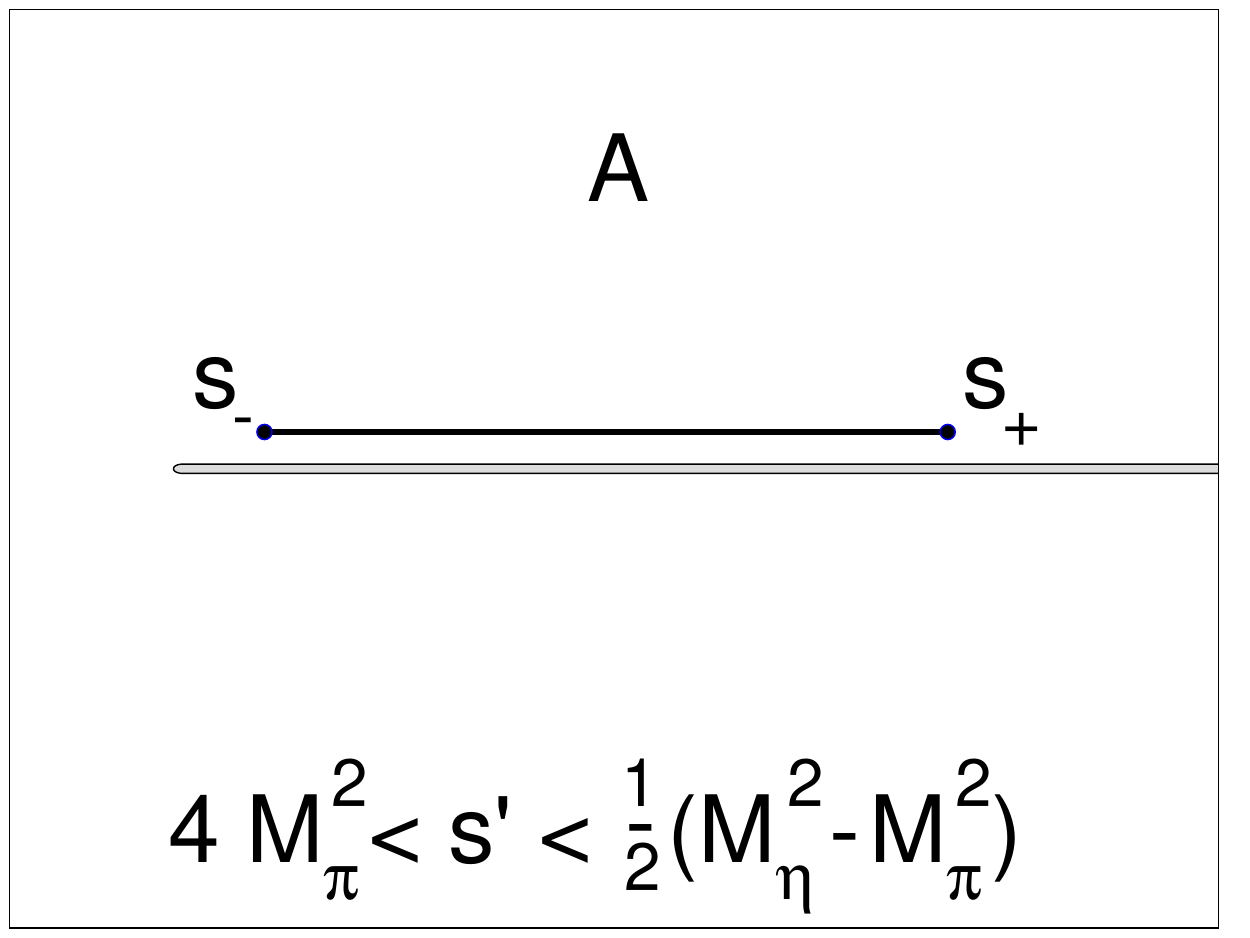}\hspace{0.3cm}\includegraphics[width=4.5cm]{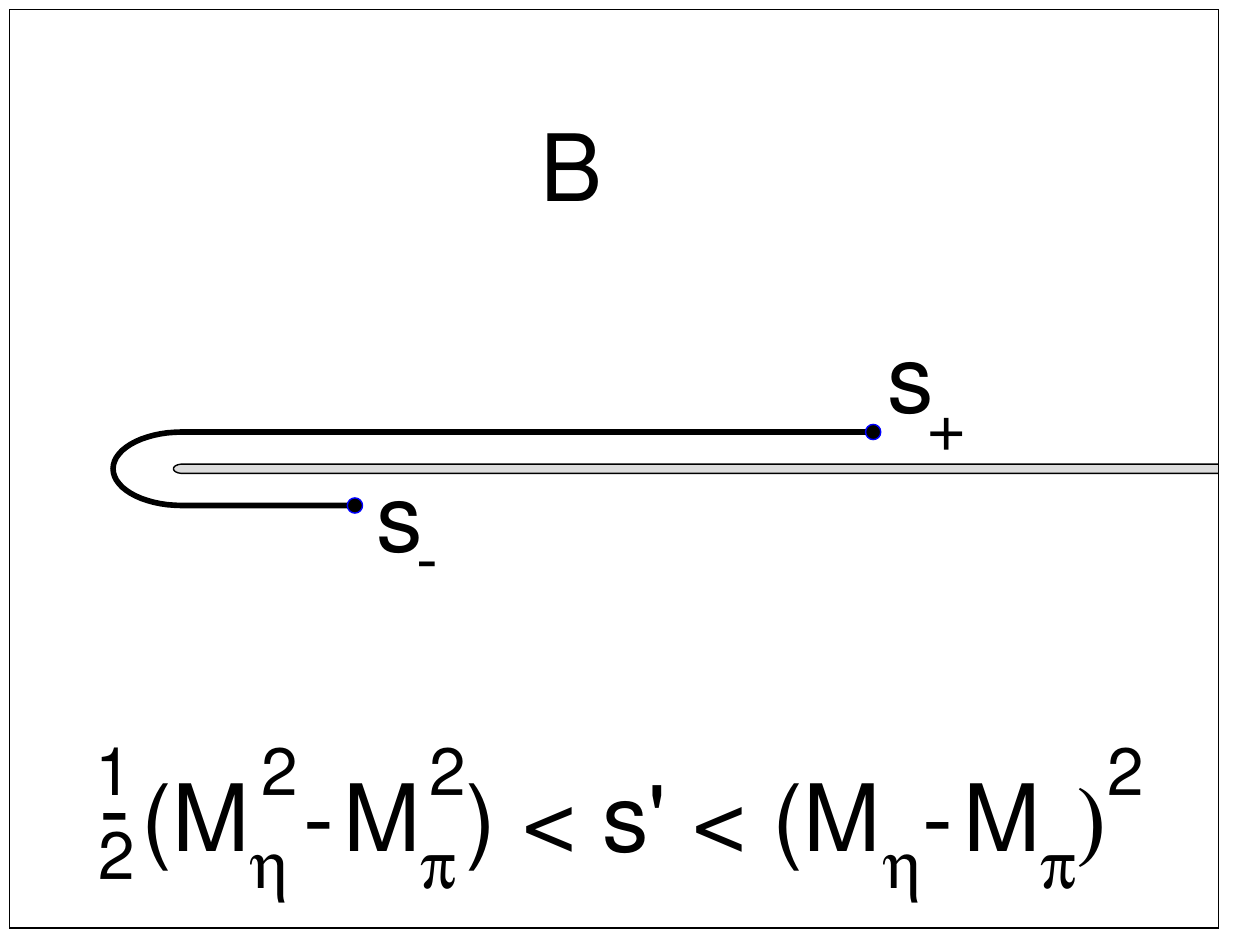}\hspace{0.3cm}\includegraphics[width=4.5cm]{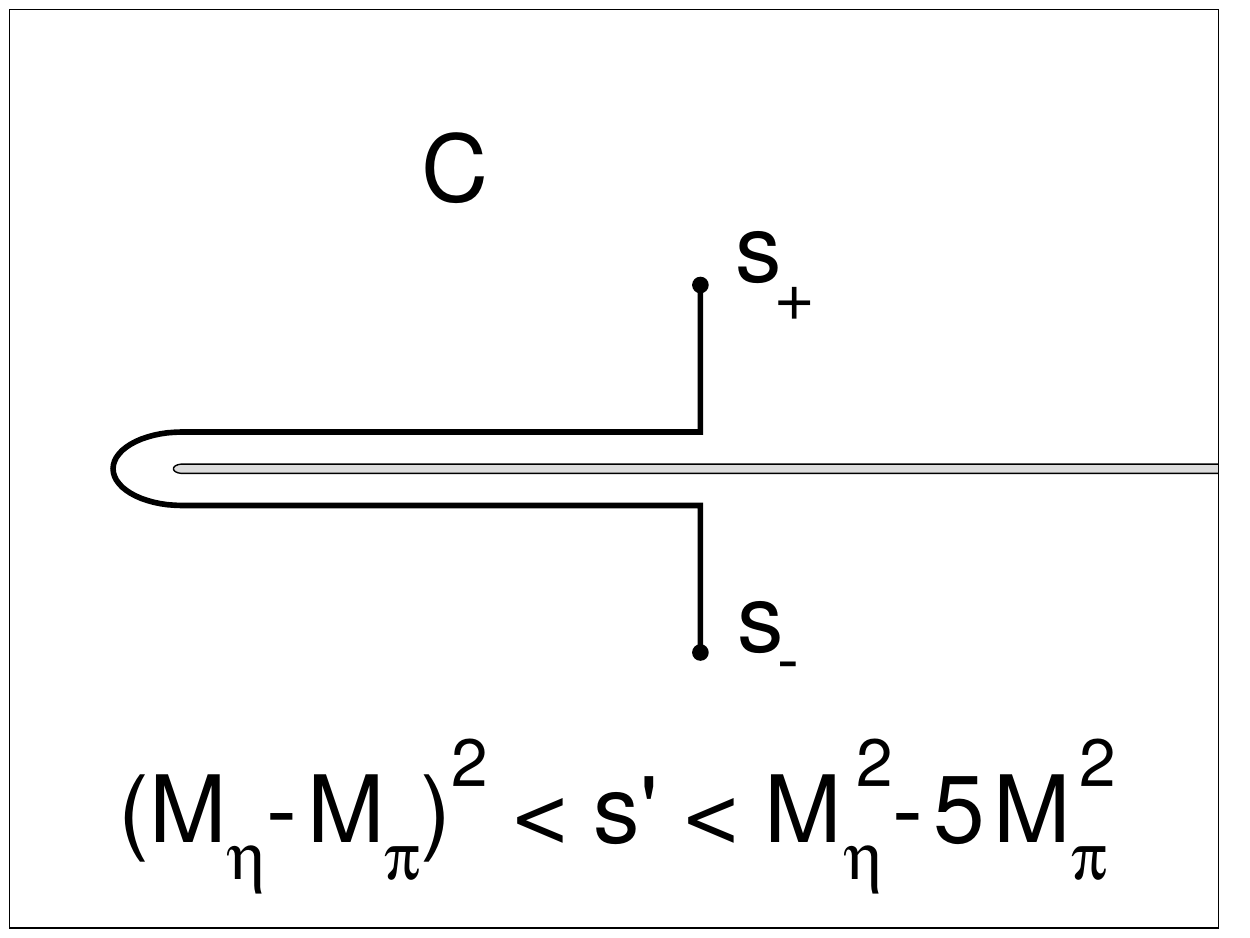}\\
\vspace{0.3cm}\includegraphics[width=4.5cm]{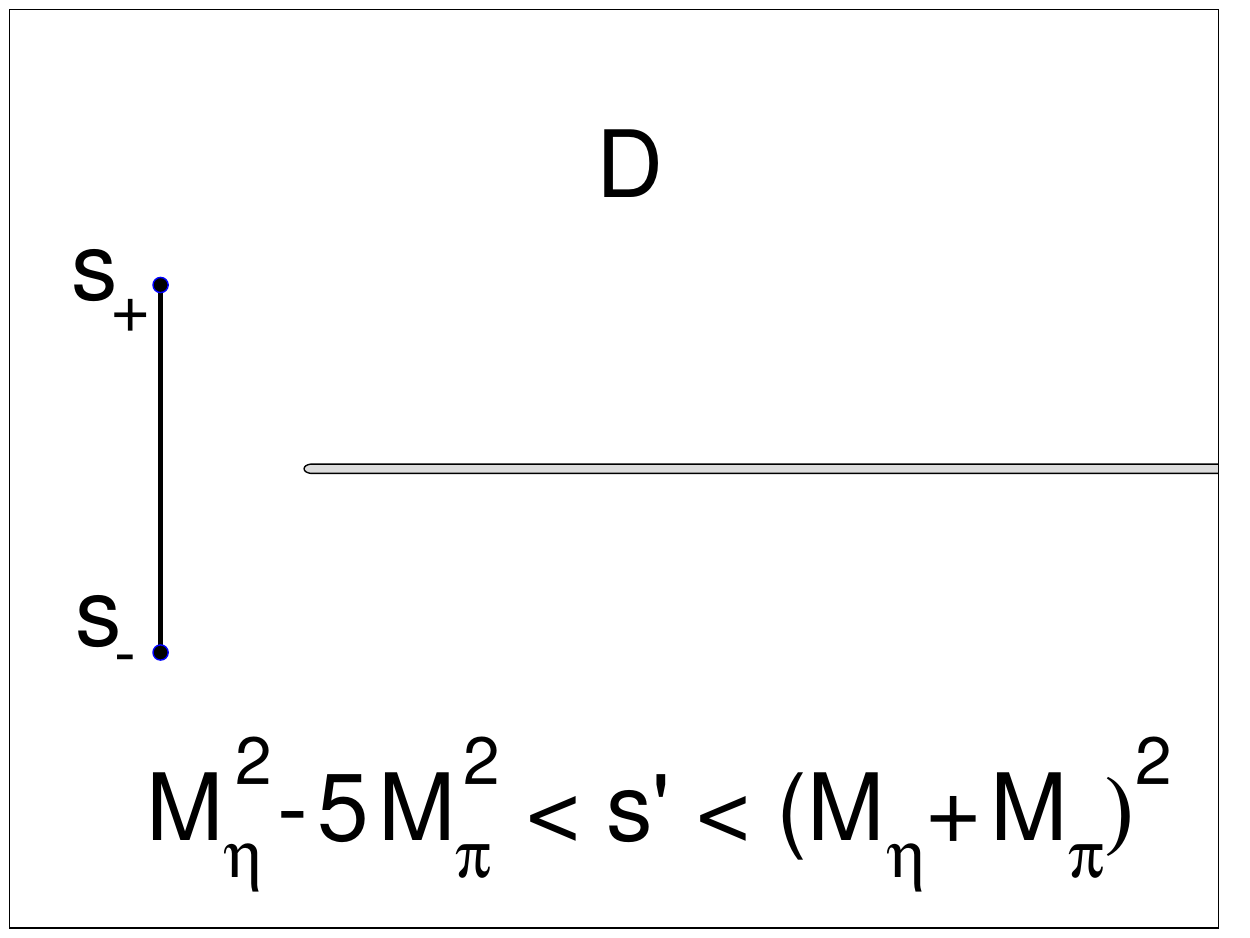}\hspace{0.3cm}\includegraphics[width=4.5cm]{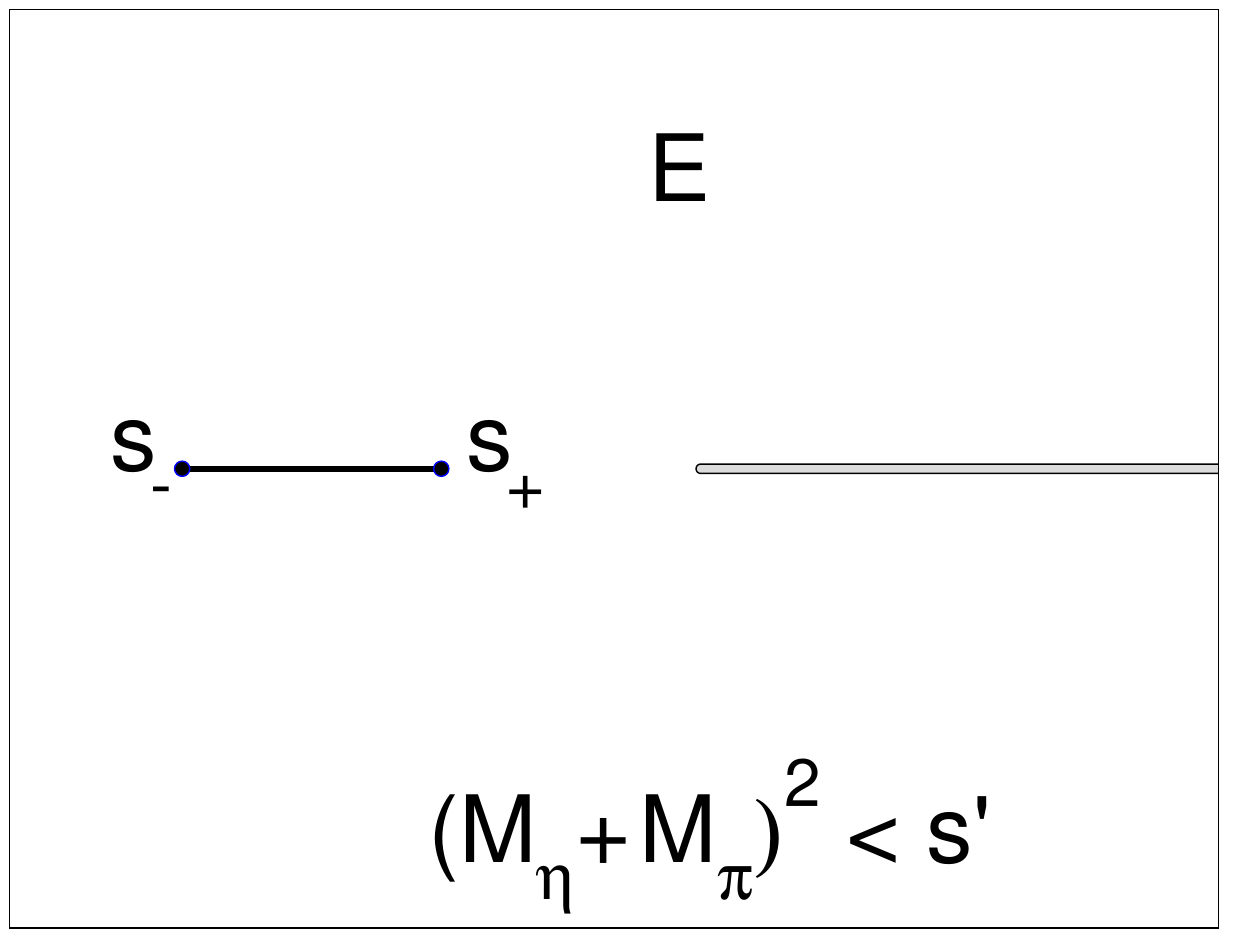}\caption{Integration path used in the evaluation of the angular averages.  \label{fig:path}}
\end{figure*} 

In the dispersion integral~\eqref{eq:MI}, the functions $\disc M_I(s')$ are needed only for $4M_\pi^2\leq s'<\infty$ and the same thus applies to $\hat M_I(s')$ and $\aav{z^n M_I}(s')$. The  amplitudes are evaluated with the physical masses,\footnote{Appendices A and B concern the isospin limit. To simplify the notation we again drop the bar and use the symbols $M_\pi$, $M_\eta$ for the masses in the isospin limit.} but to specify the angular averages, we need to replace $M_\eta$ in Eq.~\eqref{eq:kappa} by a complex variable $M$:
\begin{eqnarray}\label{eq:kappac}s_0&=&\mbox{$\frac{1}{3}$}M^2+M_\pi^2\;,\\
\kappa(s) \al =\al \sqrt{1 - 4 \mpi^2/s} \sqrt{s^2-2s(M^2+M_\pi^2)+(M^2-\mpi^2)^2}  \;. \nonumber \end{eqnarray} 
The analytic continuation in $M^2$ starts from a real value below $9M_\pi^2$, where the integral in~\eqref{eq:angularAverage} runs along the straight line
\begin{equation}\label{eq:stilde}\tilde{s}=\mbox{$\frac{3}{2}$}s_0-\mbox{$\frac{1}{2}$}s'+ \mbox{$\frac{1}{2}$} z\,\kappa(s')\;,\quad-1\leq z\leq 1\end{equation} 
which connects the points
\begin{equation}\label{eq:spm}s_\pm(s')= \mbox{$\frac{3}{2}$}s_0-\mbox{$\frac{1}{2}$}s'\pm\mbox{$\frac{1}{2}$} \kappa(s')\end{equation}
with one another. This integral needs to be continued in $M^2$, approaching the physical mass with $M^2=M_\eta^2+i\,\delta$, $\delta\rightarrow0$.  The problems arising in the evaluation of the angular averages are best understood by starting with a large value of $s'$ and then gradually lowering it. 

In the following discussion, four values of $s'$ play a special role and we introduce corresponding symbols $s_1<s_2<s_3<s_4$ to simplify the notation:
\begin{eqnarray} s_1\al = \al \mbox{$\frac{1}{2}$}(M_\eta^2-M_\pi^2)\;,\quad s_2=(M_\eta-M_\pi)^2\;, \nonumber \\
 s_3 \al = \al M_\eta^2-5M_\pi^2\;,\quad s_4=(M_\eta+M_\pi)^2\;.\end{eqnarray}
For $s'>s_4$, both $s_+(s')$ and $s_-(s')$ are real and  
smaller than $4M_\pi^2$, so that the integral over $z$ runs along the real axis, to the
left of the branch cut (Fig.~\ref{fig:path} E). In this case, the situation is essentially the same as for $\pi\pi$ scattering, where elastic unitarity also implies a representation of the form~\eqref{eq:MI} -- \eqref{eq:angularAverage}, except that $M_\eta$ is replaced by $M_\pi$ (the expression for $\kappa(s')$  then simplifies to $\kappa(s')=s'-4M_\pi^2$, so that the integration extends over the interval $4M_\pi^2-s'\leq \tilde{s}\leq 0$ of the negative real axis). 

If $s'$ falls below $s_4$, the term $\kappa(s')$ becomes imaginary: The integration runs along a line that is parallel to the imaginary axis (Fig.~\ref{fig:path}D). Note that, in view of the square roots, $\kappa(s')$ is defined only up to a sign. Since the integrand in~\eqref{eq:angularAverage} only contains the product $z\,\kappa(s')$, the average $\aav{z^n M_I}(s')$ picks up the factor $(-1)^n$ if $\kappa(s')$ changes sign. The expressions for the functions $\hat M_I(s')$ in~\eqref{eq:mhat}, however, remain invariant. Hence the representation for the discontinuities is independent of the sign chosen for $\kappa(s')$. In Fig.~\ref{fig:path}, we have chosen the sign such that $\Im \kappa(s')\geq0$.
 
 The straight line from $s_-(s')$ to $s_+(s')$ crosses the real axis at $\tilde{s}=\frac{1}{2}(3s_0-s')$. As long as $s'$ stays above $s_3$, this point is to the left of the branch cut, so that the path of integration avoids the singularity, but if $s'$ falls below that value, there is a problem: The straight line connecting $s_-(s')$ with $s_+(s')$ then crosses the singularity.
The problem would not arise if $M_\eta$ were smaller than $3M_\pi$: The quantity $M_\eta^2-5M_\pi^2$ would then stay below $4M_\pi^2$, so that, in the entire range over which $s'$ varies, the integral in~\eqref{eq:angularAverage} stays away from the branch cut. The very fact that the $\eta$ does decay into three pions, however, implies that $M_\eta$ is larger than $3M_\pi$: The straight path of integration in~\eqref{eq:angularAverage} necessarily runs across the singularity generated by the interaction among one of the pion pairs. 

The way out is to deform the path of integration. The right hand side of~\eqref{eq:angularAverage} represents an integral of the analytic function $M_I(\tilde{s})$ over its argument:
\begin{equation}\label{eq:C}\aav{z^n M_I}(s') =\kappa(s')^{-(n+1)}\,\rule{0em}{1em}^C\hspace{-0.6em}\int\!\! d\tilde{s} \,(2\tilde{s}+s'-3s_0)^nM_I(\tilde{s})\;.\end{equation}
Any path $C$ that connects the same two points $s_-(s')$ and $s_+(s')$ and does not leave the analyticity domain of $M_I(\tilde{s})$ yields the same value for the integral. If $M^2$ is equal to $M_\pi^2$, the straight path is adequate. For larger values of $M^2$, a coherent definition of elastic unitarity is obtained by (i) starting at $M^2=M_\pi^2$ with the straight path of integration that connects $s_-(s')$ with $s_+(s')$ and (ii) continuing the result analytically in $M^2$ to the physical value. Note that, since the end points depend on $M^2$, the path $C$ necessarily changes in the course of the analytic continuation. It must be chosen in such a way that, when $M^2$ increases from $M_\pi^2$ to $M_\eta^2$, the path remains within the analyticity domain, i.e.~stays away from the singularity.\footnote{For an evaluation of the angular integration that requires only a one-dimensional grid in the complex plane (along an elliptic curve), see Refs.~\cite{Kambor+1996,Colangelo:2015kha,Schneider-PhD}.}  

To apply this prescription to the case where $s'$ is in the range  $s_2< s' < s_3$, we first note that, in this range, the point $s_-(s')$ is located in the lower half plane while $s_+(s')$ is in the upper half plane. For the initial path belonging to $M_\eta=M_\pi$, the straight line connecting the two corresponds to a fixed, negative value of the real part: $\Re \tilde{s}=s_a$, with $s_a=\frac{1}{2}(4M_\pi^2-s')$. A path that stays within the analyticity domain when $M_\eta$ is increased can easily be given: a straight line from $s_-(s')$ to $s_a$, followed by a straight line from there to $s_+(s')$. For $s'>s_3$, this path can be deformed into the straight line from $s_-(s')$ to $s_+(s')$ without passing through the singularity (Fig.~\ref{fig:path} D), but for lower values of $s'$, the intersection with the real axis must stay to the left of $\tilde{s}=4M_\pi^2$.  In the numerical work, we are using the path shown in Fig.~\ref{fig:path} C, which can be reached from the one specified 
above with a deformation that avoids the singularity. It consists of (i) a vertical segment from $s_-(s')$ to the lower rim of the branch cut, (ii) a horseshoe that runs along the lower rim to the left, encircles the branch point at $\tilde{s}=4M_\pi^2$ and then runs to the right, along the upper rim of the cut and (iii) a vertical segment that ends at $s_+(s')$.

The explicit expression~\eqref{eq:kappac} for $\kappa(s')$ contains square roots. The standard convention for the numerical evaluation of square roots of complex numbers uses the first sheet, where Im$\sqrt{z}$ is of the same sign as Im$z$. With $M^2=M_\eta^2+i\,\delta$, the imaginary part of the quantity of which we need to take the square root, however, goes through zero in the interior of the region D, at the point $s'=M_\eta^2-M_\pi^2$. Accordingly, evaluating $\kappa(s')$ numerically, the result makes a jump: $\kappa(s')$ changes sign there. As discussed above, this does not matter for the integral, but for the plots shown in Fig.~\ref{fig:path} it does: There, we have fixed the sign of $\kappa(s')$ with continuity, such that $s_+$ and $s_-$ move continuously when  $s'$ is changed. To implement this choice numerically, the expression~\eqref{eq:kappac} for $\kappa(s')$ is to be used only for $s'<M_\eta^2-M_\pi^2$, while above that point, the sign of $\kappa(s')$ must be changed. 

Finally, for $4M_\pi^2 <s'<s_2$, the two end points of the path both approach the cut when $\delta$ tends to zero. The value $s'=s_2$ corresponds to the lower limit of the range considered in the preceding paragraph. For the path specified there, the vertical segments shrink to zero in that limit, but the horseshoe remains: The end points $s_-(s')$ and $s_+(s')$ are located on the lower and upper rims of the cut, respectively. The path connects them, making a detour around the branch point. When $s'$ is lowered, $\Re \kappa(s')$ takes positive values, so that $\Re s_-(s')$ is smaller than $\Re s_+(s')$:  The horseshoe becomes asymmetric (Fig.~\ref{fig:path} B). As $s'$ passes through the value $s_1$, the point $s_-(s')$ moves from the lower rim to the upper one, so that the entire path then runs along the upper rim (Fig.~\ref{fig:path} A). When $s'$ drops to $4M_\pi^2$, the term $\kappa(s')$ vanishes, so that the path shrinks to a point. 

This completes the specification of the angular averages. We emphasize that, in the above procedure, the complex parameter $M$ merely serves to determine the proper path of integration. Once this path is identified, the limit $\delta\rightarrow 0$ can be taken -- the numerical evaluation of the angular averages only involves the physical masses.

\subsection{Contribution from the horseshoe}

The representation~\eqref{eq:C} involves inverse powers of $\kappa(s')$. In the limit $\delta\rightarrow0$, $\kappa(s')$ vanishes at   $s'=4M_\pi^2$, $s'=s_2$ and $s'=s_4$, so that the integrands of the dispersion integrals~\eqref{eq:DROmega} become singular at these points. The first zero sits at the lower end of region A. The path of integration shrinks to zero there, $s'_\pm\rightarrow s_1=\frac{1}{2}(M_\eta^2-M_\pi^2)$. Indeed the formula~\eqref{eq:angularAverage} shows that the angular averages tend to a finite limit,  proportional to the value of the amplitude there:  $\langle z^n M_I\rangle(4M_\pi^2)=c_nM_I(s_1)$, with $ c_0=1,c_1=0,c_2=\frac{1}{3}$.   The third zero sits at the boundary between the regions D and E, where the path also shrinks to a point: $s_ \pm\rightarrow \bar{s}_3\equiv -M_\pi(M_\eta-M_\pi)$. The angular averages tend to a finite limit, given by an analogous formula: $\langle z^n M_I\rangle(s_3)=c_nM_I(\bar{s}_3)$.

For the zero at $s'=s_2$, however, the situation is not that simple. This value is at the boundary between the regions B and C. For values of $s'$ in these regions, it is convenient to decompose the path into three segments: (i) a straight line from $s_-(s')$ to the point $\bar{s}=\frac{1}{2}\{s_+(s')+s_-(s')\}$ on the lower rim of the real axis, (ii) a symmetric horseshoe and (iii) a straight line connecting the point $\bar{s}$  on the upper rim with $s_+(s')$. In the limit $\delta\rightarrow 0$, the segments (i) and (iii) run either along the real axis or parallel to the imaginary axis, depending on whether $s'$ is below or above $s_2$. When $s'$ approaches $s_2$, these segments shrink to a point. Their contribution to the angular average  is analogous to those encountered in the preceding paragraph and stays finite when the limit $\delta\rightarrow 0$ is taken. In the present case, the limiting value of the contribution from the segments (i) and (iii) is proportional to the difference between the values 
of the amplitude at the upper and lower rims of the cut, $M_I^+(\bar{s})-M_I^-(\bar{s})=2 i \,\disc M_I(\bar{s})$.
 
The pi\`{e}ce de r\'{e}sistance is the contribution from the horseshoe, which involves moments of the discontinuity across the cut:
\begin{equation}\label{eq:horseshoe}H_I^n(s')=i\,2^{n+1}\hspace{-0.3em}\int_{4M_\pi^2}^{\bar{s}}\hspace{-1mm}d\tilde{s}\,(\tilde{s}-\bar{s})^n\,\disc M_I(\tilde{s}) \;.\end{equation}
The corresponding contributions to the quantities $\hat{M}_0,\hat{M}_1, \hat{M_2}$ read
 \begin{equation}\hat{M}_0^H(s') =\frac{H_0(s')}{\kappa(s')}\;,\quad
 \hat{M}_1^H(s')=\frac{H_1(s')}{\kappa(s')^3}\;,\quad
 \hat{M}_2^H(s') =\frac{H_2(s')}{\kappa(s')}\;,\end{equation}
where $H_0(s)$, $H_1(s)$, $H_2(s)$ represent linear combination of these moments:
\begin{eqnarray}H_0(s')\al=\al  \mbox{$\frac{2}{3}$}H_0^0(s')+2(s'-s_0)H_1^0(s')+\mbox{$\frac{2}{3}$}H_1^1(s')+\mbox{$\frac{20}{9}$}H_2^0(s')\;,\nonumber\\
H_1(s')\al=\al  3H_0^1(s')+\mbox{$\frac{9}{2}$}(s'-s_0)H_1^1(s')+\mbox{$\frac{3}{2}$}H_1^2(s')-5H_2^1(s')\;, \nonumber\\
H_2(s')\al=\al  H_0^0(s')-\mbox{$\frac{3}{2}$}(s'-s_0)H_1^0(s')-\mbox{$\frac{1}{2}$}H_1^1(s')+\mbox{$\frac{1}{3}$}H_2^0(s')\;.\nonumber\\
\end{eqnarray}
In the limit $\delta\rightarrow 0$, the function $\kappa(s')$ is proportional to $\sqrt{s_2-s'}$. Since there is no reason for the functions $H_0(s')$, $H_1(s')$, $H_2(s')$ to vanish at $s'=s_2$, the integrands of the dispersion relation~\eqref{eq:DROmega} are singular there. In the case of $\hat{M}_0(s')$ and $\hat{M}_2(s')$, the singularity is integrable, but for $\hat{M}_1(s')$ this is not the case. 

A horseshoe occurs in the angular averages only if $s'$  is in the interval $s_1< s'< s_3$. Moreover, at the endpoints of that interval, the functions $H_I^n(s')$ vanish. In the case of $M_1(s)$, the contribution from the horseshoe thus takes the form
\begin{eqnarray}\label{eq:Htheta}M_1^H(s) \al=\al\Omega_1(s)s\int_{s_1}^{s_3} \hspace{-0.3em}ds'\frac{ \phi(s')H_1(s')}{(s'-s-i \epsilon)\,\kappa(s')^3}\;, \nonumber \\
\phi(s')\al = \al\frac{\sin\delta_1(s')}{\pi^2 s' |\Omega_1(s')|}\;.\end{eqnarray}   
Here, the imaginary part of the mass plays the role of a regulator: For positive values of $\delta$, the function $\kappa(s')$ does not have any zeros on the real axis, so that the dispersion integral is perfectly well-defined, but the limit $\delta\rightarrow 0$ cannot be interchanged with this integral. 

The regulator is needed only in the immediate vicinity of the singularity. Only one of the three zeros of $\kappa(s')$ is in the range relevant here -- in the corresponding square root, the regulator must be retained, but in the remainder, the limit can be taken: $\kappa(s')$ can be replaced by
\begin{eqnarray}\kappa(s') \al = \al \bar{\kappa}(s')\,\sqrt{(M-M_\pi)^2-s'}\;, \nonumber\\
\bar{\kappa}(s') \al = \al \sqrt{1-4M_\pi^2/s'}\sqrt{s_4-s'}\;,\end{eqnarray}
without changing the limiting value of the integral. Also, if $\phi(s')H_1(s')$ is replaced by $\phi(s')H_1(s')-\phi(s_2)H_1(s_2)$, the limit can be interchanged with the integration. The operation, however, generates fictitious logarithmic singularities at $s'=s_1,s_3$ because the modified integrand is discontinuous there. Since the function $H_1(s)$ vanishes at the endpoints, the integral~\eqref{eq:Htheta} does not contain such singularities. The artefact is avoided if the subtracted term is multiplied with a factor $h(s')$ that is equal to 1 at $s'=s_2$, but vanishes at $s'=s_1,s_3$. The singular part of the integral then boils down to
\begin{equation}\label{eq:G}G(s)=\int_{s_1}^{s_3}ds'\frac{h(s')}{(s'-s-i\epsilon)\,(a-s')^\frac{3}{2}}\;,\quad 
a=(M-M_\pi)^2\;,\end{equation}
and the contribution from the horseshoe to the amplitude $M_1(s)$ can be represented as
\begin{eqnarray}M_1^H(s) =\Omega_1(s)s \al \al \left\{ \int_{s_1}^{s_3} \hspace{-0.3em}ds'\,\frac{\bar{\phi(}s')H_1(s')-h(s')\bar{\phi}(s_2)H_1(s_2)}{(s'-s-i \epsilon)\,(s_2-s')^\frac{3}{2}} \right. \nonumber\\
\al \al~~~~~+  \left. \bar{\phi}(s_2)H_1(s_2)G(s)\right\}
 \;, \end{eqnarray}   
with $\bar{\phi}(s')\equiv \bar{\kappa}(s')^{-3/2}\phi(s')$. 

The profile of the factor $h(s')$ is irrelevant. We find it convenient to work with a parabola, $h(s')=(s'-s_1)(s_3-s')/(s_2-s_1)(s_3-s_2)$ -- for this choice, the function $G(s)$ can be given explicitly. Moreover, the integrand in~\eqref{eq:G} is then analytic in the lower half of the $s'$-plane. Hence the path of integration can be moved away from the real axis into the lower half-plane without changing the value of the integral. The limit $\delta\rightarrow 0$ can then be interchanged with the integration: On the real axis, $G(s)$ does approach a finite limit and we now remove the regularization. 

The dispersion relation only involves real values of $s$. The representation~\eqref{eq:G} shows that the function $G(s)$ admits a unique analytic continuation into the upper half of the $s$-plane. In view of the branch points at $s=s_1$ and $s=s_3$, the continuation into the lower half-plane is ambiguous. We identify the first sheet with the values reached by continuing analytically across the interval $s_1< s<s_3$ of the real axis, while the second sheet corresponds to continuation to the left of $s_1$. The difference between the values of $G(s)$ on the first and second sheets is given by $-2\pi\hspace{0.05em} i\hspace{0.05em} h(s) (s_2-s)^{-3/2}$. Since $G(s)$ is regular on the first sheet, it must be singular on the second: The singularity $\propto1/\kappa(s')^3$ encountered if the angular average is evaluated with $M=M_\eta$ sits on the second sheet. The $i\epsilon$-prescription implies that the value of $G(s)$ on the first sheet is relevant -- the presence of a singularity on the second sheet does not 
affect it. 

\begin{figure}\centering\includegraphics[width=8cm]{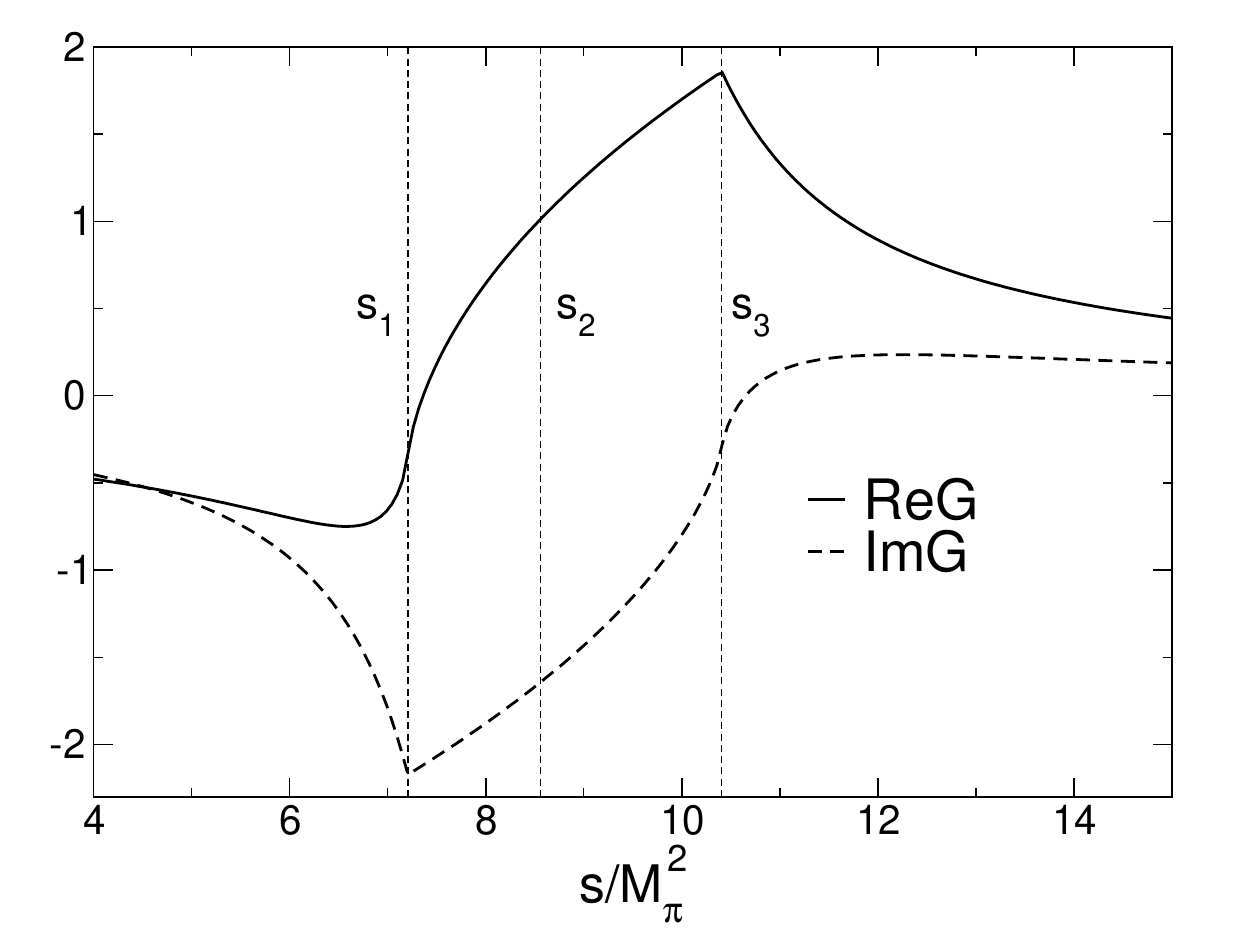}\caption{Kernel relevant for the contributions from the horseshoe.  \label{fig:G}}
\end{figure}

The integral~\eqref{eq:G} can be done explicitly. For real values of $s$ below $s_1$, the result reads:
\begin{eqnarray}\label{eq:G1}G(s)=\frac{2h(s)}{w^3}\left(\mathrm{arctanh}\frac{w_1}{w}- 
\mathrm{arctanh}\frac{w_3}{w}\right) \nonumber\\
-2\left(\frac{1}{w_1}-\frac{1}{w_3}\right)\left(\frac{1}{w^2}-\frac{1}{w_1w_3}\right)\;, \end{eqnarray}
with $w\equiv\sqrt{s_2-s}$, $w_1=\sqrt{s_2-s_1}$ , $w_3=\sqrt{s_2-s_3}$. In this expression, the branch point at $s=s_1$ is described by the term $\mathrm{arctanh}(w_1/w)$.  As long as $s$ stays below $s_2$, both $w$ and $w_1$ are real and positive -- the branch point occurs at the place where the ratio $w_1/ w$ passes through 1. On the first sheet, the analytic continuation of the function $\mathrm{arctanh}(w_1/w)$ yields a constant imaginary part equal to $\pi/2$. When $s$ crosses the point $s=s_2$, the real part of the function $\mathrm{arctanh}(w_1/w)$ goes through zero. The expansion in powers of $w$ starts with
\begin{equation}\label{eq:arctan}\mathrm{arctanh}\frac{w_1}{w}=\frac{i\pi}{2}+\frac{w}{w_1}+O(w^3)\;.\end{equation}
This formula also holds for $\mathrm{arctanh}(w_3/w)$, where $w_1$ is replaced by $w_3$.  Inserting the expansion in the first bracket of~\eqref{eq:G1}, the leading terms cancel, but those linear in $w$ do not -- they generate a simple pole at $s=s_2$, with a residue proportional to $h(s_2)=1$. The second bracket, however, contains exactly the same pole with the opposite sign, so that the explicit expression for $G(s)$ is indeed singularity free. Accordingly, the numerical representation in Fig.~\ref{fig:G} does not show any trace of a singularity at the point $s=s_2$. 
The complications concerning the behaviour in the vicinity of the zeros of $\kappa(s')$ also require extra work in the iterative procedure used to determine the fundamental solutions. For a detailed discussion, we refer to~\cite{Walker1998,Lanz-PhD}. The numerical evaluation of the dispersion relations is carried out on a lattice of points. An interpolation between these is required to calculate the integrands relevant for the next step of the iteration. At those places where the integrand varies rapidly -- in the vicinity of the threshold, for instance -- the lattice must be fine enough to arrive at an accurate result for the principal value integral. Remnants of the difficulties encountered can be seen in Fig.~\ref{fig:fundamental solution}: The fundamental solution belonging to $\alpha_0$ shows a small wiggle near $s= 4M_\pi^2$ in $M_0(s)$ (as well as in the amplitude $M_n(s)$ relevant for the neutral channel) and the plot of the component $M_1(s)$ reveals a spike at the point $s=(M_\eta-M_\pi)^2\simeq 8.
6 M_\pi^2$, which is also due to the limited accuracy of the numerical evaluation. On the other hand, in the vicinity of the third zero of $\kappa(s)$, $s=(M_\eta+M_\pi)^2\simeq 24.3 M_\pi^2$, our results do not indicate numerical deficits. 

For the determination of the quark mass ratio $Q$, we need  the integral over the square of the amplitude over the entire physical region. This integral is not sensitive to the numerical shortcomings mentioned above. The $M$-distribution in the neutral channel, however, is affected. As seen in Fig.~\ref{fig:s-distribution}, that distribution is nearly flat and high resolution is needed to resolve the structure of the cusps generated by the final state interaction $\pi^0\pi^0\rightarrow\pi^+\pi^-\rightarrow\pi^0\pi^0$. The prediction shown in that figure relies on the fundamental solutions of the integral equations obtained by Gasser and Rusetsky \cite{Gasser+2018}.  
 
\section{Representation of the transition amplitude at one loop of \chpt}\label{sec:DR NLO}
\subsection{Elastic unitarity}
In the present appendix, we show that, together with unitarity, the Taylor invariants $H_0,H_1,H_2,H_3$ uniquely determine the one-loop representation of Chiral Perturbation Theory. Since the chiral expansion of the phase shifts only starts at $O(p^2)$:
\begin{eqnarray}\delta_0^\mathrm{LO}(s)\al=\al \frac{2s-M_\pi^2}{32\pi F_\pi^2}\sigma(M_\pi,s)\;,\quad \delta_1^\mathrm{LO}(s)= \frac{s-4M_\pi^2}{96\pi F_\pi^2}\sigma(M_\pi,s)\;, \nonumber \\
\delta_2^\mathrm{LO}(s)\al=\al  -\frac{s-2M_\pi^2}{32\pi F_\pi^2}\sigma(M_\pi,s)\;,\quad
\sigma(M,s)\equiv\sqrt{1-4M^2/s}\;,\nonumber\\\end{eqnarray}
the discontinuities represent contributions of NLO. Accordingly, the factor $M_I(s)$+$\hat M_I(s)$ in~\eqref{eq:discMI} is needed only to leading order:  The representation of the amplitude at tree level in~\eqref{eq:MLO} suffices.  We may, for instance, decompose it into isospin
components with 
\begin{equation}\label{eq:MILO}
M_0^\mathrm{LO}(s)= T(s)\;,\quad M_1^\mathrm{LO}(s)=0\;,\quad M_2^\mathrm{LO}(s)=0\;.
\end{equation}
As $T(s)$ is linear in $s$, the angular averages~\eqref{eq:angularAverage} are trivial. The discontinuity then takes the form $p(s)\sigma(s)$, where $p(s)$ is a real polynomial. Hence the result may be expressed in terms of the scalar loop integral
\begin{equation}\Jbar(M,s)=\frac{1}{16\pi^2}\left\{\sigma(M,s)\ln\frac{\sigma(M,s)-1}{\sigma(M,s)+1}+2\right\}\;, \end{equation}
for which the discontinuity is proportional to $\sigma(s)$: $\disc \Jbar(s)=\sigma(s)/16 \pi$. 
The result is of the form $M_I(s)=p_I(s)+M_I^{\pi\pi}(s)$, where $p_0(s)$, $p_1(s)$, $p_2(s)$ are polynomials and the
 explicit expressions for the contributions generated by the discontinuities read 
\begin{eqnarray}\label{eq:MNLO}
M_0^{\pi\pi}(s) \al=\al \frac{1}{6F_\pi^2\Delta_{\eta\pi}}( 2s - M_\pi^2
)(6s+3M_\eta^2-11M_\pi^2)\Jbar(s)  \;,\nonumber\\
M_1^{\pi\pi}(s) \al=\al \frac{1}{4 F_\pi^2 \Delta_{\eta\pi}} (s-4 M_\pi^2)\,\Jbar(s) \;,\\
M_2^{\pi\pi}(s) \al=\al \frac{1}{4 F_\pi^2\Delta_{\eta\pi}} (s- 2 M_\pi^2) ( 3s-3M_\eta^2-M_\pi^2)\, \Jbar(s) 
 \;.\nonumber
\end{eqnarray}
with $\Delta_{\eta\pi}\equiv M_\eta^2-M_\pi^2$. Indeed, one readily verifies that the terms proportional to $\Jbar(s)$ reproduce the  contributions from the pion loops in the representation of Gasser and Leutwyler~\cite{Gasser+1985a}, but their formulae  
contain further contributions, generated by loops involving kaon or $\eta$ propagators. In the above form of the one-loop representation, these contributions are accounted for only in polynomial approximation. The corresponding full expressions can be obtained in the same way, extending elastic unitarity to the reactions $\pi\pi\leftrightarrow K\bar{K}$ and $\pi\pi\leftrightarrow \pi\eta$: Extended elastic unitarity determines the one-loop representation of the transition amplitude in terms of $F_\pi$ and the meson masses up to a polynomial. 
\subsection{Branch cuts generated by kaons and $\eta$-mesons}  
The one-loop representation is unique only up to terms of higher order. The neglected higher order contributions generate uncertainties, but the error estimates attached to the one-loop amplitude cover this source. We use a variant that differs from the one given in~\cite{Gasser+1985a} by a polynomial of NNLO, because we prefer to work with a representation that is manifestly independent of the running scale used in the renormalization of the loop graphs (for the one given in~\cite{Gasser+1985a}, a change of scale affects the amplitude by a polynomial  of NNLO). Since the one-loop representation plays a central role in our analysis and the numbers obtained with it are not completely independent of the way in which the higher order contributions are handled, we explicitly specify the one we are working with.

Loops involving kaons and $\eta$-mesons yield additional contributions:
\begin{equation}\label{eq:MSU3}M_I(s)=P_I(s)+M_I^{\pi\pi}(s)+M_I^{K\bar{K}}(s)+ M_I^{\eta\eta}(s)+M_I^{\eta\pi}(s)
+O(p^4)\;.\end{equation}
The explicit expressions for those from the discontinuities due to $K\bar{K}$ intermediate states read~\cite{Gasser+1985a}:
\begin{eqnarray}M_0^{K\bar{K}}(s)\al=\al-\frac{18s(s-M_\eta^2 -M_\pi^2)+(3M_\eta^2+M_\pi^2)^2}{12F_\pi^2\Delta_{\eta\pi}}\;\bar{J}(M_K,s)\nonumber\\
\al\al-\frac{3s}{8F_\pi^2}\frac{3s-4M_K^2}{s-4M_K^2}\left\{\bar{J}(M_K,s)-\frac{1}{8\pi^2}\right\}\;,\\
M_1^{K\bar{K}}(s)\al=\al\frac{s-4M_K^2}{8F_\pi^2\Delta_{\eta\pi}}\,\bar{J}(M_K,s)\;,\nonumber\\
M_2^{K\bar{K}}(s)\al=\al-\frac{(3s-3M_\eta^2-M_\pi^2)(3s-4M_K^2)}{8F_\pi^2\Delta_{\eta\pi}}\,\bar{J}(M_K,s)\;.\nonumber\end{eqnarray}
The branch cuts from $\eta\pi$ intermediate states only show up in $M_0$ and $M_2$. They are proportional to $M_\pi^2$ and hence very small:
\begin{eqnarray}M_0^{\eta\pi}(s)\al=\al \frac{M_\pi^2(6s+3M_\eta^2-11M_\pi^2)}{9F_\pi^2\Delta_{\eta\pi}}\,\bar{J}(M_\eta,M_\pi,s)\;,\\
M_2^{\eta\pi}(s)\al=\al -\frac{M_\pi^2(3s-3M_\eta^2-M_\pi^2)}{6F_\pi^2\Delta_{\eta\pi}}\,\bar{J}(M_\eta,M_\pi,s)\nonumber\end{eqnarray}
The discontinuity  due to $\eta\eta$ intermediate states is also proportional to $M_\pi^2$ and only contributes to $M_0$:
\begin{equation}M_0^{\eta\eta}(s)=-\frac{M_\pi^2 }{2F_\pi^2}\bar{J}(M_\eta,s)\,.\end{equation}
\subsection{Polynomial part}
In the framework of \chpt, the polynomials $P_0(s)$, $P_1(s)$, $P_2(s)$ occurring in~\eqref{eq:MSU3} are determined by the LECs of the effective Lagrangian. In the normalization we are working with, only one of the LECs occurring in the representation of the transition amplitude to one loop, $L_3$, cannot be expressed in terms of the meson masses $M_\pi$, $M_K$, $M_\eta$ and the decay constants $F_\pi$, $F_K$. The decomposition into isospin components can be chosen such that the polynomial part of $M_1(s)$ is proportional to $sL_3$,  while the amplitude $M_2(s)$ is proportional to $s^2$ with a coefficient that only involves the masses and $F_\pi$. The explicit expressions for the polynomial part then read
\begin{eqnarray}P_0(s)\al=\al T(s)\left\{1+\frac{8M_\pi^2}{3\,\Delta_{\eta\pi}}\Delta_\mathrm{F} \right\}+\frac{2(3s-8M_\pi^2)}{3\,\Delta_{\eta\pi}}\Delta_\mathrm{GMO} \nonumber \\
\al \al +\frac{k_0+k_1\,s+k_2\,s^2}{192\pi^2F_\pi^2\Delta_{\eta\pi}^2}\;\\
P_1(s)\al=\al-\frac{4\,L_3\,s}{F_\pi^2\Delta_{\eta\pi}}\;,\quad
P_2(s)=  \frac{s^2}{64\pi^2F_\pi^2\Delta_{\eta\pi}}  \left\{4\ln \frac{M_K^2}{M_\pi^2}+1\right\}\;.\nonumber\end{eqnarray}
The constants $\Delta_\mathrm{F}$ and $\Delta_\mathrm{GMO}$ are specified in Eq.~\eqref{eq:DGMODF} and the coefficients $k_0$, $k_1$, $k_2$ exclusively contain the meson masses:
\begin{eqnarray}k_0\al=\al 
-12(M_\eta^6+23M_\eta^4M_\pi^2+ M_\eta^2M_\pi^4-M_\pi^6)\ln\frac{M_K^2}{M_\pi^2}\nonumber\\
\al \al + 6M_\pi^2(40M_\eta^4-5M_\eta^2M_\pi^2+M_\pi^4)\ln\frac{M_\eta^2}{M_\pi^2} \nonumber\\
\al \al-3(M_\eta^6+17M_\eta^4M_\pi^2-21 M_\eta^2M_\pi^4+3M_\pi^6)\;,\nonumber\\
k_1\al=\al 24(6M_\eta^4-2M_\eta^2M_\pi^2+M_\pi^4)\ln\frac{M_K^2}{M_\pi^2} \nonumber\\
\al \al -18 M_\eta^2(6M_\eta^2-M_\pi^2)\ln\frac{M_\eta^2}{M_\pi^2}+ 36 M_\eta^2(M_\eta^2-M_\pi^2)\;,\nonumber\\
 k_2\al=\al 4(M_\eta^2-M_\pi^2)\left\{5\ln\frac{M_K^2}{M_\pi^2}-1\right\}\;.
\end{eqnarray} 
\subsection{Dalitz plot distribution of $\mathbf\eta\to\pi^+\pi^-\pi^0$ at one loop}
\label{sec:Dalitz plot distribution}

The present appendix concerns the structure of the DKM-amplitude in the Coulomb region, where the left panel of Fig.~\ref{fig:K} shows a spike. As discussed in Sec.~\ref{sec:Self-energy}, the phenomenon has to do with the fact that the amplitude contains several branch cuts in that region (recall that a further singularity, the one generated by the Coulomb attraction between the charged pions in the final state, is removed). We stick to the line $t_c=u_c=\frac{1}{2}(M_\eta^2+2M_{\pi^+}^2+M_{\pi^0}^2-s)$ and analyze the expansion of $M_c^\mathrm{DKM}(s_c,t_c,u_c)$ around the point $s_c=4M_{\pi^+}^2$, in powers of  $\sigma$.

The contributions from the pionic $s$-channel branch cuts are proportional to the scalar loop integrals $\bar{J}(M_{\pi^+},s_c)$ and $\bar{J}(M_{\pi^0},s_c)$, respectively, while those in the $t$- and $u$-channels are accounted for with the loop integrals $\bar{J}(M_{\pi^0},M_{\pi^+},t_c)$ and  $\bar{J}(M_{\pi^0},M_{\pi^+},u_c)$. In the region of interest, these integrals are complex, while all other contributions to the amplitude are real. Only $\bar{J}(M_{\pi^+},s_c)$ is singular at $s_c=4M_{\pi^+}^2$: The expansion starts with 
 \begin{eqnarray}\label{eq:Jbar}
 \al \al \bar{J}(M_{\pi^+},s_c) =  \frac{1}{8\pi^2 }\left\{1+\frac{i}{2} \sigma+O(\sigma^2)\right\}\;, \nonumber\\
\al \al \sigma = \sqrt{1-4M_{\pi^+}^2/s_c}\;.\end{eqnarray}
Since all other terms admit a Taylor series expansion that exclusively contains even powers of $\sigma$, the expansion of the amplitude starts with
\begin{equation}M^\mathrm{DKM}_c(s_c,t_c,u_c)=a+i (b+c\,\sigma)+O(\sigma^2)\;,\end{equation}
where $a,b,c$ are real. The constant $b$ stems from the imaginary parts of the leading terms in the expansion of $\bar{J}(M_{\pi^0},s_c)$, $\bar{J}(M_{\pi^0},M_{\pi^+},t_c)$ and  $\bar{J}(M_{\pi^0},M_{\pi^+},u_c)$, while $c$ comes from the imaginary part of $\bar{J}(M_{\pi^+},s_c)$.

The expansion of the isospin symmetric amplitude is of the same form:
\begin{equation}\Mtilde^\mathrm{GL}_c(s_c,t_c,u_c)=\tilde{a}+i (\tilde{b}+\tilde{c}\,\sigma)+O(\sigma^2)\;.\end{equation}
In this case, the term $\tilde{c}$ arises from the expansion of the function $\bar{J}(M_{\pi},\tilde{s}_c)$
 \begin{equation}\bar{J}(M_{\pi},\tilde{s}_c)=\frac{1}{8\pi^2 }\left\{1+\frac{i}{2}\kappa\, \sigma+O(\sigma^2)\right\}\;.\end{equation}
The constant $\kappa$ stems from the boundary preserving map. In the region under consideration, this map barely makes a difference: $\kappa = 0.97$ is close to unity.
 
The numerical values of the leading coefficients are quite similar: $a=0.656$, $\tilde{a}=0.638$, but the non-leading ones are very different: $b = -0.091$, $\tilde{b}= -0.133$ $c= 0.048$, $\tilde{c}=0.218$. The difference arises because the self-energy of the $\pi^+$ splits the $s$-channel branch cut of the isospin symmetric amplitude into two distinct singularities and only one of these contributes to $c$, while $\tilde{c}$ stems from the isospin limit of the sum of the two contributions. For the square of the ratio of the two amplitudes, this gives
\begin{equation}|K_c(s_c,t_c,u_c)|^2= 1.032 +0.13 \,\sigma+O(\sigma^2)\;.\end{equation}
Indeed, the spike seen in the left panel of Fig.~\ref{fig:Dcbar} is well described by this expression. 
 
\section{Kinematic map for $\eta \to 3\pi^0$}
\label{sec:Kinematic map neutral channel}

A map that takes boundary and center of the physical Dalitz plot of $\eta\to3\pi^0$ onto boundary and center of the isospin symmetric phase space is readily obtained along the lines described in Sec.~\ref{sec:Kinematic map charged channel}. The analog of~\eqref{eq:fg} and~\eqref{eq:g} reads
\begin{equation}\label{eq:fgneutral}s=f_n[s_n]\;,\quad \tau=g_n[s_n]\tau_n\;,\quad g_n[s_n]=\frac{\tau^\mathrm{max}(f_n[s_n])}{\tau_n^\mathrm{max}(s_n)}\;,\end{equation}
where $\tau^\mathrm{max}_n(s_n)$ is obtained from~\eqref{eq:taucmax} with the substitutions $s_c\to s_n$, $\tau_c\to \tau_n$, $M_{\pi^+}\to M_{\pi^0}$. Again adopting the convention $M_\pi= M_{\pi^+}$, the explicit expression for $f_n[s_n]$ becomes
 \begin{eqnarray}\label{eq:pnqn}f_n[s_n]\al=\al s_n+p_n(s_n-4M_{\pi^0}^2) \nonumber\\
\al \al +q_n(s_n-4M_{\pi^0}^2)(s_n-(M_\eta-M_{\pi^0})^2)\;,\nonumber\\
  p_n\al=\al -\frac{(M_{\pi^+}-M_{\pi^0})(2M_\eta-M_{\pi^+}-M_{\pi^0})}{(M_\eta-3M_{\pi^0})(M_\eta+M_{\pi^0})}\;,\\
  q_n\al=\al-\frac{3(M_{\pi^+}-M_{\pi^0}) (M_\eta+M_{\pi^+}+4M_{\pi^0})}{(M_\eta+3M_{\pi^0})(M_\eta-3M_{\pi^0})^2(M_\eta+M_{\pi^0})}\;.\nonumber\end{eqnarray}  
Hence the map analogous to~\eqref{eq:sctcuctilde}, 
\begin{eqnarray}\label{eq:sntnuntilde}\tilde{s}_n\al=\al f_n[s_n]\;,\nonumber\\
\tilde{t}_n\al=\al \mbox{$\frac{1}{2}$}\{3s_0-f_n[s_n]+(t_n-u_n)g_n[s_n]\}\;,\\
\tilde{u}_n\al=\al \mbox{$\frac{1}{2}$}\{3s_0-f_n[s_n]-(t_n-u_n)g_n[s_n]\}\;,\nonumber\end{eqnarray}
does preserve boundary and center of the Dalitz plot, but is not suitable for our purpose, because the amplitude obtained with it,
\begin{equation}M_n^\prime(s_n,t_n,u_n)=M_n(\tilde{s}_n,\tilde{t}_n,\tilde{u}_n)\;,\end{equation}
is not symmetric under the exchange of all three Mandelstam variables -- a characteristic property of the transition into three identical particles.

The problem arises because the relation~\eqref{eq:fgneutral} does not treat $s$ on equal footing with $t$ and $u$.  
As far as the comparison with the experimental results for the $Z$-distribution or the rate is concerned, crossing symmetry is not an issue, because these quantities only involve the integral over the angle $\varphi$ in Eq.~\eqref{eq:Z phi}, but the amplitude $M_n^\prime(s_n,t_n,u_n)$ itself and the Dalitz plot distribution obtained from it are not acceptable. We correct for that by taking the mean of the three images obtained with crossing: The map
\begin{eqnarray}\label{eq:Mntilde}\Mtilde_n(s_n,t_n,u_n)  = \mbox{$\frac{1}{3}$}\{ \al \al M_n^\prime(s_n,t_n,u_n) +M_n^\prime(t_n,u_n,s_n)  \nonumber\\
\al \al  +M_n^\prime(u_n,s_n,t_n)\}\;.\end{eqnarray}
does preserve crossing symmetry as well as boundary and center of the Dalitz plot. We make use of it when comparing our dispersive solutions with experiment in the neutral channel. In particular, the functions $\Mtilde_n^\mathrm{GL}(s_n,t_n,u_u)$, $\Mtilde_n(s_n,t_n,u_n)$ occurring in Eqs.~\eqref{eq:Kn} and~\eqref{eq:Mphys} are obtained in this way from the one-loop and dispersive representations $M_n^\mathrm{GL}(s,t,u)$, $M_n(s,t,u)$, respectively.

\section{Gaussian errors}\label{sec:Gaussian errors}

For definiteness, we describe the calculation of the Gaussian errors for our central solution. In this case, the discrepancy function  $\chi^2_\mathrm{tot}$ depends on five real parameters, which can be identified with the subtraction constants $\beta_0$, $\gamma_0$, $\delta_0$, $\beta_1$, $\gamma_1$  (since only the relative size matters, $\alpha_0$ is determined by these and by the normalization constant $H_0$, which we keep fixed at the one-loop value). The analysis is independent of the choice of independent variables -- we could just as well express the discrepancy function in terms of, say, the real parts of $K_1$, \ldots, $K_5$.  We leave the number of independent variables open and denote them by $x_1$, $x_2$, \ldots 
  
The Gaussian approximation exploits the fact that, in the vicinity of the
minimum, the discrepancy function $\chi^2_\mathrm{tot}$ can be approximated
by the truncated Taylor series in the variables $\Delta
x_i=x_i-x_i\,\rule[-0.3em]{0.05em}{1em}_{\,\mathrm{min}} $:
\newpage
\begin{equation}\chi^2_\mathrm{tot}= \chi^2_\mathrm{tot}\,\rule[-0.6em]{0.05em}{1.3em}_{\,\mathrm{min}}  +\sum_{i,k} D^{ik}\,\Delta x_i \Delta x_k+\ldots\;,\quad
D^{ik}\equiv \frac{1}{2}\frac{\partial^2\chi_\mathrm{tot}}{\partial x_i\partial x_k}\,\rule[-1em]{0.05em}{2.2em}_{\,\mathrm{min}} \;.  \end{equation}
The probability distribution in the space of the variables $x_1$,  $x_2$, \ldots then takes the form
\begin{equation} dp=N\exp\left\{ -\frac{1}{2}\sum_{i,k} D^{ik}\Delta x_i\Delta x_k\right\}dx_1dx_2 \cdots \;.\end{equation}
Accordingly, the mean values are given by
\begin{equation}\label{eq:mean values}\langle x_i\rangle =x_i\,\rule[-0.3em]{0.05em}{1em}_{\,\mathrm{min}} \;,\quad \langle \Delta x_i\Delta x_k \rangle=C_{ik} \;,\end{equation}
where $C_{ik}$ is the matrix inverse of $D^{ik}$. In particular, the Gaussian errors in the variables $x_i$ are given by the square root of the diagonal elements of the matrix $C_{ik}$. For $\mathrm{fitK\chi_6}$, for instance, the Gaussian errors in the Taylor invariants are given by 
\begin{equation}\label{eq:DeltaG}\DeltaG\beta_0= 1.2\,,\hspace{0.3em}\DeltaG\gamma_0= 7.3\,,\hspace{0.3em}\DeltaG\delta_0=17.4\,,\hspace{0.3em}\DeltaG\beta_1= 2.3\,,\hspace{0.3em}\DeltaG\gamma_1= 10.8\,. \end{equation}

Note that the errors are correlated and this must be accounted for when calculating the uncertainties in the various quantities of physical interest.  The correlations concern the off-diagonal elements of the matrix $C_{ik}$. Table~\ref{table:Correlations} lists the entries of the normalized  correlation matrix $\bar{C}_{ik}=C_{ik}/\sqrt{\rule{0em}{0.8em}C_{ii}\,C_{kk}}$ for fitK$\chi_6$. It shows, for instance, that the results for $\delta_0$ and $\gamma_1$ are strongly correlated with $\gamma_0$ and $\beta_1$, respectively.   
\begin{table}[thb]\centering
\begin{tabular}{l|rrrrr}
&$\beta_0$\rule{0.4em}{0em}&$\gamma_0$\rule{0.4em}{0em}&$\delta_0$\rule{0.4em}{0em}&$\beta_1$\rule{0.4em}{0em}&$\gamma_1$\rule{0.4em}{0em}\\
\hline
$\beta_0$&1\rule{0.5em}{0em}&      $-$0.18  &  0.28 &  $-$0.98&   0.82\\
$\gamma_0$&$-$0.18&    1\rule{0.5em}{0em}&      $-$0.95 &  0.11&   0.31\\
$\delta_0$&0.28&   $-$0.95&   1\rule{0.5em}{0em}&     $-$0.27&   $-$0.09\\
$\beta_1$&$-$0.98&   0.11&   $-$0.27&   1\rule{0.5em}{0em}&      $-$0.89\\
$\gamma_1$&0.82 &   0.31&    $-$0.09&   $-$0.89&    1\rule{0.5em}{0em}\\
\hline
\end{tabular}
\caption{\label{table:Correlations}Correlations among the subtractions constants for the central solution.} 
\end{table}
\section{Sensitivity to the $\mathbf\pi\pi$ phase shifts}\label{sec:Sensitivity to phase shifts}

As discussed in Sec.~\ref{sec:Phase shifts},  the Roy solutions of~\cite{Colangelo2001} are characterized by the values of the phase shifts at 800 MeV. We vary them independently in the range given in~\eqref{eq:deltanum}.  In addition, in order to study the sensitivity of our results to the high energy tail of the dispersion integrals, we consider the change occurring if the dispersion integral for the dominating component  is chopped off at 1 GeV. Because of the narrow resonance $f_0(980)$, the phase shift $\delta_0(s)$ rapidly passes through $\pi$ around 1 GeV. For our central input, the factor $\sin\delta_0(s)$ occurring in~\eqref{eq:dmu} thus passes through zero there, takes negative values above 1 GeV and then returns to zero at 1.7 GeV. Chopping the integral off amounts to replacing the factor $\sin\delta_0(s)$ by zero. The corresponding variation of the phase shift follows the central representation for $\delta_0(s)$ only up to the energy at which it reaches the value $\delta_0(s)=\pi$ and remains 
at that value from there on. 

In order to estimate the sensitivity to the uncertainties in $\delta_0(s)$, we evaluate the various quantities of interest for the two different configurations obtained by identifying $\delta_0(s)$ with the lower or upper boundary of the band shown in Fig.~\ref{fig:phaseshifts}, while $\delta_1(s)$ and $\delta_2(s)$ are kept fixed at the central values. We take half of the difference between the two results as an estimate for the error due to uncertainties in the low-energy behaviour of $\delta_0(s)$. The same procedure is applied to the variations in $\delta_1(s)$ and $\delta_2(s)$, as well as to the one in the contributions from above 1 GeV. The net uncertainty due to the noise in the $\pi\pi$ phase shifts is obtained by adding the four individual errors in quadrature. 

For the central solution, this leads to the following error estimates:
\begin{equation}\label{eq:Deltapipi}  \Deltapipi\beta_0= 0.24\,,\hspace{0.3em}\Deltapipi\gamma_0= 6.6,,\hspace{0.3em}\Deltapipi\delta_0=2.6\,,\hspace{0.3em}\Deltapipi\beta_1=0.23\,,\hspace{0.3em}\Deltapipi\gamma_1= 2.1\,. \end{equation}

The comparison with~\eqref{eq:DeltaG} shows that the errors generated by the noise in the phase shifts are significantly smaller than the Gaussian errors -- except for $\gamma_0$, where they are of comparable size. Table~\ref{table:fitK} lists the full uncertainties obtained by adding all errors in quadrature. The numbers indicate that, in the case of the subtraction constants, the error budget is dominated by the contributions from the Gaussian errors and from the uncertainties in the input used for the phase shifts -- those associated with the isospin breaking corrections  barely matter.  
%
\section{Sampling data in the neutral channel\label{sec:Binning}}
\begin{figure}[thb]\centering\includegraphics[width=7cm]{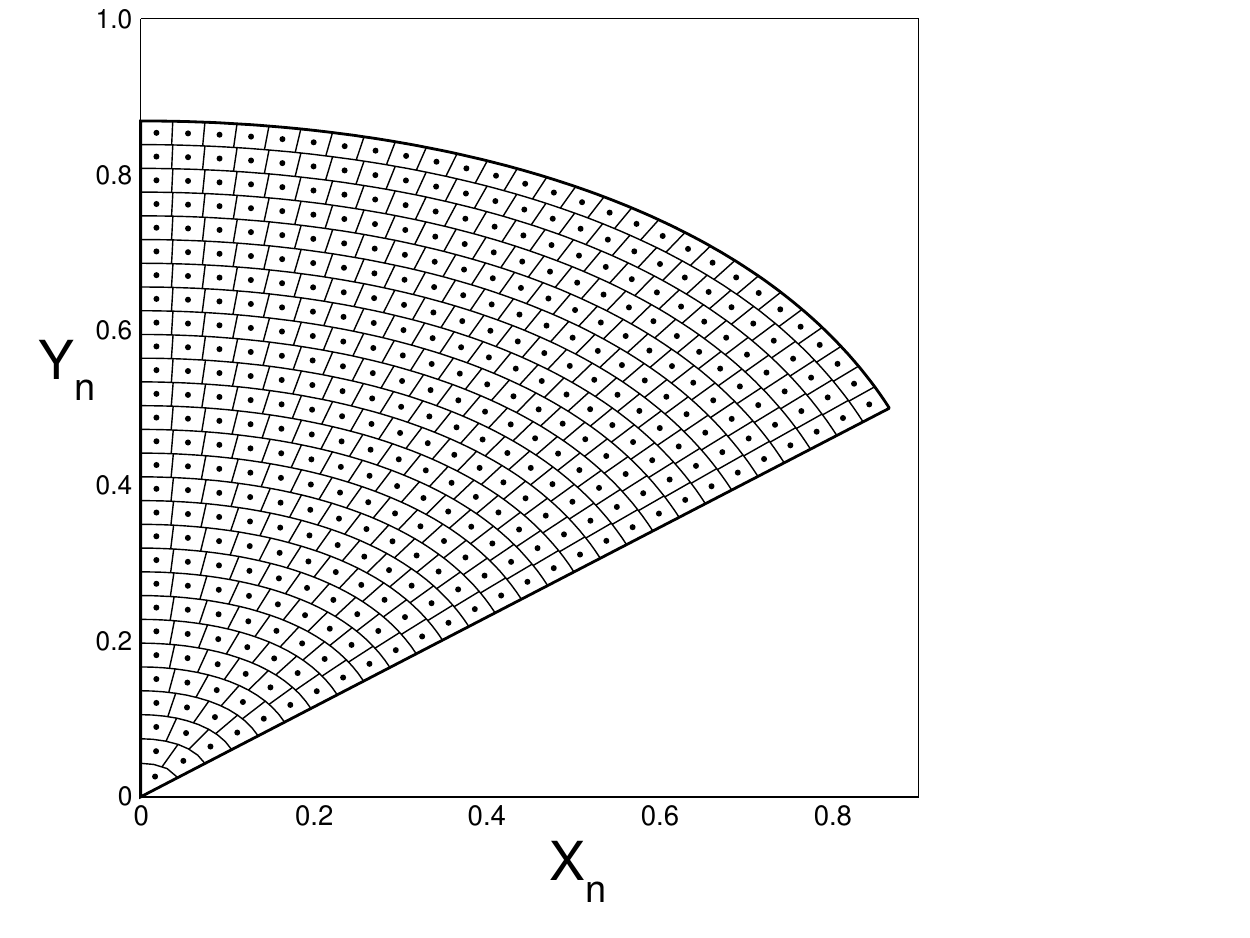}\caption{Binning the physical region of the decay $\eta\to 3\pi^0$. The bins are bounded by lines of constant $\lambda$ and lines of constant $\varphi$. The dots mark their center of gravity.\label{fig:bin28}}
\end{figure}

Binning data on the decay $\eta\to 3\pi^0$ in the standard Dalitz plot variables $X_n$ and $Y_n$ is in conflict with Bose statistics -- only one sextant of phase space contains independent events, but some of the bins necessarily reach out of this sextant. Fig.~\ref{fig:bin28}  shows an alternative sampling of the data that does respect the fact that the three pions in the final state are indistinguishable. 
It is obtained by binning in the variables $\lambda$, $\varphi$, which are defined by 
\begin{figure*}[thb]
 \includegraphics[width=5.5cm]{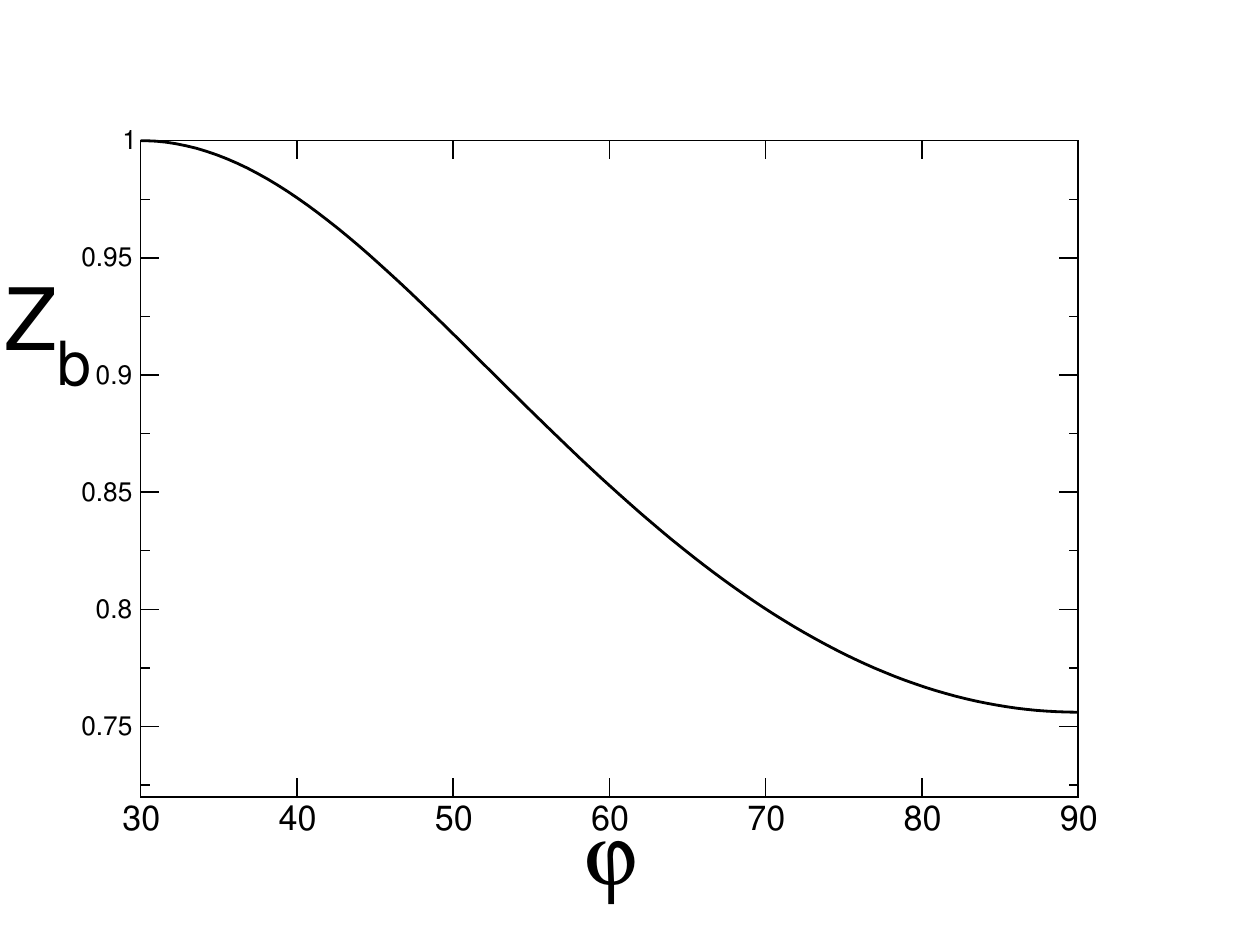}\hspace{0.5cm}\includegraphics[width=5.5cm]{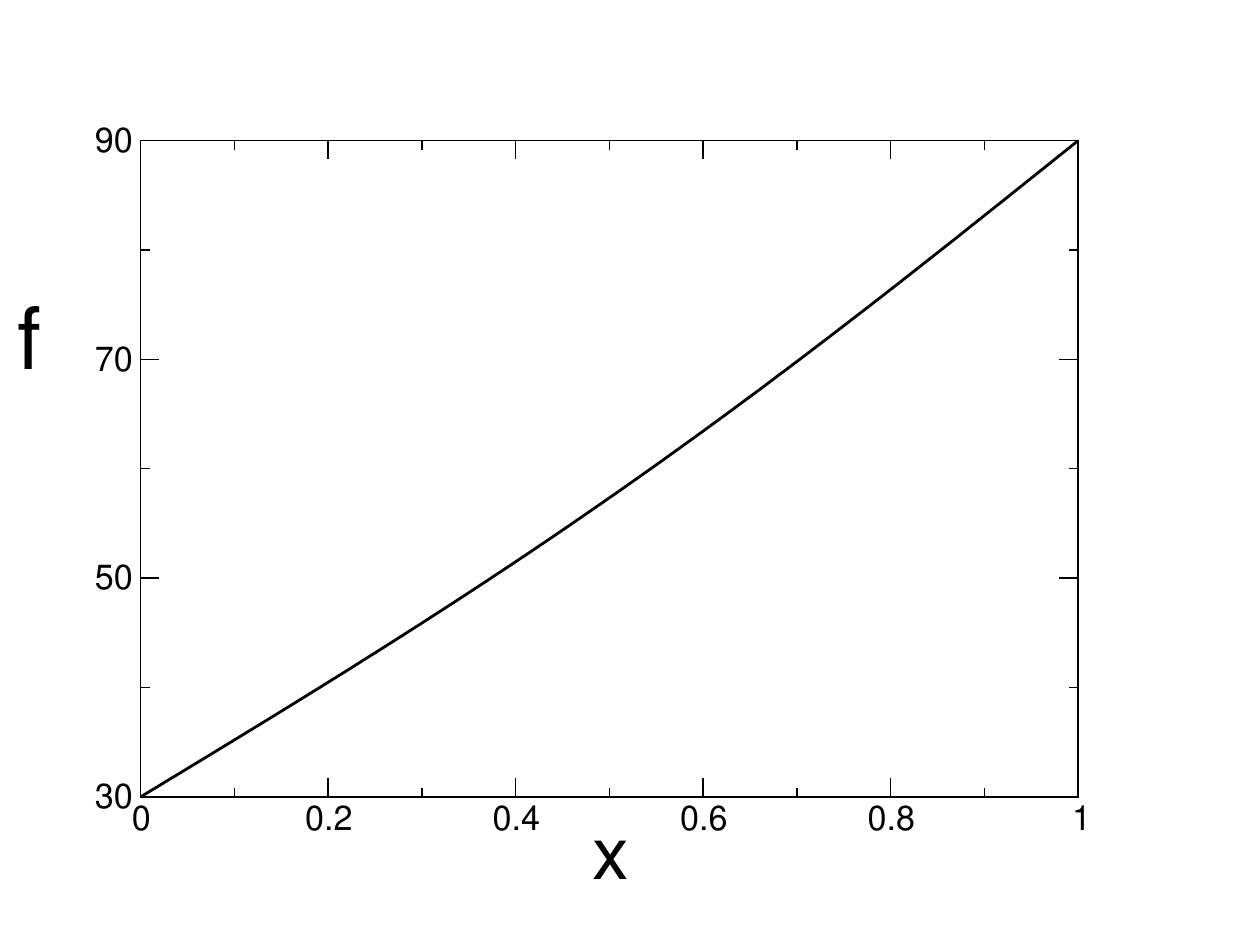}\hspace{0.5cm}\includegraphics[width=5.5cm]{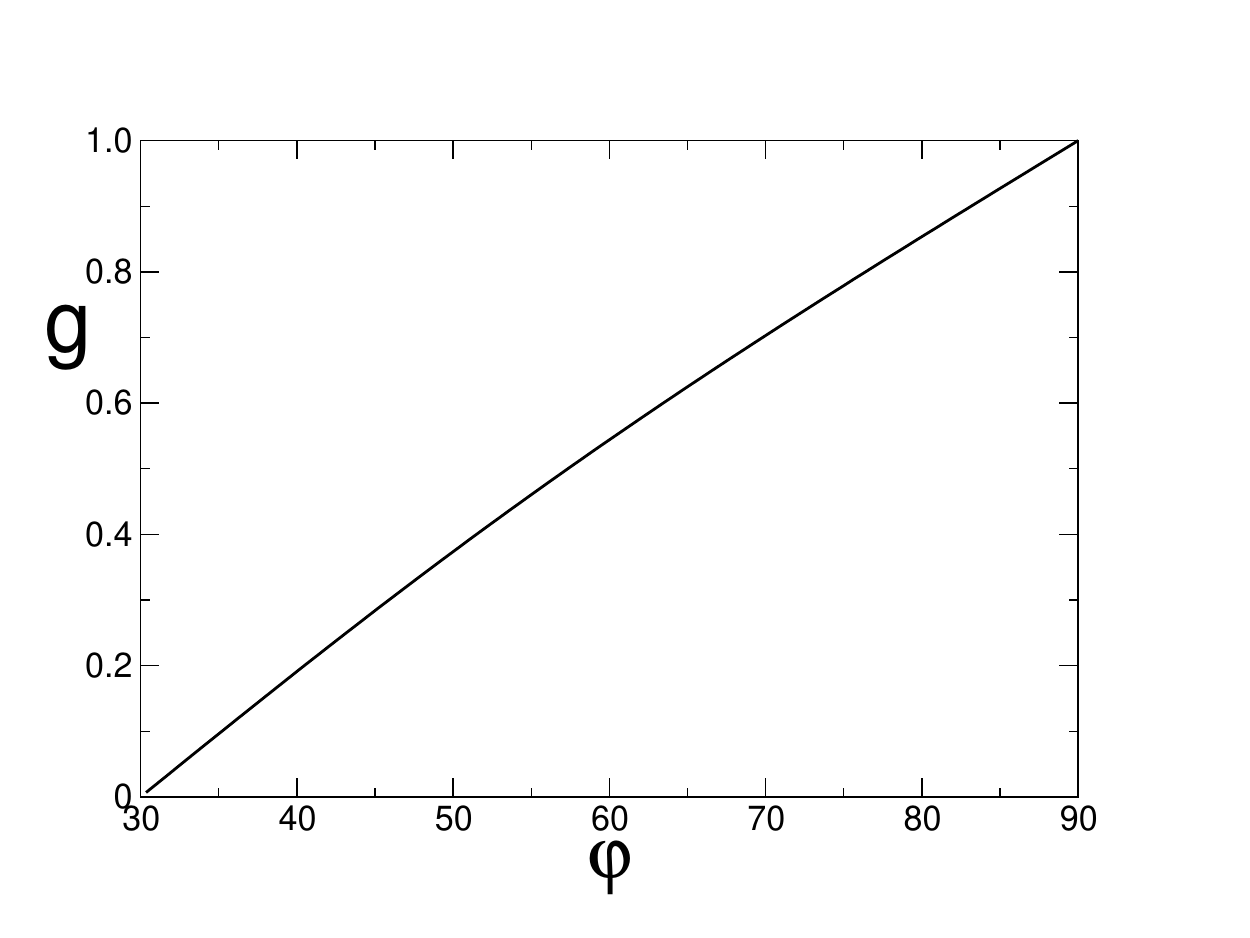}\caption{$Z_b(\varphi)$ describes the boundary of the physical region, $f(\varphi)$ represents the fraction of the area spanned by the events in the interval $\frac{\pi}{6}<\varphi'\leq \varphi$. The plot of the inverse, $g(x)$, differs from the one for $f(\varphi)$ only in the interchange of the horizontal and vertical axes. \label{fig:Zb}}
\end{figure*}

\begin{equation}\label{eq:coordinate trafo} X_n=\lambda \sqrt{Z_b(\varphi)}\cos\varphi \;,\quad Y_n=\lambda \sqrt{Z_b(\varphi)}\sin\varphi\;.\end{equation}
The function $Z_b(\varphi)$ represents the value of $Z\equiv X_n^2+Y_n^2$ at the boundary of the physical region, which depends on the angle $\varphi=\arctan Y_n/X_n$.  Fig.~\ref{fig:Zb} shows that $Z_b(\varphi)$ decreases from the maximum at $\varphi=\frac{1}{6}\pi$ to the minimum at $\varphi=\frac{1}{2}\pi$. A constant value of $\lambda$ corresponds to a curve that represents a shrunk version of the boundary, while a fixed value of $\varphi$ corresponds to a ray emanating from the origin. In these coordinates, the area element becomes
\begin{equation}\label{eq:dXdY}dX_n dY_n = Z_b(\varphi) \lambda\, d\lambda\, d\varphi\;.\end{equation}
The binning divides the range $0 \leq \lambda \leq 1$ up into a set of curved bands:
\begin{equation}\label{eq:lambdan}\Lambda(n-1)\leq\lambda\leq\Lambda(n)\;\quad   n = 1,\ldots\,, n_\mathrm{max}\;.\end{equation}
Band \#$n$ is divided into $n$ bins -- the sextant contains altogether $\frac{1}{2}n_\mathrm{max}(n_\mathrm{max}+1)$ bins  (the figure corresponds to $n_\mathrm{max}=28$ and 406 bins). For the bins to be of the same size, the area of the band must be proportional to $n$. This determines the binning in the variable $\lambda$:
\begin{equation}\label{eq:Lambdan}\Lambda(n) = \sqrt{\frac{n(n+1)}{n_\mathrm{max}(n_\mathrm{max}+1)}}\;.\end{equation} 
The requirement that the bins are of the same size determines the binning in the variable $\varphi$ as well. The first $m$ bins of band \#$n$ must cover the fraction $m/n$ of the area of this band.  We denote the area spanned by the events in the range $\frac{\pi}{6}\leq \varphi'\leq\varphi$ by
 \begin{equation}\label{eq:Farea} F(\varphi)=\frac{1}{2}\int_{\frac{\pi}{6}}^{\varphi} \hspace{-0.5em}d\varphi' \,Z_b(\varphi')\;.\end{equation}
The fraction of the area of a band that contains the events in the above range is given by
\begin{equation}\label{eq:farea} f(\varphi)=F(\varphi)/F(\mbox{$\frac{\pi}{2}$})\;.\end{equation}
The binning in the variable $\varphi$ must therefore satisfy the condition
\begin{equation}\label{eq:phimn}f[\phi(n,m)]=m/n\;,\end{equation}
where $\phi(n,m)$ is the value of $\varphi$ at the upper end of bin  \#$m$ in band \#$n$.
To solve this equation, the function $f$ needs to be inverted. We denote the inverse by $g$: $g[f(x)]\equiv x$.
In this notation, the explicit expression for the quantity $\phi(n,m)$ reads $\phi(n,m)=g(m/n)$. Hence the binning in the variable $\varphi$ is given by
\begin{equation}\label{eq:gmn}g[(m-1)/n] \leq\varphi\leq g[m/n]\;,\quad m=1,\ldots\,n\;.\end{equation}

\end{appendices}

\providecommand{\href}[2]{#2}\begingroup\raggedright\endgroup

\end{document}